\documentclass[preprint,12pt,authoryear]{elsarticle}

\usepackage{amssymb}
\usepackage{amsmath}
\usepackage[table,xcdraw]{xcolor}
\usepackage{amsfonts}
\usepackage{array}
\usepackage{colortbl}
\usepackage{nicefrac}
\usepackage{tikz}
\usetikzlibrary{arrows.meta, positioning}
\usepackage{lmodern}
\usepackage{microtype}
\newcolumntype{C}[1]{>{\centering\arraybackslash}p{#1}}
\usepackage{algorithm}
\usepackage{algorithmic}
\usepackage{booktabs}

\usepackage{float}
\usepackage{multirow}
\usepackage{url}
\usepackage[colorlinks=true, urlcolor=blue, linkcolor=blue, citecolor=blue]{hyperref}
\journal{International Journal of Forecasting}

\begin{document}
\sloppy
\begin{frontmatter}

%% Title, authors and addresses

%% use the tnoteref command within \title for footnotes;
%% use the tnotetext command for theassociated footnote;
%% use the fnref command within \author or \affiliation for footnotes;
%% use the fntext command for theassociated footnote;
%% use the corref command within \author for corresponding author footnotes;
%% use the cortext command for theassociated footnote;
%% use the ead command for the email address,
%% and the form \ead[url] for the home page:
%% \title{Title\tnoteref{label1}}
%% \tnotetext[label1]{}
%% \author{Name\corref{cor1}\fnref{label2}}
%% \ead{email address}
%% \ead[url]{home page}
%% \fntext[label2]{}
%% \cortext[cor1]{}
%% \affiliation{organization={},
%%            addressline={}, 
%%            city={},
%%            postcode={}, 
%%            state={},
%%            country={}}
%% \fntext[label3]{}

\title{Adaptive COVID-19 Trajectory Forecasting Using MAB-Inspired Ensemble Weighting}

%% Authors with explicit affiliation labels
\author[gsu]{Hamed Karami\corref{cor1}}
\ead{hkarami1@student.gsu.edu}

\author[unito]{Javier Redondo Anton}
\ead{javier.redondoanton@unito.it}

\author[asu]{Geunsoo Jang}
\ead{gjang12@asu.edu}

\author[asu]{K. Selcuk Candan}
\ead{candan@asu.edu}

\author[gsu]{Gerardo Chowell\corref{cor1}}
\ead{gchowell@gsu.edu}

\cortext[cor1]{Corresponding authors}

\affiliation[gsu]{organization={Georgia State University},
            city={Atlanta},
            state={GA},
            country={USA}}

\affiliation[unito]{organization={Universit\`a degli Studi di Torino},
            city={Turin},
            country={Italy}}

\affiliation[asu]{organization={Arizona State University},
            city={Tempe},
            state={AZ},
            country={USA}}

%% Abstract
\begin{abstract}
\noindent\textbf{Background:} Forecasting epidemic trajectories is critical for public health decision-making but remains challenging because no single model is consistently reliable across different epidemic phases and forecasting settings. While simple unweighted ensembles are often strong and difficult to outperform, they cannot adapt to shifts in the relative performance of individual component models. We investigate Multi-Armed Bandit (MAB) algorithms as adaptive weighting strategies for epidemic forecasting ensembles when model performance changes over time.

\noindent\textbf{Methods:} We evaluated three adaptive weighting strategies, UCB, EXP3, and $\varepsilon$-greedy, across three U.S. COVID-19 waves using fixed short-window and growing calibration windows. Deterministic and stochastic ensemble variants were considered. The forecasting pool included 10 component models: SIR, SEIR, GLM, Gompertz, Richards, ARIMA, random walk with drift, simple exponential smoothing, Holt's linear trend method, and exponential growth. Adaptive ensembles were compared with individual models and three benchmark methods: a naive persistence forecast, an unweighted ensemble, and an inverse-WIS weighted ensemble. Forecast performance was assessed using RMSE, weighted interval score (WIS), 95\% prediction-interval coverage, and mean 95\% prediction-interval width.

\noindent\textbf{Results:} EXP3Stoch, EXP3Det, and EPSStoch achieved the lowest mean forecast WIS across waves, calibration windows, and forecast horizons. These gains were most evident for probabilistic performance, with adaptive methods improving WIS and, in several settings, prediction-interval coverage more clearly than RMSE. This indicates that adaptive weighting primarily improved the balance between forecast accuracy and uncertainty quantification rather than uniformly reducing point forecast error. Simple benchmarks, including InverseWIS and the unweighted ensemble, remained competitive in several configurations. Although 95\% prediction-interval coverage remained below the nominal level on average, stochastic adaptive ensembles generally provided better coverage than many deterministic methods and simple benchmarks.

\noindent\textbf{Conclusions:} MAB-inspired adaptive weighting offers a flexible approach for combining heterogeneous epidemic forecasting models when component-model performance changes across epidemic phases. In this national U.S. analysis, the adaptive ensembles were most useful for improving probabilistic forecast quality, while simple averaging and calibration-performance-based weighting remained strong competitors. These findings suggest that MAB-inspired ensembles are best viewed as a complementary tool rather than a universal replacement for simpler ensemble strategies, particularly in settings where forecast uncertainty is substantial and model skill is time-varying.

\end{abstract}

%% Keywords
\begin{keyword}
Epidemic forecasting \sep Ensemble methods \sep Multi-Armed Bandit \sep Adaptive ensemble weighting \sep COVID-19 \sep Uncertainty quantification
\end{keyword}

\end{frontmatter}

%% Add \usepackage{lineno} before \begin{document} and uncomment 
%% following line to enable line numbers
%% \linenumbers

%% main text

\section{Introduction}\label{sec:intro}

Mathematical modeling and forecasting have become central tools for managing large disease outbreaks~\citep{cao2020mathematical,usikalua2025mathematical,karami2026comparative,karami2025bayesianfitforecast}. Public health officials use forecasts to make key decisions, such as planning hospital capacity, issuing health warnings, changing travel rules, and deciding when to use control measures like social distancing or vaccination~\citep{dashtbali2021compartmental,lutz2019applying,desai2019real,ordu2021novel}. During the COVID-19 pandemic, forecasts played an important role in guiding decisions about resource allocation and social distancing policies~\citep{ray2020ensemble,usikalua2025mathematical}. Forecasting models have also been used to study mpox, helping to predict how it spreads and to assess the effects of control strategies~\citep{bakare2025time,biggerstaff2022improving}. Accurate forecasting remains difficult because epidemics are complex and constantly changing~\citep{gandon2016forecasting}. Disease spread can shift rapidly due to new variants or changes in human behavior, such as shifts in risk perception or public policies~\citep{vardavas2021modeling}. These rapid changes create unstable conditions where small differences can lead to large changes in epidemic trends~\citep{chowell2017fitting}. As a result, although forecasting is necessary, it involves substantial uncertainty, and no single model can be relied upon to consistently capture the future course of an epidemic. These challenges are closely related to model calibration and parameter identifiability, which can strongly affect uncertainty quantification and predictive performance in epidemic models~\citep{jang2026comparative,karami2026parameter}.

A major challenge in forecasting epidemics is structural uncertainty, meaning that no single model works well throughout an entire outbreak~\citep{silk2022uncertainty,desai2019real}. Scientists use many types of models, from mechanistic ones like SIR and SEIR to purely statistical approaches, and each is built on different assumptions about how the disease spreads and how people behave~\citep{manfredi2013modeling,hens2012modeling}. As a result, a model that fits the early, fast-growing phase of an epidemic might perform poorly once interventions change people's contact patterns~\citep{wang2020epidemiological}. Likewise, a statistical model that works well during a steady decline might fail to detect a new wave sparked by a more contagious variant. This variability in performance is well documented~\citep{barbaglia2023testing}; for instance, a comparative study of Bayesian and frequentist methods has shown that different modeling frameworks often yield divergent predictions depending on the data context and epidemic phase~\citep{karami2026comparative}. Because the ``best'' model changes as conditions evolve, relying on a single model is risky~\citep{susser1996choosing}. If its assumptions no longer match real-world dynamics, forecasts can quickly become misleading, leaving public health officials with unreliable guidance precisely when timely and accurate information is most needed~\citep{ioannidis2022forecasting,desai2019real,sterman1988skeptic}.

A common way to address this model uncertainty is ensemble forecasting, where predictions from many different models are combined into one~\citep{parker2013ensemble,lindstrom2015bayesian,bannick2020ensemble,wu2021ensemble,chowell2021ensemble}. The idea is that each model captures different aspects of epidemic dynamics, some focus on transmission mechanisms, others on aggregate growth patterns, so combining them helps balance individual weaknesses. Studies, including~\citet{chowell2021ensemble}, show that ensembles often outperform single models and produce more stable forecasts across diverse outbreak settings. Recent infectious-disease forecasting studies further support the use of ensemble frameworks that combine complementary model classes to improve forecast accuracy and uncertainty quantification~\citep{yoon2026enhancing}. Notably, large-scale forecasting hubs such as the CDC COVID-19 Forecast Hub, the COVID-19 Forecasting Initiative at Reich Lab, and RIVM’s ensemble-based surveillance systems have demonstrated that ensemble approaches can substantially improve operational forecasting reliability when model performance varies over time~\citep{cramer2022united,ray2020ensemble,van2022prediction,cramer2022evaluation}. Recent work has also shown that ensemble size and composition can affect the stability and performance of combined real-time COVID-19 forecasts~\citep{becker2025influence}. Simple ensembles, such as unweighted model averaging, are popular because they are easy to implement and can perform surprisingly well~\citep{reich2019collaborative}. However, unweighted ensembles implicitly assume equal model reliability, even when some models consistently outperform others~\citep{graefe2015limitations}. When poor-performing models receive the same weight as stronger ones, overall forecast performance may degrade~\citep{yamana2017individual,viboud2018rapidd}. This limitation has motivated the development of weighted ensemble methods such as Bayesian Model Averaging (BMA), stacking, superensembles, and quantile-based regression approaches~\citep{wilson2007calibrated,divina2018stacking,krishnamurti2016review,taillardat2016calibrated}. Yet these approaches typically rely on retrospective training archives and update weights slowly, limiting their effectiveness in highly non-stationary epidemic settings where model skill can shift rapidly~\citep{hamill2004ensemble}.

To overcome these limitations, researchers have increasingly explored reinforcement learning and sequential decision-making methods for epidemic response, including applications to testing, vaccination, and non-pharmaceutical intervention policies~\citep{warren2025integrating,zhang2025adaptive}. However, to the best of our knowledge, RL-inspired methods such as Multi-Armed Bandit (MAB) algorithms have not been systematically applied to the problem of dynamically combining epidemic forecasting models. A key advantage of MAB algorithms is their balance between exploitation, favoring models that have recently performed well, and exploration, preserving the possibility of shifting weight toward models that become more useful as epidemic conditions change~\citep{robbins1952sequential,auer2002finite,bubeck2012regret,lai1985asymptotically,sutton1998reinforcement}. In our setting, losses for all component models can be evaluated during the calibration period, so the algorithms are used as adaptive weighting rules inspired by MAB learning rather than as a pure bandit-feedback system. This makes them a natural framework for updating ensemble weights when model performance varies over time.

In this study, we implement three adaptive weighting strategies: the Upper Confidence Bound (UCB) algorithm, EXP3, and $\varepsilon$-greedy~\citep{auer2002finite,sutton1998reinforcement}. We evaluate these methods using national U.S. COVID-19 incidence data across three distinct epidemic waves, applying them to a 10-model pool spanning mechanistic, phenomenological, and statistical forecasts: SIR, SEIR, GLM, Gompertz, Richards, ARIMA, RWDrift, SES, Holt, and ExpGrowth~\citep{miyama2022phenomenological,karami2026comparative,dhahbi2022forecasting,hyndman2008automatic,hyndman2018forecasting}. The analysis uses both fixed short-window and growing calibration windows and evaluates forecasts across short and longer horizons. This design allows us to test whether adaptive weighting can improve forecast performance relative to individual models, simple averaging, a naive persistence forecast, and a calibration-performance-based weighting baseline. We focus especially on WIS, prediction-interval coverage, and interval width, since reliable uncertainty estimates are central to epidemic forecasting. To the best of our knowledge, this study represents one of the first systematic evaluations of MAB-inspired ensemble strategies for epidemic forecasting across multiple pandemic waves.

The remainder of this paper is organized as follows. Section \ref{sec:data} describes the epidemiological data utilized in this study, including the characteristics of the three COVID-19 waves analyzed. Section \ref{sec:model} details the 10 base models employed to generate the underlying forecasts. Section \ref{sec:method} presents the ensemble forecasting framework and formally defines the three MAB algorithms and calibration strategies. Section \ref{sec:metric} outlines the performance metrics used to evaluate both point prediction accuracy and probabilistic forecast quality. Section \ref{sec:res} reports the comparative performance of the ensemble methods across all forecasting scenarios. Finally, Section \ref{sec:discussion} discusses the implications of our findings and concludes with broader perspectives on adaptive ensemble forecasting for epidemics.

\section{Data}\label{sec:data}

\begin{figure}[H]
\centering
\includegraphics[width=0.6\linewidth]{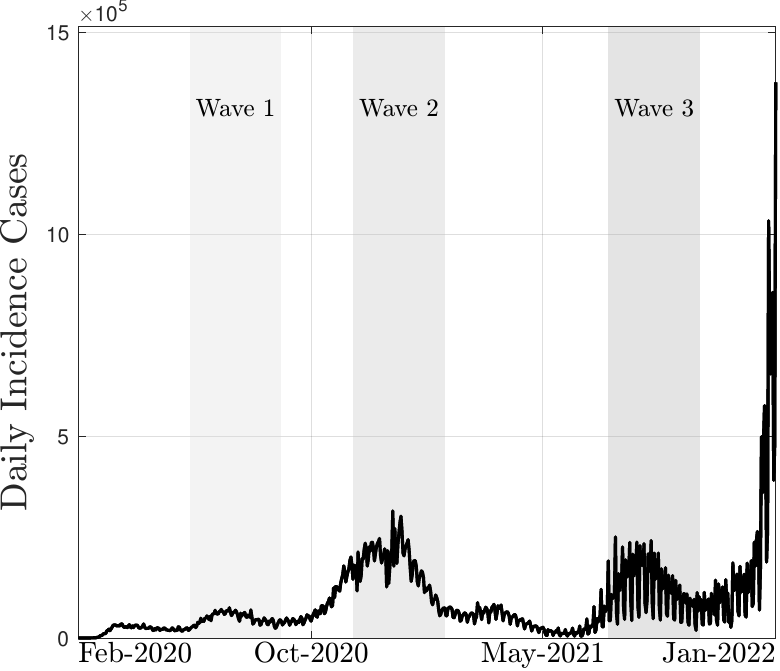}
\caption{Daily COVID-19 cases in the United States from February 2020 to January 2022, showing the three major epidemic waves analyzed in this study. Shaded gray areas indicate Wave 1 (June--September 2020), Wave 2 (November 2020--February 2021), and Wave 3 (July--October 2021). The waves differ in peak size, growth pattern, and duration, providing varied conditions for evaluating adaptive ensemble forecasting methods.}
\label{fig:pandemic_overview_USA}
\end{figure}

We evaluated the MAB ensemble forecasting framework using daily COVID-19 incidence data from the United States, spanning from February 25, 2020 onward, as shown in Figure~\ref{fig:pandemic_overview_USA}. The daily incidence series was constructed from cumulative confirmed case counts obtained from a publicly available COVID-19 forecasting data repository associated with prior ensemble sub-epidemic modeling work~\citep{chowell2022ensemble}, by first differencing the cumulative counts, with the first cumulative observation used as the first daily count. The analysis focuses on three major national epidemic waves: Wave 1 (June--September 2020), Wave 2 (November 2020--February 2021), and Wave 3 (July--October 2021). These periods were selected to represent distinct phases of the U.S. COVID-19 epidemic with different peak sizes, growth patterns, and levels of day-to-day variability. The wave periods were fixed as part of the evaluation design before comparing the forecasting methods and were not selected based on the performance of any model or ensemble method. Wave 1 corresponds to an early summer wave with a lower peak and relatively gradual rise and decline. Wave 2 shows a larger winter wave with sustained high incidence. Wave 3 captures a later wave with pronounced daily fluctuations and several sharp spikes. These differences provide a useful evaluation setting for ensemble methods, since the relative performance of forecasting models may change across waves and epidemic phases.

Because this study is retrospective, all forecasts were generated from the same finalized incidence series rather than from a sequence of real-time data releases. This design ensures that all forecasting methods are evaluated on an identical data record, but it does not reproduce the changes in the reported data over time that would occur in a prospective forecasting exercise. In real-time applications, recent COVID-19 observations may be revised as reports are updated, which could affect model calibration, recent loss estimates, and therefore the adaptive weights assigned by the MAB algorithms. We therefore interpret the results as a retrospective assessment of adaptive ensemble weighting under a fixed finalized data record, rather than as a direct replication of real-time operational forecasting with reporting revisions.

\begin{figure}[H]
\centering
\includegraphics[width=\linewidth]{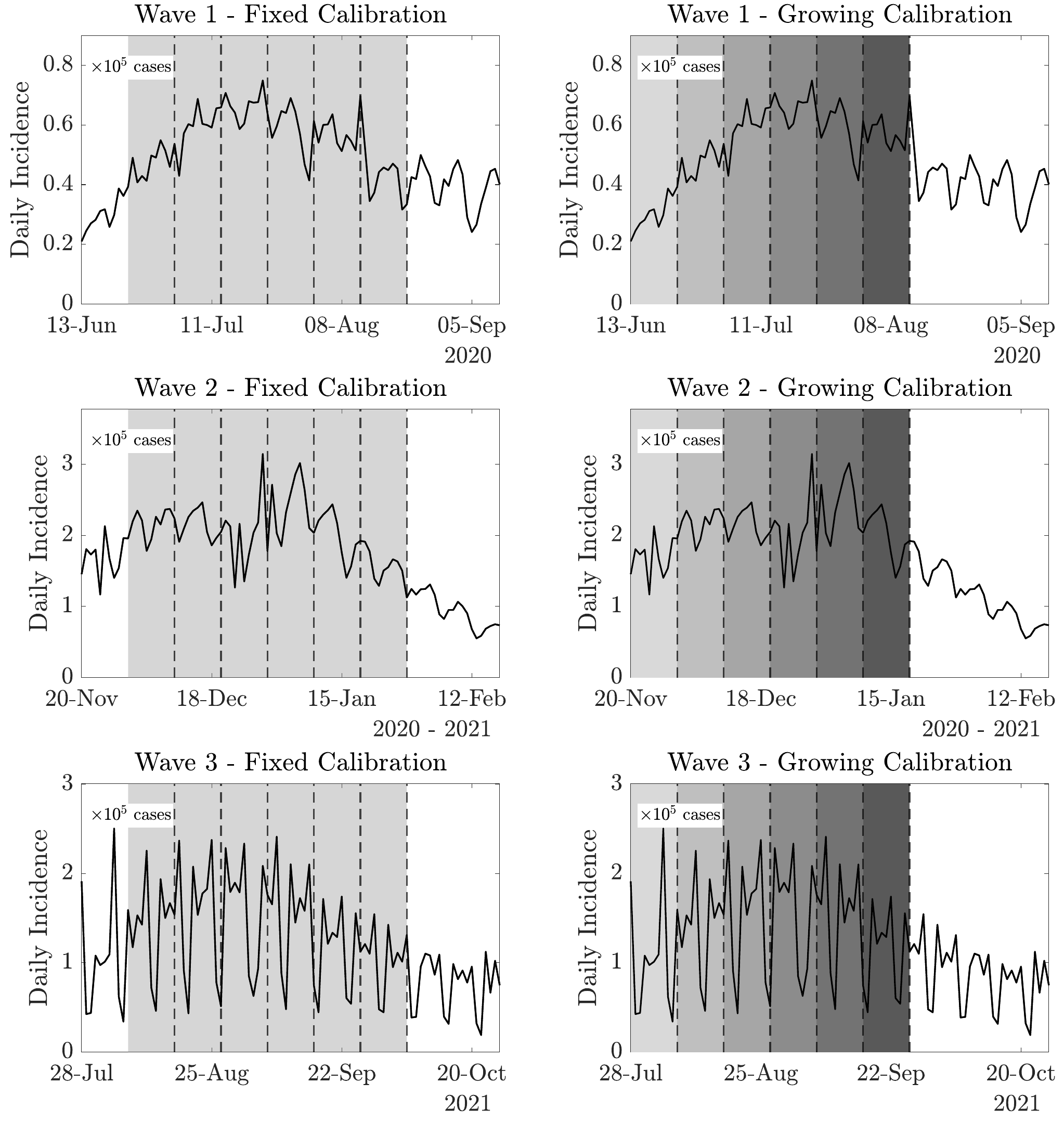}
\caption{Fixed and growing calibration strategies for evaluating ensemble forecasts. Left panels show the fixed calibration approach: six 10-day calibration windows (gray shading) starting at different times and spaced 10 days apart. Right panels show the growing calibration approach: six nested windows (progressively darker gray) starting at the same time but with calibration periods increasing from 30 to 80 days in 10-day steps. Black dashed lines mark the boundary between calibration and forecast periods. The fixed strategy serves as a short-window stress test across different epidemic phases, while the growing strategy examines performance when more calibration data are available.}
\label{fig:analysis_strategy_visualization_USA}
\end{figure}

To assess ensemble performance under different amounts of available calibration data, we used two complementary calibration strategies, shown in Figure~\ref{fig:analysis_strategy_visualization_USA}. In the fixed calibration approach, we use a 10-day calibration window and shift it forward in time within each epidemic wave, using six time points spaced 10 days apart. This setting evaluates forecasts based only on recent local epidemic dynamics and allows us to assess how performance changes when the forecast begins during the growth, peak, or decline phase of a wave. In the growing calibration approach, we keep the starting point fixed and gradually increase the length of the calibration period from 30 to 80 days in 10-day steps, providing a longer-history setting for model fitting. Together, the fixed and growing designs evaluate complementary forecasting conditions.

The forecast horizons also vary by calibration strategy. For the fixed calibration approach, we use 5-day and 10-day forecast horizons. For the growing calibration approach, we use 10-day and 30-day forecast horizons. These horizons are short enough to be relevant for operational epidemic forecasting, while still allowing us to assess forecast accuracy, interval coverage, and robustness across changing epidemic conditions.

\section{Model}\label{sec:model}

In this section, we describe the 10 base models used to construct the ensemble: SIR, SEIR, GLM, Gompertz, Richards, ARIMA, RWDrift, SES, Holt, and ExpGrowth. These models represent complementary forecasting approaches. SIR and SEIR are mechanistic compartmental models that describe transmission through epidemiological states. GLM, Gompertz, and Richards are phenomenological growth models, while ARIMA, RWDrift, SES, Holt, and ExpGrowth provide empirical time-series forecasts.

We selected this model pool to capture different aspects of epidemic behavior. Mechanistic models provide epidemiological structure through parameters such as transmission and recovery rates, while phenomenological and statistical models can flexibly fit or extrapolate observed incidence patterns without requiring detailed assumptions about transmission mechanisms. This diversity is important because the relative performance of individual models may change across epidemic phases and forecast horizons.

The SIR and SEIR models are widely used in infectious disease modeling and provide interpretable descriptions of transmission dynamics. The GLM, Gompertz, Richards, and ExpGrowth models capture different forms of epidemic growth, including sub-exponential, asymmetric, flexible sigmoidal, and early exponential patterns. The statistical time-series models provide complementary empirical forecasts and serve as useful benchmarks for assessing whether adaptive ensemble weighting adds value beyond standard extrapolation methods.

This 10-model pool is used to evaluate the MAB ensemble framework under a heterogeneous set of candidate forecasts. In practice, the set of component models can be modified depending on the forecasting goal, available data, and computational resources. The ensemble algorithms then update model weights based on calibration-period performance.

\begin{figure}[H]
\centering
\begin{tikzpicture}[
    workflowbox/.style={
        rectangle,
        draw=black!70,
        line width=0.4pt,
        fill=white,
        rounded corners=1pt,
        text width=0.82\textwidth,
        minimum height=1.18cm,
        align=center,
        inner xsep=7pt,
        inner ysep=5pt,
        font=\footnotesize
    },
    workflowarrow/.style={-{Latex[length=2.2mm,width=1.5mm]}, line width=0.45pt, draw=black!75},
    node distance=3.5mm
]
\node[workflowbox] (data) {\textbf{Observed incidence data}\\[2pt]
Daily U.S. COVID-19 incidence data are divided into calibration and forecast periods.};
\node[workflowbox, below=of data] (windows) {\textbf{Calibration-window design}\\[2pt]
Fixed short-window calibration and growing calibration windows are evaluated across three epidemic waves.};
\node[workflowbox, below=of windows] (bootstrap) {\textbf{Bootstrap uncertainty generation}\\[2pt]
For each calibration window, $B$ noisy bootstrap realizations of the observed incidence series are generated.};
\node[workflowbox, below=of bootstrap] (models) {\textbf{Fit 10-model pool}\\[2pt]
Each bootstrap realization is used to fit SIR, SEIR, GLM, Gompertz, Richards, ARIMA, RWDrift, SES, Holt, and ExpGrowth.};
\node[workflowbox, below=of models] (trajectories) {\textbf{Generate predictive trajectories}\\[2pt]
Each fitted model produces calibration and forecast trajectories for every bootstrap replicate.};
\node[workflowbox, below=of trajectories] (ensembles) {\textbf{Construct ensembles and benchmarks}\\[2pt]
Adaptive ensembles are constructed using EXP3, $\varepsilon$-greedy, and UCB variants; benchmarks include Unweighted, Naive, and InverseWIS.};
\node[workflowbox, below=of ensembles] (evaluation) {\textbf{Predictive summaries and evaluation}\\[2pt]
Predictive medians and 95\% prediction intervals are obtained from bootstrap quantiles and evaluated using RMSE, WIS, 95\% PI coverage, and mean 95\% PI width.};

\draw[workflowarrow] (data) -- (windows);
\draw[workflowarrow] (windows) -- (bootstrap);
\draw[workflowarrow] (bootstrap) -- (models);
\draw[workflowarrow] (models) -- (trajectories);
\draw[workflowarrow] (trajectories) -- (ensembles);
\draw[workflowarrow] (ensembles) -- (evaluation);
\end{tikzpicture}
\caption{Workflow of the 10-model forecasting and uncertainty-quantification framework. For each calibration window, bootstrap realizations of the observed incidence series are generated and the model pool is fitted. Predictive medians and 95\% prediction intervals are obtained from the empirical distribution of bootstrap trajectories, and forecasts are evaluated using RMSE, WIS, 95\% PI coverage, and mean 95\% PI width.}
\label{fig:workflow_uq}
\end{figure}
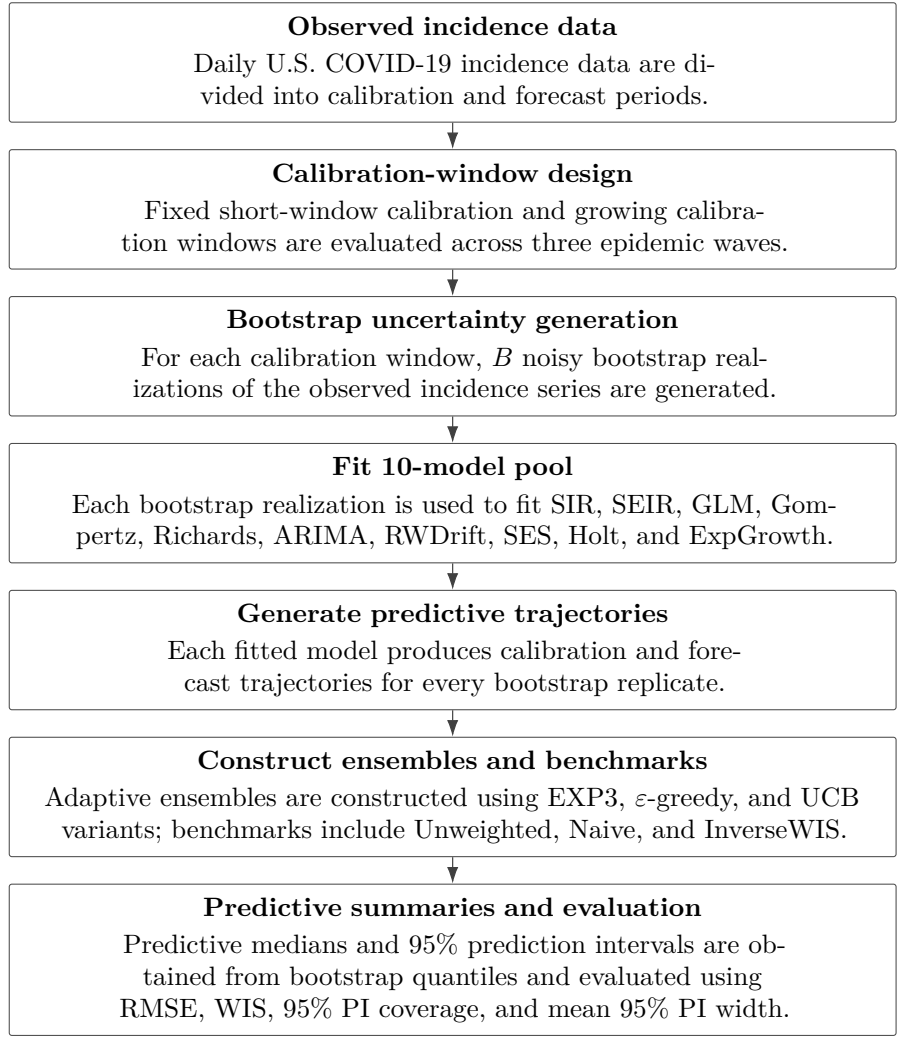

Before presenting the individual component models, we summarize the full forecasting and evaluation workflow in Figure~\ref{fig:workflow_uq}. The workflow shows how the daily incidence data are divided into calibration and forecast periods, how bootstrap realizations are generated, how the 10-model pool is fitted, and how adaptive ensembles and benchmark forecasts are constructed and evaluated.

\subsection{SIR Model}

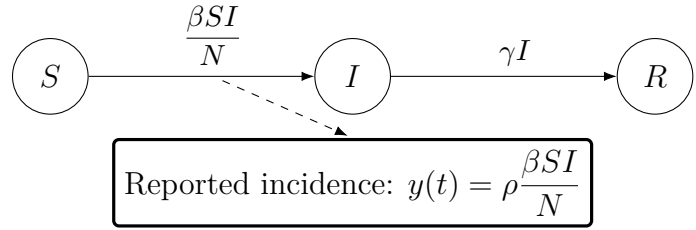
\begin{figure}[H]
\centering
\begin{tikzpicture}[>=Latex]
\tikzstyle{comp}=[circle, draw, minimum size=10mm, inner sep=0pt]
\tikzstyle{lab}=[font=\small, align=center]

\node[comp] (S) at (0,0) {$S$};
\node[comp] (I) at (4,0) {$I$};
\node[comp] (R) at (8,0) {$R$};

\draw[->] (S) -- (I)
  node[lab, pos=0.55, above] {$\dfrac{\beta S I}{N}$}
  coordinate[pos=0.55] (infflow);
\draw[->] (I) -- (R)
  node[lab, pos=0.55, above] {$\gamma I$};

\node[draw, very thick, rounded corners=2pt, fill=white,
      minimum width=26mm, minimum height=7mm,
      anchor=south] (C) at (4,-2) {Reported incidence: $y(t)=\rho\dfrac{\beta S I}{N}$};

\draw[dashed, ->, shorten >=2pt, shorten <=4pt] (infflow) -- (C.north);

\end{tikzpicture}
\caption{Compartmental diagram of the SIR model with reported incidence. Circles represent the compartments for susceptible ($S$), infectious ($I$), and recovered ($R$) individuals. Solid arrows indicate transitions between compartments, while the dashed arrow indicates the observation process associated with newly reported infections.}
\label{fig:sir}
\end{figure}

The SIR model provides a foundational framework for describing the transmission dynamics of infectious diseases within a closed population. In this model, the population is partitioned into three epidemiological states: susceptible ($S$), infectious ($I$), and recovered ($R$). Additionally, we denote by $C$ the cumulative number of reported cases over time. The model dynamics are governed by the following system of differential equations:
\begin{equation}
\begin{gathered}
\dfrac{dS}{dt} = -\dfrac{\beta SI}{N}, \quad
\dfrac{dI}{dt} = \dfrac{\beta SI}{N} - \gamma I, \quad
\dfrac{dR}{dt} = \gamma I, \quad
\dfrac{dC}{dt} = \rho\dfrac{\beta SI}{N},
\end{gathered}
\end{equation}
where $\beta$ denotes the transmission rate, $\gamma$ the recovery rate, $\rho$ the reporting rate, and $N$ the total population size. The initial conditions are specified as
\begin{equation}
S(0) = N - I_0, \quad I(0) = I_0, \quad R(0) = 0, \quad C(0) = I_0,
\end{equation}
where $I_0$ represents the initial number of infections, which are also included in the cumulative count of reported cases. The compartmental structure and observation process for the SIR model are shown in Figure~\ref{fig:sir}.

\subsection{SEIR Model}

To capture the delay between infection and the onset of infectiousness, the SEIR model extends the SIR framework by introducing an additional compartment, $E$, representing individuals who have been exposed to the pathogen but are not yet infectious. The model equations are given by
\begin{equation}
\begin{gathered}
\dfrac{dS}{dt} = -\dfrac{\beta SI}{N}, \quad
\dfrac{dE}{dt} = \dfrac{\beta SI}{N} - \kappa E, \quad
\dfrac{dI}{dt} = \kappa E - \gamma I, \quad \\
\dfrac{dR}{dt} = \gamma I, \quad
\dfrac{dC}{dt} = \kappa\rho E,
\end{gathered}
\end{equation}
where $\kappa$ denotes the rate at which exposed individuals progress to the infectious stage (i.e., the inverse of the incubation period). The initial conditions are defined as
\begin{equation}
S(0) = N - I_0, \quad E(0) = 0, \quad I(0) = I_0, \quad R(0) = 0, \quad C(0) = I_0.
\end{equation}

\begin{figure}[H]
\centering
\begin{tikzpicture}[>=Latex]
\tikzstyle{comp}=[circle, draw, minimum size=10mm, inner sep=0pt]
\tikzstyle{lab}=[font=\small, align=center]

\node[comp] (S) at (0,0) {$S$};
\node[comp] (E) at (3.5,0) {$E$};
\node[comp] (I) at (7,0) {$I$};
\node[comp] (R) at (10.5,0) {$R$};

\draw[->] (S) -- (E)
  node[lab, pos=0.55, above] {$\dfrac{\beta S I}{N}$};
\draw[->] (E) -- (I)
  node[lab, pos=0.55, above] {$\kappa E$}
  coordinate[pos=0.55] (progflow);
\draw[->] (I) -- (R)
  node[lab, pos=0.55, above] {$\gamma I$};

\node[draw, very thick, rounded corners=2pt, fill=white,
      minimum width=28mm, minimum height=7mm,
      anchor=south] (C) at (5.25,-2) {Reported incidence: $y(t)=\kappa\rho E(t)$};

\draw[dashed, ->, shorten >=2pt, shorten <=4pt] (progflow) -- (C.north);

\end{tikzpicture}
\caption{Compartmental diagram of the SEIR model with reported incidence. Circles represent the compartments for susceptible ($S$), exposed ($E$), infectious ($I$), and recovered ($R$) individuals. Solid arrows indicate transitions between compartments, while the dashed arrow indicates the observation process associated with newly reported infections.}
\label{fig:seir}
\end{figure}
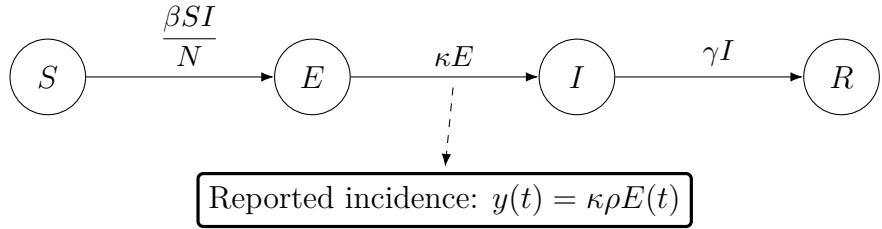

The compartmental structure and observation process for the SEIR model are shown in Figure~\ref{fig:seir}.

\subsection{Phenomenological Growth Models}

\paragraph{Generalized Logistic Model (GLM)}

The GLM extends the classical logistic growth framework by introducing a deceleration parameter that modulates the rate at which the epidemic curve slows as it approaches its upper bound. The model is expressed as
\begin{equation}
\dfrac{dC}{dt} =\rho r\, C^{p}\!\left(1 - \dfrac{C}{K}\right),
\end{equation}
where $r$ denotes the intrinsic growth rate, $p$ the deceleration parameter that controls the sub-exponential growth behavior, $K$ the carrying capacity, and $\rho$ the reporting rate. Here, $K$ corresponds to the (asymptotic) maximum cumulative number of reported cases.

\paragraph{Gompertz Model}

The Gompertz model characterizes epidemic growth by assuming that the relative growth rate of cumulative cases decreases exponentially over time. This formulation captures asymmetric epidemic curves, where the early growth phase is faster than the decline phase. The model is defined as
\begin{equation}
\dfrac{dC}{dt} = \rho r\, C\, e^{-b t},
\end{equation}
where $r$ denotes the initial growth rate, $b$ the rate at which the growth rate decays exponentially, and $\rho$ the reporting rate. 

\paragraph{Richards Model}

The Richards model generalizes both the logistic and Gompertz models by introducing a shape parameter that allows for flexible curvature and asymmetry in the epidemic trajectory. The model is expressed as
\begin{equation}
\dfrac{dC}{dt} = \rho r\, C \left[1 - \left(\dfrac{C}{K}\right)^{\alpha}\right],
\end{equation}
where $r$ is the intrinsic growth rate, $K$ the carrying capacity, $\alpha$ a shape parameter that governs the degree of asymmetry of the epidemic curve, and $\rho$ the reporting rate. 

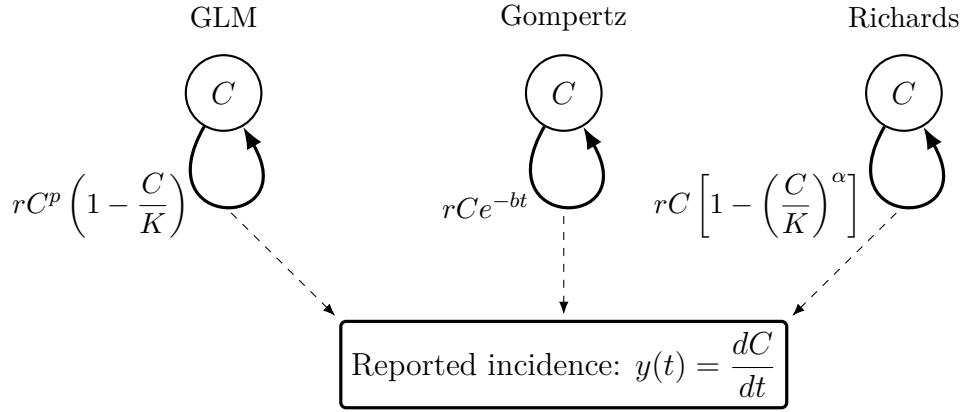
\begin{figure}[H]
\centering
\begin{tikzpicture}[>=Latex]
\tikzstyle{comp}=[circle, draw, minimum size=10mm, inner sep=0pt, thick]
\tikzstyle{lab}=[font=\small, align=center]

%---------------- GLM ----------------%
\node[lab] (GLM_Title) at (0, 1) {GLM};
\node[comp] (GLM_C) at (0,0) {$C$};
\draw[->, very thick] (GLM_C) to[out=240, in=300, looseness=8]
    node[lab, pos=0.5, left=3mm] {$rC^p\left(1-\dfrac{C}{K}\right)$}
    coordinate[pos=0.5] (GLM_flow)
    (GLM_C);

%---------------- Gompertz ----------------%
\node[lab] (GOM_Title) at (4.5, 1) {Gompertz};
\node[comp] (GOM_C) at (4.5,0) {$C$};
\draw[->, very thick] (GOM_C) to[out=240, in=300, looseness=8]
    node[lab, pos=0.5, left=3mm] {$rCe^{-bt}$}
    coordinate[pos=0.5] (GOM_flow)
    (GOM_C);

%---------------- Richards ----------------%
\node[lab] (RICH_Title) at (9, 1) {Richards};
\node[comp] (RICH_C) at (9,0) {$C$};
\draw[->, very thick] (RICH_C) to[out=240, in=300, looseness=8]
    node[lab, pos=0.5, left=4mm] {$rC\left[1-\left(\dfrac{C}{K}\right)^\alpha\right]$}
    coordinate[pos=0.5] (RICH_flow)
    (RICH_C);

%---------------- Observation box ----------------%
\node[draw, very thick, rounded corners=2pt, fill=white,
      minimum width=56mm, minimum height=7mm,
      anchor=north] (Obs) at (4.5,-3.0)
      {Reported incidence: $y(t)=\dfrac{dC}{dt}$};

%---------------- Dashed arrows from growth flows ----------------%
\draw[dashed, ->, shorten >=2pt, shorten <=4pt] (GLM_flow) -- (Obs.north west);
\draw[dashed, ->, shorten >=2pt, shorten <=4pt] (GOM_flow) -- (Obs.north);
\draw[dashed, ->, shorten >=2pt, shorten <=4pt] (RICH_flow) -- (Obs.north east);

\end{tikzpicture}
\caption{Diagrams of the three phenomenological cumulative growth models. The circles represent the cumulative case count $C(t)$ in each model. Solid self-loops indicate the growth process $dC/dt$ with model-specific functional forms. Dashed arrows indicate the reported incidence generated by the rate of change of cumulative cases.}
\label{fig:growth_models}
\end{figure}

All three phenomenological models are initialized with the same initial condition,
\begin{equation}
C(0) = C_0,
\end{equation}
where $C_0$ denotes the initial number of reported cases in the given data. The cumulative-case structure and observation process for these three phenomenological growth models are summarized in Figure~\ref{fig:growth_models}.

\subsection{Statistical and Time-Series Models}

In addition to the mechanistic and phenomenological models above, we include five statistical and time-series forecasting models: ARIMA, random walk with drift (RWDrift), simple exponential smoothing (SES), Holt's linear trend method, and exponential growth (ExpGrowth). These models are defined directly on the daily incidence $D_t$ and provide complementary empirical forecasts that do not require explicit compartmental transmission assumptions.

\paragraph{ARIMA}

The ARIMA model represents daily incidence using autoregressive and differencing components. In this study, we use a nonseasonal ARIMA$(1,1,0)$ specification,
\begin{equation}
\Delta D_t = c + \phi_1 \Delta D_{t-1} + \epsilon_t,
\end{equation}
where $\Delta D_t = D_t-D_{t-1}$, $c$ is a constant term, $\phi_1$ is the autoregressive coefficient, and $\epsilon_t$ is assumed to be a white-noise error term. We chose ARIMA(1,1,0) as a simple base model. The differencing step helps account for changes in the level of daily incidence over time, and the single autoregressive term allows the model to use recent incidence patterns without adding too many parameters for the short calibration windows.

\paragraph{Random Walk with Drift (RWDrift)}

The random walk with drift model extrapolates the most recent incidence value using a linear drift term. Its forecast form is
\begin{equation}
\hat D_{m+h} = D_m + h d,
\end{equation}
where $D_m$ is the last observed incidence value in the calibration window, $h$ is the forecast horizon, and $d$ denotes the estimated local drift.

\paragraph{Simple Exponential Smoothing (SES)}

The SES model represents the incidence series through a smoothed level component,
\begin{equation}
\ell_t = \alpha D_t + (1-\alpha)\ell_{t-1},
\end{equation}
where $\ell_t$ is the level at time $t$ and $\alpha$ is the smoothing parameter. The resulting forecast is constant at the final level,
\begin{equation}
\hat D_{m+h} = \ell_m.
\end{equation}

\paragraph{Holt's Linear Trend Method}

Holt's method extends SES by including both a level and a trend component:
\begin{equation}
\ell_t = \alpha D_t + (1-\alpha)(\ell_{t-1}+b_{t-1}),
\end{equation}
\begin{equation}
b_t = \beta(\ell_t-\ell_{t-1}) + (1-\beta)b_{t-1},
\end{equation}
where $b_t$ is the trend component and $\alpha,\beta$ are smoothing parameters. The forecast is
\begin{equation}
\hat D_{m+h} = \ell_m + h b_m.
\end{equation}

\paragraph{Exponential Growth (ExpGrowth)}

The ExpGrowth model extrapolates recent multiplicative growth in daily incidence. A simple form is
\begin{equation}
\hat D_{m+h} = D_m \exp(hr),
\end{equation}
where $r$ denotes the recent exponential growth rate and $h$ is the forecast horizon.

\section{Methodology}\label{sec:method}

Let $\mathbf{D} = [D_1, D_2, \ldots, D_m]^\top$ denote the observed daily incidence data during a calibration period of length $m$ days. We consider an ensemble of $K$ base forecasting models, denoted by ${\Phi_1,\Phi_2,\ldots,\Phi_K}$, with $K=10$ in the final analysis. Each fitted model produces daily-incidence predictions during the calibration period and forecasts over a horizon of $h$ days beyond the calibration period, denoted by $\hat D_{m+1},\hat D_{m+2},\ldots,\hat D_{m+h}$.

The base models are fitted or constructed using model-specific procedures. The mechanistic and phenomenological models, SIR, SEIR, GLM, Gompertz, and Richards, are parameterized models whose parameters are estimated from the calibration data. The statistical and time-series models, ARIMA, RWDrift, SES, Holt, and ExpGrowth, are fitted directly to the daily incidence series using the procedures described below. For comparison and ensemble construction, all model forecasts are expressed as daily incidence.

For the mechanistic and phenomenological models, we employ a multi-start constrained optimization procedure, given in Algorithm~\ref{alg1}, to estimate model-specific parameters $\hat{\theta}_k$.

\begin{algorithm}
\caption{Parameter estimation for mechanistic and phenomenological base models}
\label{alg1}
\begin{algorithmic}[1]
\REQUIRE Incidence data $\mathbf{D} = [D_1,D_2,\ldots,D_m]^\top$, model $\Phi_k$, initial guess $\theta_0$, parameter bounds $[\theta^L,\theta^U]$, number of random starts $n_{\text{starts}}$
\ENSURE Estimated parameters $\hat{\theta}_k$ for model $k$
\STATE Initialize $\hat{\theta}_k \leftarrow \theta_0$ and $f_{\min} \leftarrow \infty$
\FOR{$i = 1$ to $n_{\text{starts}}$}
    \IF{$i = 1$}
        \STATE Set $\theta^{(0)} \leftarrow \theta_0$
    \ELSE
        \STATE Draw $\theta^{(0)} \leftarrow \theta^L + \mathcal{U}(0,1)(\theta^U-\theta^L)$
    \ENDIF
    \STATE Define $f(\theta) = \sum_{t=1}^{m} \left(D_t-\Phi_k(\theta,t)\right)^2$
    \STATE Solve $\theta^* = \arg\min_{\theta} f(\theta)$ subject to $\theta^L \leq \theta \leq \theta^U$
    \IF{$f(\theta^*) < f_{\min}$}
        \STATE Set $\hat{\theta}_k \leftarrow \theta^*$
        \STATE Set $f_{\min} \leftarrow f(\theta^*)$
    \ENDIF
\ENDFOR
\RETURN $\hat{\theta}_k$
\end{algorithmic}
\end{algorithm}

The optimization is carried out using the interior-point method (\emph{fmincon} in MATLAB) with 20 random starting points to reduce sensitivity to local minimum.

For the five statistical and time-series models, we use model-specific fitting rules rather than the constrained optimization in Algorithm~\ref{alg1}. ARIMA is fitted directly to the calibration incidence series as a nonseasonal ARIMA$(1,1,0)$ model. RWDrift estimates a local drift from recent incidence increments; specifically, with $q=\min(7,m-1)$, the drift is
\begin{equation}
d=\frac{1}{q}\sum_{j=m-q+1}^{m}(D_j-D_{j-1}),
\end{equation}
and the forecast is $\hat D_{m+h}=D_m+hd$. For SES, the smoothing parameter $\alpha$ is selected from a fixed grid by minimizing the calibration squared error, and forecasts are set equal to the final smoothed level. For Holt's linear trend method, the smoothing parameters $\alpha$ and $\beta$ are similarly selected over a fixed grid by minimizing calibration squared error, and forecasts use the final estimated level and trend. ExpGrowth estimates recent multiplicative growth from the last up to seven log-incidence ratios using a pseudo-count $c=1$,
\begin{equation}
g_j=\log\left(\frac{D_j+c}{D_{j-1}+c}\right),
\end{equation}
and extrapolates incidence using the average recent log-growth rate. All statistical and time-series forecasts are truncated at zero when needed to avoid negative incidence values. These model-specific procedures generate fitted calibration-period predictions and forecast trajectories on the same daily-incidence scale as the mechanistic and phenomenological models.

Predictive uncertainty was quantified using $B=300$ noise-perturbed bootstrap realizations for each calibration window. The bootstrap samples were generated by adding multiplicative Gaussian noise with a fixed relative noise level of 0.2 to the observed calibration incidence series. We used 0.2 to introduce moderate variation around the observed incidence values while keeping the bootstrap series close to the original epidemic trajectory. This perturbation was used only to create repeated noisy realizations of the calibration data for uncertainty propagation, not to modify the held-out forecast observations. For each bootstrap realization, all component models were refit or recomputed, and the resulting trajectories were combined by each ensemble method. Predictive medians and 95\% prediction intervals were then obtained from the empirical distribution of the bootstrap trajectories. No model-specific analytical prediction intervals were used.

\subsection{Ensemble Weight Adaptation via Multi-Armed Bandit Algorithms}

The MAB ensemble framework treats each base model as a candidate forecasting arm. During calibration, the algorithms compare recent prediction errors and update model weights over time. Because this is a retrospective forecast evaluation, calibration-period losses are available for all component models at each time point. Thus, although the algorithms retain the selection and weighting structure of MAB methods, our implementation is best interpreted as a MAB-inspired adaptive weighting procedure in a full-information setting rather than as a strict bandit-feedback formulation. This allows the ensemble weights to shift toward models that are performing well as epidemic conditions change.

After fitting or constructing the base model forecasts, we build ensembles using three MAB algorithms: EXP3, $\varepsilon$-greedy, and UCB. Each algorithm assigns time-varying weights $w_k^{(t)}$ for $k = 1, \ldots, K$, where $K=10$ is the number of base models. Let $\hat D_{k,t}$ denote the calibration-period prediction from model $k$ at time $t$.

\paragraph{Exponential-weight algorithm for Exploration and Exploitation (EXP3)}

The EXP3 algorithm (Algorithm~\ref{alg2}) updates model weights using exponential rewards. It is designed for settings where model performance can change unpredictably over time. The learning rate $\eta$ controls how quickly the algorithm adapts, balancing exploration and exploitation. Following the recommendation in~\citet{auer2002nonstochastic}, we set $\eta = \sqrt{\frac{2\log K}{K m}}$.

\begin{algorithm}
\caption{EXP3 ensemble algorithm}
\label{alg2}
\begin{algorithmic}[1]
\REQUIRE Incidence data $\mathbf{D} = [D_1, D_2, \ldots, D_m]^\top$, calibration predictions $\{\hat D_{k,t}: k=1,\ldots,K,\; t=1,\ldots,m\}$, learning rate $\eta$
\ENSURE Time-varying model weights
\STATE Initialize weights: $w_k^{(1)} \leftarrow \frac{1}{K}$ for all $k=1,\ldots,K$
\FOR{$t = 1$ to $m$}
\STATE Normalize weights: $p_k^{(t)} \leftarrow \frac{w_k^{(t)}}{\sum_{j=1}^K w_j^{(t)}}$
\STATE Sample model: $k_t \sim \text{Categorical}(p_1^{(t)}, p_2^{(t)}, \ldots, p_K^{(t)})$
\STATE Compute losses: $\ell_k^{(t)} = (\hat D_{k,t} - D_t)^2$ for $k = 1, \ldots, K$
\STATE Normalize losses to rewards:
\STATE \hspace{2em} $r_k^{(t)} = 1 - \frac{\ell_k^{(t)} - \min_j \ell_j^{(t)}}{\max_j \ell_j^{(t)} - \min_j \ell_j^{(t)}}$
\STATE Compute estimated reward: $\tilde{r}_{k_t}^{(t)} = \frac{r_{k_t}^{(t)}}{p_{k_t}^{(t)}}$, and $\tilde{r}_k^{(t)} = 0$ for $k \neq k_t$
\STATE Compute unnormalized updated weights: $\bar w_k \leftarrow w_k^{(t)} e^{\eta \tilde{r}_k^{(t)}}$ for all $k=1,\ldots,K$
\STATE Normalize updated weights: $w_k^{(t+1)} \leftarrow \frac{\bar w_k}{\sum_{j=1}^K \bar w_j}$ for all $k=1,\ldots,K$
\ENDFOR
\RETURN $\text{Weights } \{\mathbf{w}_k\}_{k=1}^K, \quad \mathbf{w}_k = (w_k^{(1)}, \ldots, w_k^{(m)})$
\end{algorithmic}
\end{algorithm}

\paragraph{$\varepsilon$-Greedy}

The $\varepsilon$-greedy algorithm (Algorithm~\ref{alg3}) uses a simple exploration strategy: with probability $\varepsilon$, a model is selected uniformly at random, and with probability $1-\varepsilon$, the model with the lowest cumulative error is selected. We use a fixed exploration rate $\varepsilon = \sqrt{K/m}$ during the calibration period of length $m$. In contrast to declining-$\varepsilon$ formulations, where $\varepsilon(t)=\sqrt{K/t}$ decreases over time~\citep{sutton1998reinforcement}, we adopt the fixed version for implementation simplicity and stability across bootstrap samples.

\begin{algorithm}
\caption{$\varepsilon$-greedy ensemble algorithm}
\label{alg3}
\begin{algorithmic}[1]
\REQUIRE Incidence data $\mathbf{D} = [D_1,D_2,\ldots,D_m]^\top$, calibration predictions $\{\hat D_{k,t}: k=1,\ldots,K,\; t=1,\ldots,m\}$, exploration rate $\varepsilon$
\ENSURE Time-varying model weights
\STATE Initialize cumulative errors: $E_k \leftarrow 0$ for all $k=1,\ldots,K$
\STATE Initialize selection counts: $N_k \leftarrow 0$ for all $k=1,\ldots,K$
\STATE Initialize weights: $w_k^{(0)} \leftarrow 0$ for all $k=1,\ldots,K$
\FOR{$t = 1$ to $m$}
\STATE Pick $u_t \sim \mathcal{U}(0,1)$
\IF{$u_t < \varepsilon$}
\STATE Select $k_t \sim \text{Uniform}\{1,2,\ldots,K\}$ \COMMENT{Explore}
\ELSE
\STATE Select $k_t$ uniformly at random from $\arg\min_k E_k$ \COMMENT{Exploit}
\ENDIF
\STATE Update selection count: $N_{k_t} \leftarrow N_{k_t}+1$
\STATE Update weights: $w_k^{(t)} \leftarrow \frac{N_k}{t}$ for all $k=1,\ldots,K$
\STATE Compute losses: $\ell_k^{(t)} = (\hat D_{k,t} - D_t)^2$ for all $k = 1,\ldots,K$
\STATE Update cumulative errors: $E_k \leftarrow E_k + \ell_k^{(t)}$ for all $k=1,\ldots,K$
\ENDFOR
\RETURN $\text{Weights } \{\mathbf{w}_k\}_{k=1}^K, \quad \mathbf{w}_k = (w_k^{(1)}, \ldots, w_k^{(m)})$
\end{algorithmic}
\end{algorithm}

\paragraph{Upper Confidence Bound (UCB)}

The UCB ensemble algorithm (Algorithm~\ref{alg4}) selects models by combining average calibration loss with an exploration term. Because calibration-period predictions are available for all component models, cumulative losses are updated for all models at each calibration time. Accordingly, the average-loss term is normalized by elapsed calibration time, while the exploration term and final pseudo-weights use the number of times each model has been selected. We set the exploration parameter to $c=2$, a commonly used choice in UCB-type algorithms~\citep{auer2002finite}.

\begin{algorithm}
\caption{UCB ensemble algorithm}
\label{alg4}
\begin{algorithmic}[1]
\REQUIRE Incidence data $\mathbf{D} = [D_1,D_2,\ldots,D_m]^\top$, calibration predictions $\{\hat D_{k,t}: k=1,\ldots,K,\; t=1,\ldots,m\}$, exploration parameter $c$
\ENSURE Time-varying model weights 
\STATE Initialize cumulative errors: $E_k \leftarrow 0$ for all $k=1,\ldots,K$
\STATE Initialize selection counts: $N_k \leftarrow 0$ for all $k=1,\ldots,K$
\STATE Initialize weights: $w_k^{(0)} \leftarrow 0$ for all $k=1,\ldots,K$
\FOR{$t = 1$ to $m$}
\FOR{$k = 1$ to $K$}
\IF{$N_k = 0$}
\STATE $\mathrm{UCB}_k^{(t)} \leftarrow +\infty$
\ELSE
\STATE $\mathrm{UCB}_k^{(t)} \leftarrow -\dfrac{E_k}{\max(t-1,1)} + \sqrt{\dfrac{c\log(\max(t,2))}{N_k}}$
\ENDIF
\ENDFOR
\STATE Select $k_t$ uniformly at random from $\arg\max_{k} \mathrm{UCB}_k^{(t)}$
\STATE Update selection count: $N_{k_t} \leftarrow N_{k_t}+1$
\STATE Update weights: $w_k^{(t)} \leftarrow \dfrac{N_k}{t}$ for all $k=1,\ldots,K$
\STATE Compute losses: $\ell_k^{(t)} = (\hat D_{k,t} - D_t)^2$ for all $k=1,\ldots,K$
\STATE Update cumulative errors: $E_k \leftarrow E_k+\ell_k^{(t)}$ for all $k=1,\ldots,K$
\ENDFOR
\RETURN $\{\mathbf{w}_k\}_{k=1}^K$, where $\mathbf{w}_k=(w_k^{(1)},\ldots,w_k^{(m)})$
\end{algorithmic}
\end{algorithm}

\subsection{Deterministic and Stochastic Ensemble Generation}

For each bootstrap iteration $b=1,\ldots,B$, each base model produces a component trajectory $\hat D_k^{(b)}(t)$, $k=1,\ldots,K$, over the calibration and forecast periods. The adaptive weighting algorithm is run separately within each bootstrap, so the resulting weights are bootstrap-specific.

\paragraph{Deterministic ensembles}

The deterministic variants combine component trajectories by weighted averaging. For EXP3, let $w_k^{(t,b)}$ denote the EXP3 weight for model $k$ at calibration time $t$ in bootstrap realization $b$. During the forecast period, we use the average calibration weight,
\[
\bar w_k^{(b)} = \frac{1}{m}\sum_{s=1}^{m} w_k^{(s,b)}.
\]
Thus, over the combined calibration and forecast period, the weights used in the ensemble are
\[
\tilde w_k^{(t,b)} =
\begin{cases}
w_k^{(t,b)}, & 1\leq t \leq m,\\
\bar w_k^{(b)}, & t > m.
\end{cases}
\]
The deterministic EXP3 ensemble trajectory is
\[
\hat D_{\mathrm{ens,det}}^{(b)}(t)
=
\sum_{k=1}^{K} \tilde w_k^{(t,b)} \hat D_k^{(b)}(t).
\]
For $\varepsilon$-greedy and UCB, the deterministic variants use a bootstrap-specific pseudo-weight vector. Let $N_k^{(b)}$ denote the number of times model $k$ is selected during calibration in bootstrap iteration $b$. The pseudo-weight for model $k$ is defined as
\[
\pi_k^{(b)} = \frac{N_k^{(b)}}{\sum_{j=1}^{K} N_j^{(b)}}.
\]
This fixed pseudo-weight vector is then used over both calibration and forecast times:
\[
\hat D_{\mathrm{ens,det}}^{(b)}(t)
=
\sum_{k=1}^{K} \pi_k^{(b)} \hat D_k^{(b)}(t).
\]

\paragraph{Stochastic ensembles}

The stochastic variants use the bootstrap-specific weights as sampling probabilities rather than forming a weighted average. Sampling is performed independently at each time point within each bootstrap realization. Thus, the stochastic ensemble value at time $t$ is the prediction from the model sampled for that time point, rather than a single component-model trajectory selected for the entire forecast horizon. For EXP3, the selected model index at time $t$ in bootstrap realization $b$ is sampled as
\[
k_t^{(b)} \sim \mathrm{Categorical}(\tilde{\mathbf w}^{(t,b)}),
\]
while for $\varepsilon$-greedy and UCB, it is sampled from the bootstrap-specific pseudo-weight vector,
\[
k_t^{(b)} \sim \mathrm{Categorical}(\boldsymbol{\pi}^{(b)}).
\]
The stochastic ensemble forecast at time $t$ is then given by the prediction of the component model sampled for that time point:
\[
\hat D_{\mathrm{ens,stoch}}^{(b)}(t)
=
\hat D_{k_t^{(b)}}^{(b)}(t).
\]

For both deterministic and stochastic variants, the predictive median is computed pointwise across bootstrap ensemble trajectories. The 95\% prediction interval is given by the pointwise 2.5th and 97.5th percentiles. These are empirical bootstrap intervals; model-specific analytical prediction intervals are not used.

\subsection{Benchmark Ensembles and Baselines}

In addition to the adaptive MAB ensembles, we consider three benchmark methods: an unweighted ensemble, a naive persistence forecast, and an inverse-WIS weighted ensemble. These benchmarks provide reference models for assessing whether adaptive weighting improves forecast performance beyond simple averaging, persistence, and calibration-performance-based weighting.

\paragraph{Unweighted ensemble}

The unweighted ensemble assigns equal weight to all $K$ base models. For bootstrap $b$, the ensemble trajectory is
\[
\hat D_{\mathrm{unw}}^{(b)}(t)
=
\frac{1}{K}\sum_{k=1}^{K}\hat D_k^{(b)}(t).
\]
This benchmark separates the effect of adaptive weighting from the benefit of combining a diverse set of component models.

\paragraph{Naive persistence forecast}

The naive benchmark assumes that the most recent observed incidence remains unchanged over the forecast horizon. Thus, for $h\geq 1$,
\[
\hat D_{\mathrm{naive}}(m+h)=D_m.
\]
For bootstrap-based evaluation, the same persistence rule is applied to each noisy bootstrap realization. This benchmark provides a simple reference for short-term epidemic forecasting.

\paragraph{Inverse-WIS weighted ensemble}

We also consider a calibration-performance-based weighted ensemble. For each component model, its calibration-period WIS is computed, and models with lower WIS receive larger weights. The inverse-WIS weight for model $k$ is defined as
\[
v_k =
\frac{1/\mathrm{WIS}_k}{\sum_{j=1}^{K} 1/\mathrm{WIS}_j}.
\]
The corresponding ensemble trajectory is
\[
\hat D_{\mathrm{invWIS}}^{(b)}(t)
=
\sum_{k=1}^{K} v_k^{(b)} \hat D_k^{(b)}(t).
\]
This benchmark tests whether the MAB algorithms provide improvement beyond a direct weighting rule based on past calibration performance.

\section{Performance Metrics}\label{sec:metric}

We assess predictive performance using four complementary metrics that evaluate point forecast accuracy, probabilistic calibration, and interval sharpness~\citep{gneiting2007scoring, bracher2021evaluating}. Let $\mathcal{T} = \{t_1,\dots,t_N\}$ denote the set of evaluation time points, $D_t$ the observed incidence at time $t$, and $\hat D_t$ the corresponding predictive median.

\subsection{Point Prediction Accuracy}

To quantify point forecast error, we employ the Root Mean Squared Error (RMSE). Because point-forecast rankings can depend on the chosen error measure, we report RMSE as the point-accuracy metric, consistent with the squared-error loss used in model fitting~\citep{kolassa2020best}.

\begin{equation}
\label{eq:rmse}
\mathrm{RMSE}
=
\sqrt{
\frac{1}{N}
\sum_{t\in\mathcal{T}}
\left(\hat D_t-D_t\right)^2
}.
\end{equation}
Lower RMSE indicates better point forecast accuracy, with larger deviations penalized quadratically.

\subsection{Probabilistic Calibration and Sharpness}

To evaluate the predictive distribution, we use the Weighted Interval Score (WIS), a proper scoring rule that decomposes performance into sharpness and penalties for lack of coverage~\citep{bracher2021evaluating}. We compute WIS using the interval levels
\[
\mathcal{A}=\{0.02,0.05,0.10,0.20,\ldots,0.90\}.
\]
For each $\alpha\in\mathcal{A}$, let $L_t^\alpha$ and $U_t^\alpha$ denote the $\alpha/2$ and $1-\alpha/2$ predictive quantiles, respectively. The Interval Score (IS) is defined as
\begin{equation}
\label{eq:IS}
\mathrm{IS}_{\alpha,t}
=
U_t^\alpha-L_t^\alpha
+
\frac{2}{\alpha}(L_t^\alpha-D_t)\mathbf{1}\{D_t<L_t^\alpha\}
+
\frac{2}{\alpha}(D_t-U_t^\alpha)\mathbf{1}\{D_t>U_t^\alpha\}.
\end{equation}
With $J=|\mathcal{A}|=11$, the pointwise WIS is
\begin{equation}
\label{eq:wis_t}
\mathrm{WIS}_t
=
\frac{1}{J+\frac{1}{2}}
\left[
\frac{1}{2}|D_t-\hat D_t|
+
\sum_{\alpha\in\mathcal{A}}
\frac{\alpha}{2}\mathrm{IS}_{\alpha,t}
\right].
\end{equation}
We report the mean WIS over all evaluation time points,
\begin{equation}
\label{eq:mean_wis}
\mathrm{WIS}
=
\frac{1}{N}
\sum_{t\in\mathcal{T}}
\mathrm{WIS}_t.
\end{equation}
Lower WIS indicates better probabilistic forecast performance, combining both accuracy and interval sharpness.

To evaluate prediction-interval calibration, we examine the empirical coverage of the $95\%$ prediction interval (PI). Let $L_t$ and $U_t$ denote the 2.5th and 97.5th predictive quantiles. Coverage is reported as the percentage of observations falling within these bounds:
\begin{equation}
\label{eq:coverage}
\mathrm{95\%~PI~Coverage}
=
\frac{100}{N}
\sum_{t\in\mathcal{T}}
\mathbf{1}\{L_t \leq D_t \leq U_t\}.
\end{equation}
A calibrated model should achieve coverage close to the nominal level of $95\%$; deviations indicate either overconfidence, corresponding to intervals that are too narrow, or underconfidence, corresponding to intervals that are too wide. We also report the mean width of the $95\%$ PI,
\begin{equation}
\label{eq:width}
\mathrm{Mean~95\%~PI~Width}
=
\frac{1}{N}
\sum_{t\in\mathcal{T}}
(U_t-L_t),
\end{equation}
to quantify interval sharpness.

\section{Results}\label{sec:res}

We evaluated 19 forecasting methods across 12 U.S. forecast configurations, defined by three epidemic waves, two calibration designs, and two forecast horizons. The methods included 10 base models (SIR, SEIR, GLM, Gompertz, Richards, ARIMA, RWDrift, SES, Holt, and ExpGrowth) and nine ensemble or benchmark methods (EXP3Det, EXP3Stoch, EPSDet, EPSStoch, UCBDet, UCBStoch, Unweighted, Naive, and InverseWIS). Forecast performance was assessed using RMSE, WIS, 95\% PI coverage, and mean 95\% PI width. We emphasize WIS as the primary probabilistic forecast score, while using RMSE as a point-forecast comparison.

\subsection{Overall forecasting performance across configurations}

\begin{figure}[H]
\centering
\includegraphics[width=\linewidth]{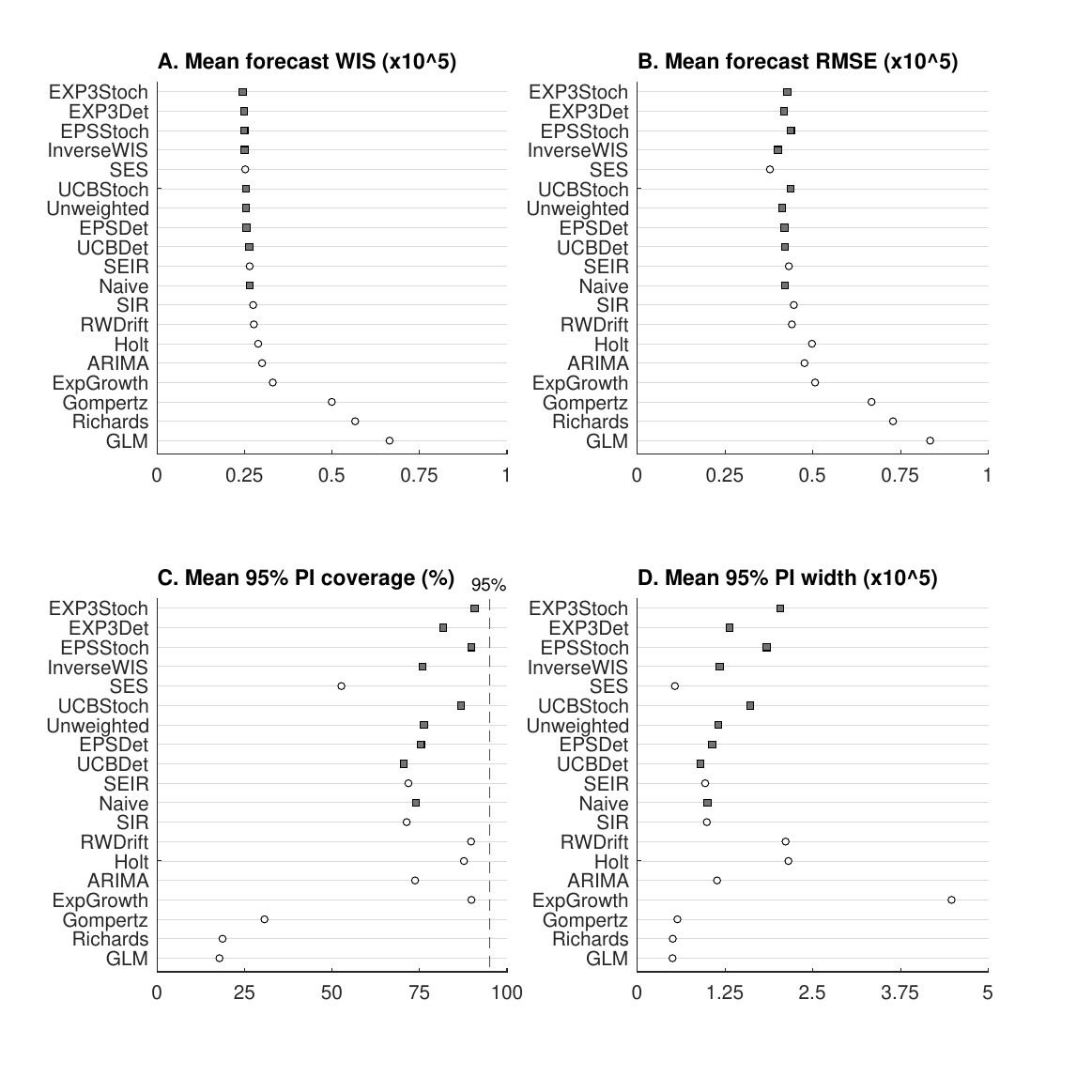}
\caption{Overall forecast performance across the 12 forecast configurations. Points show mean forecast-period RMSE, WIS, empirical 95\% prediction-interval coverage, and mean 95\% prediction-interval width for the 10 base models and nine ensemble/comparison methods. Methods are ordered by mean WIS.}
\label{fig:forecast_performance_summary}
\end{figure}

Figure~\ref{fig:forecast_performance_summary} summarizes mean forecast-period performance across the 12 U.S. forecast configurations. The methods are ordered by mean WIS. EXP3Stoch, EXP3Det, and EPSStoch achieved the lowest overall mean WIS values, indicating strong average probabilistic performance. However, the separation from InverseWIS, SES, and the Unweighted ensemble was modest, showing that simple benchmarks and statistical models remained competitive.

The RMSE panel shows a different pattern. SES was particularly strong for point accuracy, and InverseWIS also had competitive average RMSE. Thus, adaptive methods showed clearer gains for WIS and coverage than for RMSE. The coverage and interval-width panels further show that stochastic adaptive ensembles had higher 95\% PI coverage than many deterministic methods and simple benchmarks, but coverage remained below the nominal 95\% level on average for all methods.

\begin{table}[H]
\centering
\scriptsize
\caption{Summary of forecast-period performance across the 12 forecast configurations. For each method and configuration, forecast-period metrics were first averaged over forecast days within each evaluation window and then over evaluation windows. The values in this table average these configuration-level summaries equally across the 12 configurations. The final three columns report the number of configurations in which each method achieved lower WIS than the corresponding baseline.}
\label{tab:forecast_summary}
\resizebox{\textwidth}{!}{%
\begin{tabular}{lrrrrccc}
\hline
\multirow{2}{*}{Method}
& \multirow{2}{*}{Mean WIS}
& \multirow{2}{*}{Mean RMSE}
& \multirow{2}{*}{95\% PI Coverage (\%)}
& \multirow{2}{*}{Mean 95\% PI Width}
& \multicolumn{3}{c}{Lower WIS than} \\
\cline{6-8}
& & & & & Naive & Unweighted & InverseWIS \\
\hline
EXP3Stoch             & 24438.7 & 42717.5 & 90.8 & 203881.8 & 6/12 & 6/12 & 6/12 \\
EXP3Det               & 24772.5 & 41816.7 & 81.8 & 131528.6 & 7/12 & 9/12 & 7/12 \\
EPSStoch              & 24897.6 & 43746.9 & 89.8 & 184367.5 & 6/12 & 5/12 & 6/12 \\
InverseWIS            & 24915.4 & 40029.0 & 75.8 & 117441.2 & 8/12 & 6/12 & -- \\
SES (best individual model in aggregate) & 25151.8 & 37793.8 & 52.6 & 53586.4  & 7/12 & 6/12 & 6/12 \\
UCBStoch              & 25313.4 & 43720.3 & 86.8 & 160659.3 & 6/12 & 4/12 & 5/12 \\
Unweighted            & 25381.2 & 41220.3 & 76.3 & 115519.4 & 7/12 & --   & 6/12 \\
EPSDet                & 25477.3 & 41882.5 & 75.4 & 106816.1 & 7/12 & 4/12 & 5/12 \\
UCBDet                & 26306.1 & 42128.3 & 70.5 & 90209.6  & 6/12 & 6/12 & 5/12 \\
Naive                 & 26450.2 & 42087.5 & 73.9 & 100105.2 & --   & 5/12 & 4/12 \\
\hline
\end{tabular}%
}
\end{table}

Table~\ref{tab:forecast_summary} provides the numerical summary corresponding to Figure~\ref{fig:forecast_performance_summary}. EXP3Stoch ranked first by mean WIS, followed by EXP3Det and EPSStoch. InverseWIS and the Unweighted ensemble remained close competitors, and SES was the strongest base model in this aggregate summary. These results indicate that adaptive weighting can improve probabilistic performance in selected settings, but it does not uniformly dominate simpler alternatives.

\subsection{Relative WIS improvement against benchmark methods}

\begin{figure}[H]
\centering
\includegraphics[width=\linewidth]{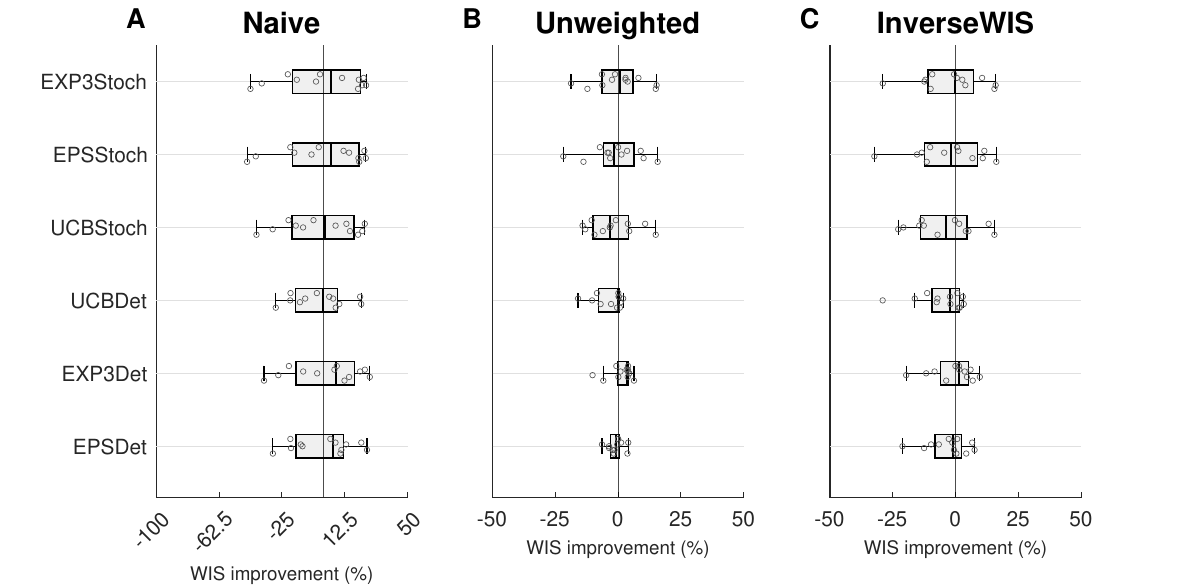}
\caption{WIS percent change of the six adaptive ensemble methods relative to the Naive, Unweighted, and InverseWIS baselines across the 12 forecast configurations. Positive values indicate lower WIS, and therefore better probabilistic forecast performance, than the corresponding baseline.}
\label{fig:wis_percent_change}
\end{figure}

Figure~\ref{fig:wis_percent_change} compares the six adaptive methods against the Naive, Unweighted, and InverseWIS benchmarks using WIS percent change across the same 12 U.S. forecast configurations. Positive values indicate lower WIS than the corresponding benchmark within the same configuration. The paired comparisons show that improvements are configuration-dependent. EXP3Det most consistently improved on the Unweighted ensemble, but some fixed short-window settings favored simple benchmarks or statistical models.

These comparisons support a measured interpretation of the adaptive methods. Adaptive weighting can help when recent calibration performance provides useful information about which component forecasts to emphasize, but it does not uniformly outperform Naive, Unweighted, or InverseWIS. In the national U.S. aggregate, simple ensemble benchmarks remain strong competitors.

\subsection{Configuration-level winners}

Table~\ref{tab:configuration_winners} complements the aggregate summaries by showing which method wins in each individual wave--calibration--horizon configuration. WIS is used as the primary probabilistic criterion, and RMSE is shown as the point-forecast comparison.

\begin{table}[H]
\centering
\small
\caption{Configuration-level winners across the 12 U.S. forecasting settings. For each configuration, the table reports the lowest-WIS method, the second- and third-lowest WIS methods, and the lowest-RMSE method. Lower values indicate better performance.}
\label{tab:configuration_winners}
\resizebox{\textwidth}{!}{%
\begin{tabular}{lllll}
\toprule
Wave & Setting & Top WIS method & Second/third WIS methods & RMSE winner \\
\midrule
Wave 1 & Fixed 5d & SES (5,606) & Naive (6,533); ARIMA (7,238) & SES (9,148) \\
Wave 1 & Fixed 10d & SES (6,355) & Naive (6,780); ARIMA (8,082) & SES (10,495) \\
Wave 1 & Growing 10d & EXP3Det (5,026) & UCBStoch (5,028); EPSStoch (5,044) & EXP3Stoch (8,845) \\
Wave 1 & Growing 30d & EXP3Det (7,239) & Unweighted (7,246); EPSDet (7,407) & EXP3Stoch (12,178) \\
Wave 2 & Fixed 5d & SES (15,934) & ARIMA (17,741); RWDrift (18,356) & ExpGrowth (28,350) \\
Wave 2 & Fixed 10d & ARIMA (19,189) & SES (21,929); Naive (21,980) & ARIMA (34,896) \\
Wave 2 & Growing 10d & Holt (20,036) & ARIMA (20,558); InverseWIS (21,112) & RWDrift (33,739) \\
Wave 2 & Growing 30d & EXP3Det (27,430) & InverseWIS (27,816); Unweighted (28,563) & EXP3Stoch (49,505) \\
Wave 3 & Fixed 5d & SES (40,138) & EPSStoch (44,949); EXP3Stoch (45,349) & SES (58,773) \\
Wave 3 & Fixed 10d & SES (37,231) & UCBStoch (40,977); EPSStoch (41,787) & SES (55,572) \\
Wave 3 & Growing 10d & Holt (33,506) & EXP3Stoch (34,456); EPSStoch (36,561) & Holt (55,839) \\
Wave 3 & Growing 30d & Holt (32,192) & EXP3Det (33,834); EXP3Stoch (34,162) & InverseWIS (55,686) \\
\bottomrule
\end{tabular}
}
\end{table}

SES was the most frequent WIS winner in fixed-window settings, while EXP3Det won several growing-window WIS configurations. Holt and ARIMA also won selected configurations. RMSE winners were not always the same as WIS winners, confirming that the best method depends on wave, calibration design, forecast horizon, and metric.

\subsection{Sensitivity to MAB constants}

To assess robustness to the MAB parameter choices, we repeated the U.S. ensemble construction across the same parameter grid used in the state-level analysis. Table~\ref{tab:mab_sensitivity_USA} reports the resulting average forecast performance across the 12 U.S. configurations. The results show that performance can vary with the parameter choice, but adaptive methods remain competitive across a range of settings.

\begin{table}[H]
\centering
\scriptsize
\caption{Sensitivity of U.S. adaptive ensemble performance to MAB constants. Metrics are averaged across the 12 U.S. forecast configurations using the same aggregation scheme as in the main results. Lower WIS and RMSE indicate better performance, while coverage closer to 95\% indicates better interval calibration. For EXP3, $\eta_0=\sqrt{2\log(K)/(Km)}$.}
\label{tab:mab_sensitivity_USA}
\resizebox{\textwidth}{!}{%
\begin{tabular}{llrrrr}
\toprule
Method & Tested value & Mean WIS & Mean RMSE & 95\% PI Coverage (\%) & Mean 95\% PI Width \\
\midrule
\multirow{5}{*}{EXP3Det} & $0.5\eta_0$ & 25171.2 & 41221.9 & 78.6 & 116606.2 \\
 & $\eta_0$ & 24574.3 & 41565.2 & 82.5 & 135156.3 \\
 & \textbf{$2\eta_0$} & \textbf{24008.4} & \textbf{41868.6} & \textbf{87.5} & \textbf{167482.8} \\
 & $5\eta_0$ & 25058.9 & 43572.9 & 89.2 & 195237.0 \\
 & $10\eta_0$ & 25447.2 & 44847.1 & 88.7 & 195315.6 \\
\addlinespace
\multirow{5}{*}{EXP3Stoch} & $0.5\eta_0$ & 24454.2 & 42614.7 & 90.7 & 211595.1 \\
 & $\eta_0$ & 24430.5 & 43006.3 & 91.0 & 208077.5 \\
 & \textbf{$2\eta_0$} & \textbf{24100.0} & \textbf{42104.6} & \textbf{90.7} & \textbf{207659.8} \\
 & $5\eta_0$ & 25068.7 & 43964.5 & 91.2 & 216867.1 \\
 & $10\eta_0$ & 25342.7 & 45153.4 & 90.9 & 214871.9 \\
\addlinespace
\multirow{5}{*}{EPSDet} & $\sqrt{K/m}$ & 25555.2 & 41807.6 & 74.3 & 103911.9 \\
 & $0.1$ & 24831.7 & 40597.7 & 72.7 & 98692.6 \\
 & $0.2$ & 24675.8 & 40317.9 & 72.9 & 100620.0 \\
 & \textbf{$0.5$} & \textbf{24568.1} & \textbf{40362.1} & \textbf{76.2} & \textbf{108371.1} \\
 & $1$ & 24961.1 & 41300.4 & 79.8 & 122605.0 \\
\addlinespace
\multirow{5}{*}{EPSStoch} & $\sqrt{K/m}$ & 24873.0 & 43576.3 & 90.0 & 185508.5 \\
 & $0.1$ & 24048.8 & 41900.1 & 86.4 & 151949.2 \\
 & $0.2$ & 23908.6 & 42093.3 & 88.6 & 162610.4 \\
 & \textbf{$0.5$} & \textbf{23849.9} & \textbf{42363.6} & \textbf{90.2} & \textbf{184281.0} \\
 & $1$ & 24642.2 & 42754.7 & 91.2 & 215732.3 \\
\addlinespace
\multirow{5}{*}{UCBDet} & $c=0.5$ & 26306.1 & 42128.3 & 70.5 & 90209.6 \\
 & $c=1$ & 26306.1 & 42128.3 & 70.5 & 90209.6 \\
 & \textbf{$c=2$} & \textbf{26306.1} & \textbf{42128.3} & \textbf{70.5} & \textbf{90209.6} \\
 & $c=5$ & 26306.1 & 42128.3 & 70.5 & 90209.6 \\
 & $c=10$ & 26306.1 & 42128.3 & 70.5 & 90209.6 \\
\addlinespace
\multirow{5}{*}{UCBStoch} & \textbf{$c=0.5$} & \textbf{25262.3} & \textbf{43682.4} & \textbf{86.8} & \textbf{160167.7} \\
 & $c=1$ & 25322.4 & 43844.8 & 87.7 & 160203.4 \\
 & $c=2$ & 25356.3 & 43928.0 & 86.7 & 159498.1 \\
 & $c=5$ & 25330.6 & 43827.5 & 85.7 & 159121.5 \\
 & $c=10$ & 25397.7 & 43944.4 & 86.9 & 159507.5 \\
\bottomrule
\end{tabular}
}
\end{table}

The sensitivity analysis shows that the numerically best setting varies by algorithm and variant. In this replay, $\varepsilon=0.5$ gave the lowest mean WIS among the tested adaptive settings, while EXP3 variants remained competitive across several learning-rate multipliers. UCBDet was insensitive to the tested values of $c$, and UCBStoch varied only modestly. For UCBDet, the tested values of $c$ did not change the deterministic model-selection sequence, because differences in squared calibration loss on the raw incidence scale dominated the exploration term. We therefore keep the predefined parameter choices for the primary analysis and interpret this table as a sensitivity check rather than as a replacement for the main results.

\subsection{Representative forecast trajectories}

\begin{figure}[H]
\centering
\includegraphics[width=\linewidth]{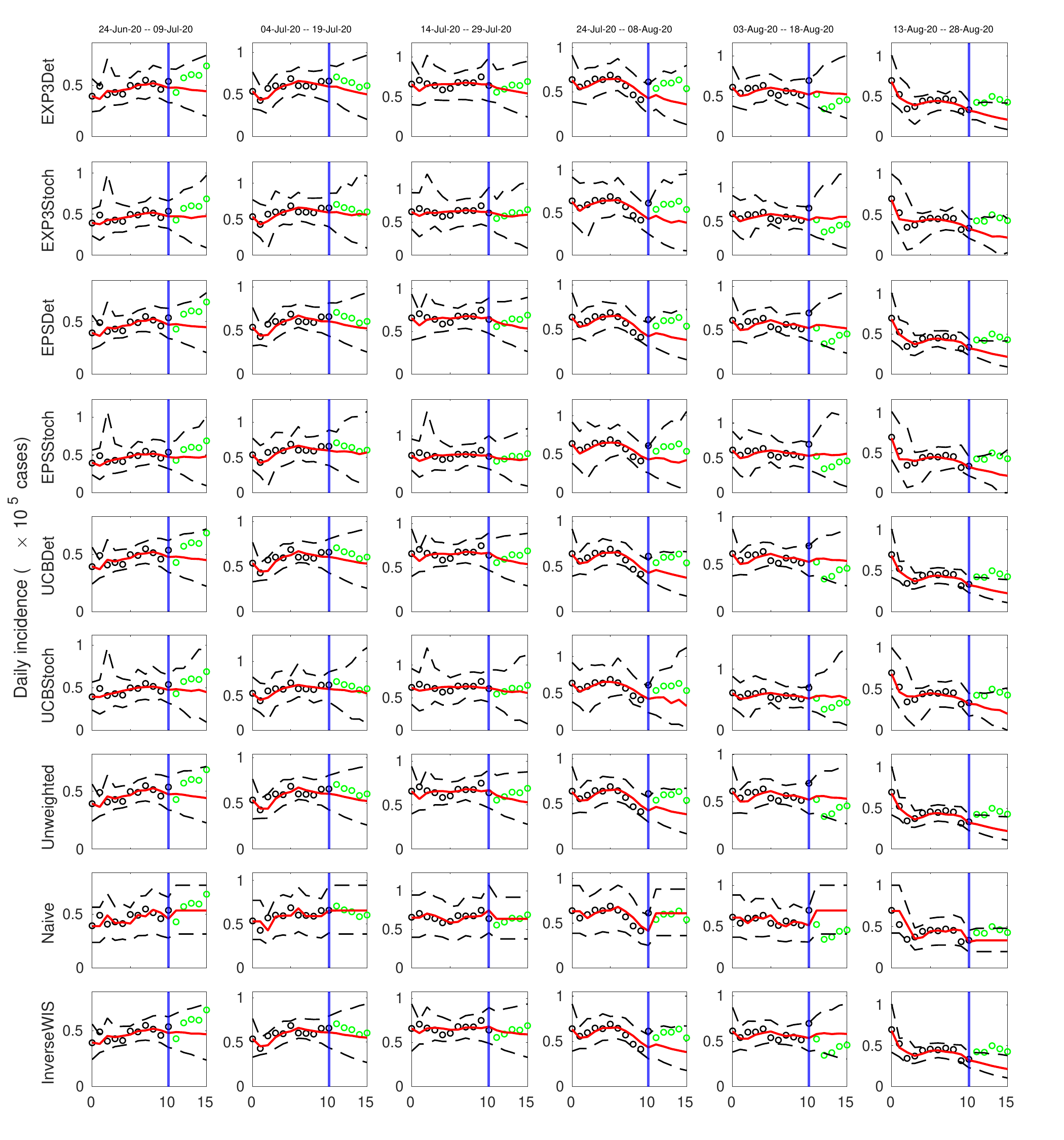}
\caption{Representative U.S. ensemble and benchmark forecasts for Wave 1 under the fixed calibration setting with a short forecast horizon. Rows show adaptive ensemble methods and comparison benchmarks constructed from the 10-model pool. Solid trajectories show predictive medians, dashed curves show 95\% prediction intervals, black points show calibration observations, green points show forecast observations, and the vertical blue line marks the boundary between calibration and forecast periods.}
\label{fig:representative_ensemble_forecast}
\end{figure}

Figure~\ref{fig:representative_ensemble_forecast} shows detailed forecast trajectories and uncertainty intervals for a representative fixed Wave 1 short-window setting. This panel illustrates how predictive medians and interval widths differ across adaptive ensembles and benchmark methods. It should be interpreted as an example that supports the aggregate findings, not as a replacement for the full configuration-level evaluation in Figure~\ref{fig:forecast_performance_summary}, Figure~\ref{fig:wis_percent_change}, and Tables~\ref{tab:forecast_summary}--\ref{tab:mab_sensitivity_USA}.

\subsection{Supplementary Alabama case-study analysis}

The Alabama analysis is supplementary rather than part of the primary U.S. analysis. It uses the same forecasting pipeline and is reported in the Supplementary Material as a state-level comparison. Overall, the Alabama results were consistent with the main message of the U.S. analysis: adaptive ensemble methods remained competitive with simple benchmarks and individual models, and their strongest advantages were seen in probabilistic performance rather than in uniformly lower RMSE. The Alabama case study also reinforces the configuration-dependent nature of the results, showing that the relative ranking of adaptive methods, simple ensembles, and individual time-series models can vary across epidemic scale, wave, calibration design, and forecast horizon.

\section{Discussion}\label{sec:discussion}

This study evaluated MAB-inspired ensemble methods for COVID-19 forecasting using a 10-model pool that spans mechanistic, phenomenological, and statistical forecasting approaches. The results show that adaptive weighting can improve probabilistic forecast performance and prediction-interval coverage in the national U.S. aggregate, particularly when uncertainty is propagated through bootstrap trajectories and stochastic ensemble construction. However, the advantage over simple benchmark ensembles was modest and configuration-dependent, indicating that adaptive weighting should be viewed as a complement to, rather than a uniform replacement for, simple averaging and calibration-performance weighting.

The paired WIS percent-change analysis shows that the benefit of adaptive weighting is configuration dependent. Adaptive ensembles improved WIS in several settings, but selected configurations favored simple benchmark approaches. This pattern is consistent with the role of national aggregation: averaging across heterogeneous local outbreaks can produce a smoother signal for which simple ensemble averaging and inverse-WIS weighting are strong competitors. In contrast, adaptive weighting may be more advantageous when the relative skill of individual component models changes sharply over time.

A key implication is that ensemble forecasts should be evaluated using interval-based scores rather than point accuracy alone. RMSE describes the predictive median, but WIS, 95\% PI coverage, and mean 95\% PI width jointly characterize whether forecasts are accurate, calibrated, and sharp. Wider intervals may increase coverage but are not necessarily preferable if they are insufficiently sharp; WIS provides a way to summarize this trade-off. The comparison across the main performance summaries therefore supports using WIS as the main forecast score while retaining coverage and interval width as diagnostic summaries.

The conclusions of this study should also be interpreted in light of the selected component-model pool. Although we expanded the original pool to include mechanistic, phenomenological, and statistical time-series models, the results remain conditional on these 10 candidate models. A broader or more specialized model pool could change the relative benefit of adaptive weighting, especially if the component models produce more or less diverse forecasts. This interpretation is consistent with recent work showing that the size and composition of real-time COVID-19 forecasting ensembles can influence aggregate forecast performance~\citep{becker2025influence}. This is consistent with recent work on infectious-disease ensemble nowcasting, where the performance of weighted ensemble schemes depended on the available component models and on how model weights were distributed across them~\citep{amaral2025post}. Thus, our findings support the usefulness of MAB-inspired adaptive weighting for this model pool and forecasting design, but they should not be interpreted as a universal ranking of ensemble strategies across all possible epidemic model pools.

Several limitations remain. First, the evaluation is retrospective and uses finalized national U.S. COVID-19 incidence data. In prospective forecasting, real-time reporting revisions can affect model fitting, recent loss estimates, adaptive ensemble weights, and forecast evaluation; these changes in the reported data over time were not explicitly modeled here. Second, although the analysis uses the same broad setting as the U.S. COVID-19 Forecasting Hub, we did not directly compare our ensembles with Hub submissions. Such a comparison would require aligning training windows, forecast targets, submission dates, and real-time data versions. Third, performance may differ for regional outbreaks, other pathogens, or settings with sparse or delayed reporting. Finally, although adaptive ensembles improved interval coverage in many settings, coverage often remained below the nominal 95\% level, likely reflecting the high variability of daily cases and the limits of a bootstrap-based uncertainty procedure.

A further limitation is that the adaptive weights are learned from calibration-period reconstruction errors rather than from a fully rolling forecast evaluation. This choice was made because the fixed calibration windows are short, and repeatedly refitting mechanistic and phenomenological models on even shorter partial windows would lead to unstable parameter estimates. Therefore, the calibration losses used for weight adaptation should be interpreted as measures of how well each component model reconstructs the available calibration window, not as independent out-of-sample forecast errors. The final forecast evaluation, however, is performed only on the held-out forecast period.

Overall, the results suggest that MAB-inspired adaptive weighting provides a useful framework for combining heterogeneous epidemic forecasting models. Its clearest benefit in this study was not uniformly lower point forecast error, but improved WIS and better forecast-period interval coverage relative to several simple benchmarks. These findings support the use of adaptive ensembles as practical tools for improving forecast reliability when model performance changes across epidemic waves, calibration windows, and forecast horizons.

\section*{Code availability}

\noindent All code needed to reproduce the U.S. analysis are available in the
\href{https://github.com/hkarami-GSU/Adaptive-COVID-19-Trajectory-Forecasting-Using-MAB-Inspired-Ensemble-Weighting}{GitHub repository}.
The processed incidence data are provided in \texttt{code\_release\_USA/data/incidence\_cases.xlsx}.

\section*{CRediT authorship contribution statement}

\noindent\textbf{Hamed Karami:} Conceptualization, Methodology, Software, Formal analysis, Investigation, Data curation, Visualization, Writing -- original draft. \textbf{Javier Redondo Anton:} Methodology, Validation, Writing -- review \& editing. \textbf{Geunsoo Jang:} Validation, Writing -- review \& editing. \textbf{K. Selcuk Candan:} Conceptualization, Methodology, Supervision, Writing -- review \& editing. \textbf{Gerardo Chowell:} Conceptualization, Methodology, Supervision, Project administration, Writing -- review \& editing.

\section*{Declaration of competing interest}

\noindent The authors declare that they have no known competing financial interests or personal relationships that could have appeared to influence the work reported in this paper.

\section*{Acknowledgments}

This work was supported by NSF grants DBI 2412115 and 2622265 as part of the US NSF Center for Analysis and Prediction of Pandemic Expansion (APPEX). Additional support was provided by NSF ACED grant 2435886.

%% Bibliography using BibTeX
\bibliographystyle{elsarticle-harv} 
\bibliography{mybib}
\clearpage
\section*{Supplementary Material}

% Reset and prefix numbering for supplementary material.
\setcounter{section}{0}
\setcounter{subsection}{0}
\setcounter{subsubsection}{0}
\setcounter{figure}{0}
\setcounter{table}{0}
\setcounter{equation}{0}
\renewcommand{\thesection}{S\arabic{section}}
\renewcommand{\thesubsection}{S\arabic{section}.\arabic{subsection}}
\renewcommand{\thesubsubsection}{S\arabic{section}.\arabic{subsection}.\arabic{subsubsection}}
\renewcommand{\thefigure}{S\arabic{figure}}
\renewcommand{\thetable}{S\arabic{table}}
\renewcommand{\theequation}{S\arabic{equation}}
\renewcommand{\theHsection}{supp.\arabic{section}}
\renewcommand{\theHsubsection}{supp.\arabic{section}.\arabic{subsection}}
\renewcommand{\theHsubsubsection}{supp.\arabic{section}.\arabic{subsection}.\arabic{subsubsection}}
\renewcommand{\theHfigure}{supp.\arabic{figure}}
\renewcommand{\theHtable}{supp.\arabic{table}}
\renewcommand{\theHequation}{supp.\arabic{equation}}

\newcommand{\fixedcalibration}{a 10-day fixed calibration period}
\newcommand{\growingcalibration}{a growing calibration period of 30, 40, 50, 60, 70, and 80 days}
\newcommand{\basemodels}{SIR, SEIR, GLM, Gompertz, Richards, ARIMA, RWDrift, SES, Holt, and ExpGrowth}
\newcommand{\comparisonmethods}{EXP3Det, EXP3Stoch, EPSDet, EPSStoch, UCBDet, UCBStoch, Unweighted, Naive, and InverseWIS}
\newcommand{\adaptivemethods}{EXP3Det, EXP3Stoch, EPSDet, EPSStoch, UCBDet, and UCBStoch}
\newcommand{\basepanelcaption}[3]{Forecasting performance for Wave #1 under #2 and a #3-day forecast horizon across six evaluation windows. Rows show the expanded 10-model pool: \basemodels. Solid trajectories show predictive medians and dashed bounds show 95\% prediction intervals.}
\newcommand{\comparisonpanelcaption}[3]{Forecasting performance for Wave #1 under #2 and a #3-day forecast horizon across six evaluation windows. Rows show the comparison methods \comparisonmethods, constructed from the expanded 10-model pool. Dashed bounds show 95\% prediction intervals.}
\newcommand{\weightpanelcaption}[2]{Evolution of adaptive ensemble weights for Wave #1 under #2 across six evaluation windows. Panels are shown only for the adaptive methods \adaptivemethods; lines represent weights assigned to the 10 base models. Unweighted, Naive, and InverseWIS are comparison baselines and are not included in this weight-evolution figure.}
\newcommand{\metricstablecaption}[3]{Average calibration and forecasting performance for Alabama in Wave #1 under #2 and a #3-day forecast horizon, averaged across six evaluation windows. Reported measures are RMSE, WIS, coverage of the 95\% prediction interval, and mean width of the 95\% prediction interval. Rows show the 10 base models and the nine ensemble/comparison methods.}
\section{U.S. analysis}

\subsection{U.S. first wave}
\subsubsection{Fixed Calibration Period}
\paragraph{Five-Day Forecasting Horizon}

\begin{figure}[H]
\centering
\includegraphics[width=\linewidth]{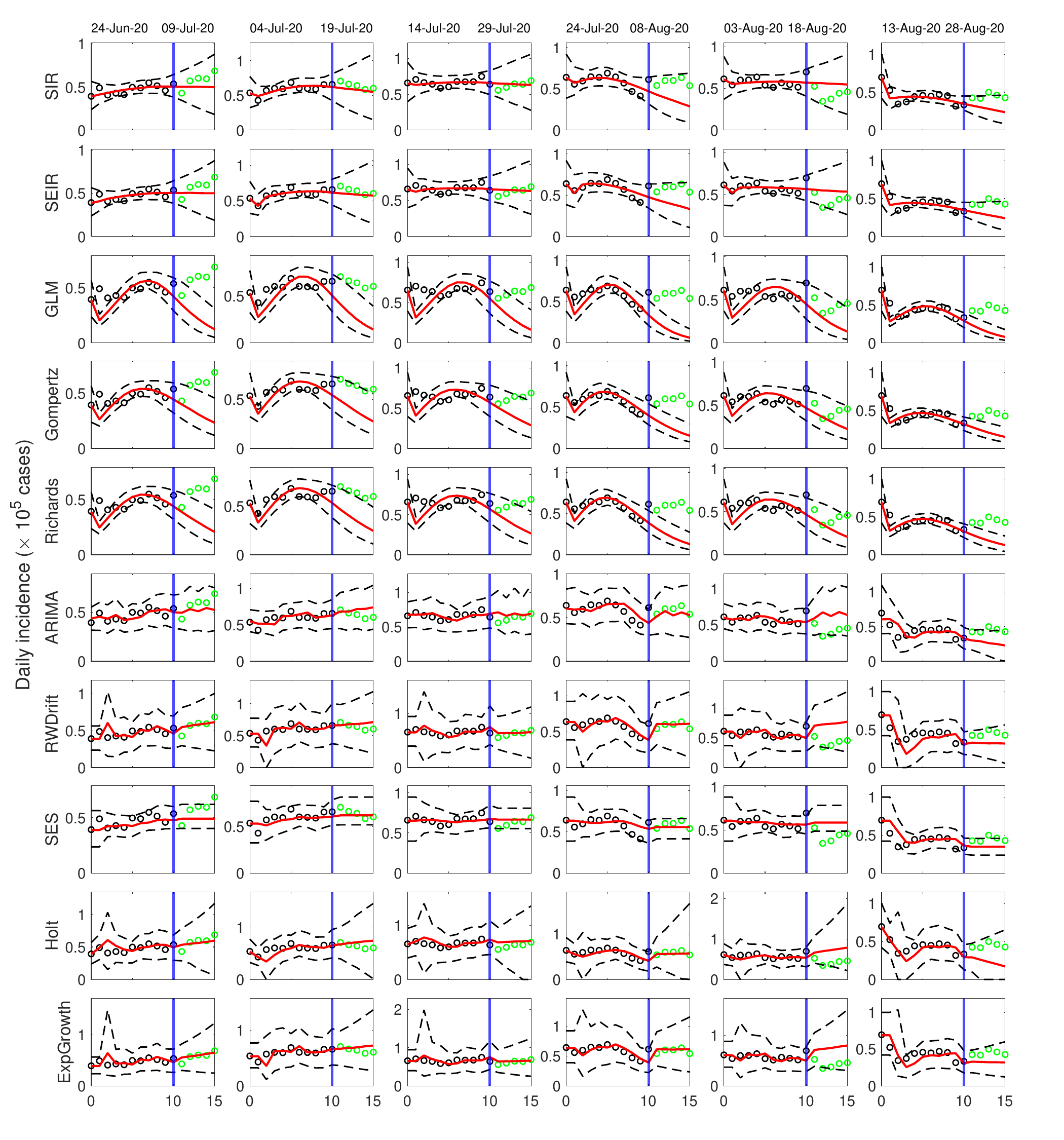}
\caption{\basepanelcaption{1}{\fixedcalibration}{5}}
\label{fig:usa_base_fixed_wave1_fcst5}
\end{figure}
\begin{figure}[H]
\centering
\includegraphics[width=\linewidth]{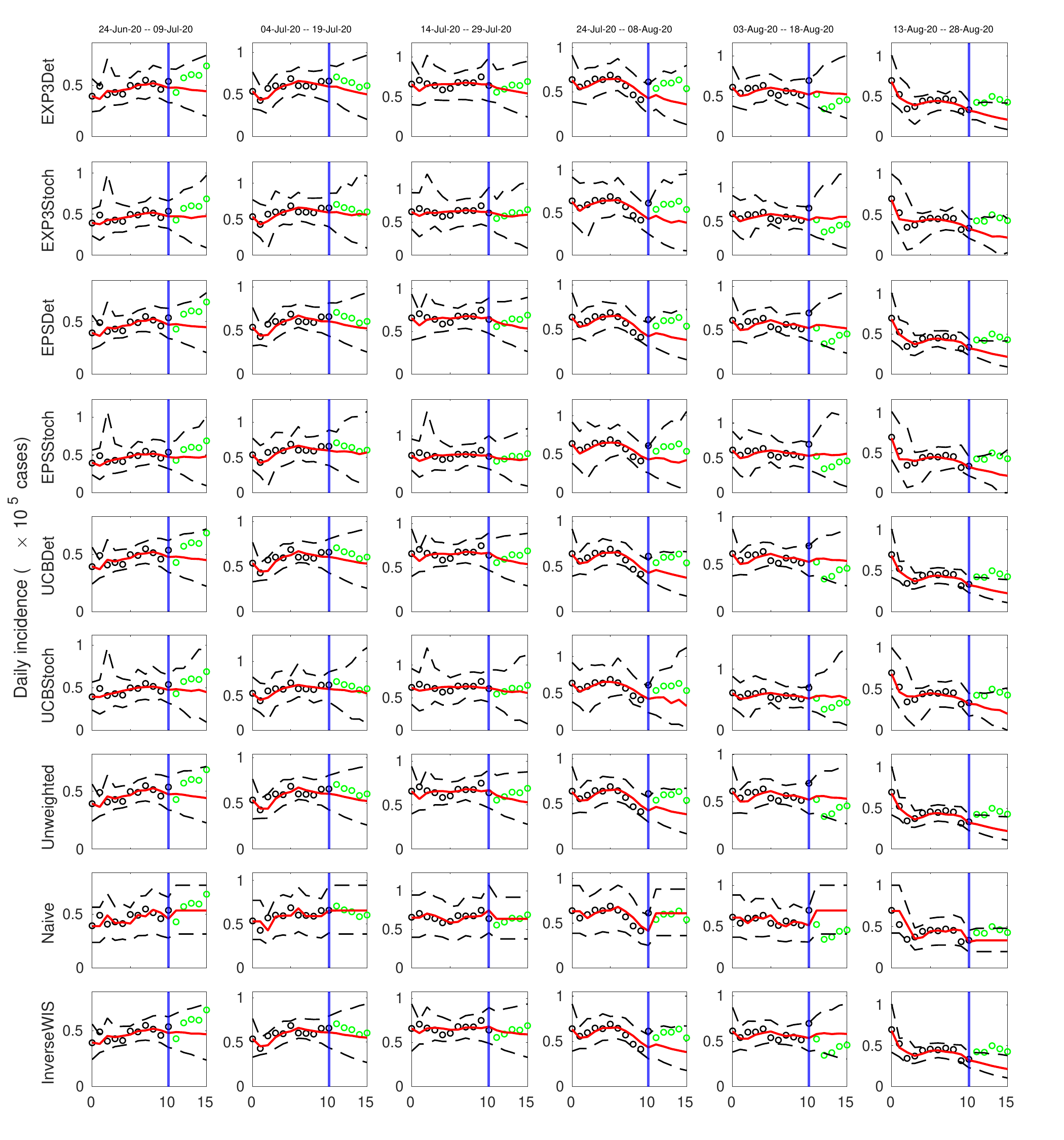}
\caption{\comparisonpanelcaption{1}{\fixedcalibration}{5}}
\label{fig:usa_ensemble_fixed_wave1_fcst5}
\end{figure}
\begin{table}[H]
\centering
\scriptsize
\resizebox{\textwidth}{!}{%
\begin{tabular}{lrrrrrrrr}
\toprule
& \multicolumn{4}{c}{Calibration} & \multicolumn{4}{c}{Forecasting} \\
\cmidrule(lr){2-5} \cmidrule(lr){6-9}
Model & RMSE & WIS & 95\% PI Coverage (\%) & Mean 95\% PI Width & RMSE & WIS & 95\% PI Coverage (\%) & Mean 95\% PI Width \\
\midrule
SIR & 5908.10 & 2980.38 & 95.0\% & 22630.57 & 13090.69 & 7886.81 & 93.3\% & 54128.64 \\
SEIR & 6092.75 & 3026.25 & 93.3\% & 23144.60 & 12179.25 & 7342.39 & 90.0\% & 50378.33 \\
GLM & 12260.87 & 7168.37 & 61.7\% & 22288.93 & 34944.81 & 26140.37 & 23.3\% & 32535.62 \\
Gompertz & 9214.60 & 5044.18 & 73.3\% & 21671.18 & 26147.95 & 17983.47 & 43.3\% & 34172.94 \\
Richards & 9895.09 & 5521.06 & 68.3\% & 21629.42 & 28065.84 & 19954.18 & 26.7\% & 33079.67 \\
ARIMA & 8030.64 & 4174.74 & 98.3\% & 36239.33 & 12280.62 & 7237.72 & 86.7\% & 51792.15 \\
RWDrift & 10329.88 & 5646.74 & 96.7\% & 58021.26 & 11222.05 & 7463.34 & 100.0\% & 70015.33 \\
SES & 7870.89 & 4109.96 & 88.3\% & 30423.67 & 9147.79 & 5606.21 & 76.7\% & 25591.88 \\
Holt & 9887.50 & 5153.23 & 100.0\% & 52719.78 & 13673.24 & 8776.38 & 96.7\% & 95039.49 \\
ExpGrowth & 10192.58 & 5529.05 & 96.7\% & 59713.65 & 11616.77 & 7729.04 & 96.7\% & 74185.96 \\
EXP3Det & 5985.55 & 3027.47 & 98.3\% & 33063.40 & 14208.09 & 7891.99 & 86.7\% & 51165.21 \\
EXP3Stoch & 5878.55 & 3186.56 & 98.3\% & 42350.40 & 13203.56 & 7933.41 & 96.7\% & 72206.21 \\
EPSDet & 6116.93 & 3044.63 & 96.7\% & 28030.09 & 13879.87 & 7835.27 & 83.3\% & 47256.69 \\
EPSStoch & 6150.99 & 3316.75 & 98.3\% & 44133.91 & 12889.18 & 7837.12 & 96.7\% & 72666.14 \\
UCBDet & 6020.58 & 2990.12 & 95.0\% & 25250.97 & 13701.61 & 7829.43 & 80.0\% & 44301.27 \\
UCBStoch & 5654.53 & 3165.42 & 98.3\% & 43489.58 & 13672.63 & 7907.50 & 96.7\% & 73865.71 \\
Unweighted & 6181.21 & 3072.92 & 93.3\% & 26326.79 & 13830.75 & 7836.30 & 83.3\% & 44632.38 \\
Naive & 7367.58 & 3960.13 & 98.3\% & 42921.97 & 10396.62 & 6532.95 & 90.0\% & 49174.20 \\
InverseWIS & 5684.48 & 2834.66 & 96.7\% & 24716.69 & 13401.74 & 7884.24 & 80.0\% & 44278.15 \\
\bottomrule
\end{tabular}%
}
\caption{Average calibration and forecasting performance for the U.S. in Wave 1 under \fixedcalibration\, and a 5-day forecast horizon, averaged across six evaluation windows. Reported measures are RMSE, WIS, coverage of the 95\% prediction interval, and mean width of the 95\% prediction interval. Rows show the 10 base models and the nine ensemble/comparison methods.}
\label{tab:usa_metrics_fixed_wave1_fcst5}
\end{table}
\paragraph{Ten-Day Forecasting Horizon}

\begin{figure}[H]
\centering
\includegraphics[width=\linewidth]{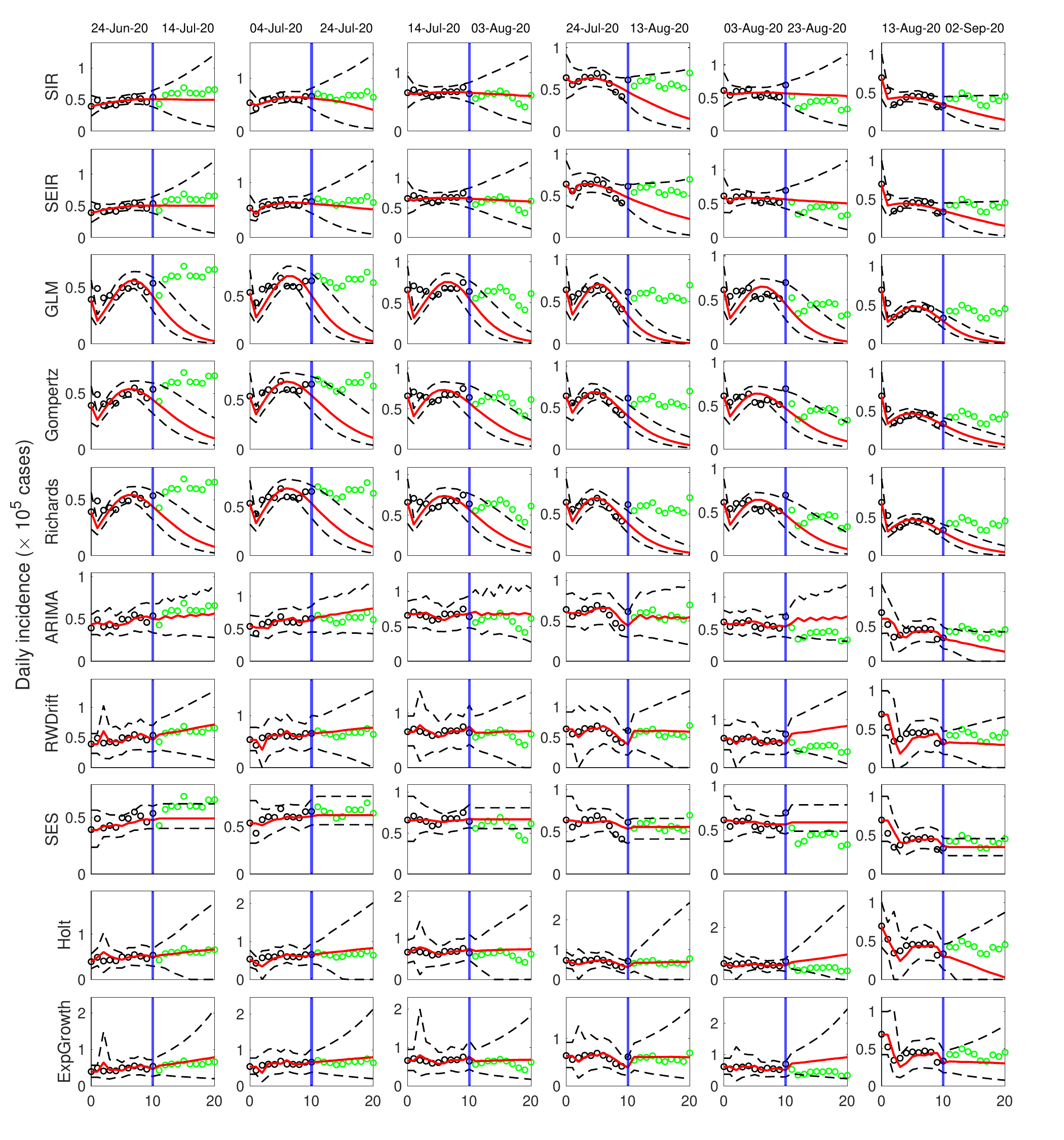}
\caption{\basepanelcaption{1}{\fixedcalibration}{10}}
\label{fig:usa_base_fixed_wave1_fcst10}
\end{figure}
\begin{figure}[H]
\centering
\includegraphics[width=\linewidth]{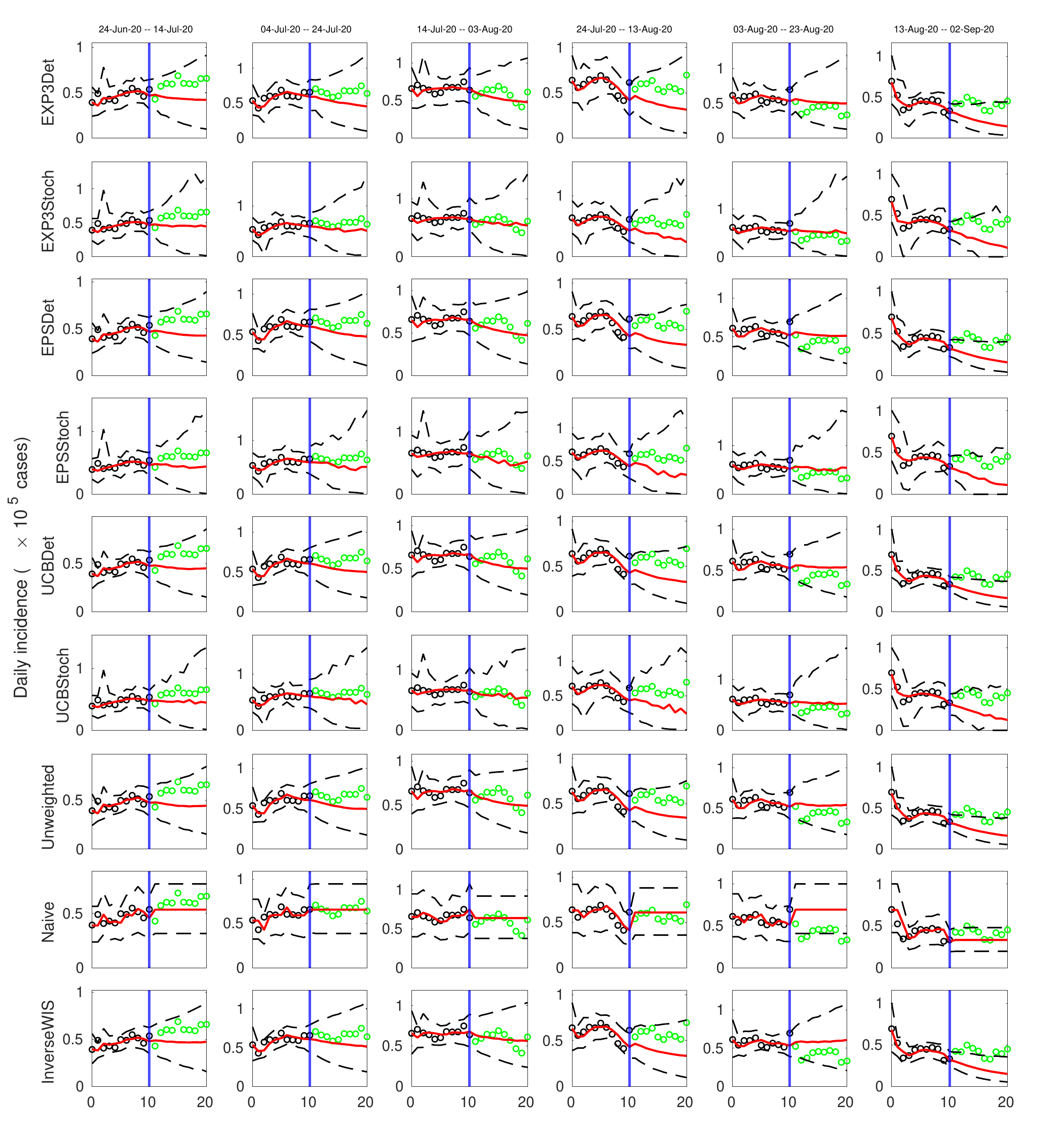}
\caption{\comparisonpanelcaption{1}{\fixedcalibration}{10}}
\label{fig:usa_ensemble_fixed_wave1_fcst10}
\end{figure}
\begin{table}[H]
\centering
\scriptsize
\resizebox{\textwidth}{!}{%
\begin{tabular}{lrrrrrrrr}
\toprule
& \multicolumn{4}{c}{Calibration} & \multicolumn{4}{c}{Forecasting} \\
\cmidrule(lr){2-5} \cmidrule(lr){6-9}
Model & RMSE & WIS & 95\% PI Coverage (\%) & Mean 95\% PI Width & RMSE & WIS & 95\% PI Coverage (\%) & Mean 95\% PI Width \\
\midrule
SIR & 5908.10 & 2980.38 & 95.0\% & 22630.57 & 17222.98 & 9683.97 & 96.7\% & 70964.13 \\
SEIR & 6092.75 & 3026.25 & 93.3\% & 23144.60 & 15201.86 & 8754.06 & 95.0\% & 67157.17 \\
GLM & 12260.87 & 7168.37 & 61.7\% & 22288.93 & 42511.11 & 35092.50 & 11.7\% & 24293.52 \\
Gompertz & 9214.60 & 5044.18 & 73.3\% & 21671.18 & 34341.48 & 25519.29 & 25.0\% & 31033.29 \\
Richards & 9895.09 & 5521.06 & 68.3\% & 21629.42 & 36204.60 & 28031.45 & 13.3\% & 28777.27 \\
ARIMA & 8030.64 & 4174.74 & 98.3\% & 36239.33 & 14129.62 & 8082.36 & 91.7\% & 58104.63 \\
RWDrift & 10329.88 & 5646.74 & 96.7\% & 58021.26 & 13511.39 & 8639.05 & 100.0\% & 89980.80 \\
SES & 7870.89 & 4109.96 & 88.3\% & 30423.67 & 10495.00 & 6354.97 & 71.7\% & 25591.88 \\
Holt & 9887.50 & 5153.23 & 100.0\% & 52719.78 & 17670.60 & 11308.05 & 98.3\% & 134414.71 \\
ExpGrowth & 10192.58 & 5529.05 & 96.7\% & 59713.65 & 14724.83 & 9403.92 & 98.3\% & 104543.58 \\
EXP3Det & 6094.95 & 3074.03 & 96.7\% & 34353.47 & 16488.37 & 9197.50 & 91.7\% & 64233.02 \\
EXP3Stoch & 5983.27 & 3265.73 & 98.3\% & 43008.73 & 16301.87 & 9746.81 & 98.3\% & 93090.72 \\
EPSDet & 6152.63 & 3078.28 & 96.7\% & 28564.96 & 16011.08 & 8838.79 & 91.7\% & 56616.10 \\
EPSStoch & 5900.68 & 3272.39 & 100.0\% & 43896.97 & 17193.58 & 9882.89 & 98.3\% & 92361.19 \\
UCBDet & 6021.42 & 2990.17 & 95.0\% & 25253.95 & 15686.58 & 8721.37 & 85.0\% & 53080.01 \\
UCBStoch & 5774.50 & 3150.90 & 98.3\% & 42437.64 & 16035.26 & 9503.37 & 98.3\% & 92126.59 \\
Unweighted & 6182.25 & 3072.96 & 93.3\% & 26328.96 & 15830.66 & 8684.80 & 86.7\% & 53076.02 \\
Naive & 7367.58 & 3960.13 & 98.3\% & 42921.97 & 11390.55 & 6779.76 & 91.7\% & 49174.20 \\
InverseWIS & 5684.62 & 2834.71 & 96.7\% & 24718.02 & 15754.73 & 8865.24 & 85.0\% & 53895.65 \\
\bottomrule
\end{tabular}%
}
\caption{Average calibration and forecasting performance for the U.S. in Wave 1 under \fixedcalibration\, and a 10-day forecast horizon, averaged across six evaluation windows. Reported measures are RMSE, WIS, coverage of the 95\% prediction interval, and mean width of the 95\% prediction interval. Rows show the 10 base models and the nine ensemble/comparison methods.}
\label{tab:usa_metrics_fixed_wave1_fcst10}
\end{table}
\begin{figure}[H]
\centering
\includegraphics[width=\linewidth]{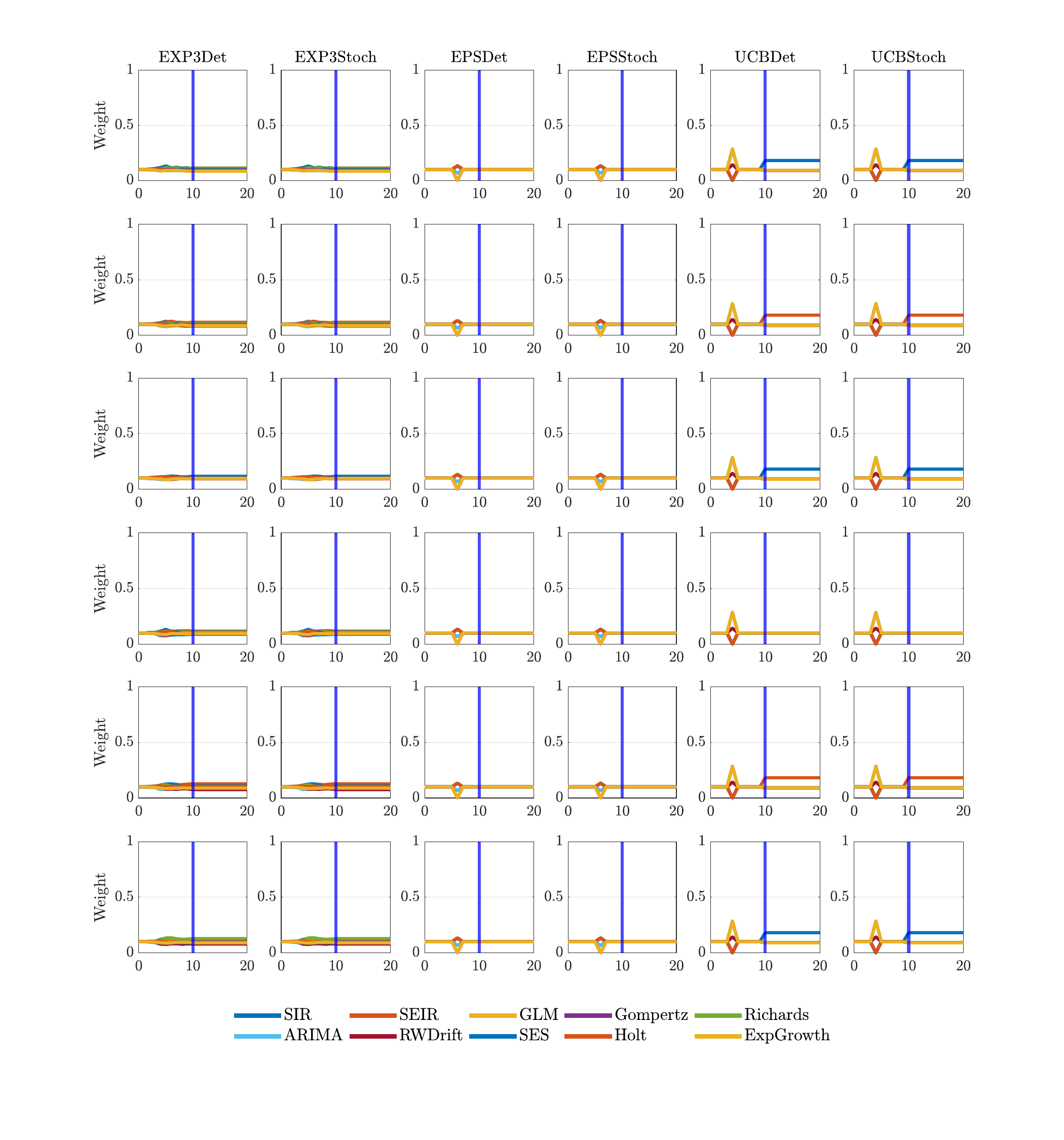}
\caption{\weightpanelcaption{1}{\fixedcalibration}}
\label{fig:usa_weights_fixed_wave1}
\end{figure}
\subsubsection{Growing Calibration Period}
\paragraph{Ten-Day Forecasting Horizon}

\begin{figure}[H]
\centering
\includegraphics[width=\linewidth]{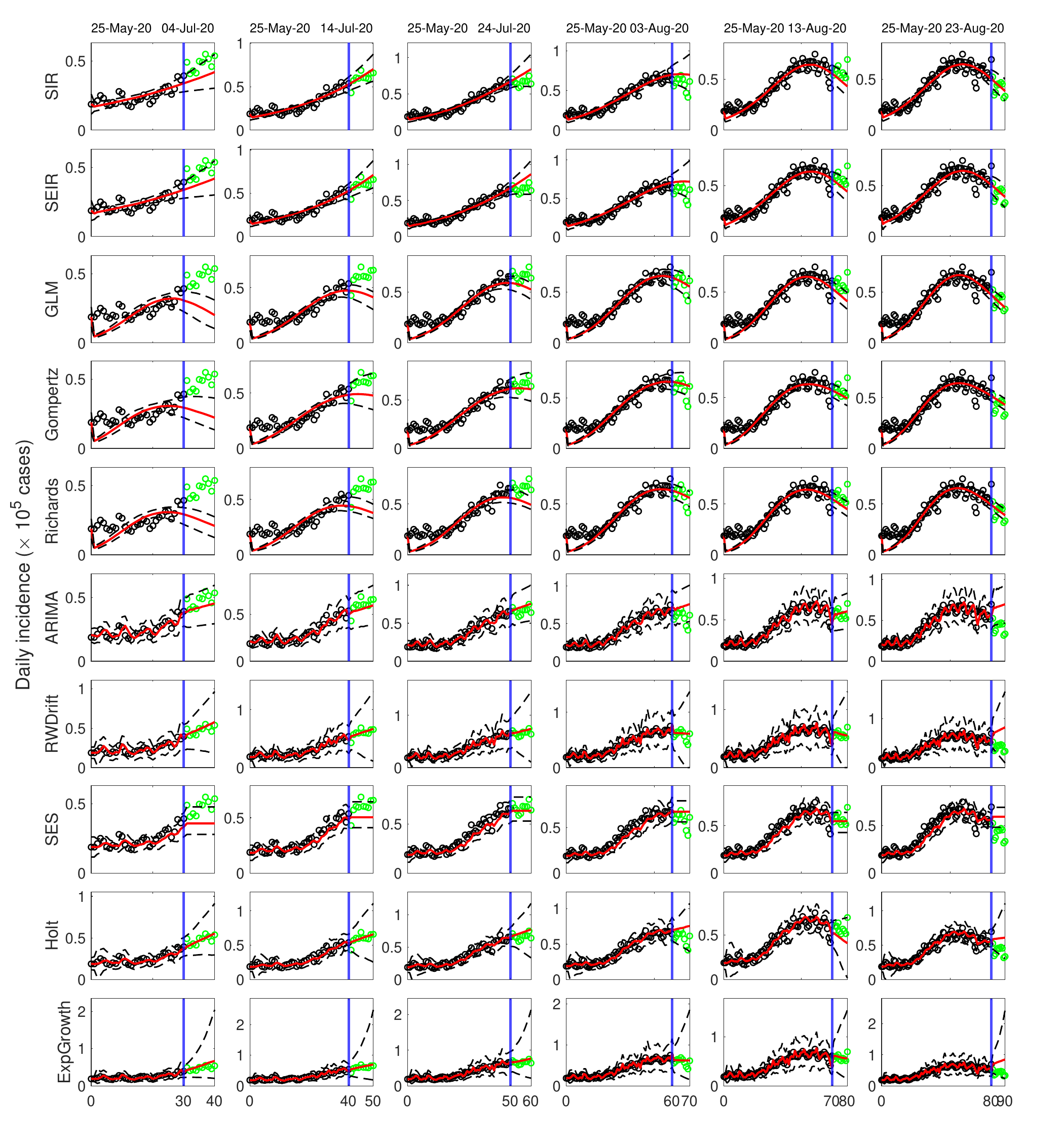}
\caption{\basepanelcaption{1}{\growingcalibration}{10}}
\label{fig:usa_base_growing_wave1_fcst10}
\end{figure}
\begin{figure}[H]
\centering
\includegraphics[width=\linewidth]{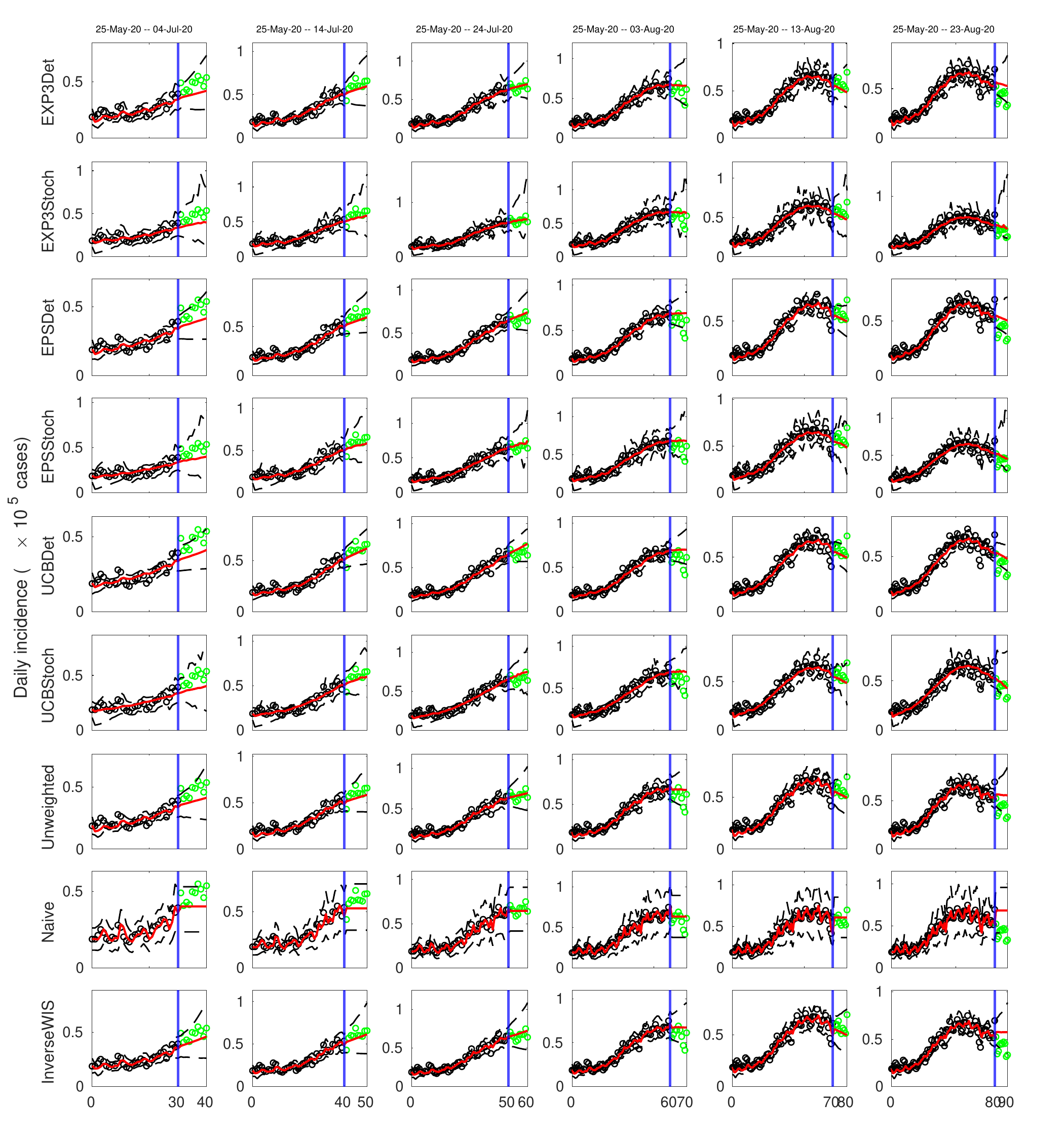}
\caption{\comparisonpanelcaption{1}{\growingcalibration}{10}}
\label{fig:usa_ensemble_growing_wave1_fcst10}
\end{figure}
\begin{table}[H]
\centering
\scriptsize
\resizebox{\textwidth}{!}{%
\begin{tabular}{lrrrrrrrr}
\toprule
& \multicolumn{4}{c}{Calibration} & \multicolumn{4}{c}{Forecasting} \\
\cmidrule(lr){2-5} \cmidrule(lr){6-9}
Model & RMSE & WIS & 95\% PI Coverage (\%) & Mean 95\% PI Width & RMSE & WIS & 95\% PI Coverage (\%) & Mean 95\% PI Width \\
\midrule
SIR & 4985.21 & 2891.63 & 59.5\% & 7462.08 & 9954.53 & 5481.51 & 75.0\% & 24585.00 \\
SEIR & 4995.21 & 2897.46 & 60.1\% & 7579.29 & 10355.20 & 5616.02 & 76.7\% & 25329.62 \\
GLM & 8022.07 & 5084.02 & 44.5\% & 6462.28 & 13108.03 & 9376.61 & 36.7\% & 16177.87 \\
Gompertz & 7805.68 & 4848.25 & 50.2\% & 6864.17 & 11314.44 & 7117.13 & 66.7\% & 19519.44 \\
Richards & 8205.12 & 5197.42 & 45.3\% & 6772.03 & 13858.75 & 10185.64 & 33.3\% & 13407.66 \\
ARIMA & 5976.20 & 3092.74 & 87.1\% & 20644.20 & 11592.14 & 6820.59 & 85.0\% & 39688.45 \\
RWDrift & 6299.45 & 3297.92 & 99.0\% & 32508.95 & 10989.54 & 7723.04 & 98.3\% & 89611.76 \\
SES & 6218.41 & 3471.07 & 65.9\% & 13134.63 & 11191.24 & 6908.94 & 63.3\% & 22187.36 \\
Holt & 5370.89 & 2796.73 & 92.6\% & 20296.74 & 11121.15 & 6177.84 & 93.3\% & 51538.97 \\
ExpGrowth & 6316.44 & 3318.34 & 100.0\% & 33444.94 & 12293.21 & 8756.81 & 98.3\% & 113303.71 \\
EXP3Det & 4618.44 & 2337.93 & 87.7\% & 16125.14 & 9139.14 & 5025.54 & 90.0\% & 37960.57 \\
EXP3Stoch & 4513.99 & 2302.82 & 97.8\% & 23614.00 & 8844.66 & 5075.90 & 95.0\% & 61102.41 \\
EPSDet & 4550.96 & 2373.50 & 75.2\% & 10860.97 & 9388.49 & 5167.51 & 83.3\% & 29134.64 \\
EPSStoch & 4528.44 & 2268.78 & 95.9\% & 20764.17 & 9176.01 & 5043.93 & 95.0\% & 48211.76 \\
UCBDet & 4545.38 & 2417.49 & 72.8\% & 9991.32 & 9470.81 & 5213.68 & 76.7\% & 25017.66 \\
UCBStoch & 4623.26 & 2303.52 & 92.5\% & 17446.81 & 9158.73 & 5028.12 & 88.3\% & 38298.75 \\
Unweighted & 4787.15 & 2519.63 & 76.1\% & 12178.02 & 9532.15 & 5230.46 & 88.3\% & 34731.06 \\
Naive & 5090.63 & 2649.06 & 97.5\% & 28344.83 & 11191.50 & 6660.33 & 90.0\% & 45849.90 \\
InverseWIS & 4500.93 & 2283.66 & 84.1\% & 13713.63 & 9255.03 & 5099.87 & 88.3\% & 38624.77 \\
\bottomrule
\end{tabular}%
}
\caption{Average calibration and forecasting performance for the U.S. in Wave 1 under \growingcalibration\, and a 10-day forecast horizon, averaged across six evaluation windows. Reported measures are RMSE, WIS, coverage of the 95\% prediction interval, and mean width of the 95\% prediction interval. Rows show the 10 base models and the nine ensemble/comparison methods.}
\label{tab:usa_metrics_growing_wave1_fcst10}
\end{table}
\paragraph{Thirty-Day Forecasting Horizon}

\begin{figure}[H]
\centering
\includegraphics[width=\linewidth]{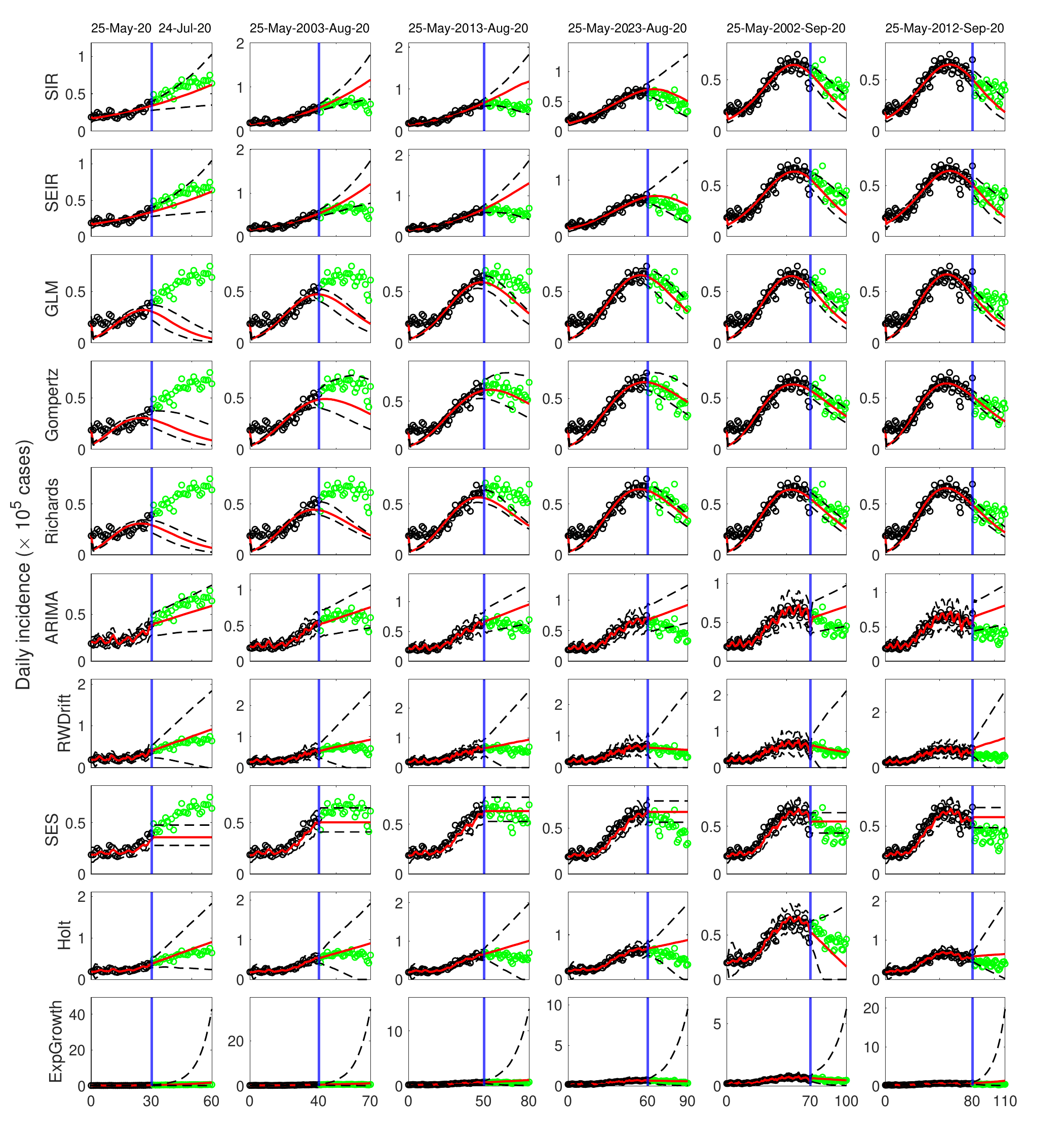}
\caption{\basepanelcaption{1}{\growingcalibration}{30}}
\label{fig:usa_base_growing_wave1_fcst30}
\end{figure}
\begin{figure}[H]
\centering
\includegraphics[width=\linewidth]{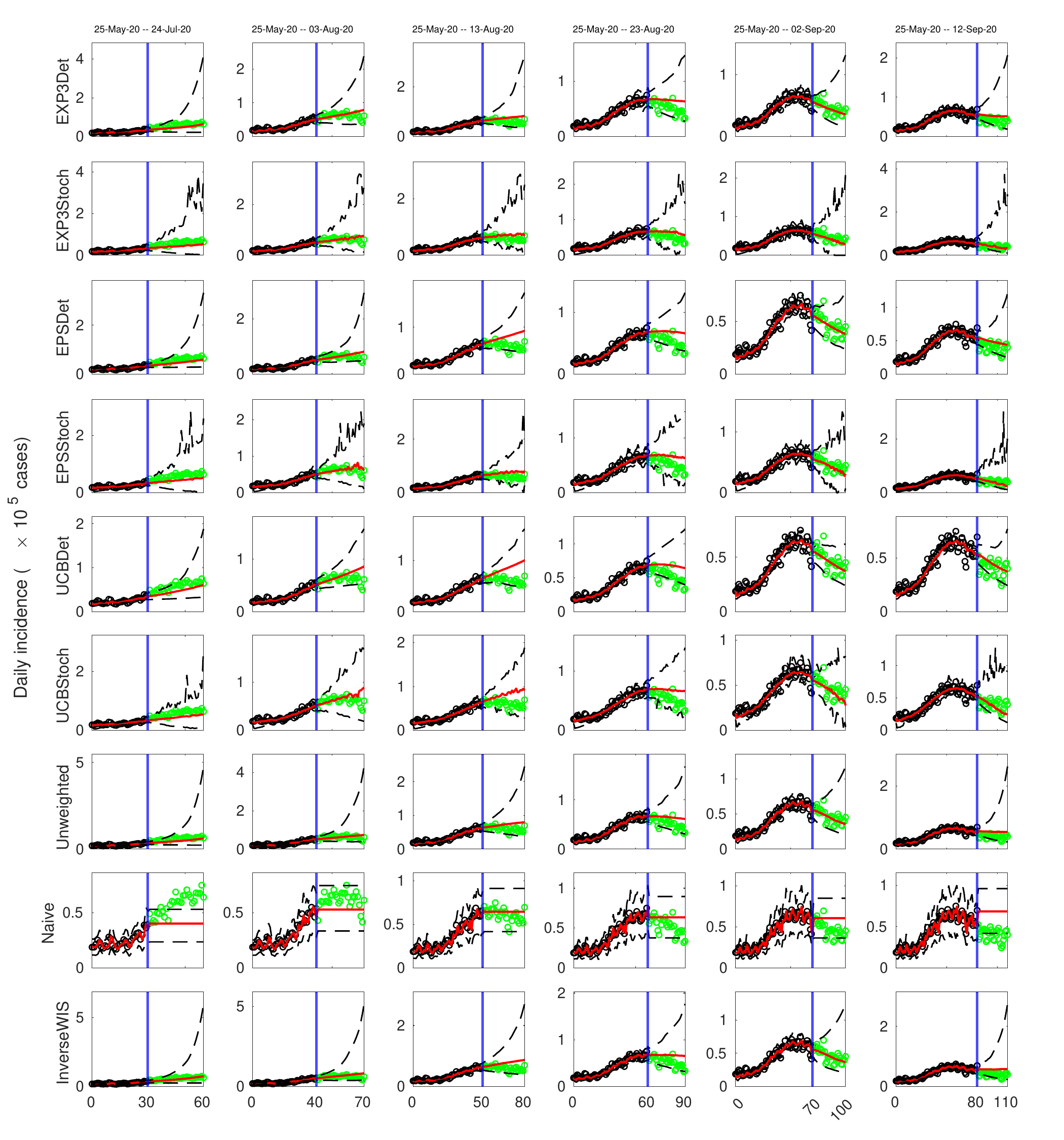}
\caption{\comparisonpanelcaption{1}{\growingcalibration}{30}}
\label{fig:usa_ensemble_growing_wave1_fcst30}
\end{figure}
\begin{table}[H]
\centering
\scriptsize
\resizebox{\textwidth}{!}{%
\begin{tabular}{lrrrrrrrr}
\toprule
& \multicolumn{4}{c}{Calibration} & \multicolumn{4}{c}{Forecasting} \\
\cmidrule(lr){2-5} \cmidrule(lr){6-9}
Model & RMSE & WIS & 95\% PI Coverage (\%) & Mean 95\% PI Width & RMSE & WIS & 95\% PI Coverage (\%) & Mean 95\% PI Width \\
\midrule
SIR & 4985.21 & 2891.63 & 59.5\% & 7462.08 & 19917.10 & 10755.75 & 75.0\% & 44349.38 \\
SEIR & 4995.21 & 2897.46 & 60.1\% & 7579.29 & 21542.73 & 11339.69 & 75.0\% & 45494.04 \\
GLM & 8022.07 & 5084.02 & 44.5\% & 6462.28 & 21722.04 & 17119.24 & 27.2\% & 15779.00 \\
Gompertz & 7805.68 & 4848.25 & 50.2\% & 6864.17 & 15731.48 & 10635.87 & 63.3\% & 23970.51 \\
Richards & 8205.12 & 5197.42 & 45.3\% & 6772.03 & 20525.93 & 16562.22 & 26.1\% & 12058.43 \\
ARIMA & 5976.20 & 3092.74 & 87.1\% & 20644.20 & 22071.04 & 13304.54 & 65.0\% & 46688.34 \\
RWDrift & 6299.45 & 3297.92 & 99.0\% & 32508.95 & 20473.76 & 13201.38 & 99.4\% & 154012.42 \\
SES & 6218.41 & 3471.07 & 65.9\% & 13134.63 & 16019.33 & 10753.68 & 45.6\% & 22187.36 \\
Holt & 5370.89 & 2796.73 & 92.6\% & 20296.74 & 20917.84 & 10556.31 & 97.8\% & 100269.14 \\
ExpGrowth & 6316.44 & 3318.34 & 100.0\% & 33444.94 & 31816.46 & 23274.61 & 99.4\% & 594300.71 \\
EXP3Det & 4639.07 & 2343.18 & 88.0\% & 16168.59 & 13182.42 & 7239.06 & 96.7\% & 90849.76 \\
EXP3Stoch & 4544.38 & 2307.37 & 97.3\% & 23576.62 & 12177.67 & 7700.73 & 98.9\% & 131883.44 \\
EPSDet & 4566.18 & 2378.92 & 75.5\% & 10854.31 & 14100.53 & 7406.74 & 88.9\% & 62302.33 \\
EPSStoch & 4530.65 & 2280.99 & 95.3\% & 20492.75 & 12909.11 & 7474.86 & 97.8\% & 95547.68 \\
UCBDet & 4544.13 & 2417.50 & 72.8\% & 9992.68 & 14629.12 & 7751.45 & 85.0\% & 46069.58 \\
UCBStoch & 4643.79 & 2318.11 & 92.6\% & 17506.64 & 15067.27 & 7687.13 & 97.2\% & 72301.78 \\
Unweighted & 4787.14 & 2519.55 & 76.1\% & 12178.08 & 12902.62 & 7246.09 & 94.4\% & 92044.66 \\
Naive & 5090.63 & 2649.06 & 97.5\% & 28344.83 & 16751.07 & 9984.61 & 75.0\% & 45849.90 \\
InverseWIS & 4500.80 & 2283.68 & 84.1\% & 13714.89 & 14019.53 & 8011.37 & 93.9\% & 105443.09 \\
\bottomrule
\end{tabular}%
}
\caption{Average calibration and forecasting performance for the U.S. in Wave 1 under \growingcalibration\, and a 30-day forecast horizon, averaged across six evaluation windows. Reported measures are RMSE, WIS, coverage of the 95\% prediction interval, and mean width of the 95\% prediction interval. Rows show the 10 base models and the nine ensemble/comparison methods.}
\label{tab:usa_metrics_growing_wave1_fcst30}
\end{table}
\begin{figure}[H]
\centering
\includegraphics[width=\linewidth]{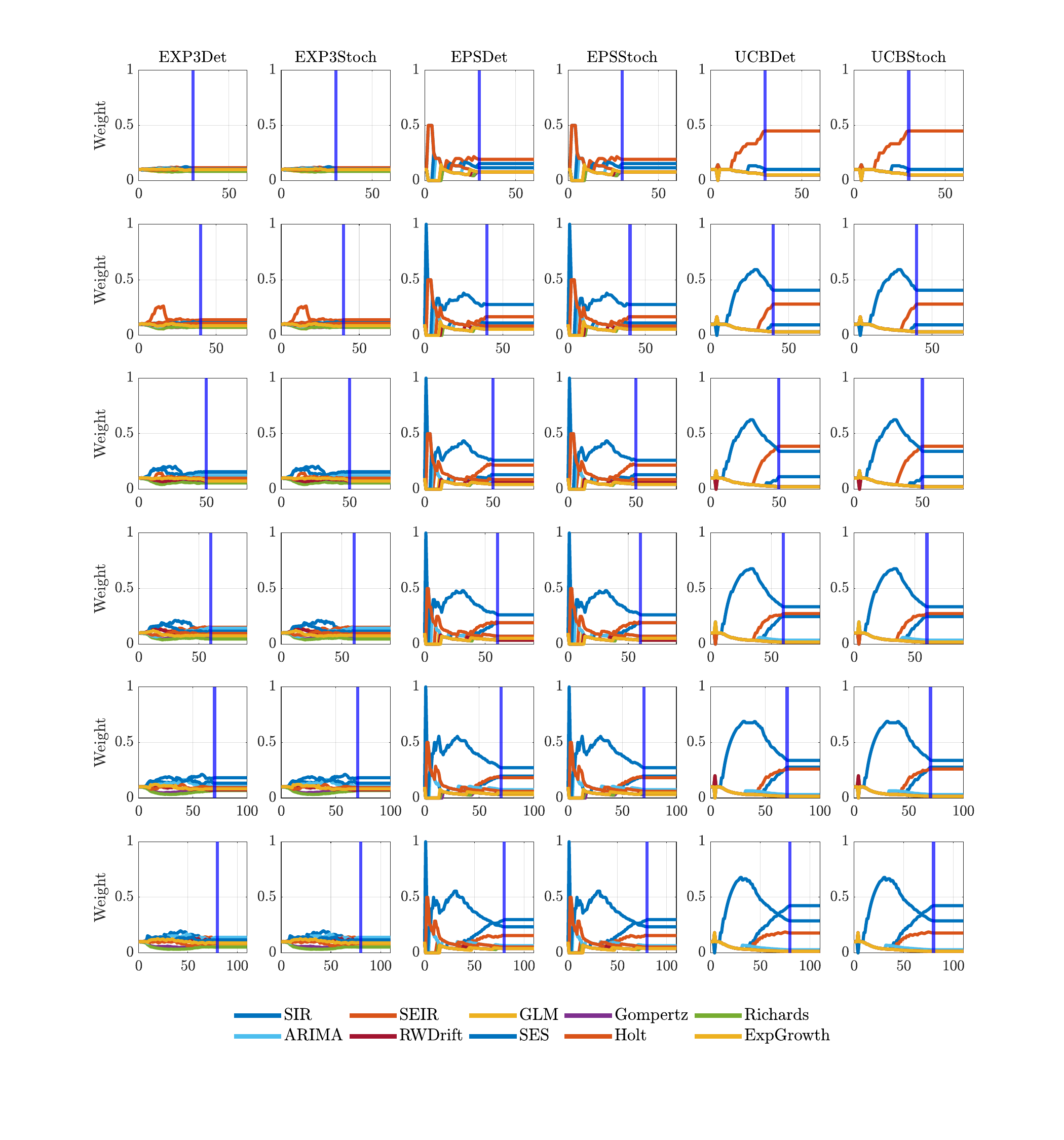}
\caption{\weightpanelcaption{1}{\growingcalibration}}
\label{fig:usa_weights_growing_wave1}
\end{figure}
\subsection{U.S. second wave}
\subsubsection{Fixed Calibration Period}
\paragraph{Five-Day Forecasting Horizon}

\begin{figure}[H]
\centering
\includegraphics[width=\linewidth]{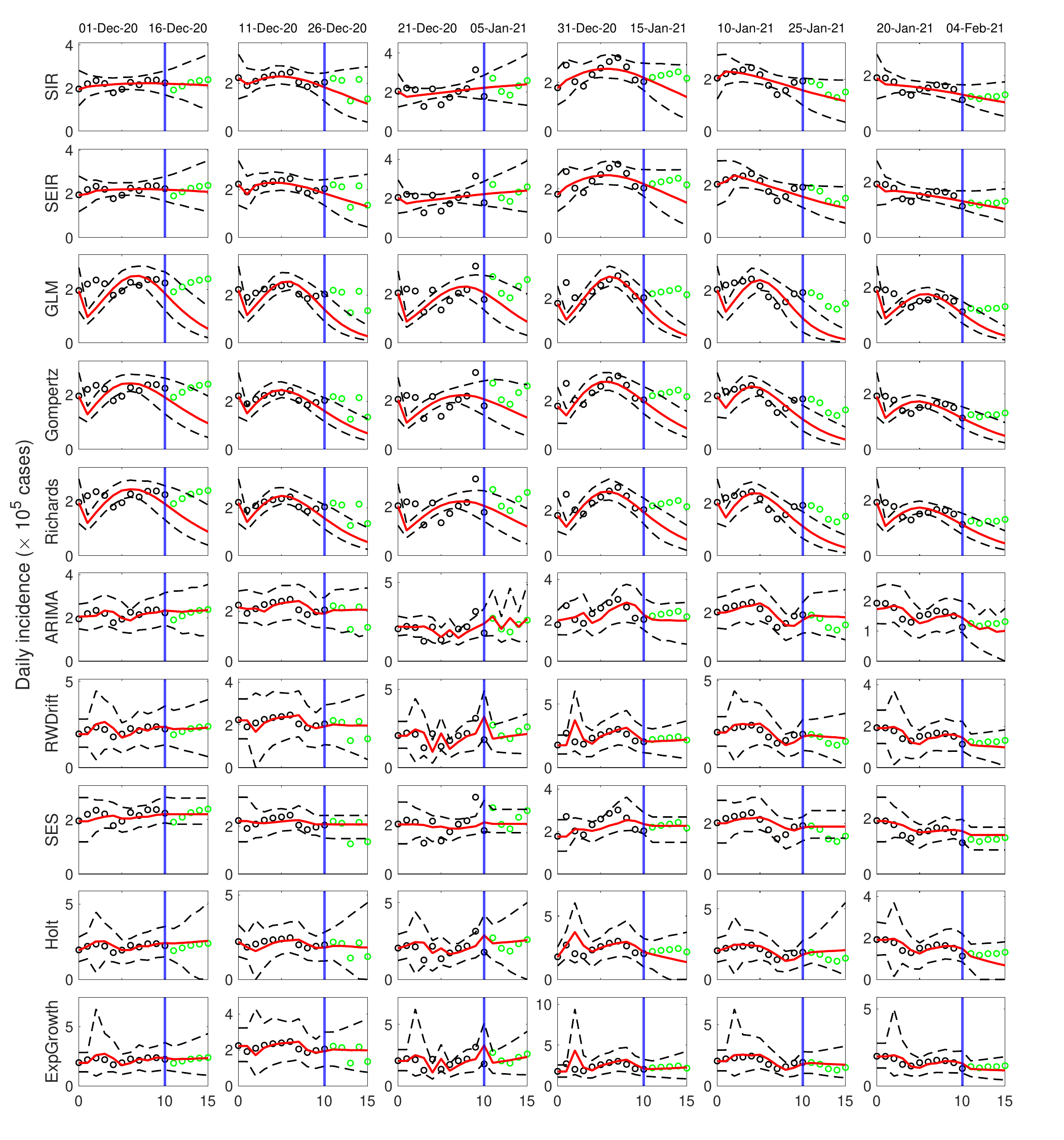}
\caption{\basepanelcaption{2}{\fixedcalibration}{5}}
\label{fig:usa_base_fixed_wave2_fcst5}
\end{figure}
\begin{figure}[H]
\centering
\includegraphics[width=\linewidth]{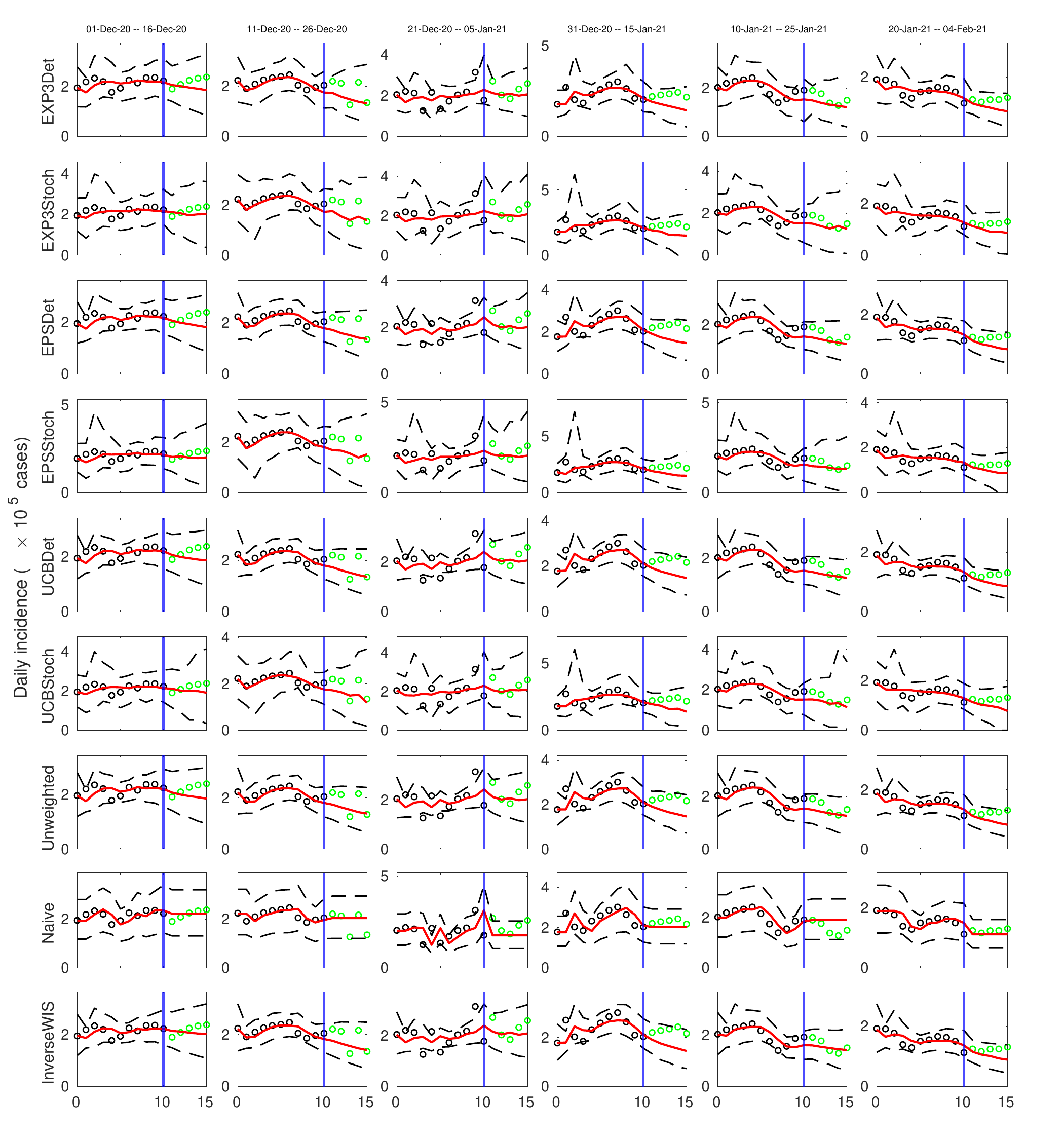}
\caption{\comparisonpanelcaption{2}{\fixedcalibration}{5}}
\label{fig:usa_ensemble_fixed_wave2_fcst5}
\end{figure}
\begin{table}[H]
\centering
\scriptsize
\resizebox{\textwidth}{!}{%
\begin{tabular}{lrrrrrrrr}
\toprule
& \multicolumn{4}{c}{Calibration} & \multicolumn{4}{c}{Forecasting} \\
\cmidrule(lr){2-5} \cmidrule(lr){6-9}
Model & RMSE & WIS & 95\% PI Coverage (\%) & Mean 95\% PI Width & RMSE & WIS & 95\% PI Coverage (\%) & Mean 95\% PI Width \\
\midrule
SIR & 28239.57 & 15433.54 & 81.7\% & 83625.77 & 35839.09 & 19563.94 & 100.0\% & 171595.17 \\
SEIR & 28741.52 & 15645.77 & 80.0\% & 85715.10 & 34977.47 & 18893.65 & 100.0\% & 163552.03 \\
GLM & 50276.13 & 29033.29 & 65.0\% & 82505.81 & 120540.39 & 90877.54 & 16.7\% & 106586.80 \\
Gompertz & 39301.02 & 22121.00 & 70.0\% & 79380.56 & 88348.88 & 59269.42 & 40.0\% & 121419.15 \\
Richards & 41817.84 & 23652.80 & 65.0\% & 79891.84 & 95323.92 & 66812.71 & 26.7\% & 113679.87 \\
ARIMA & 40104.74 & 21564.35 & 88.3\% & 130866.68 & 32559.46 & 17741.17 & 100.0\% & 192695.99 \\
RWDrift & 49765.89 & 25446.06 & 95.0\% & 213988.46 & 29407.91 & 18356.01 & 100.0\% & 225862.66 \\
SES & 36018.75 & 19628.13 & 81.7\% & 113023.80 & 28899.02 & 15934.05 & 83.3\% & 100017.79 \\
Holt & 46094.02 & 23583.71 & 96.7\% & 203352.82 & 44528.53 & 25424.71 & 100.0\% & 315582.88 \\
ExpGrowth & 52582.38 & 26350.43 & 95.0\% & 231467.93 & 28350.00 & 18390.21 & 100.0\% & 232001.03 \\
EXP3Det & 29483.99 & 15985.85 & 93.3\% & 135262.09 & 40729.88 & 21634.76 & 100.0\% & 169965.59 \\
EXP3Stoch & 28567.18 & 15865.31 & 96.7\% & 172152.03 & 38764.07 & 22370.13 & 100.0\% & 240541.39 \\
EPSDet & 29897.42 & 16089.13 & 90.0\% & 107356.56 & 41821.05 & 21914.62 & 100.0\% & 155723.37 \\
EPSStoch & 30342.74 & 16373.27 & 98.3\% & 178596.87 & 39252.25 & 22656.30 & 100.0\% & 245773.33 \\
UCBDet & 29623.56 & 16071.26 & 86.7\% & 96537.69 & 40142.82 & 21403.33 & 100.0\% & 143880.00 \\
UCBStoch & 28508.76 & 15612.94 & 98.3\% & 172257.47 & 39780.58 & 22490.74 & 100.0\% & 246590.01 \\
Unweighted & 30108.98 & 16357.11 & 86.7\% & 100475.37 & 40958.43 & 21838.57 & 96.7\% & 144939.56 \\
Naive & 34370.12 & 19351.97 & 91.7\% & 159796.19 & 35009.04 & 19271.27 & 93.3\% & 157661.27 \\
InverseWIS & 28375.73 & 15411.20 & 90.0\% & 97292.22 & 36528.07 & 19962.39 & 96.7\% & 145784.06 \\
\bottomrule
\end{tabular}%
}
\caption{Average calibration and forecasting performance for the U.S. in Wave 2 under \fixedcalibration\, and a 5-day forecast horizon, averaged across six evaluation windows. Reported measures are RMSE, WIS, coverage of the 95\% prediction interval, and mean width of the 95\% prediction interval. Rows show the 10 base models and the nine ensemble/comparison methods.}
\label{tab:usa_metrics_fixed_wave2_fcst5}
\end{table}
\paragraph{Ten-Day Forecasting Horizon}

\begin{figure}[H]
\centering
\includegraphics[width=\linewidth]{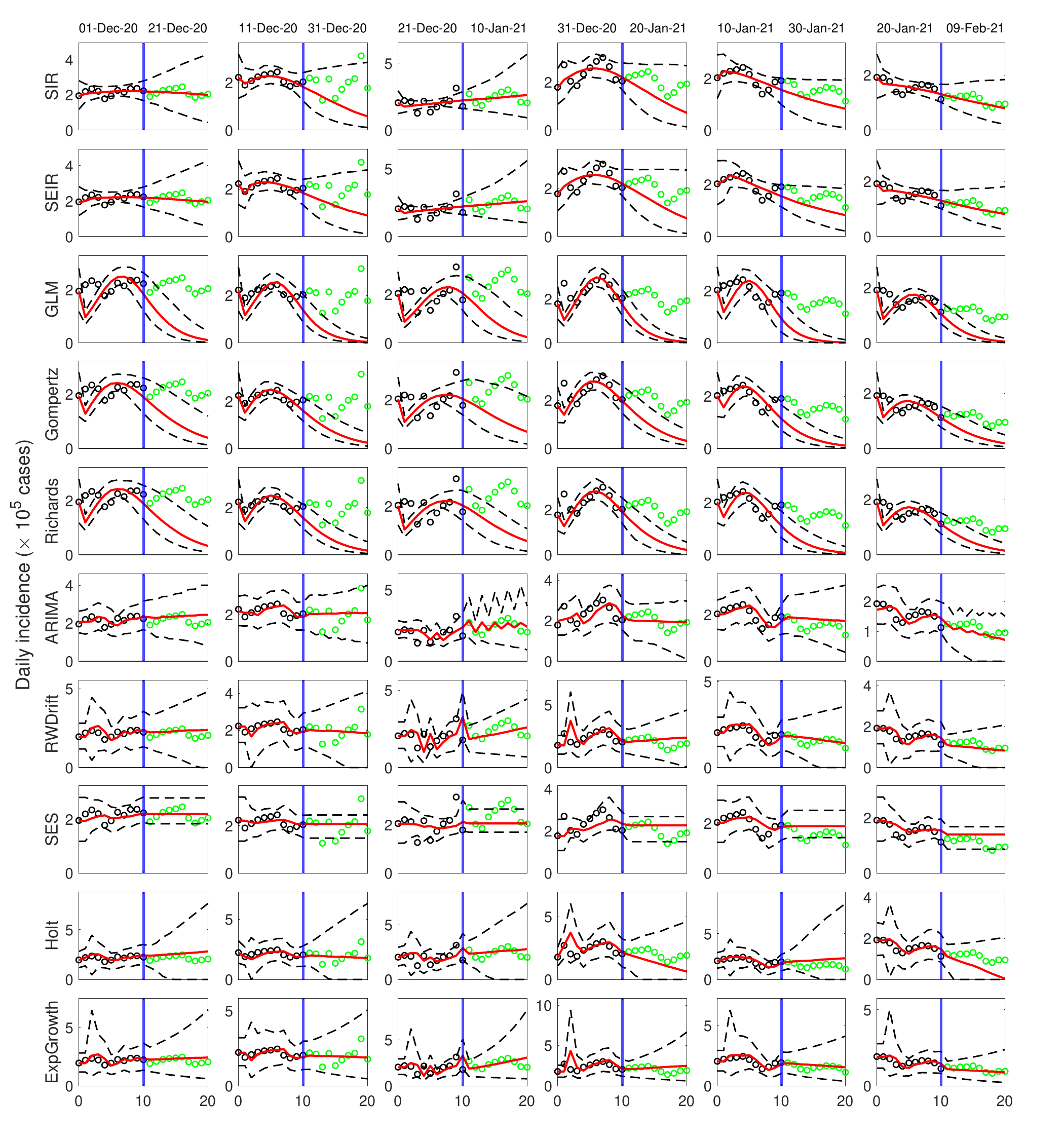}
\caption{\basepanelcaption{2}{\fixedcalibration}{10}}
\label{fig:usa_base_fixed_wave2_fcst10}
\end{figure}
\begin{figure}[H]
\centering
\includegraphics[width=\linewidth]{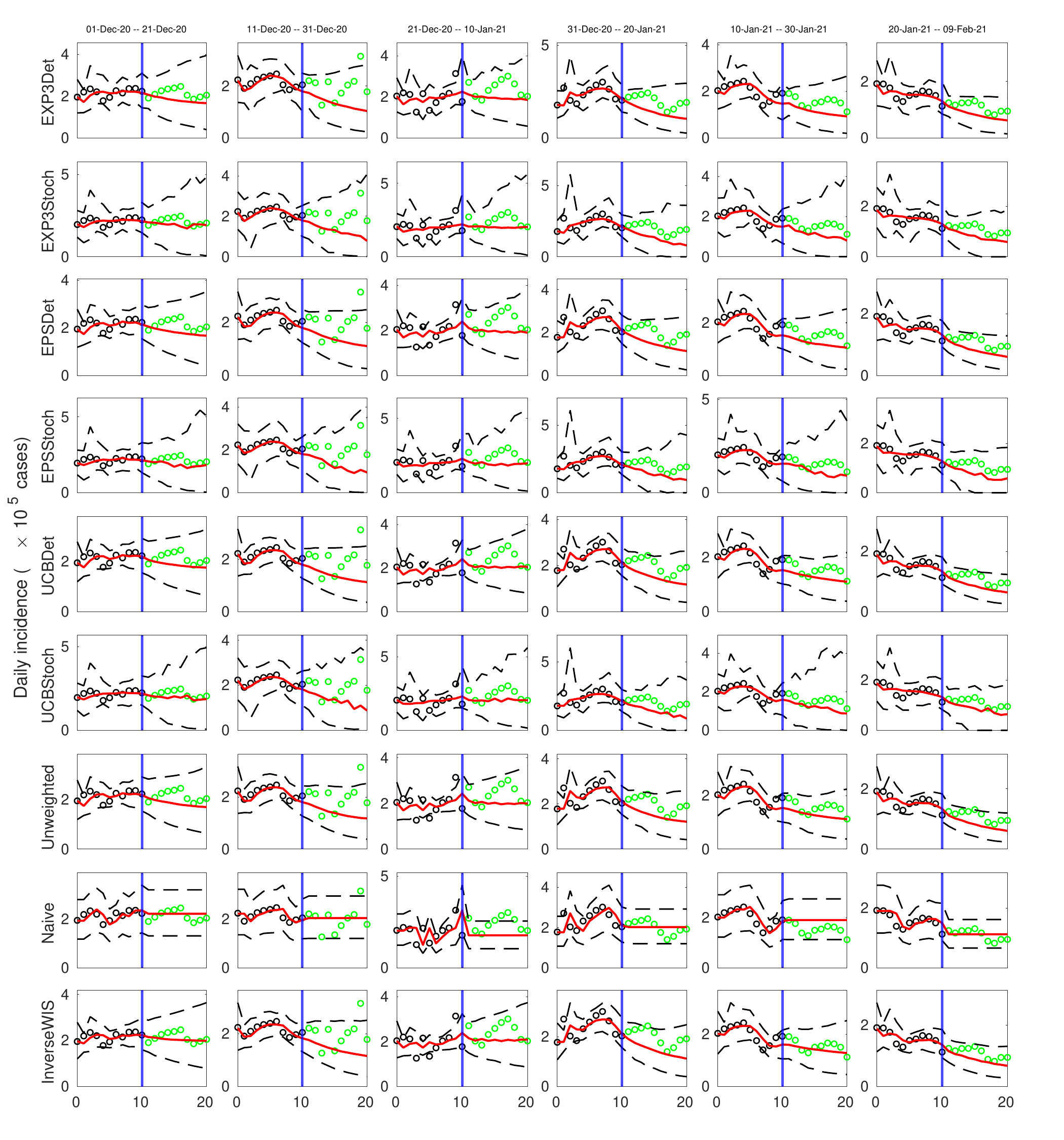}
\caption{\comparisonpanelcaption{2}{\fixedcalibration}{10}}
\label{fig:usa_ensemble_fixed_wave2_fcst10}
\end{figure}
\begin{table}[H]
\centering
\scriptsize
\resizebox{\textwidth}{!}{%
\begin{tabular}{lrrrrrrrr}
\toprule
& \multicolumn{4}{c}{Calibration} & \multicolumn{4}{c}{Forecasting} \\
\cmidrule(lr){2-5} \cmidrule(lr){6-9}
Model & RMSE & WIS & 95\% PI Coverage (\%) & Mean 95\% PI Width & RMSE & WIS & 95\% PI Coverage (\%) & Mean 95\% PI Width \\
\midrule
SIR & 28239.57 & 15433.54 & 81.7\% & 83625.77 & 50569.63 & 26933.35 & 98.3\% & 216121.17 \\
SEIR & 28741.52 & 15645.77 & 80.0\% & 85715.10 & 47561.38 & 25409.55 & 98.3\% & 207068.49 \\
GLM & 50276.13 & 29033.29 & 65.0\% & 82505.81 & 146420.68 & 121008.36 & 8.3\% & 78717.23 \\
Gompertz & 39301.02 & 22121.00 & 70.0\% & 79380.56 & 116677.57 & 85108.47 & 21.7\% & 109643.69 \\
Richards & 41817.84 & 23652.80 & 65.0\% & 79891.84 & 123637.28 & 95378.93 & 13.3\% & 95616.78 \\
ARIMA & 40104.74 & 21564.35 & 88.3\% & 130866.68 & 34895.71 & 19189.04 & 100.0\% & 221792.56 \\
RWDrift & 49765.89 & 25446.06 & 95.0\% & 213988.46 & 36577.82 & 22921.38 & 100.0\% & 285153.78 \\
SES & 36018.75 & 19628.13 & 81.7\% & 113023.80 & 40599.17 & 21929.09 & 81.7\% & 100017.79 \\
Holt & 46094.02 & 23583.71 & 96.7\% & 203352.82 & 58135.30 & 32894.57 & 100.0\% & 423001.76 \\
ExpGrowth & 52582.38 & 26350.43 & 95.0\% & 231467.93 & 38194.07 & 23600.77 & 100.0\% & 321017.82 \\
EXP3Det & 30034.22 & 16146.23 & 93.3\% & 134388.63 & 54677.05 & 27945.50 & 98.3\% & 205304.09 \\
EXP3Stoch & 29484.52 & 16191.52 & 98.3\% & 169564.58 & 56127.50 & 30117.89 & 100.0\% & 300144.94 \\
EPSDet & 29996.29 & 16006.52 & 91.7\% & 110021.35 & 51489.96 & 26276.31 & 98.3\% & 188028.53 \\
EPSStoch & 28796.31 & 16042.62 & 98.3\% & 174809.82 & 57744.67 & 30900.43 & 100.0\% & 306100.46 \\
UCBDet & 29624.80 & 16072.14 & 86.7\% & 96540.46 & 48306.26 & 25107.36 & 98.3\% & 169329.85 \\
UCBStoch & 28283.12 & 15528.44 & 98.3\% & 168011.47 & 51227.83 & 28683.87 & 100.0\% & 304302.27 \\
Unweighted & 30110.50 & 16357.81 & 86.7\% & 100480.64 & 49160.27 & 25363.14 & 96.7\% & 169037.19 \\
Naive & 34370.12 & 19351.97 & 91.7\% & 159796.19 & 40202.25 & 21980.14 & 88.3\% & 157661.27 \\
InverseWIS & 28375.27 & 15411.37 & 90.0\% & 97297.97 & 43819.75 & 23348.19 & 96.7\% & 172028.24 \\
\bottomrule
\end{tabular}%
}
\caption{Average calibration and forecasting performance for the U.S. in Wave 2 under \fixedcalibration\, and a 10-day forecast horizon, averaged across six evaluation windows. Reported measures are RMSE, WIS, coverage of the 95\% prediction interval, and mean width of the 95\% prediction interval. Rows show the 10 base models and the nine ensemble/comparison methods.}
\label{tab:usa_metrics_fixed_wave2_fcst10}
\end{table}
\begin{figure}[H]
\centering
\includegraphics[width=\linewidth]{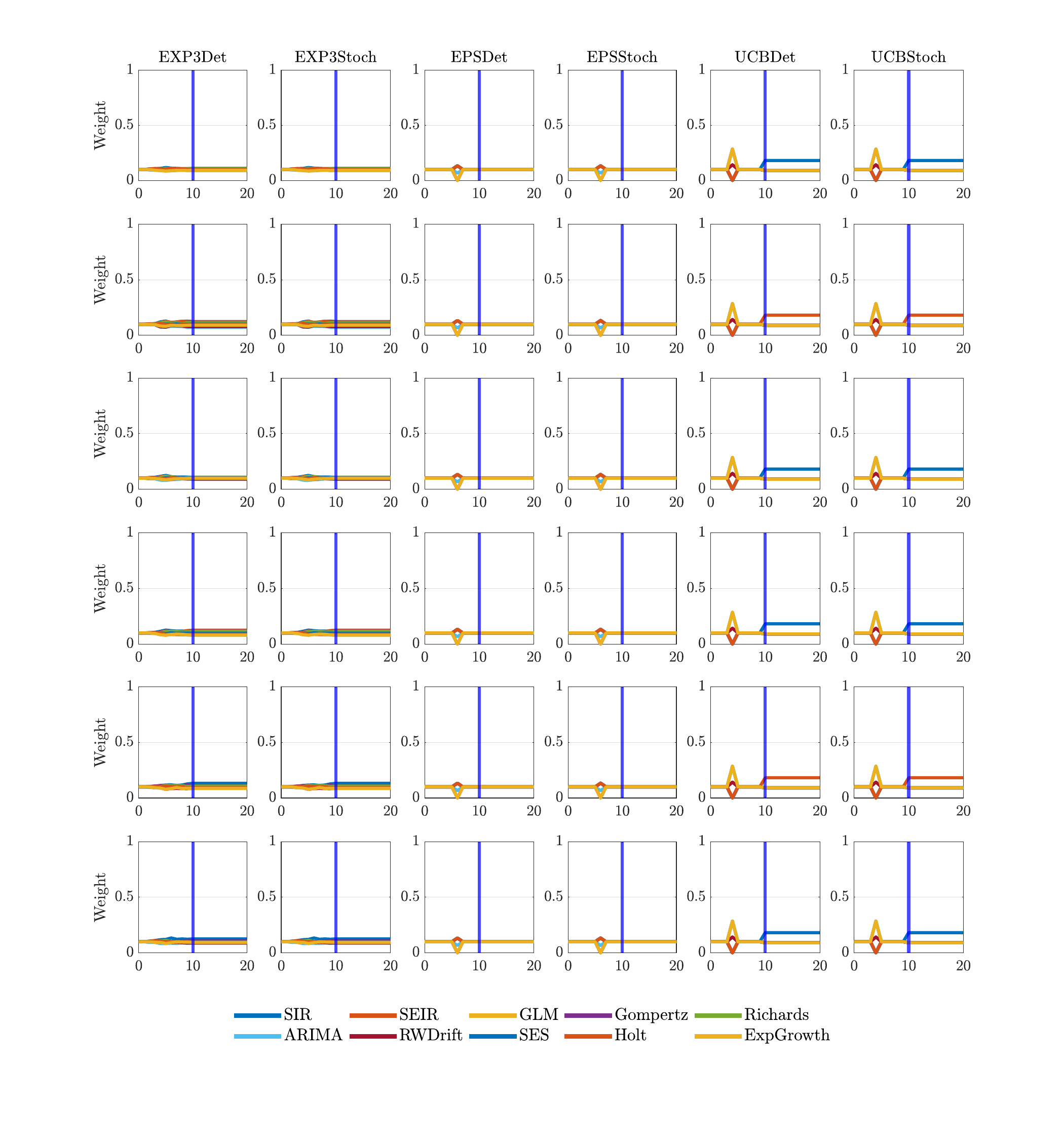}
\caption{\weightpanelcaption{2}{\fixedcalibration}}
\label{fig:usa_weights_fixed_wave2}
\end{figure}
\subsubsection{Growing Calibration Period}
\paragraph{Ten-Day Forecasting Horizon}

\begin{figure}[H]
\centering
\includegraphics[width=\linewidth]{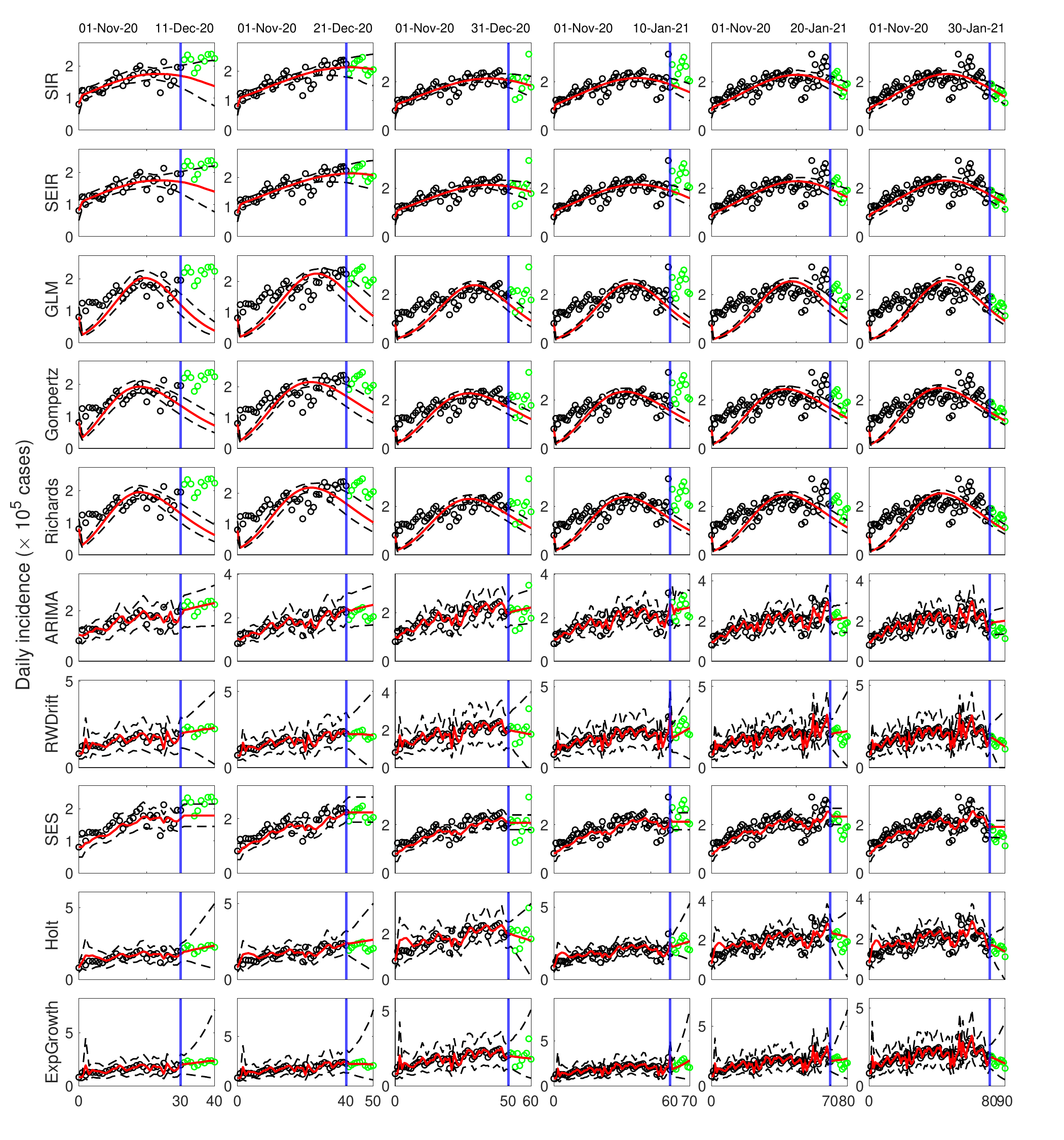}
\caption{\basepanelcaption{2}{\growingcalibration}{10}}
\label{fig:usa_base_growing_wave2_fcst10}
\end{figure}
\begin{figure}[H]
\centering
\includegraphics[width=\linewidth]{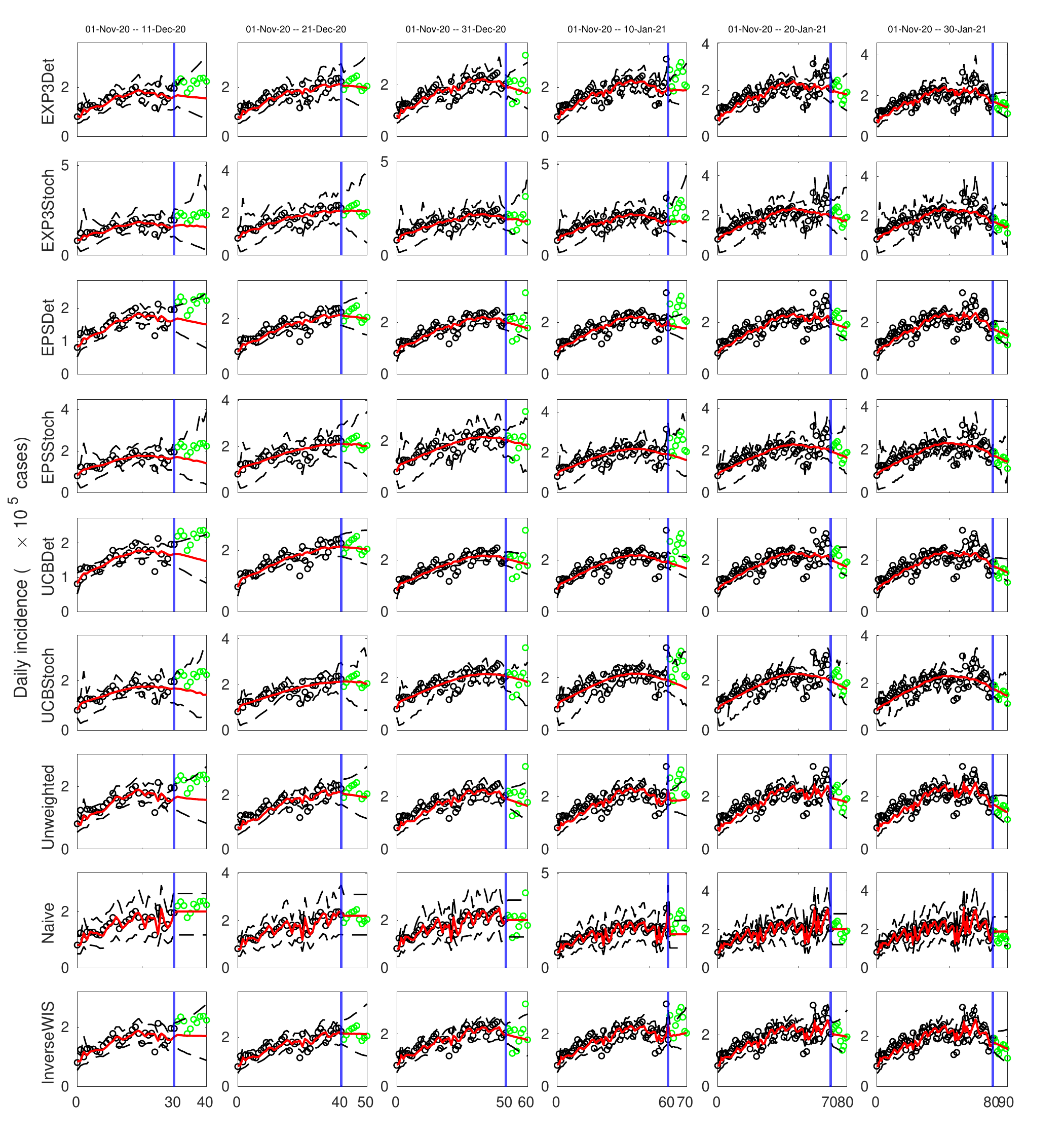}
\caption{\comparisonpanelcaption{2}{\growingcalibration}{10}}
\label{fig:usa_ensemble_growing_wave2_fcst10}
\end{figure}
\begin{table}[H]
\centering
\scriptsize
\resizebox{\textwidth}{!}{%
\begin{tabular}{lrrrrrrrr}
\toprule
& \multicolumn{4}{c}{Calibration} & \multicolumn{4}{c}{Forecasting} \\
\cmidrule(lr){2-5} \cmidrule(lr){6-9}
Model & RMSE & WIS & 95\% PI Coverage (\%) & Mean 95\% PI Width & RMSE & WIS & 95\% PI Coverage (\%) & Mean 95\% PI Width \\
\midrule
SIR & 28652.68 & 17434.72 & 46.6\% & 33578.72 & 46078.91 & 29271.94 & 61.7\% & 75664.01 \\
SEIR & 28551.40 & 17360.49 & 46.2\% & 32717.61 & 45285.84 & 28849.61 & 58.3\% & 70522.34 \\
GLM & 60863.40 & 44489.72 & 27.2\% & 34951.26 & 102377.34 & 83679.57 & 11.7\% & 59962.35 \\
Gompertz & 51293.37 & 36115.81 & 31.6\% & 34326.90 & 82067.83 & 65224.70 & 15.0\% & 54920.86 \\
Richards & 55063.32 & 39265.45 & 29.6\% & 35196.02 & 90230.39 & 73055.75 & 11.7\% & 55189.41 \\
ARIMA & 33148.85 & 16718.51 & 91.7\% & 100585.30 & 38803.15 & 20558.27 & 90.0\% & 143043.84 \\
RWDrift & 37060.57 & 18905.34 & 93.9\% & 159582.67 & 33739.23 & 22285.32 & 98.3\% & 304011.13 \\
SES & 32626.43 & 19179.28 & 55.8\% & 57022.96 & 42538.83 & 25819.55 & 55.0\% & 72143.88 \\
Holt & 35414.33 & 18074.33 & 90.9\% & 113012.00 & 36036.41 & 20036.18 & 100.0\% & 245618.07 \\
ExpGrowth & 38027.56 & 19374.39 & 93.9\% & 166406.47 & 34115.02 & 23440.07 & 100.0\% & 363159.95 \\
EXP3Det & 29343.63 & 15178.01 & 84.2\% & 84325.57 & 42814.35 & 23586.42 & 85.0\% & 135147.89 \\
EXP3Stoch & 28558.31 & 14763.00 & 94.0\% & 128856.79 & 42407.38 & 23745.06 & 100.0\% & 226483.96 \\
EPSDet & 27902.31 & 15454.27 & 63.8\% & 47409.76 & 43529.99 & 25579.19 & 75.0\% & 96653.72 \\
EPSStoch & 27425.85 & 14019.75 & 92.1\% & 112899.27 & 44791.46 & 24356.78 & 96.7\% & 191547.49 \\
UCBDet & 27204.28 & 15237.71 & 59.2\% & 40760.33 & 44344.19 & 27242.02 & 68.3\% & 84598.63 \\
UCBStoch & 27331.75 & 14229.66 & 86.5\% & 87939.32 & 45016.83 & 25498.42 & 91.7\% & 154922.37 \\
Unweighted & 30639.29 & 17120.25 & 67.2\% & 60604.18 & 43829.43 & 24682.99 & 78.3\% & 115169.75 \\
Naive & 33611.87 & 16612.03 & 93.0\% & 138428.31 & 41277.12 & 22697.63 & 86.7\% & 154775.20 \\
InverseWIS & 28205.19 & 14621.19 & 79.2\% & 68077.39 & 38966.69 & 21111.76 & 85.0\% & 127803.56 \\
\bottomrule
\end{tabular}%
}
\caption{Average calibration and forecasting performance for the U.S. in Wave 2 under \growingcalibration\, and a 10-day forecast horizon, averaged across six evaluation windows. Reported measures are RMSE, WIS, coverage of the 95\% prediction interval, and mean width of the 95\% prediction interval. Rows show the 10 base models and the nine ensemble/comparison methods.}
\label{tab:usa_metrics_growing_wave2_fcst10}
\end{table}
\paragraph{Thirty-Day Forecasting Horizon}

\begin{figure}[H]
\centering
\includegraphics[width=\linewidth]{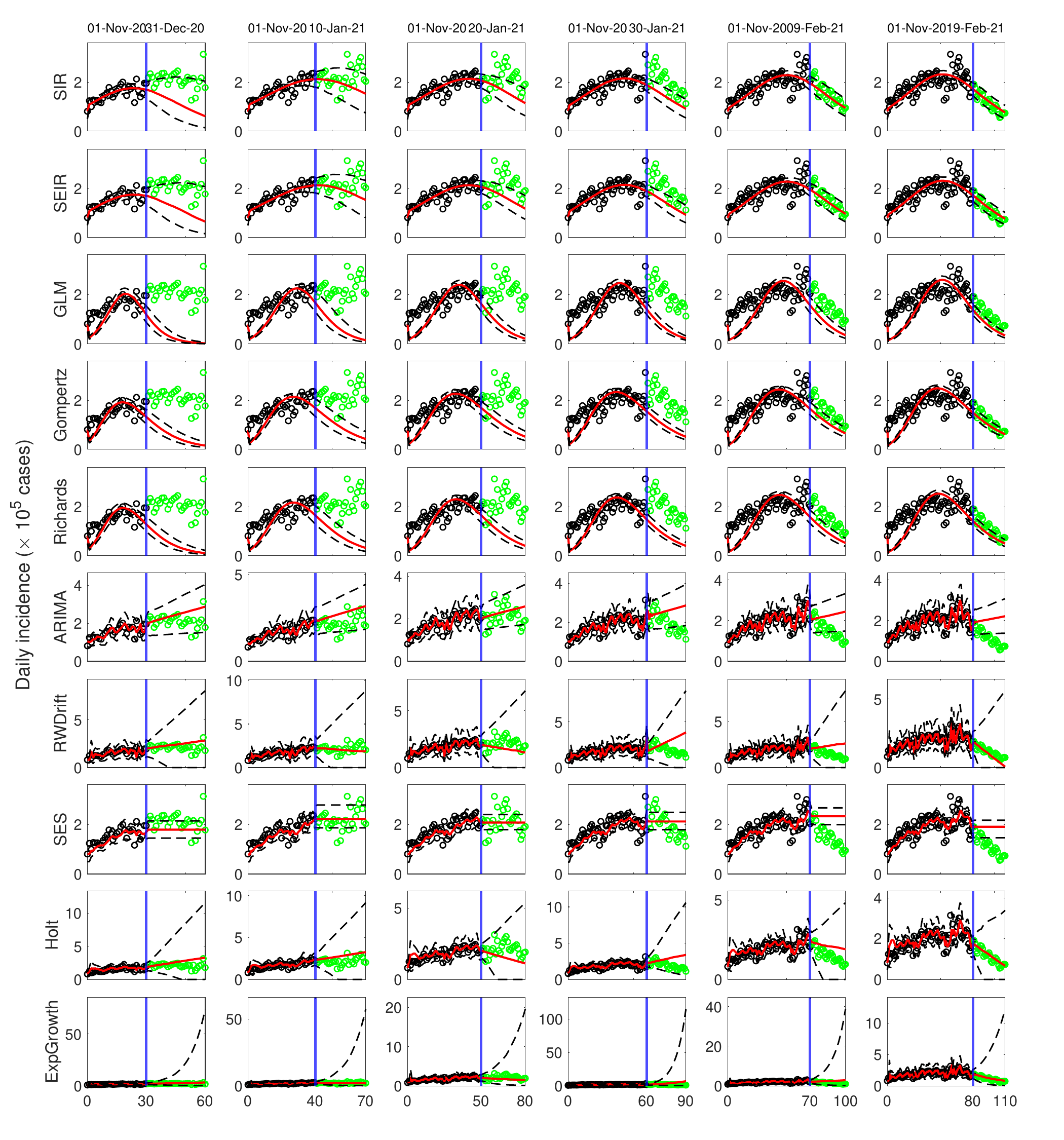}
\caption{\basepanelcaption{2}{\growingcalibration}{30}}
\label{fig:usa_base_growing_wave2_fcst30}
\end{figure}
\begin{figure}[H]
\centering
\includegraphics[width=\linewidth]{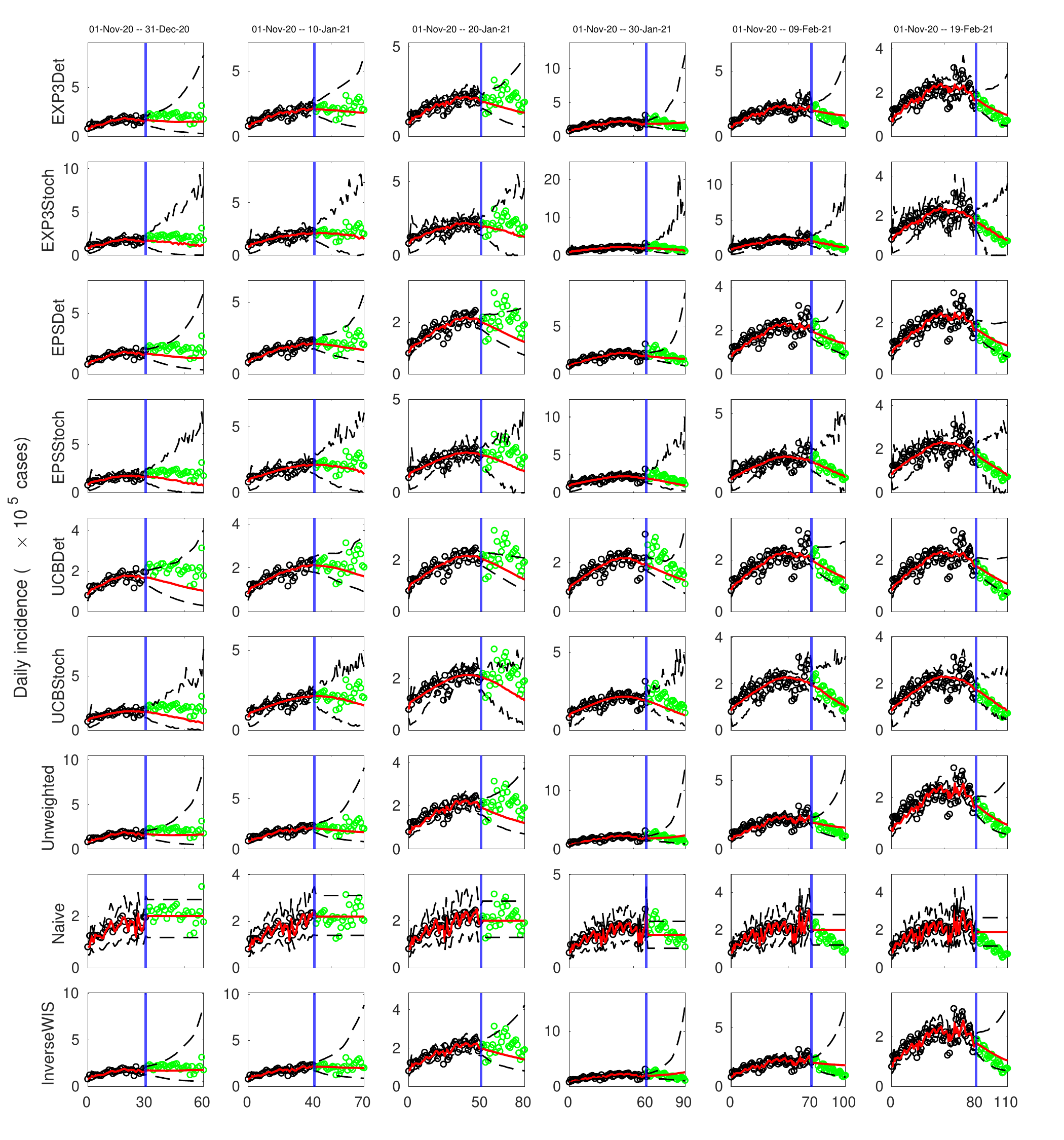}
\caption{\comparisonpanelcaption{2}{\growingcalibration}{30}}
\label{fig:usa_ensemble_growing_wave2_fcst30}
\end{figure}
\begin{table}[H]
\centering
\scriptsize
\resizebox{\textwidth}{!}{%
\begin{tabular}{lrrrrrrrr}
\toprule
& \multicolumn{4}{c}{Calibration} & \multicolumn{4}{c}{Forecasting} \\
\cmidrule(lr){2-5} \cmidrule(lr){6-9}
Model & RMSE & WIS & 95\% PI Coverage (\%) & Mean 95\% PI Width & RMSE & WIS & 95\% PI Coverage (\%) & Mean 95\% PI Width \\
\midrule
SIR & 28652.68 & 17434.72 & 46.6\% & 33578.72 & 57719.60 & 34681.33 & 60.6\% & 95386.72 \\
SEIR & 28551.40 & 17360.49 & 46.2\% & 32717.61 & 56327.80 & 33887.14 & 60.0\% & 89863.58 \\
GLM & 60863.40 & 44489.72 & 27.2\% & 34951.26 & 126075.93 & 109895.72 & 5.0\% & 43950.52 \\
Gompertz & 51293.37 & 36115.81 & 31.6\% & 34326.90 & 101255.37 & 84114.66 & 16.1\% & 47948.30 \\
Richards & 55063.32 & 39265.45 & 29.6\% & 35196.02 & 110386.93 & 93679.25 & 8.9\% & 45356.10 \\
ARIMA & 33148.85 & 16718.51 & 91.7\% & 100585.30 & 76450.39 & 43945.08 & 69.4\% & 170274.90 \\
RWDrift & 37060.57 & 18905.34 & 93.9\% & 159582.67 & 73032.13 & 43432.24 & 99.4\% & 505321.94 \\
SES & 32626.43 & 19179.28 & 55.8\% & 57022.96 & 60231.78 & 40835.40 & 41.1\% & 72143.88 \\
Holt & 35414.33 & 18074.33 & 90.9\% & 113012.00 & 71309.50 & 39753.98 & 100.0\% & 464788.67 \\
ExpGrowth & 38027.56 & 19374.39 & 93.9\% & 166406.47 & 96379.67 & 67994.49 & 100.0\% & 1619670.82 \\
EXP3Det & 29209.70 & 15136.66 & 83.7\% & 83201.56 & 51215.37 & 27429.67 & 95.6\% & 279250.39 \\
EXP3Stoch & 28388.87 & 14752.22 & 94.6\% & 128947.94 & 49504.94 & 30405.08 & 100.0\% & 442433.11 \\
EPSDet & 27984.72 & 15503.01 & 63.7\% & 47636.90 & 52726.30 & 28574.28 & 83.9\% & 185061.84 \\
EPSStoch & 27519.60 & 14046.51 & 92.3\% & 111869.32 & 55765.38 & 30635.94 & 97.8\% & 335432.79 \\
UCBDet & 27202.63 & 15238.82 & 59.1\% & 40757.61 & 54666.91 & 30968.76 & 73.9\% & 130852.60 \\
UCBStoch & 27320.81 & 14106.07 & 85.6\% & 86348.79 & 56860.61 & 31568.97 & 95.6\% & 237636.43 \\
Unweighted & 30645.49 & 17122.16 & 67.2\% & 60605.52 & 53355.29 & 28562.60 & 90.0\% & 266787.35 \\
Naive & 33611.87 & 16612.03 & 93.0\% & 138428.31 & 54348.82 & 29745.11 & 80.6\% & 154775.20 \\
InverseWIS & 28203.59 & 14622.20 & 79.6\% & 68079.27 & 52082.43 & 27815.73 & 92.8\% & 290042.89 \\
\bottomrule
\end{tabular}%
}
\caption{Average calibration and forecasting performance for the U.S. in Wave 2 under \growingcalibration\, and a 30-day forecast horizon, averaged across six evaluation windows. Reported measures are RMSE, WIS, coverage of the 95\% prediction interval, and mean width of the 95\% prediction interval. Rows show the 10 base models and the nine ensemble/comparison methods.}
\label{tab:usa_metrics_growing_wave2_fcst30}
\end{table}
\begin{figure}[H]
\centering
\includegraphics[width=\linewidth]{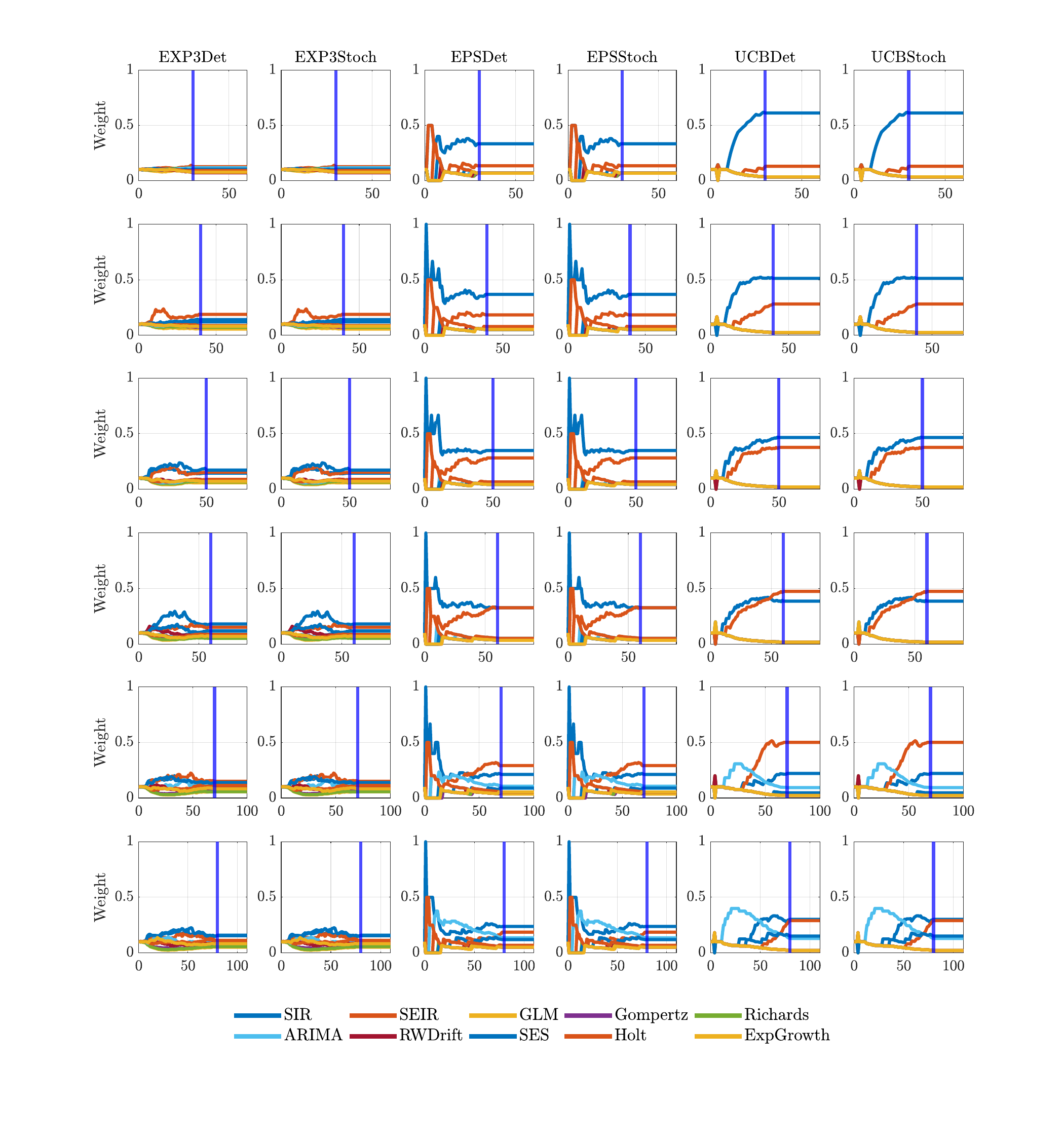}
\caption{\weightpanelcaption{2}{\growingcalibration}}
\label{fig:usa_weights_growing_wave2}
\end{figure}
\subsection{U.S. third wave}
\subsubsection{Fixed Calibration Period}
\paragraph{Five-Day Forecasting Horizon}

\begin{figure}[H]
\centering
\includegraphics[width=\linewidth]{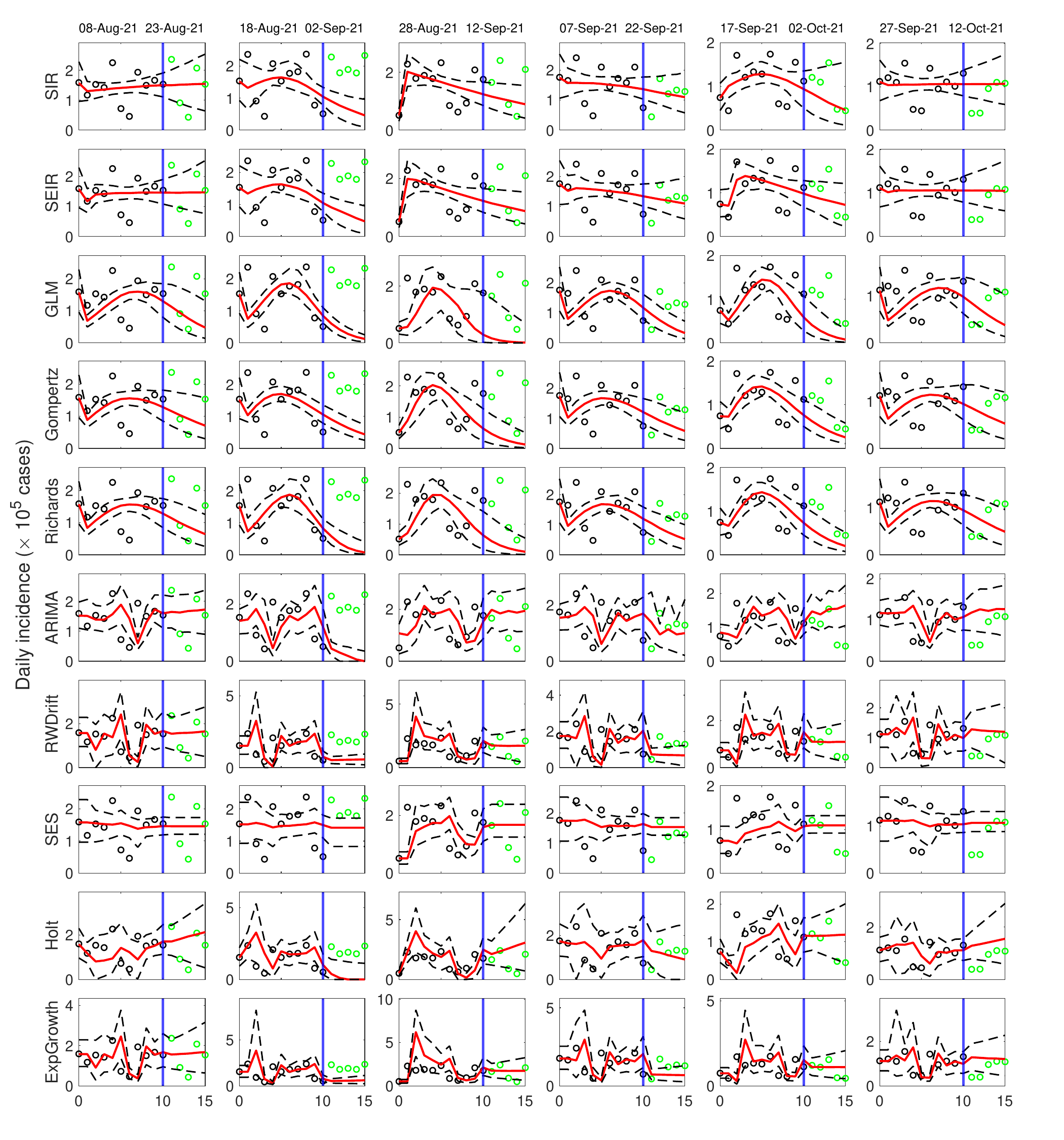}
\caption{\basepanelcaption{3}{\fixedcalibration}{5}}
\label{fig:usa_base_fixed_wave3_fcst5}
\end{figure}
\begin{figure}[H]
\centering
\includegraphics[width=\linewidth]{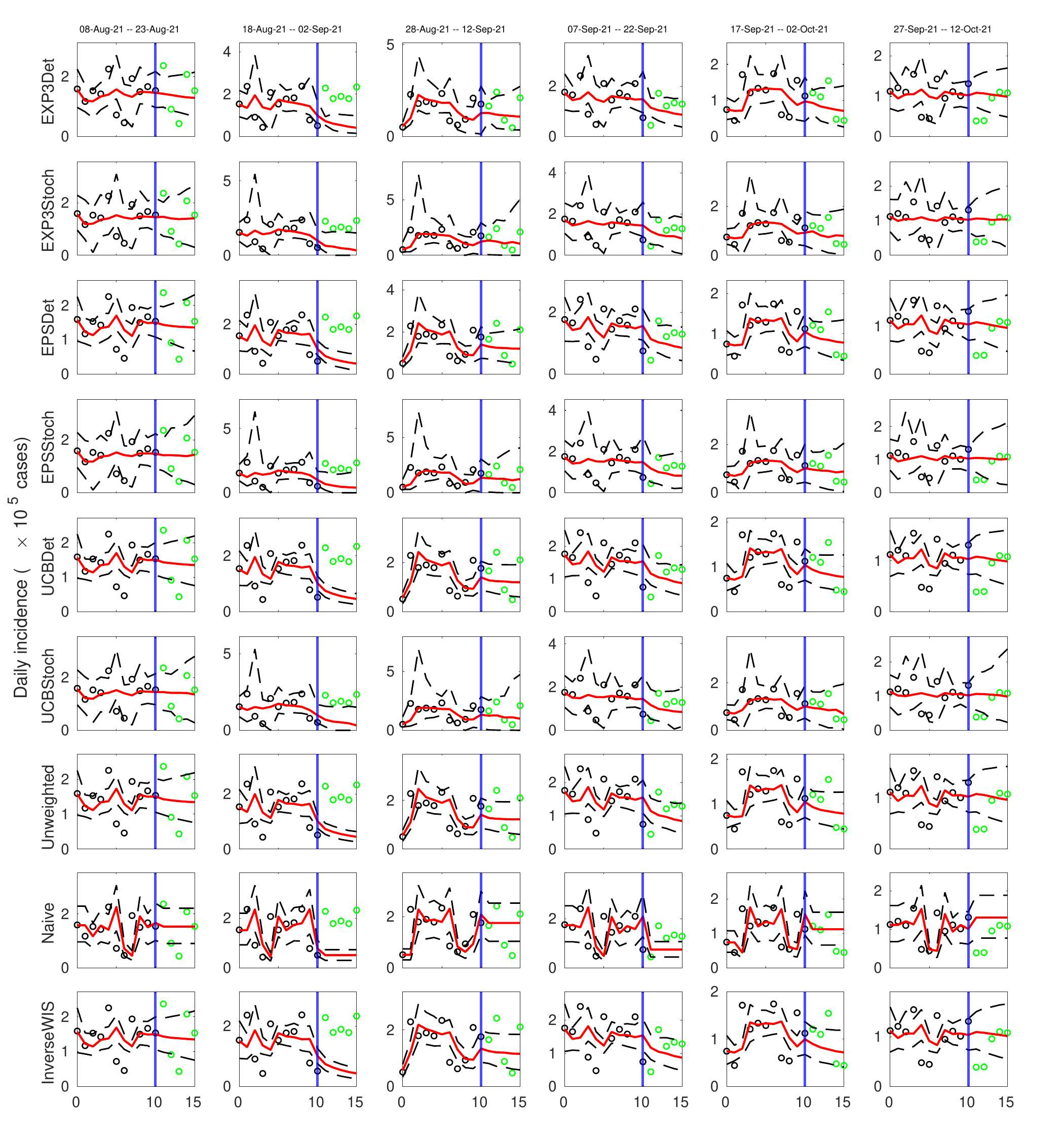}
\caption{\comparisonpanelcaption{3}{\fixedcalibration}{5}}
\label{fig:usa_ensemble_fixed_wave3_fcst5}
\end{figure}
\begin{table}[H]
\centering
\scriptsize
\resizebox{\textwidth}{!}{%
\begin{tabular}{lrrrrrrrr}
\toprule
& \multicolumn{4}{c}{Calibration} & \multicolumn{4}{c}{Forecasting} \\
\cmidrule(lr){2-5} \cmidrule(lr){6-9}
Model & RMSE & WIS & 95\% PI Coverage (\%) & Mean 95\% PI Width & RMSE & WIS & 95\% PI Coverage (\%) & Mean 95\% PI Width \\
\midrule
SIR & 54902.68 & 36082.46 & 45.0\% & 60730.97 & 71912.79 & 49002.52 & 50.0\% & 109407.74 \\
SEIR & 56280.95 & 36770.62 & 45.0\% & 62686.86 & 70465.81 & 48898.64 & 46.7\% & 100463.18 \\
GLM & 65109.09 & 42463.10 & 40.0\% & 71669.26 & 108000.33 & 81264.38 & 23.3\% & 81141.63 \\
Gompertz & 60839.38 & 41038.53 & 40.0\% & 61212.21 & 90927.48 & 67175.06 & 30.0\% & 81654.08 \\
Richards & 61461.65 & 41242.78 & 35.0\% & 64852.22 & 99667.77 & 75073.73 & 20.0\% & 75725.78 \\
ARIMA & 73537.97 & 49736.90 & 43.3\% & 93839.19 & 88281.81 & 63149.75 & 46.7\% & 115081.36 \\
RWDrift & 103285.09 & 65351.60 & 48.3\% & 140251.73 & 77670.77 & 52693.37 & 53.3\% & 141811.42 \\
SES & 63393.99 & 42216.07 & 38.3\% & 80446.33 & 58772.83 & 40138.10 & 36.7\% & 65934.05 \\
Holt & 85705.16 & 50689.26 & 56.7\% & 149606.74 & 97758.27 & 61484.80 & 56.7\% & 203792.30 \\
ExpGrowth & 115180.09 & 71062.84 & 46.7\% & 162559.12 & 77576.55 & 52643.36 & 53.3\% & 141651.92 \\
EXP3Det & 58453.83 & 35392.42 & 51.7\% & 119870.55 & 73025.92 & 49976.98 & 56.7\% & 113050.46 \\
EXP3Stoch & 57050.09 & 32465.13 & 68.3\% & 162899.36 & 74474.69 & 45349.27 & 66.7\% & 191465.31 \\
EPSDet & 60761.30 & 38592.43 & 48.3\% & 85113.30 & 72496.33 & 51365.37 & 53.3\% & 103054.83 \\
EPSStoch & 57222.53 & 32534.55 & 70.0\% & 169640.95 & 73046.83 & 44948.96 & 66.7\% & 192712.45 \\
UCBDet & 60000.25 & 39675.76 & 43.3\% & 69350.75 & 72080.46 & 52974.66 & 46.7\% & 89694.54 \\
UCBStoch & 55882.16 & 31851.29 & 73.3\% & 163229.49 & 73720.59 & 45354.84 & 70.0\% & 192427.95 \\
Unweighted & 61321.87 & 40389.83 & 45.0\% & 71646.13 & 72447.48 & 53340.78 & 50.0\% & 90153.80 \\
Naive & 78332.26 & 46700.05 & 58.3\% & 106984.85 & 78449.61 & 57035.60 & 43.3\% & 98870.32 \\
InverseWIS & 59095.72 & 39027.85 & 45.0\% & 67454.86 & 72782.59 & 53629.72 & 46.7\% & 87718.68 \\
\bottomrule
\end{tabular}%
}
\caption{Average calibration and forecasting performance for the U.S. in Wave 3 under \fixedcalibration\, and a 5-day forecast horizon, averaged across six evaluation windows. Reported measures are RMSE, WIS, coverage of the 95\% prediction interval, and mean width of the 95\% prediction interval. Rows show the 10 base models and the nine ensemble/comparison methods.}
\label{tab:usa_metrics_fixed_wave3_fcst5}
\end{table}
\paragraph{Ten-Day Forecasting Horizon}

\begin{figure}[H]
\centering
\includegraphics[width=\linewidth]{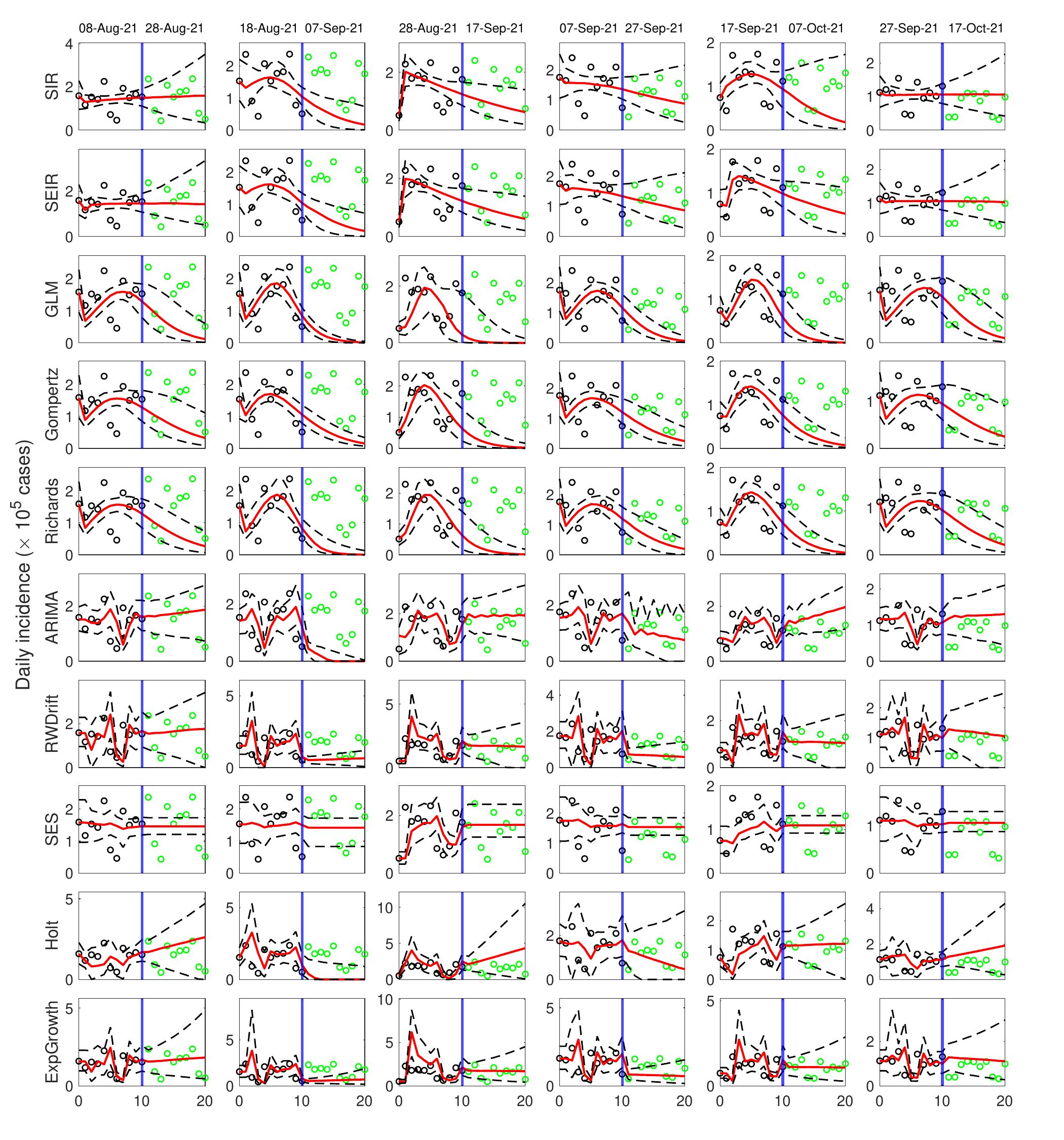}
\caption{\basepanelcaption{3}{\fixedcalibration}{10}}
\label{fig:usa_base_fixed_wave3_fcst10}
\end{figure}
\begin{figure}[H]
\centering
\includegraphics[width=\linewidth]{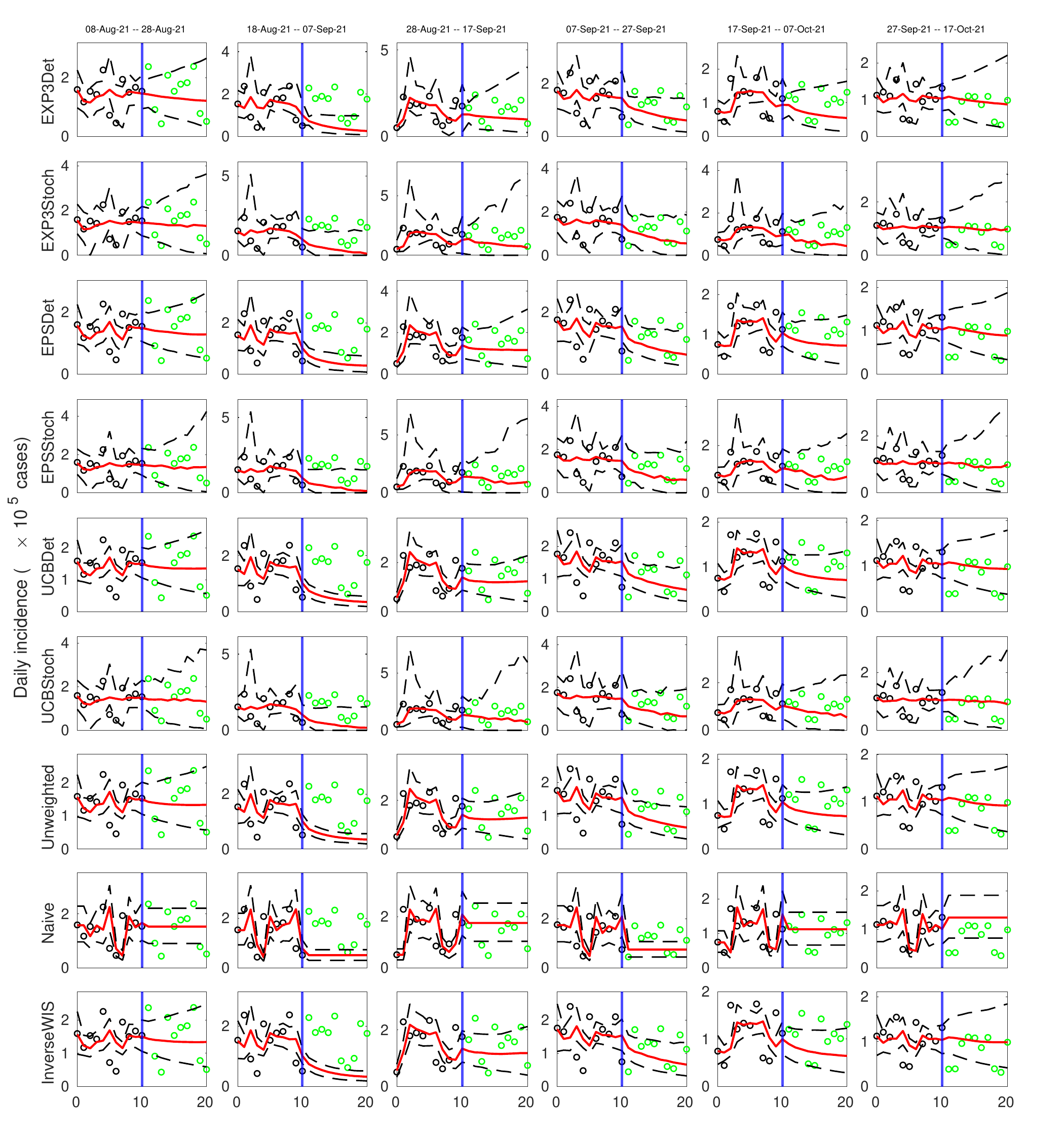}
\caption{\comparisonpanelcaption{3}{\fixedcalibration}{10}}
\label{fig:usa_ensemble_fixed_wave3_fcst10}
\end{figure}
\begin{table}[H]
\centering
\scriptsize
\resizebox{\textwidth}{!}{%
\begin{tabular}{lrrrrrrrr}
\toprule
& \multicolumn{4}{c}{Calibration} & \multicolumn{4}{c}{Forecasting} \\
\cmidrule(lr){2-5} \cmidrule(lr){6-9}
Model & RMSE & WIS & 95\% PI Coverage (\%) & Mean 95\% PI Width & RMSE & WIS & 95\% PI Coverage (\%) & Mean 95\% PI Width \\
\midrule
SIR & 54902.68 & 36082.46 & 45.0\% & 60730.97 & 75404.36 & 47473.53 & 61.7\% & 134357.70 \\
SEIR & 56280.95 & 36770.62 & 45.0\% & 62686.86 & 71404.94 & 46226.52 & 55.0\% & 122205.09 \\
GLM & 65109.09 & 42463.10 & 40.0\% & 71669.26 & 114573.63 & 90554.66 & 18.3\% & 60445.26 \\
Gompertz & 60839.38 & 41038.53 & 40.0\% & 61212.21 & 99230.51 & 74563.55 & 28.3\% & 71965.83 \\
Richards & 61461.65 & 41242.78 & 35.0\% & 64852.22 & 106612.46 & 82409.15 & 21.7\% & 61888.84 \\
ARIMA & 73537.97 & 49736.90 & 43.3\% & 93839.19 & 83692.62 & 56578.43 & 53.3\% & 130429.02 \\
RWDrift & 103285.09 & 65351.60 & 48.3\% & 140251.73 & 68986.62 & 42622.83 & 70.0\% & 179940.21 \\
SES & 63393.99 & 42216.07 & 38.3\% & 80446.33 & 55571.66 & 37231.10 & 41.7\% & 65934.05 \\
Holt & 85705.16 & 50689.26 & 56.7\% & 149606.74 & 110161.73 & 64054.18 & 75.0\% & 284636.14 \\
ExpGrowth & 115180.09 & 71062.84 & 46.7\% & 162559.12 & 69166.43 & 41983.11 & 73.3\% & 189883.16 \\
EXP3Det & 58024.30 & 35205.78 & 55.0\% & 121147.40 & 71139.83 & 44371.03 & 71.7\% & 147173.99 \\
EXP3Stoch & 56960.42 & 32370.07 & 68.3\% & 161559.63 & 76006.41 & 42202.69 & 78.3\% & 235740.77 \\
EPSDet & 60409.95 & 38331.12 & 48.3\% & 87692.72 & 68700.58 & 44068.53 & 63.3\% & 122687.60 \\
EPSStoch & 57303.87 & 32349.57 & 71.7\% & 168377.75 & 73390.09 & 41786.63 & 80.0\% & 238665.94 \\
UCBDet & 59999.18 & 39675.73 & 43.3\% & 69352.06 & 67992.11 & 45827.55 & 55.0\% & 103040.73 \\
UCBStoch & 56199.15 & 31927.75 & 70.0\% & 162295.34 & 72215.55 & 40977.44 & 78.3\% & 235641.63 \\
Unweighted & 61322.02 & 40389.62 & 45.0\% & 71648.42 & 67890.54 & 45915.43 & 55.0\% & 103345.60 \\
Naive & 78332.26 & 46700.05 & 58.3\% & 106984.85 & 70433.25 & 47382.93 & 50.0\% & 98870.32 \\
InverseWIS & 59094.37 & 39027.84 & 45.0\% & 67459.13 & 69325.50 & 47176.64 & 50.0\% & 99316.06 \\
\bottomrule
\end{tabular}%
}
\caption{Average calibration and forecasting performance for the U.S. in Wave 3 under \fixedcalibration\, and a 10-day forecast horizon, averaged across six evaluation windows. Reported measures are RMSE, WIS, coverage of the 95\% prediction interval, and mean width of the 95\% prediction interval. Rows show the 10 base models and the nine ensemble/comparison methods.}
\label{tab:usa_metrics_fixed_wave3_fcst10}
\end{table}
\begin{figure}[H]
\centering
\includegraphics[width=\linewidth]{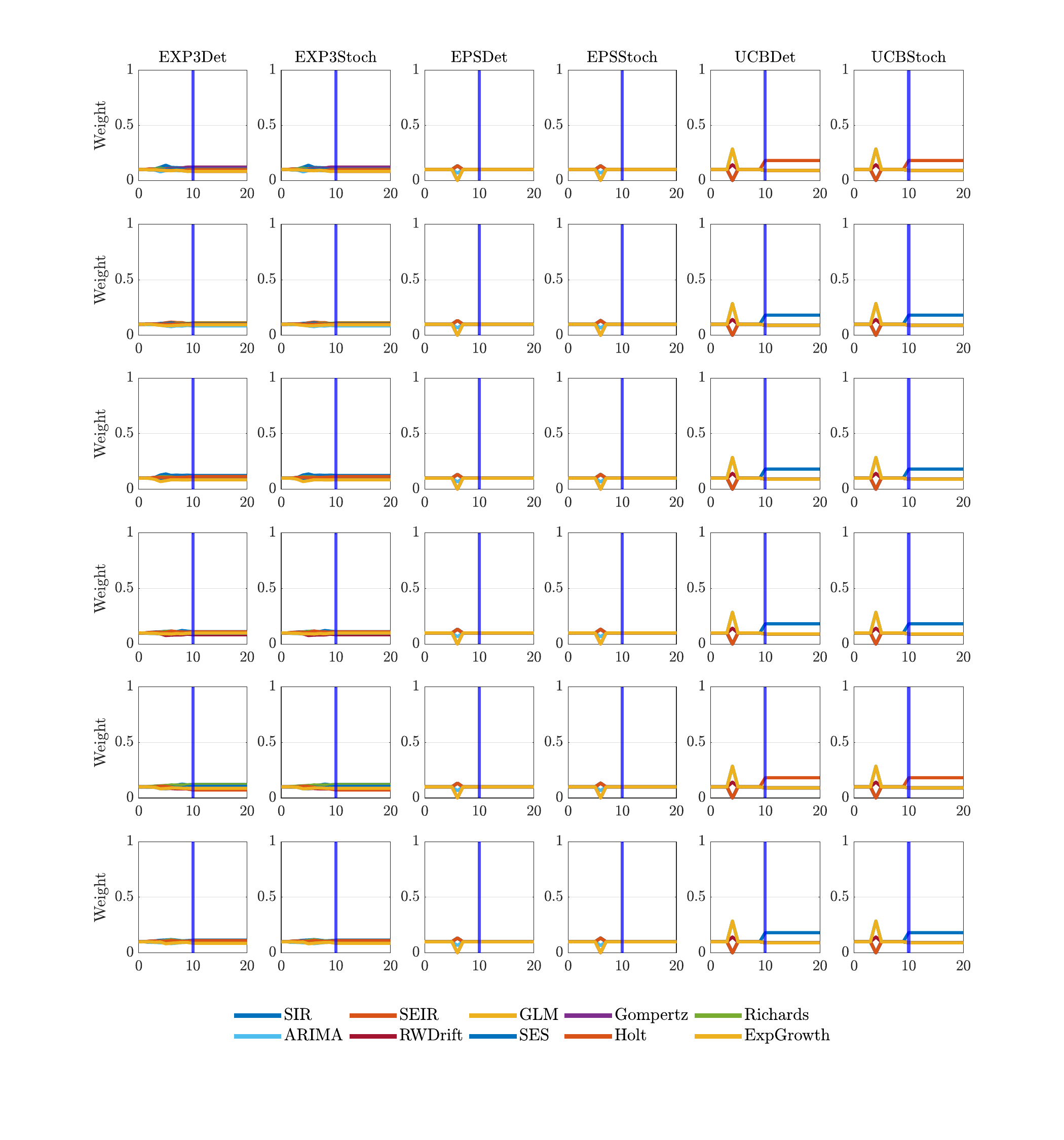}
\caption{\weightpanelcaption{3}{\fixedcalibration}}
\label{fig:usa_weights_fixed_wave3}
\end{figure}
\subsubsection{Growing Calibration Period}
\paragraph{Ten-Day Forecasting Horizon}

\begin{figure}[H]
\centering
\includegraphics[width=\linewidth]{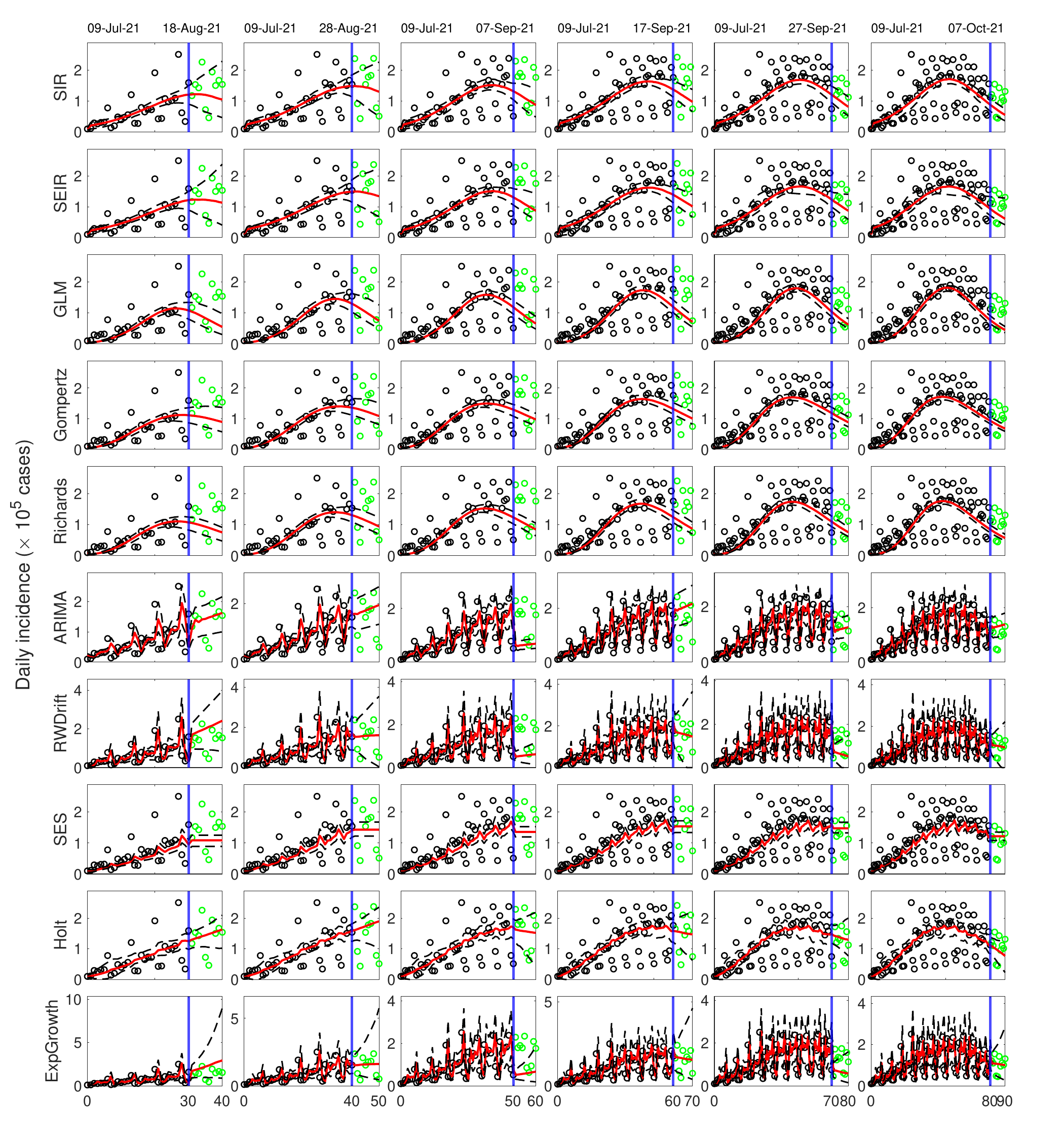}
\caption{\basepanelcaption{3}{\growingcalibration}{10}}
\label{fig:usa_base_growing_wave3_fcst10}
\end{figure}
\begin{figure}[H]
\centering
\includegraphics[width=\linewidth]{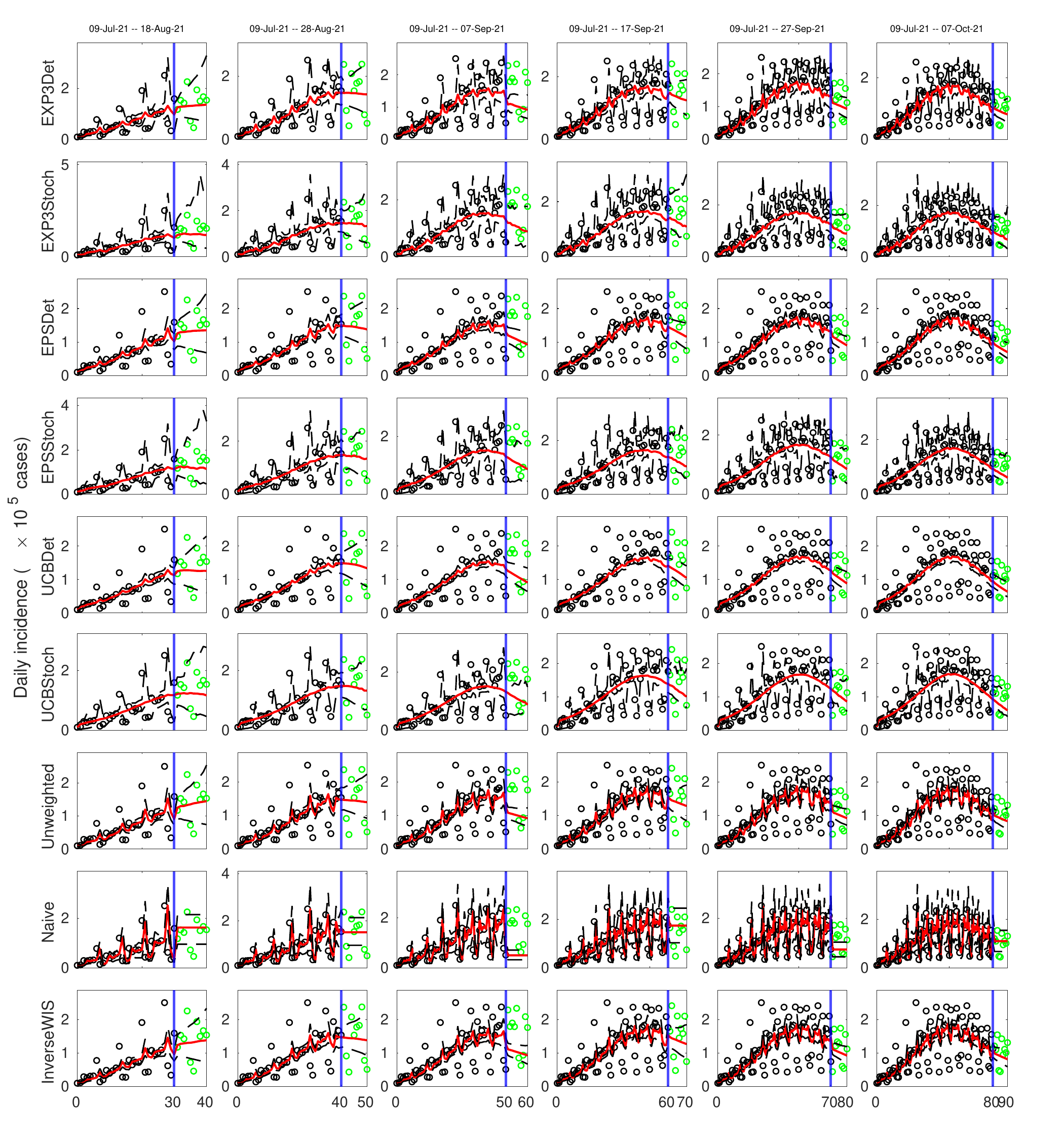}
\caption{\comparisonpanelcaption{3}{\growingcalibration}{10}}
\label{fig:usa_ensemble_growing_wave3_fcst10}
\end{figure}
\begin{table}[H]
\centering
\scriptsize
\resizebox{\textwidth}{!}{%
\begin{tabular}{lrrrrrrrr}
\toprule
& \multicolumn{4}{c}{Calibration} & \multicolumn{4}{c}{Forecasting} \\
\cmidrule(lr){2-5} \cmidrule(lr){6-9}
Model & RMSE & WIS & 95\% PI Coverage (\%) & Mean 95\% PI Width & RMSE & WIS & 95\% PI Coverage (\%) & Mean 95\% PI Width \\
\midrule
SIR & 50307.28 & 31272.04 & 40.5\% & 29713.58 & 62674.45 & 42038.48 & 36.7\% & 83183.54 \\
SEIR & 50104.63 & 31070.15 & 42.8\% & 31412.96 & 61274.90 & 39843.81 & 41.7\% & 89067.31 \\
GLM & 54164.28 & 37318.93 & 21.5\% & 20092.35 & 76968.44 & 57664.77 & 16.7\% & 46761.11 \\
Gompertz & 52097.41 & 35594.52 & 20.8\% & 19793.08 & 63147.88 & 48202.18 & 6.7\% & 44858.35 \\
Richards & 53146.76 & 36656.84 & 21.9\% & 19963.28 & 68931.27 & 53953.22 & 10.0\% & 38849.05 \\
ARIMA & 68226.34 & 41876.20 & 36.7\% & 57666.87 & 69708.02 & 45434.08 & 55.0\% & 88848.05 \\
RWDrift & 82882.66 & 50009.41 & 37.6\% & 85624.01 & 73789.58 & 45372.70 & 71.7\% & 183645.76 \\
SES & 57193.69 & 39946.40 & 3.8\% & 24024.57 & 58235.35 & 43856.96 & 18.3\% & 35643.41 \\
Holt & 51838.26 & 30655.96 & 51.1\% & 43372.16 & 55839.18 & 33505.99 & 56.7\% & 95575.40 \\
ExpGrowth & 83536.29 & 50251.49 & 40.1\% & 90388.07 & 77420.26 & 46575.28 & 70.0\% & 231372.36 \\
EXP3Det & 52907.71 & 32203.83 & 44.7\% & 57784.88 & 58923.95 & 39137.11 & 41.7\% & 88279.37 \\
EXP3Stoch & 51320.45 & 29557.26 & 60.0\% & 86501.30 & 60489.55 & 34455.75 & 68.3\% & 157825.99 \\
EPSDet & 52013.58 & 33501.97 & 36.4\% & 27097.88 & 59306.36 & 41258.87 & 31.7\% & 71774.07 \\
EPSStoch & 50397.73 & 29238.17 & 57.9\% & 77287.07 & 61213.42 & 36560.92 & 63.3\% & 140491.73 \\
UCBDet & 50746.20 & 32744.53 & 34.5\% & 23959.59 & 60246.41 & 41831.22 & 31.7\% & 72482.40 \\
UCBStoch & 50113.05 & 30269.51 & 51.0\% & 57238.07 & 61162.97 & 38890.99 & 50.0\% & 108408.80 \\
Unweighted & 54946.33 & 35676.45 & 34.3\% & 31255.22 & 58857.07 & 40686.53 & 36.7\% & 71710.54 \\
Naive & 74444.85 & 41523.21 & 53.2\% & 76095.47 & 68988.08 & 46115.29 & 50.0\% & 94300.35 \\
InverseWIS & 54202.47 & 35329.62 & 33.8\% & 29387.14 & 58725.38 & 40986.91 & 36.7\% & 67983.63 \\
\bottomrule
\end{tabular}%
}
\caption{Average calibration and forecasting performance for the U.S. in Wave 3 under \growingcalibration\, and a 10-day forecast horizon, averaged across six evaluation windows. Reported measures are RMSE, WIS, coverage of the 95\% prediction interval, and mean width of the 95\% prediction interval. Rows show the 10 base models and the nine ensemble/comparison methods.}
\label{tab:usa_metrics_growing_wave3_fcst10}
\end{table}
\paragraph{Thirty-Day Forecasting Horizon}

\begin{figure}[H]
\centering
\includegraphics[width=\linewidth]{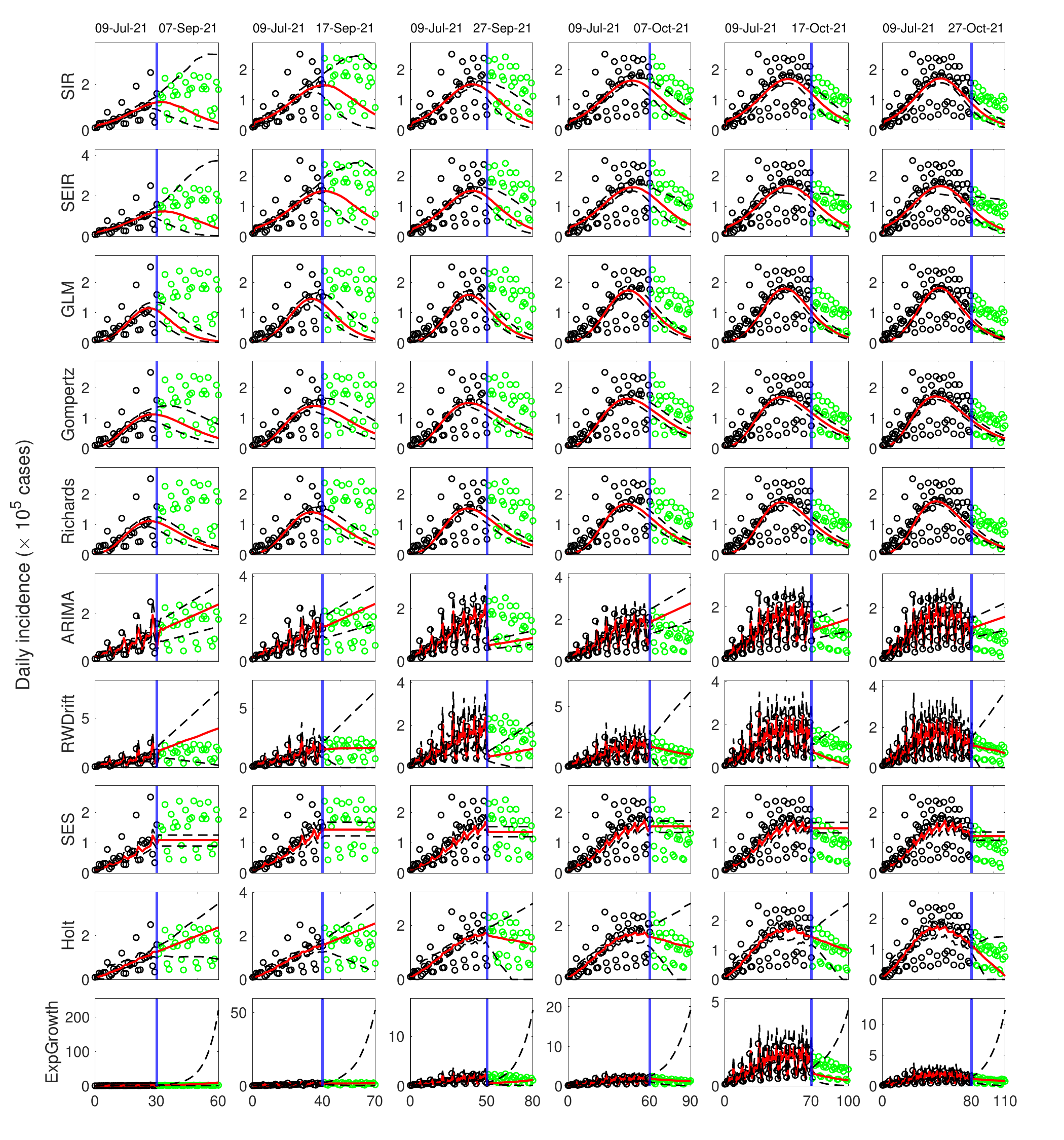}
\caption{\basepanelcaption{3}{\growingcalibration}{30}}
\label{fig:usa_base_growing_wave3_fcst30}
\end{figure}
\begin{figure}[H]
\centering
\includegraphics[width=\linewidth]{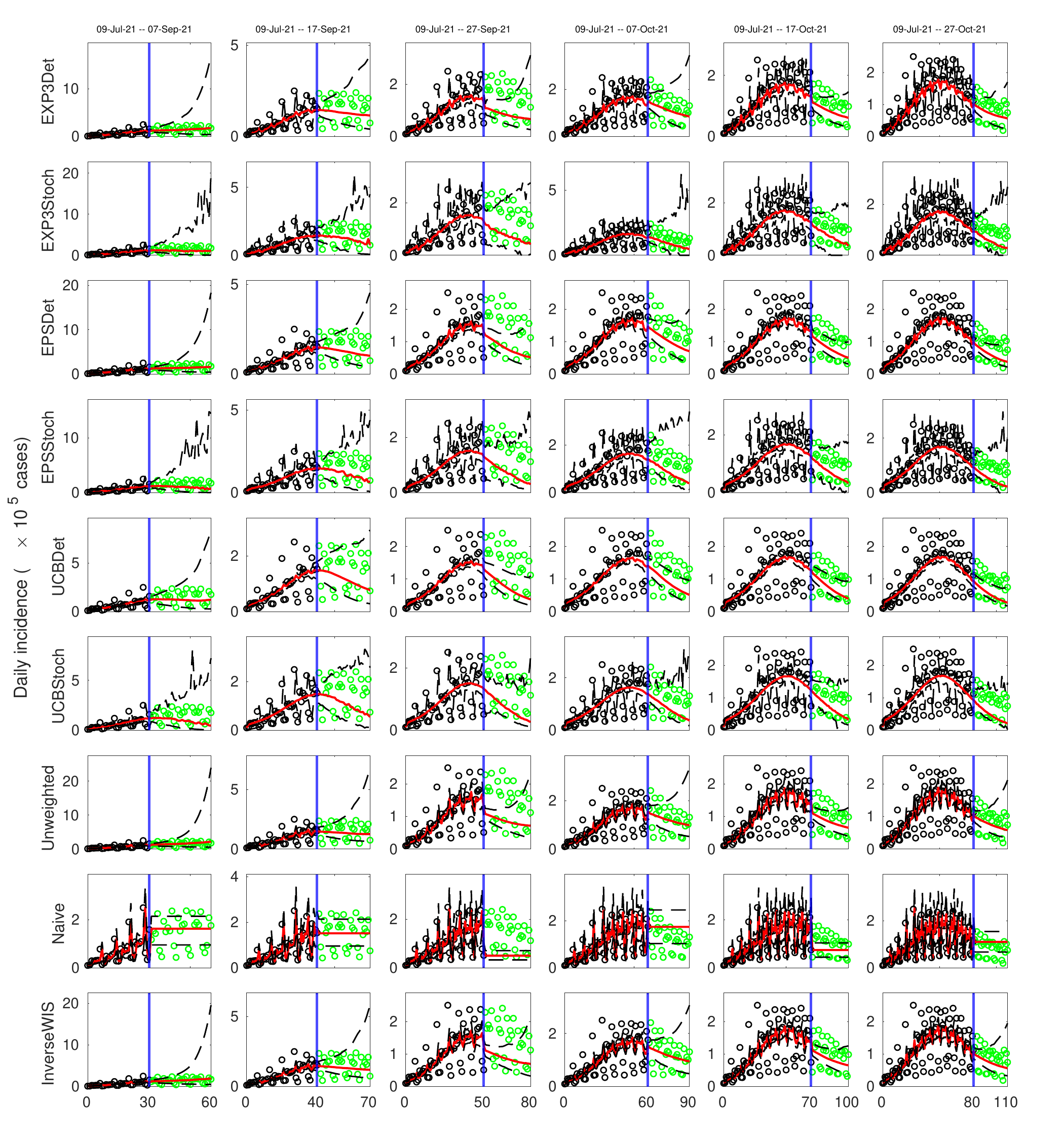}
\caption{\comparisonpanelcaption{3}{\growingcalibration}{30}}
\label{fig:usa_ensemble_growing_wave3_fcst30}
\end{figure}
\begin{table}[H]
\centering
\scriptsize
\resizebox{\textwidth}{!}{%
\begin{tabular}{lrrrrrrrr}
\toprule
& \multicolumn{4}{c}{Calibration} & \multicolumn{4}{c}{Forecasting} \\
\cmidrule(lr){2-5} \cmidrule(lr){6-9}
Model & RMSE & WIS & 95\% PI Coverage (\%) & Mean 95\% PI Width & RMSE & WIS & 95\% PI Coverage (\%) & Mean 95\% PI Width \\
\midrule
SIR & 50307.28 & 31272.04 & 40.5\% & 29713.58 & 74798.29 & 46167.11 & 46.7\% & 108750.30 \\
SEIR & 50104.63 & 31070.15 & 42.8\% & 31412.96 & 71564.67 & 41886.85 & 65.0\% & 128988.40 \\
GLM & 54164.28 & 37318.93 & 21.5\% & 20092.35 & 94158.74 & 74337.77 & 14.4\% & 35469.71 \\
Gompertz & 52097.41 & 35594.52 & 20.8\% & 19793.08 & 71725.16 & 53783.56 & 11.7\% & 44384.86 \\
Richards & 53146.76 & 36656.84 & 21.9\% & 19963.28 & 80971.13 & 63856.15 & 12.2\% & 32444.45 \\
ARIMA & 68226.34 & 41876.20 & 36.7\% & 57666.87 & 87137.73 & 57510.61 & 41.7\% & 105559.40 \\
RWDrift & 82882.66 & 50009.41 & 37.6\% & 85624.01 & 79171.41 & 46663.84 & 86.1\% & 305852.48 \\
SES & 57193.69 & 39946.40 & 3.8\% & 24024.57 & 61823.11 & 46453.05 & 16.7\% & 35643.41 \\
Holt & 51838.26 & 30655.96 & 51.1\% & 43372.16 & 60162.88 & 32191.86 & 77.8\% & 169014.06 \\
ExpGrowth & 83536.29 & 50251.49 & 40.1\% & 90388.07 & 116324.82 & 72503.74 & 88.3\% & 1387838.66 \\
EXP3Det & 52718.08 & 32234.65 & 41.8\% & 56105.35 & 56255.95 & 33834.13 & 67.8\% & 195962.72 \\
EXP3Stoch & 51235.25 & 29499.11 & 59.1\% & 85718.64 & 64307.99 & 34161.56 & 87.2\% & 293663.50 \\
EPSDet & 51997.38 & 33532.93 & 37.0\% & 26940.92 & 59139.45 & 37442.03 & 51.7\% & 163500.07 \\
EPSStoch & 50351.06 & 29252.03 & 58.0\% & 76898.55 & 67590.84 & 36685.88 & 85.0\% & 252899.21 \\
UCBDet & 50746.07 & 32744.18 & 34.5\% & 23959.21 & 64272.30 & 40802.20 & 45.0\% & 120168.44 \\
UCBStoch & 50155.51 & 30380.71 & 52.1\% & 56274.97 & 70724.25 & 40168.79 & 75.6\% & 171389.55 \\
Unweighted & 54946.90 & 35676.43 & 34.3\% & 31255.08 & 56048.88 & 35186.45 & 59.4\% & 200604.94 \\
Naive & 74444.85 & 41523.21 & 53.2\% & 76095.47 & 66612.31 & 43217.00 & 48.3\% & 94300.35 \\
InverseWIS & 54201.30 & 35329.37 & 33.8\% & 29388.68 & 55686.34 & 35092.89 & 58.3\% & 176375.56 \\
\bottomrule
\end{tabular}%
}
\caption{Average calibration and forecasting performance for the U.S. in Wave 3 under \growingcalibration\, and a 30-day forecast horizon, averaged across six evaluation windows. Reported measures are RMSE, WIS, coverage of the 95\% prediction interval, and mean width of the 95\% prediction interval. Rows show the 10 base models and the nine ensemble/comparison methods.}
\label{tab:usa_metrics_growing_wave3_fcst30}
\end{table}
\begin{figure}[H]
\centering
\includegraphics[width=\linewidth]{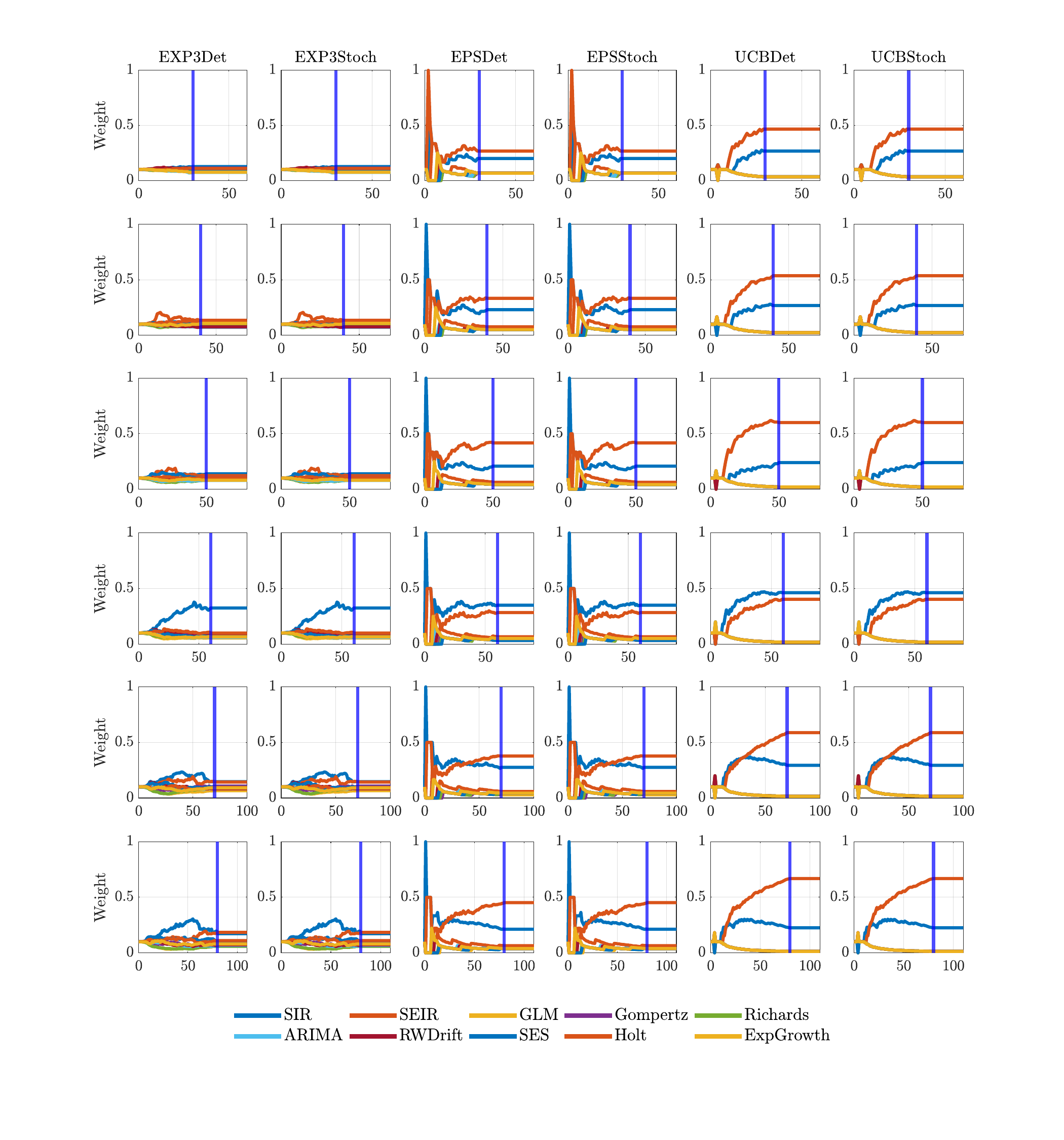}
\caption{\weightpanelcaption{3}{\growingcalibration}}
\label{fig:usa_weights_growing_wave3}
\end{figure}

\section{Supplementary Alabama analysis}

This section contains the Alabama case-study figures and tables that were used in the previous main manuscript or supplementary materials. They are retained here as supplementary material and are distinct from the U.S. national primary analysis.

\subsection{Alabama main-manuscript summary figures and sensitivity analysis}
\begin{figure}[H]
\centering
\includegraphics[width=\linewidth]{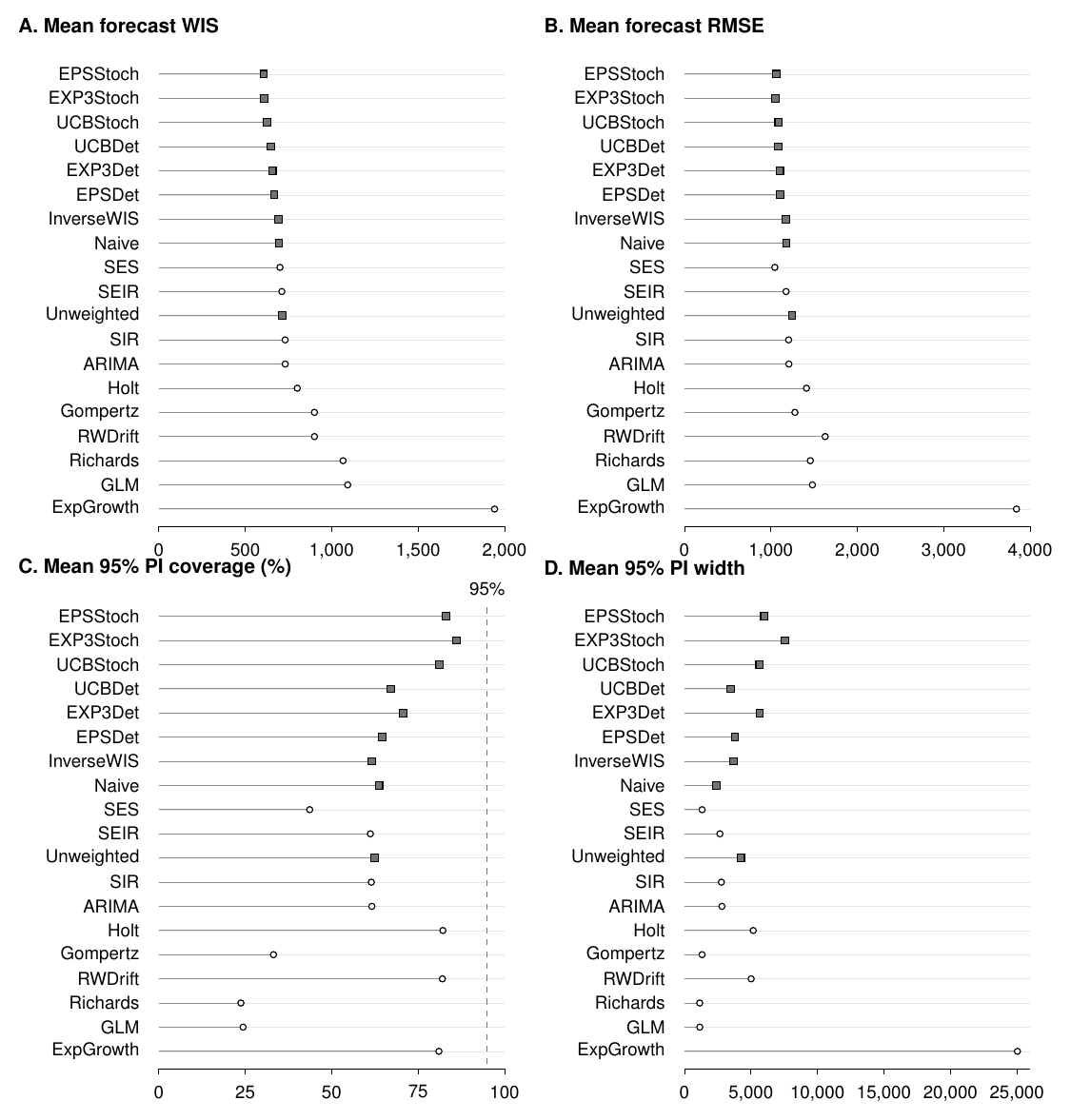}
\caption{Alabama case-study overall forecast performance across the 12 forecast configurations. Points show mean forecast-period RMSE, WIS, empirical 95\% prediction-interval coverage, and mean 95\% prediction-interval width for the 10 base models and nine ensemble/comparison methods. Methods are ordered by mean WIS.}
\label{fig:alabama_forecast_performance_summary}
\end{figure}
\begin{figure}[H]
\centering
\includegraphics[width=\linewidth]{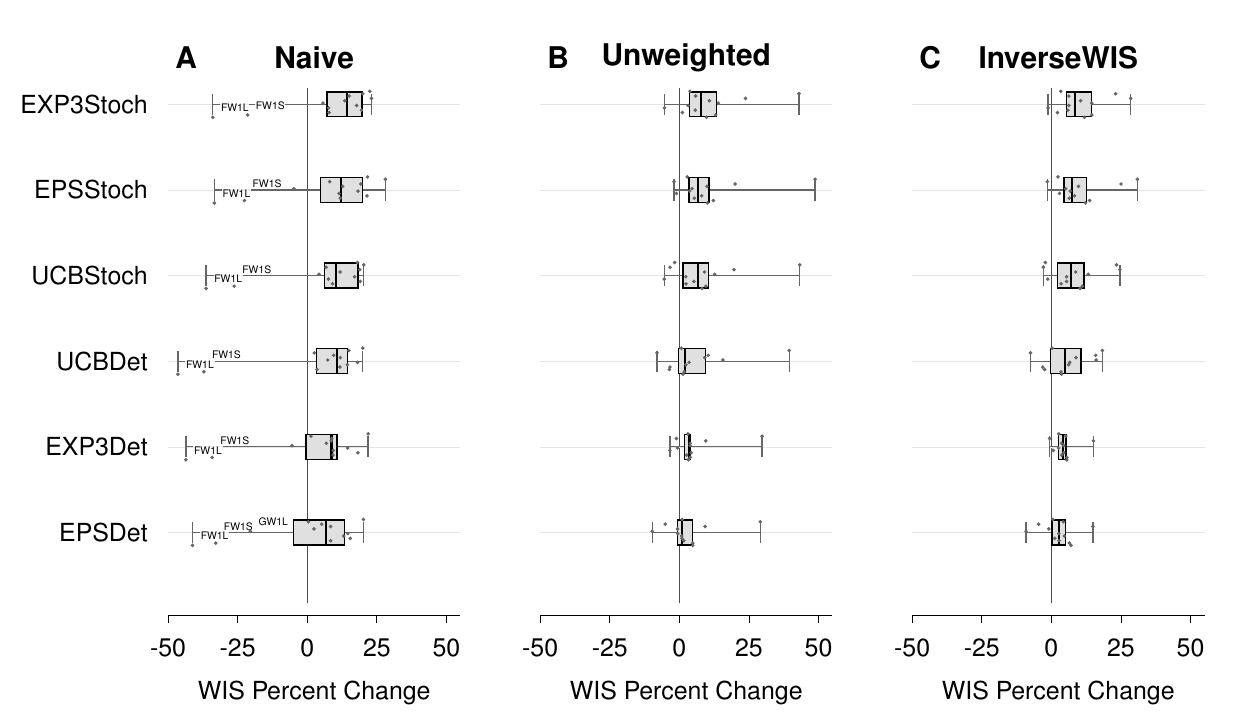}
\caption{Alabama case-study WIS percent change of the six adaptive ensemble methods relative to the Naive, Unweighted, and InverseWIS baselines across the 12 forecast configurations. Positive values indicate lower WIS, and therefore better probabilistic forecast performance, than the corresponding baseline.}
\label{fig:alabama_wis_percent_change}
\end{figure}
\begin{table}[H]
\centering
\scriptsize
\caption{Alabama sensitivity analysis for the main MAB constants. Values are averaged across all 12 forecast configurations and all evaluation windows. EXP3 values are factors of $\eta_0=\sqrt{2\log(K)/(Km)}$; EPS values are fixed exploration probabilities; UCB values are the constant $c$. Bold WIS marks the lowest WIS within each method group.}
\label{tab:alabama_mab_tuning_sensitivity_detailed}
\resizebox{\textwidth}{!}{%
\begin{tabular}{llrrrr}
\hline
Method &  Value & WIS & RMSE & 95\% PI Coverage (\%) & Mean 95\% PI Width \\
\hline
\multirow{5}{*}{EXP3Det}
& $0.5\eta_0$ & 689.8 & 1190.9 & 63.7 & 3928.4 \\
& $\eta_0$ & 657.3 & 1100.2 & 69.2 & 5273.5 \\
& $2\eta_0$ & \textbf{632.2} & 1087.0 & 79.7 & 6670.7 \\
& $5\eta_0$ & 641.0 & 1061.7 & 82.1 & 9435.4 \\
& $10\eta_0$ & 639.3 & 1049.8 & 82.2 & 6960.3 \\
\hline
\multirow{5}{*}{EXP3Stoch}
& $0.5\eta_0$ & 630.9 & 1078.9 & 85.3 & 8273.8 \\
& $\eta_0$ & \textbf{625.8} & 1072.1 & 85.0 & 7647.8 \\
& $2\eta_0$ & 626.0 & 1093.0 & 84.2 & 7680.6 \\
& $5\eta_0$ & 645.2 & 1082.0 & 85.5 & 10184.5 \\
& $10\eta_0$ & 639.9 & 1066.1 & 85.5 & 7614.3 \\
\hline
\multirow{4}{*}{EPSDet}
& $\varepsilon=0.1$ & 670.2 & 1112.7 & 60.6 & 2639.6 \\
& $\varepsilon=0.2$ & \textbf{661.8} & 1093.2 & 62.3 & 2750.5 \\
& $\varepsilon=0.5$ & 669.6 & 1121.8 & 61.9 & 3525.1 \\
& $\varepsilon=1$ & 698.8 & 1207.8 & 65.5 & 4408.8 \\
\hline
\multirow{4}{*}{EPSStoch}
& $\varepsilon=0.1$ & 656.6 & 1157.1 & 79.4 & 3879.7 \\
& $\varepsilon=0.2$ & 646.4 & 1154.5 & 81.0 & 4455.8 \\
& $\varepsilon=0.5$ & 632.1 & 1116.3 & 83.5 & 6239.3 \\
& $\varepsilon=1$ & \textbf{621.7} & 1048.6 & 85.4 & 8183.1 \\
\hline
\multirow{5}{*}{UCBDet}
& $c=0.5$ & 651.1 & 1084.0 & 67.0 & 2877.5 \\
& $c=1$ & 651.1 & 1084.0 & 67.0 & 2877.5 \\
& $c=2$ & \textbf{651.1} & 1084.0 & 67.0 & 2877.5 \\
& $c=5$ & 651.1 & 1084.0 & 67.0 & 2877.5 \\
& $c=10$ & 651.1 & 1084.0 & 67.0 & 2877.5 \\
\hline
\multirow{5}{*}{UCBStoch}
& $c=0.5$ & 629.7 & 1110.9 & 81.1 & 4824.0 \\
& $c=1$ & \textbf{627.2} & 1106.7 & 81.5 & 4823.6 \\
& $c=2$ & 630.0 & 1113.2 & 81.8 & 4788.0 \\
& $c=5$ & 628.8 & 1108.8 & 81.3 & 4719.9 \\
& $c=10$ & 629.3 & 1106.8 & 81.8 & 4782.5 \\
\hline
\end{tabular}%
}
\end{table}

\subsection{Alabama first wave}
\subsubsection{Fixed Calibration Period}
\paragraph{Five-Day Forecasting Horizon}

\begin{figure}[H]
\centering
\includegraphics[width=\linewidth]{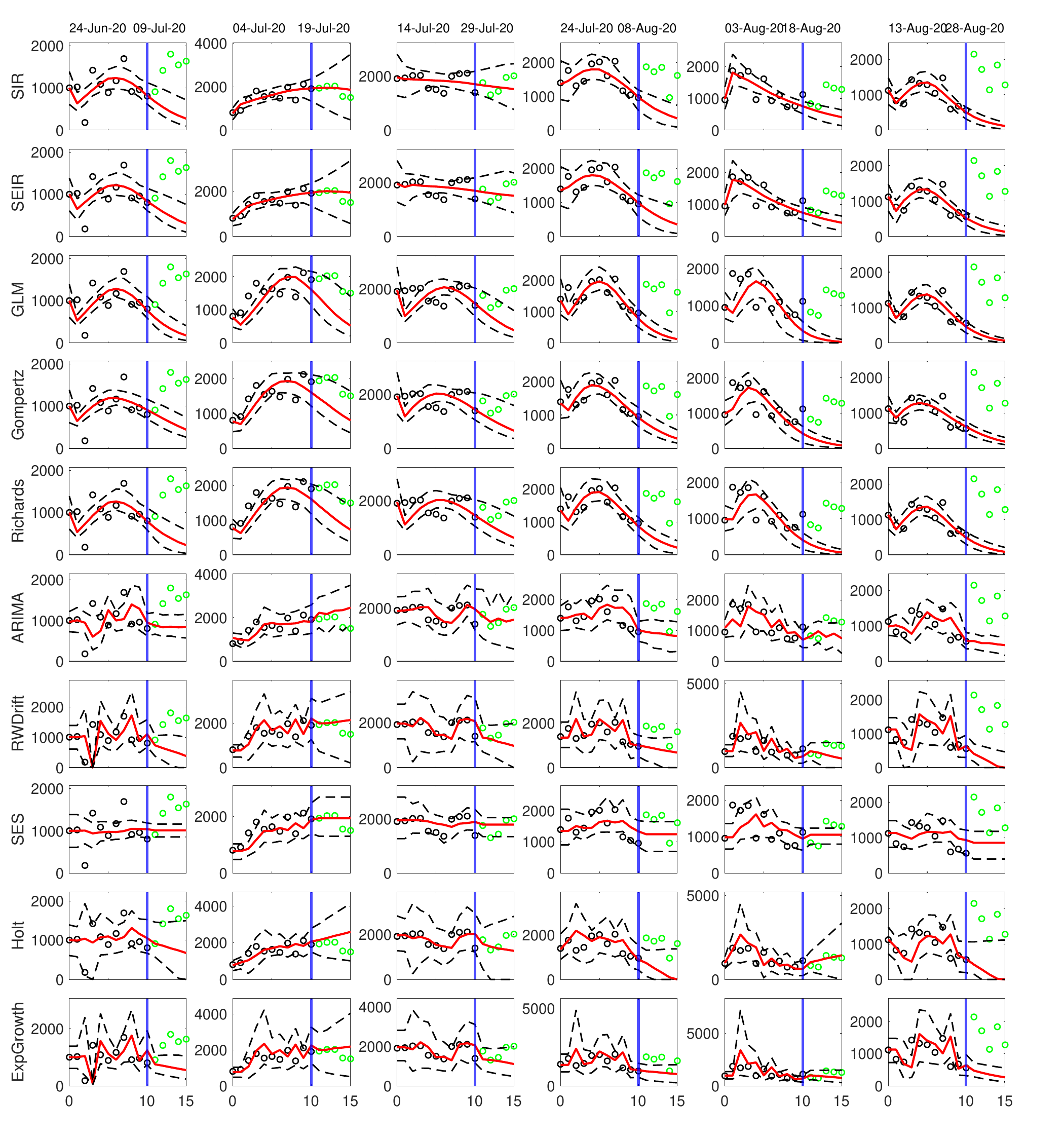}
\caption{\basepanelcaption{1}{\fixedcalibration}{5}}
\label{fig:base_fixed_wave1_fcst5}
\end{figure}
\begin{figure}[H]
\centering
\includegraphics[width=\linewidth]{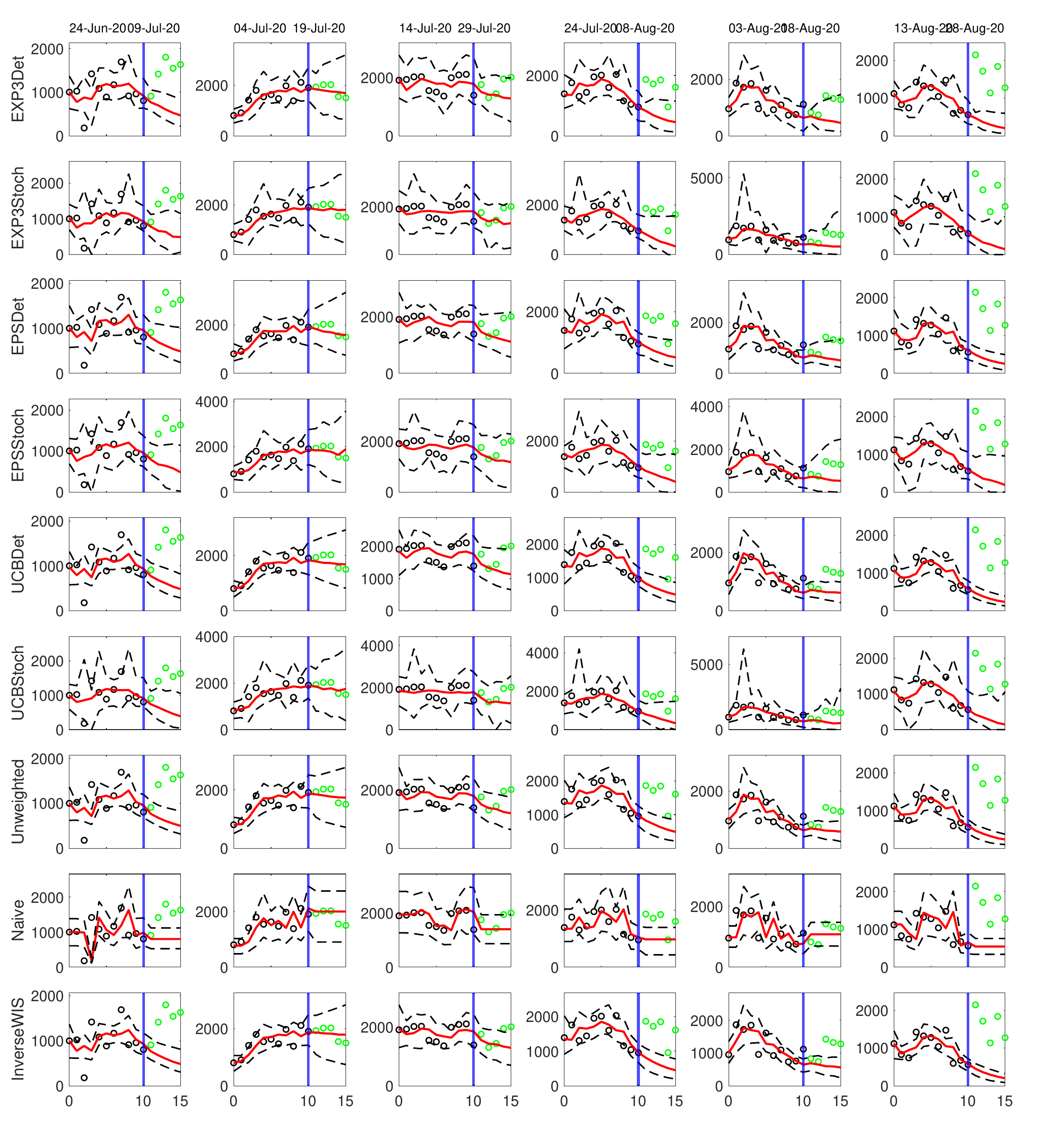}
\caption{\comparisonpanelcaption{1}{\fixedcalibration}{5}}
\label{fig:ensemble_fixed_wave1_fcst5}
\end{figure}
\begin{table}[H]
\centering
\scriptsize
\resizebox{\textwidth}{!}{%
\begin{tabular}{lrrrrrrrr}
\toprule
& \multicolumn{4}{c}{Calibration} & \multicolumn{4}{c}{Forecasting} \\
\cmidrule(lr){2-5} \cmidrule(lr){6-9}
Model & RMSE & WIS & 95\% PI Coverage (\%) & Mean 95\% PI Width & RMSE & WIS & 95\% PI Coverage (\%) & Mean 95\% PI Width \\
\midrule
SIR & 252.07 & 145.22 & 71.2\% & 597.71 & 809.54 & 653.33 & 43.3\% & 937.77 \\
SEIR & 250.60 & 143.14 & 74.2\% & 620.12 & 803.13 & 637.21 & 43.3\% & 918.52 \\
GLM & 372.93 & 222.53 & 57.6\% & 602.48 & 1166.27 & 972.35 & 16.7\% & 610.46 \\
Gompertz & 328.10 & 191.53 & 62.1\% & 575.65 & 998.82 & 802.10 & 23.3\% & 648.32 \\
Richards & 341.29 & 199.85 & 63.6\% & 603.89 & 1081.19 & 889.34 & 20.0\% & 641.02 \\
ARIMA & 347.85 & 190.55 & 74.2\% & 888.15 & 661.24 & 480.28 & 43.3\% & 1036.65 \\
RWDrift & 485.13 & 253.76 & 86.4\% & 1360.06 & 817.35 & 572.53 & 50.0\% & 1384.84 \\
SES & 357.99 & 202.76 & 71.2\% & 769.82 & 454.98 & 315.05 & 46.7\% & 761.17 \\
Holt & 379.84 & 186.68 & 89.4\% & 1308.50 & 814.54 & 534.50 & 60.0\% & 1797.22 \\
ExpGrowth & 523.35 & 270.38 & 83.3\% & 1507.56 & 731.88 & 523.90 & 50.0\% & 1234.73 \\
EXP3Det & 299.19 & 159.91 & 80.3\% & 969.74 & 769.18 & 573.24 & 50.0\% & 1060.38 \\
EXP3Stoch & 288.32 & 155.90 & 84.8\% & 1209.43 & 782.62 & 518.73 & 56.7\% & 1539.31 \\
EPSDet & 324.12 & 176.51 & 77.3\% & 854.73 & 762.78 & 567.78 & 50.0\% & 1028.72 \\
EPSStoch & 298.60 & 160.87 & 86.4\% & 1197.90 & 772.40 & 523.70 & 56.7\% & 1588.79 \\
UCBDet & 310.61 & 175.54 & 68.2\% & 684.24 & 763.06 & 585.96 & 36.7\% & 821.00 \\
UCBStoch & 288.21 & 156.41 & 89.4\% & 1354.27 & 797.49 & 539.48 & 56.7\% & 1568.47 \\
Unweighted & 312.30 & 176.31 & 69.7\% & 705.31 & 754.99 & 596.58 & 36.7\% & 804.66 \\
Naive & 418.95 & 226.27 & 78.8\% & 1047.12 & 592.84 & 427.28 & 50.0\% & 945.78 \\
InverseWIS & 294.09 & 165.15 & 69.7\% & 671.86 & 758.32 & 607.10 & 36.7\% & 783.47 \\
\bottomrule
\end{tabular}%
}
\caption{\metricstablecaption{1}{\fixedcalibration}{5}}
\label{tab:ensemble_fixed_wave1_fcst5}
\end{table}
\paragraph{Ten-Day Forecasting Horizon}

\begin{figure}[H]
\centering
\includegraphics[width=\linewidth]{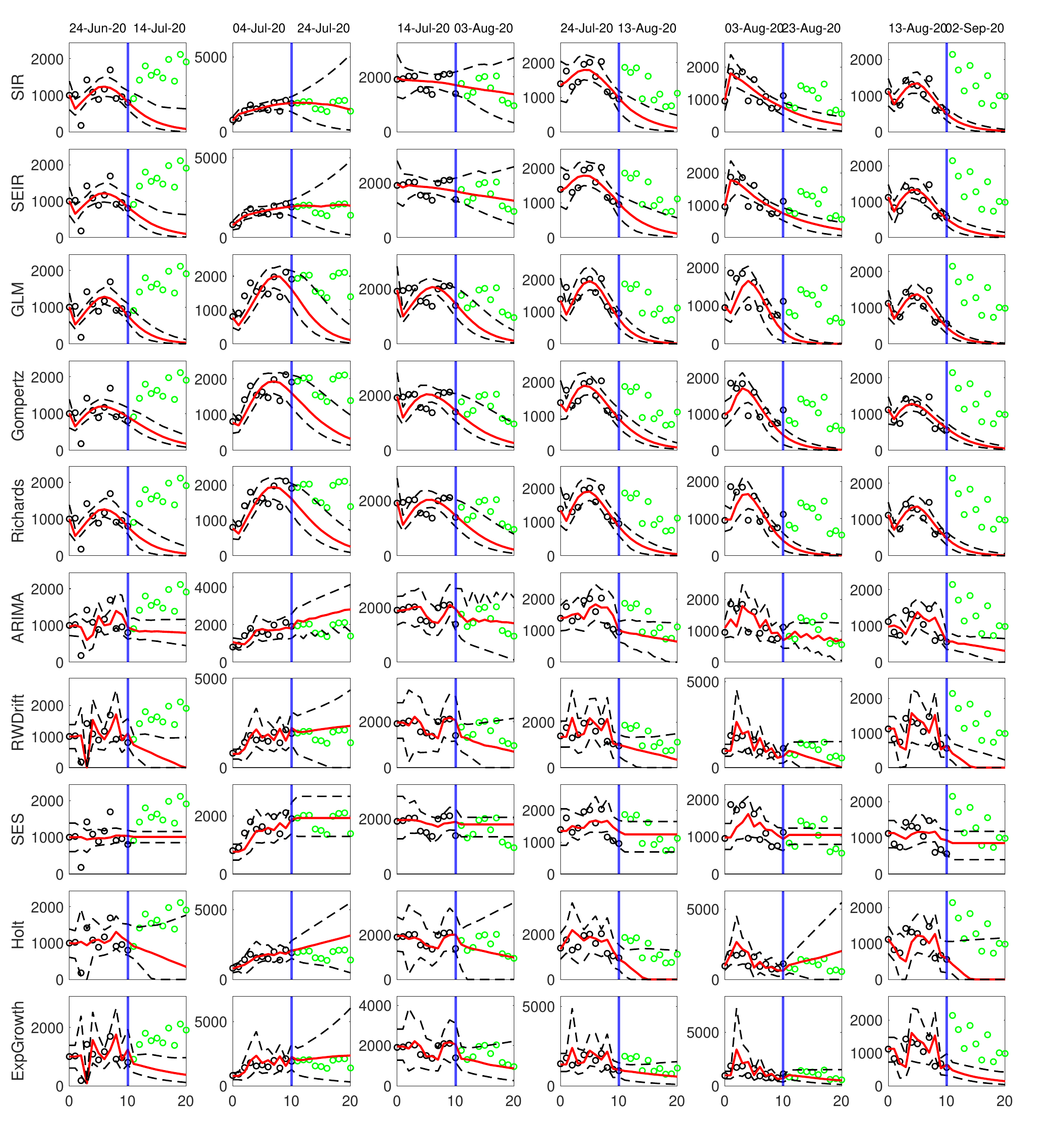}
\caption{\basepanelcaption{1}{\fixedcalibration}{10}}
\label{fig:base_fixed_wave1_fcst10}
\end{figure}
\begin{figure}[H]
\centering
\includegraphics[width=\linewidth]{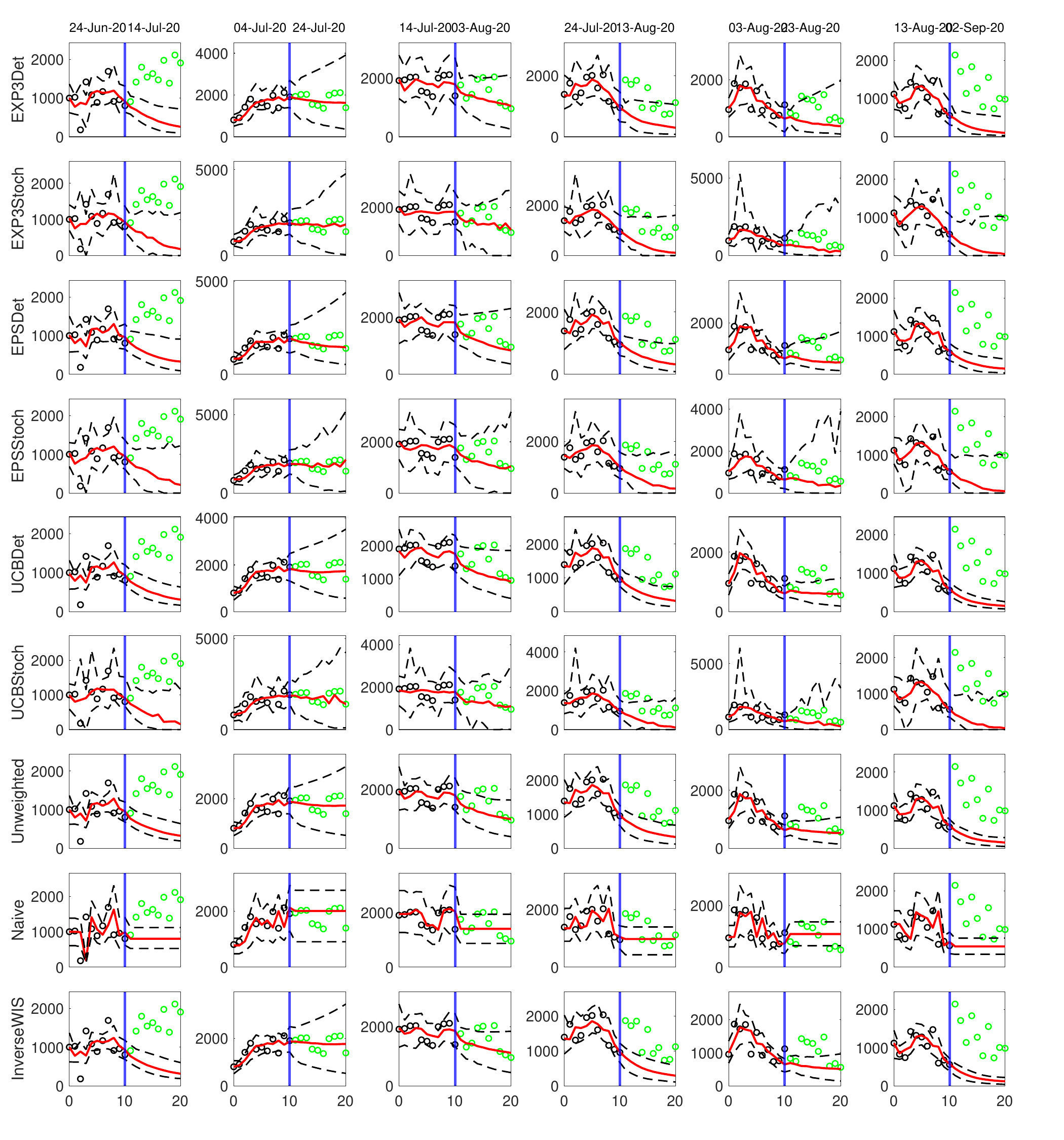}
\caption{\comparisonpanelcaption{1}{\fixedcalibration}{10}}
\label{fig:ensemble_fixed_wave1_fcst10}
\end{figure}
\begin{table}[H]
\centering
\scriptsize
\resizebox{\textwidth}{!}{%
\begin{tabular}{lrrrrrrrr}
\toprule
& \multicolumn{4}{c}{Calibration} & \multicolumn{4}{c}{Forecasting} \\
\cmidrule(lr){2-5} \cmidrule(lr){6-9}
Model & RMSE & WIS & 95\% PI Coverage (\%) & Mean 95\% PI Width & RMSE & WIS & 95\% PI Coverage (\%) & Mean 95\% PI Width \\
\midrule
SIR & 252.07 & 145.22 & 71.2\% & 597.71 & 819.28 & 661.00 & 38.3\% & 1137.26 \\
SEIR & 250.60 & 143.14 & 74.2\% & 620.12 & 812.60 & 643.28 & 38.3\% & 1094.72 \\
GLM & 372.93 & 222.53 & 57.6\% & 602.48 & 1218.08 & 1039.02 & 8.3\% & 455.79 \\
Gompertz & 328.10 & 191.53 & 62.1\% & 575.65 & 1067.92 & 863.10 & 18.3\% & 564.96 \\
Richards & 341.29 & 199.85 & 63.6\% & 603.89 & 1140.50 & 950.32 & 10.0\% & 520.90 \\
ARIMA & 347.85 & 190.55 & 74.2\% & 888.15 & 659.92 & 444.04 & 50.0\% & 1181.23 \\
RWDrift & 485.13 & 253.76 & 86.4\% & 1360.06 & 852.90 & 571.90 & 56.7\% & 1544.02 \\
SES & 357.99 & 202.76 & 71.2\% & 769.82 & 483.37 & 316.93 & 51.7\% & 761.17 \\
Holt & 379.84 & 186.68 & 89.4\% & 1308.50 & 943.75 & 572.94 & 73.3\% & 2246.87 \\
ExpGrowth & 523.35 & 270.38 & 83.3\% & 1507.56 & 728.54 & 494.92 & 58.3\% & 1489.81 \\
EXP3Det & 299.19 & 159.91 & 80.3\% & 969.74 & 763.68 & 544.41 & 55.0\% & 1251.30 \\
EXP3Stoch & 288.32 & 155.90 & 84.8\% & 1209.43 & 791.85 & 507.80 & 68.3\% & 1903.38 \\
EPSDet & 324.12 & 176.51 & 77.3\% & 854.73 & 766.56 & 535.45 & 53.3\% & 1227.45 \\
EPSStoch & 298.60 & 160.87 & 86.4\% & 1197.90 & 795.12 & 505.69 & 66.7\% & 1949.08 \\
UCBDet & 310.61 & 175.54 & 68.2\% & 684.24 & 744.76 & 555.34 & 41.7\% & 921.16 \\
UCBStoch & 288.21 & 156.41 & 89.4\% & 1354.27 & 805.28 & 516.84 & 66.7\% & 1875.96 \\
Unweighted & 312.30 & 176.31 & 69.7\% & 705.31 & 739.61 & 562.83 & 38.3\% & 885.65 \\
Naive & 418.95 & 226.27 & 78.8\% & 1047.12 & 566.38 & 379.14 & 51.7\% & 945.78 \\
InverseWIS & 294.09 & 165.15 & 69.7\% & 671.86 & 742.83 & 576.18 & 38.3\% & 874.58 \\
\bottomrule
\end{tabular}%
}
\caption{\metricstablecaption{1}{\fixedcalibration}{10}}
\label{tab:ensemble_fixed_wave1_fcst10}
\end{table}
\begin{figure}[H]
\centering
\includegraphics[width=\linewidth]{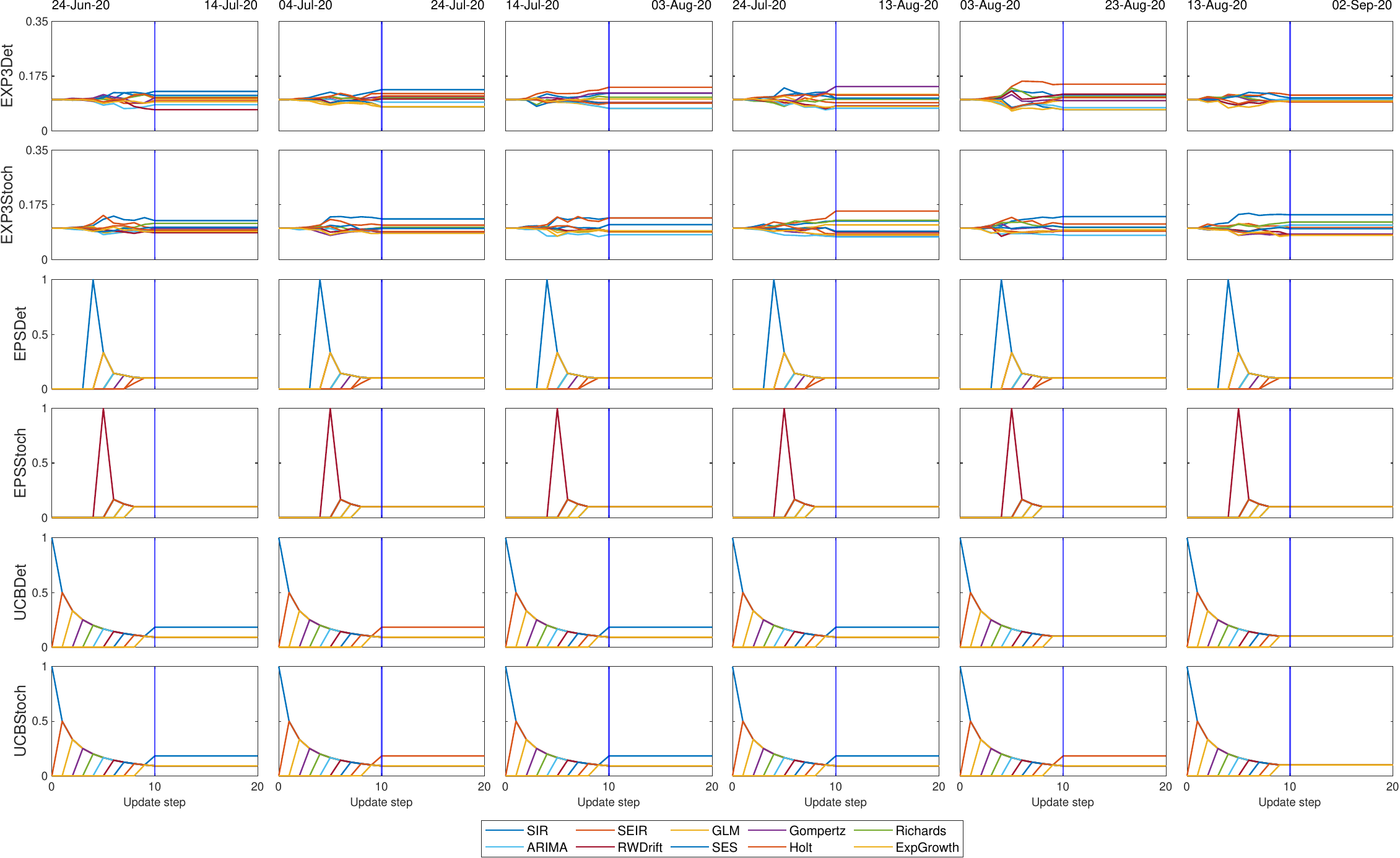}
\caption{\weightpanelcaption{1}{\fixedcalibration}}
\label{fig:weights_fixed_wave1}
\end{figure}
\subsubsection{Growing Calibration Period}
\paragraph{Ten-Day Forecasting Horizon}

\begin{figure}[H]
\centering
\includegraphics[width=\linewidth]{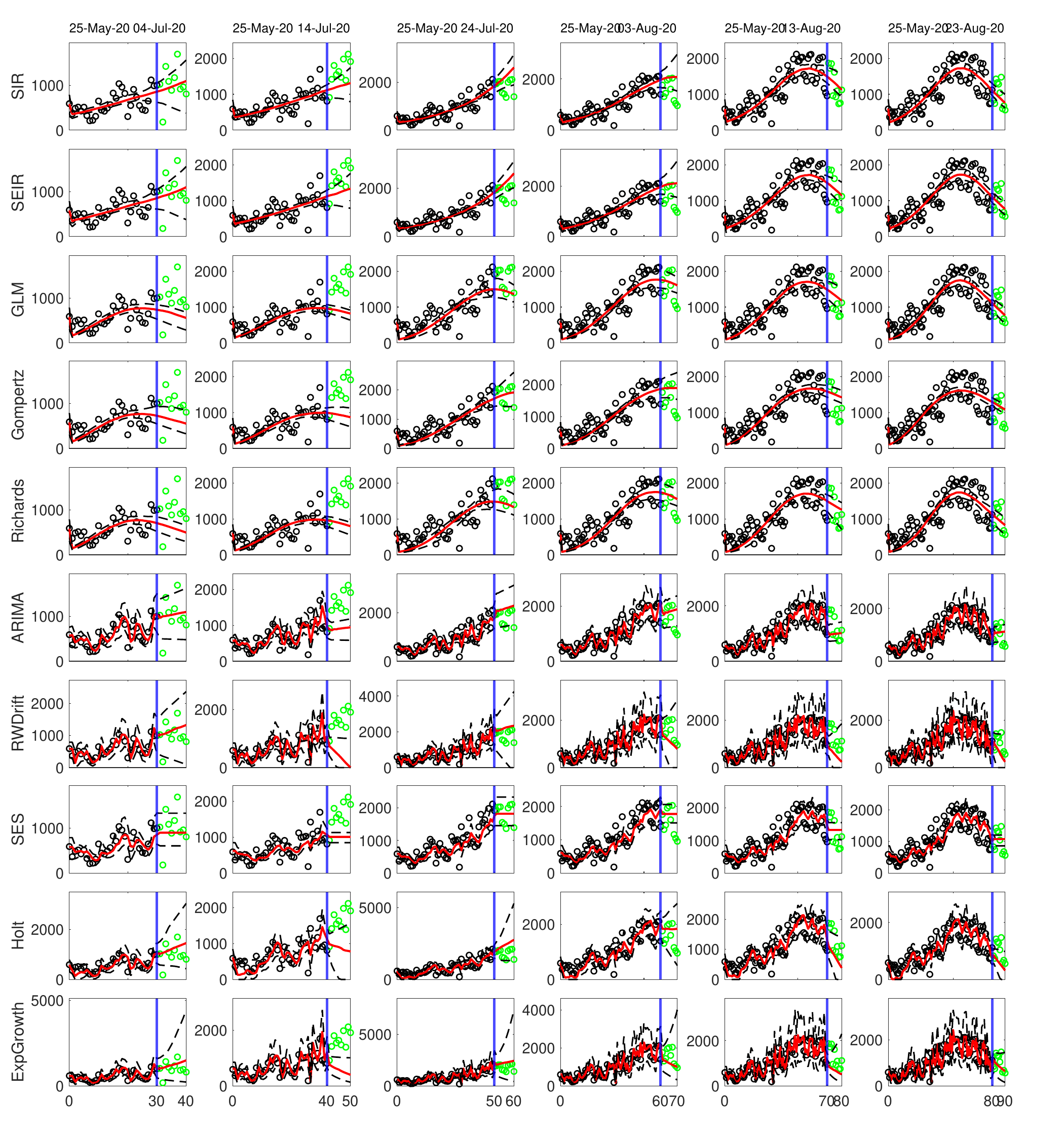}
\caption{\basepanelcaption{1}{\growingcalibration}{10}}
\label{fig:base_growing_wave1_fcst10}
\end{figure}
\begin{figure}[H]
\centering
\includegraphics[width=\linewidth]{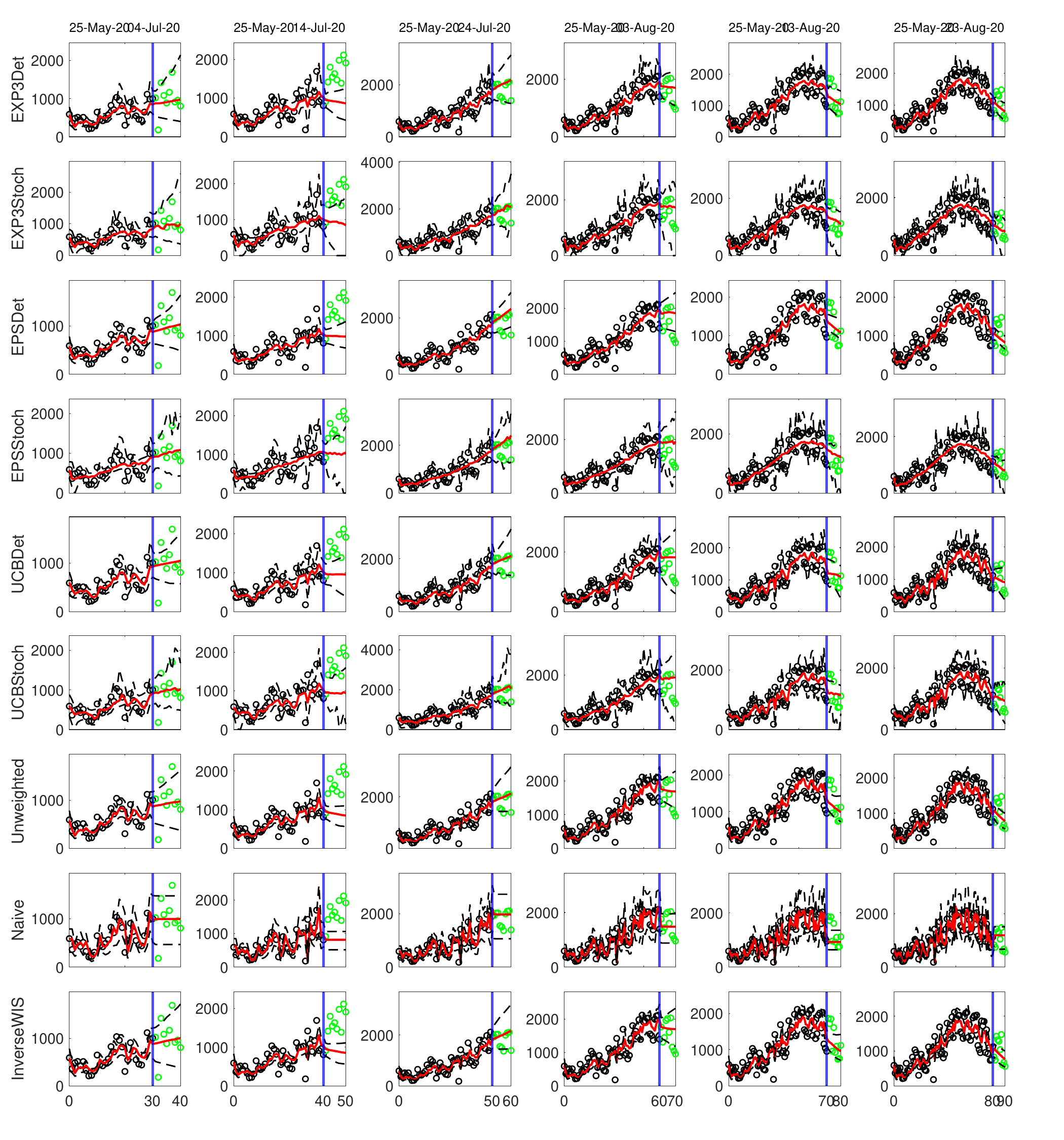}
\caption{\comparisonpanelcaption{1}{\growingcalibration}{10}}
\label{fig:ensemble_growing_wave1_fcst10}
\end{figure}
\begin{table}[H]
\centering
\scriptsize
\resizebox{\textwidth}{!}{%
\begin{tabular}{lrrrrrrrr}
\toprule
& \multicolumn{4}{c}{Calibration} & \multicolumn{4}{c}{Forecasting} \\
\cmidrule(lr){2-5} \cmidrule(lr){6-9}
Model & RMSE & WIS & 95\% PI Coverage (\%) & Mean 95\% PI Width & RMSE & WIS & 95\% PI Coverage (\%) & Mean 95\% PI Width \\
\midrule
SIR & 265.89 & 181.70 & 30.4\% & 218.96 & 466.36 & 296.40 & 43.3\% & 714.86 \\
SEIR & 265.98 & 183.50 & 25.9\% & 208.04 & 469.00 & 298.23 & 41.7\% & 697.35 \\
GLM & 291.50 & 208.66 & 25.4\% & 201.98 & 482.70 & 360.84 & 23.3\% & 435.72 \\
Gompertz & 294.59 & 213.76 & 25.0\% & 184.36 & 497.80 & 360.51 & 33.3\% & 519.87 \\
Richards & 300.13 & 215.50 & 25.7\% & 211.65 & 494.14 & 367.77 & 21.7\% & 468.69 \\
ARIMA & 307.82 & 165.75 & 64.2\% & 531.66 & 504.35 & 312.60 & 56.7\% & 891.25 \\
RWDrift & 361.15 & 183.42 & 76.8\% & 808.41 & 681.49 & 411.23 & 75.0\% & 1811.69 \\
SES & 295.85 & 181.22 & 43.4\% & 345.85 & 436.66 & 305.20 & 40.0\% & 549.59 \\
Holt & 320.44 & 176.84 & 69.6\% & 580.47 & 602.03 & 348.02 & 71.7\% & 1543.89 \\
ExpGrowth & 368.43 & 185.00 & 76.9\% & 833.74 & 611.02 & 373.08 & 76.7\% & 2021.77 \\
EXP3Det & 267.48 & 154.23 & 58.3\% & 469.98 & 453.00 & 282.31 & 60.0\% & 896.81 \\
EXP3Stoch & 267.61 & 146.49 & 76.9\% & 688.78 & 449.51 & 262.17 & 76.7\% & 1453.24 \\
EPSDet & 262.52 & 165.09 & 41.3\% & 274.42 & 457.49 & 295.77 & 41.7\% & 668.29 \\
EPSStoch & 263.26 & 148.05 & 72.4\% & 618.31 & 461.22 & 264.54 & 71.7\% & 1315.49 \\
UCBDet & 272.26 & 148.01 & 66.6\% & 512.43 & 449.49 & 267.04 & 70.0\% & 1003.95 \\
UCBStoch & 270.98 & 144.71 & 75.0\% & 631.23 & 459.22 & 267.34 & 70.0\% & 1302.50 \\
Unweighted & 271.28 & 163.73 & 44.6\% & 315.45 & 449.93 & 293.86 & 51.7\% & 771.05 \\
Naive & 334.13 & 169.82 & 72.0\% & 695.85 & 485.43 & 303.02 & 66.7\% & 974.47 \\
InverseWIS & 271.63 & 162.12 & 46.0\% & 326.60 & 452.84 & 293.00 & 51.7\% & 794.59 \\
\bottomrule
\end{tabular}%
}
\caption{\metricstablecaption{1}{\growingcalibration}{10}}
\label{tab:ensemble_growing_wave1_fcst10}
\end{table}
\paragraph{Thirty-Day Forecasting Horizon}

\begin{figure}[H]
\centering
\includegraphics[width=\linewidth]{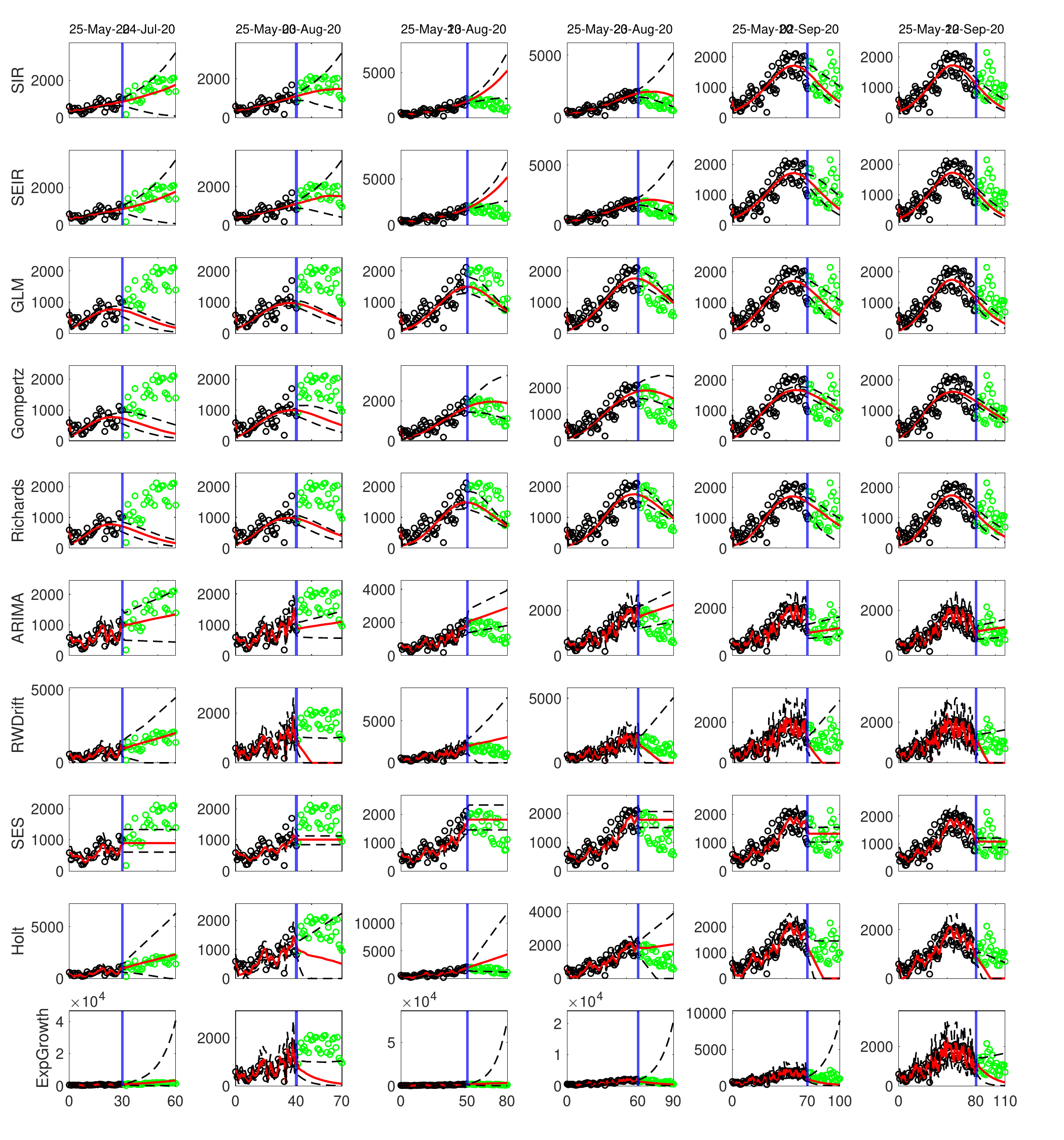}
\caption{\basepanelcaption{1}{\growingcalibration}{30}}
\label{fig:base_growing_wave1_fcst30}
\end{figure}
\begin{figure}[H]
\centering
\includegraphics[width=\linewidth]{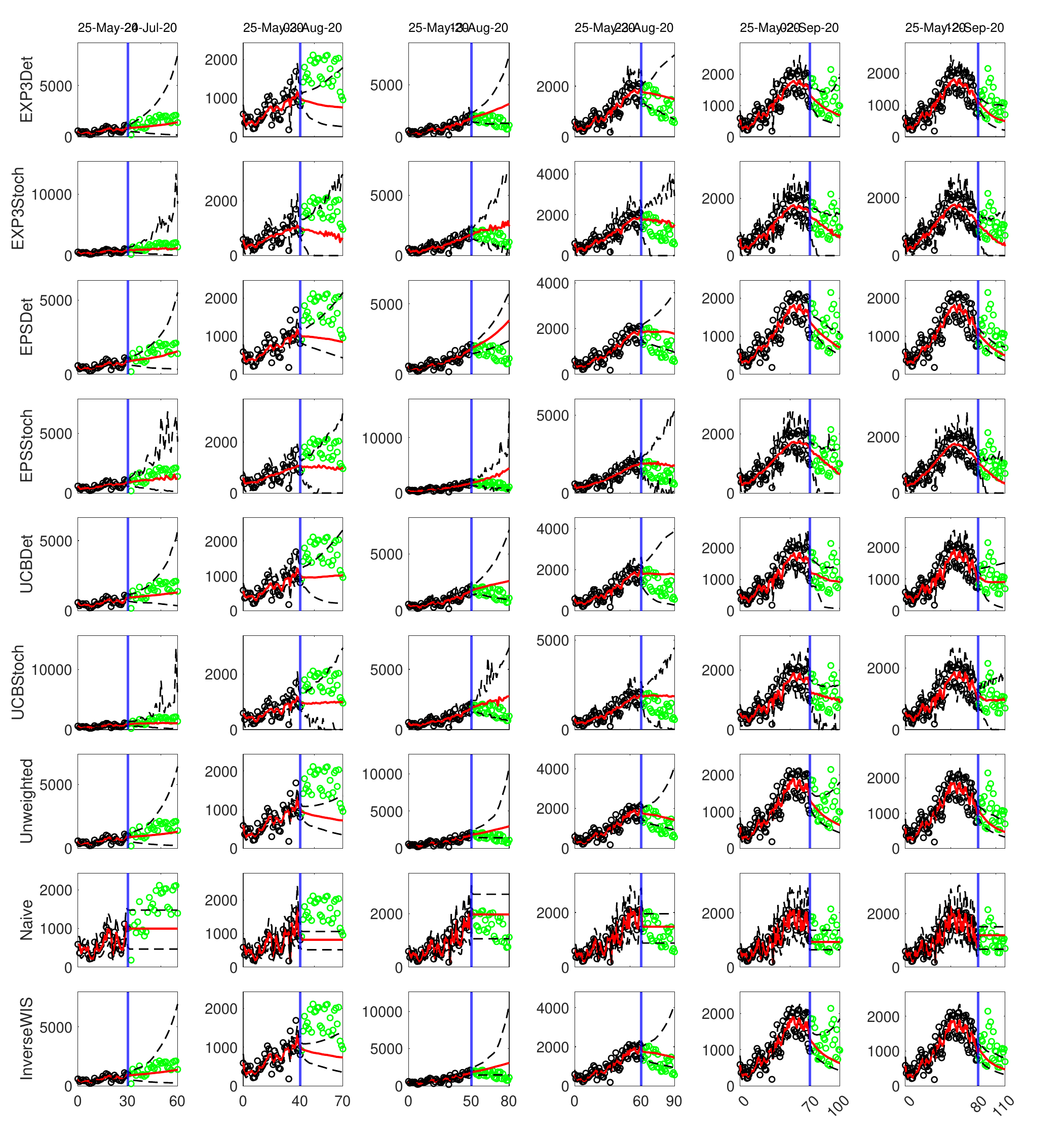}
\caption{\comparisonpanelcaption{1}{\growingcalibration}{30}}
\label{fig:ensemble_growing_wave1_fcst30}
\end{figure}
\begin{table}[H]
\centering
\scriptsize
\resizebox{\textwidth}{!}{%
\begin{tabular}{lrrrrrrrr}
\toprule
& \multicolumn{4}{c}{Calibration} & \multicolumn{4}{c}{Forecasting} \\
\cmidrule(lr){2-5} \cmidrule(lr){6-9}
Model & RMSE & WIS & 95\% PI Coverage (\%) & Mean 95\% PI Width & RMSE & WIS & 95\% PI Coverage (\%) & Mean 95\% PI Width \\
\midrule
SIR & 265.89 & 181.70 & 30.4\% & 218.96 & 842.71 & 532.12 & 47.2\% & 1467.06 \\
SEIR & 265.98 & 183.50 & 25.9\% & 208.04 & 840.97 & 545.99 & 44.4\% & 1418.29 \\
GLM & 291.50 & 208.66 & 25.4\% & 201.98 & 699.50 & 544.59 & 23.9\% & 407.80 \\
Gompertz & 294.59 & 213.76 & 25.0\% & 184.36 & 712.06 & 520.73 & 28.9\% & 677.21 \\
Richards & 300.13 & 215.50 & 25.7\% & 211.65 & 694.47 & 545.78 & 22.2\% & 430.77 \\
ARIMA & 307.82 & 165.75 & 64.2\% & 531.66 & 698.16 & 437.40 & 46.7\% & 1049.38 \\
RWDrift & 361.15 & 183.42 & 76.8\% & 808.41 & 1017.55 & 590.96 & 78.9\% & 2663.08 \\
SES & 295.85 & 181.22 & 43.4\% & 345.85 & 588.10 & 413.27 & 30.6\% & 549.59 \\
Holt & 320.44 & 176.84 & 69.6\% & 580.47 & 1031.76 & 605.42 & 70.6\% & 2690.69 \\
ExpGrowth & 368.43 & 185.00 & 76.9\% & 833.74 & 945.21 & 600.23 & 79.4\% & 8090.99 \\
EXP3Det & 267.48 & 154.23 & 58.3\% & 469.98 & 701.75 & 405.76 & 61.7\% & 1731.21 \\
EXP3Stoch & 267.61 & 146.49 & 76.9\% & 688.78 & 683.07 & 362.94 & 83.9\% & 2453.57 \\
EPSDet & 262.52 & 165.09 & 41.3\% & 274.42 & 753.67 & 463.28 & 44.4\% & 1203.33 \\
EPSStoch & 263.26 & 148.05 & 72.4\% & 618.31 & 804.98 & 402.94 & 83.3\% & 2456.87 \\
UCBDet & 272.26 & 148.01 & 66.6\% & 512.43 & 634.29 & 356.26 & 77.8\% & 1844.72 \\
UCBStoch & 270.98 & 144.71 & 75.0\% & 631.23 & 663.91 & 368.42 & 81.7\% & 2289.46 \\
Unweighted & 271.28 & 163.73 & 44.6\% & 315.45 & 695.85 & 422.15 & 53.9\% & 1585.95 \\
Naive & 334.13 & 169.82 & 72.0\% & 695.85 & 603.80 & 384.52 & 56.1\% & 974.47 \\
InverseWIS & 271.63 & 162.12 & 46.0\% & 326.60 & 702.70 & 424.64 & 53.9\% & 1636.97 \\
\bottomrule
\end{tabular}%
}
\caption{\metricstablecaption{1}{\growingcalibration}{30}}
\label{tab:ensemble_growing_wave1_fcst30}
\end{table}
\begin{figure}[H]
\centering
\includegraphics[width=\linewidth]{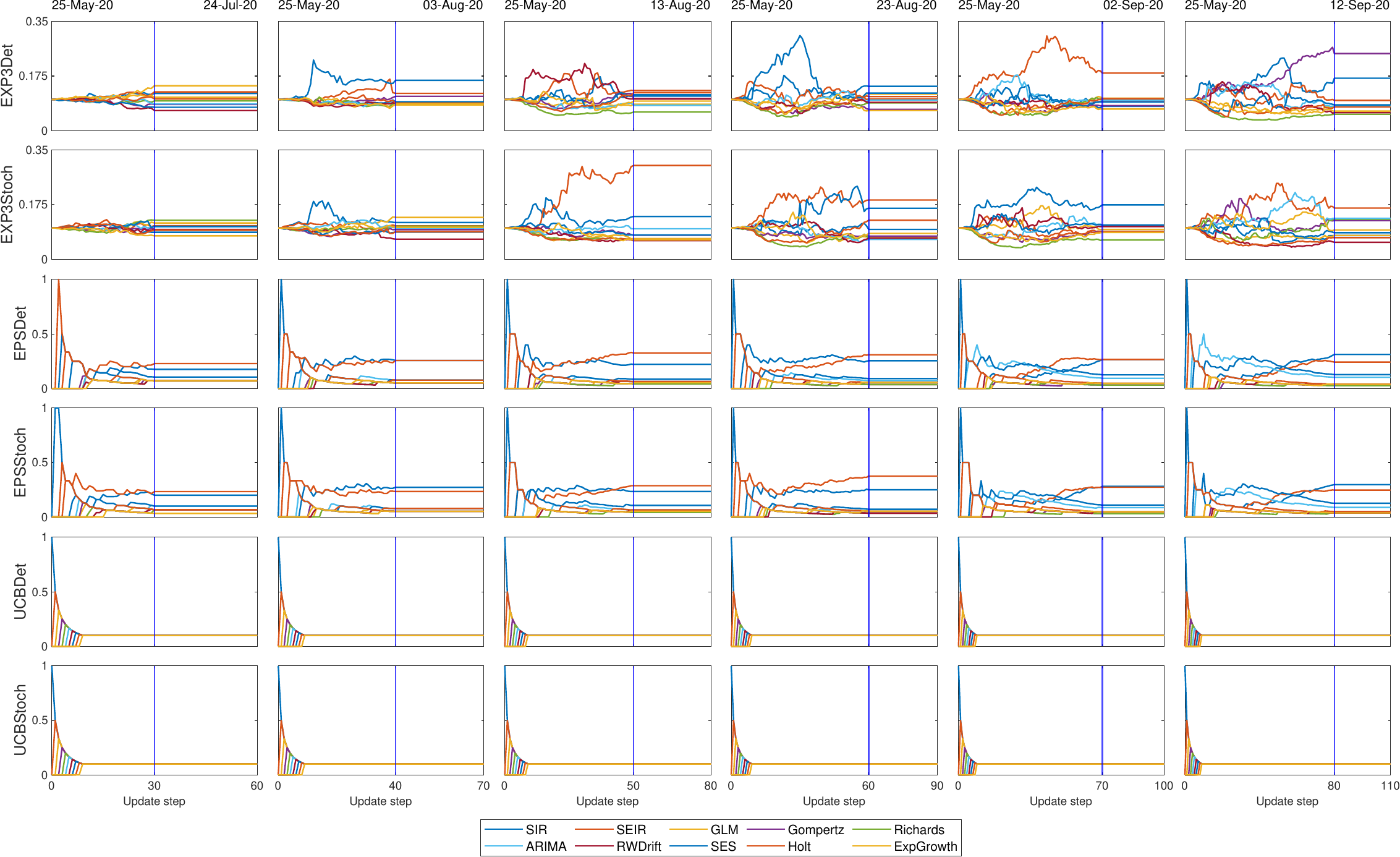}
\caption{\weightpanelcaption{1}{\growingcalibration}}
\label{fig:weights_growing_wave1}
\end{figure}
\subsection{Alabama second wave}
\subsubsection{Fixed Calibration Period}
\paragraph{Five-Day Forecasting Horizon}

\begin{figure}[H]
\centering
\includegraphics[width=\linewidth]{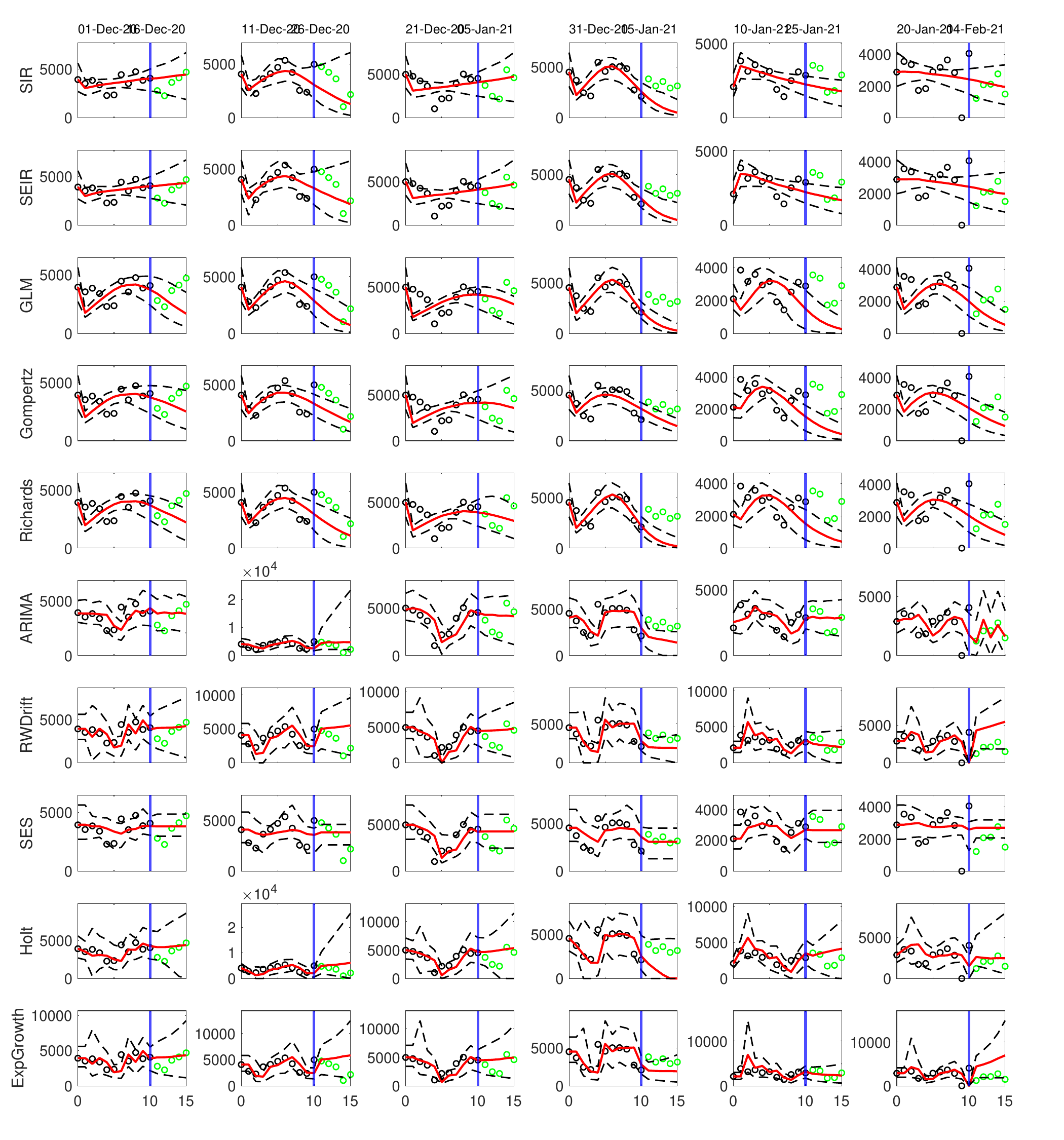}
\caption{\basepanelcaption{2}{\fixedcalibration}{5}}
\label{fig:base_fixed_wave2_fcst5}
\end{figure}
\begin{figure}[H]
\centering
\includegraphics[width=\linewidth]{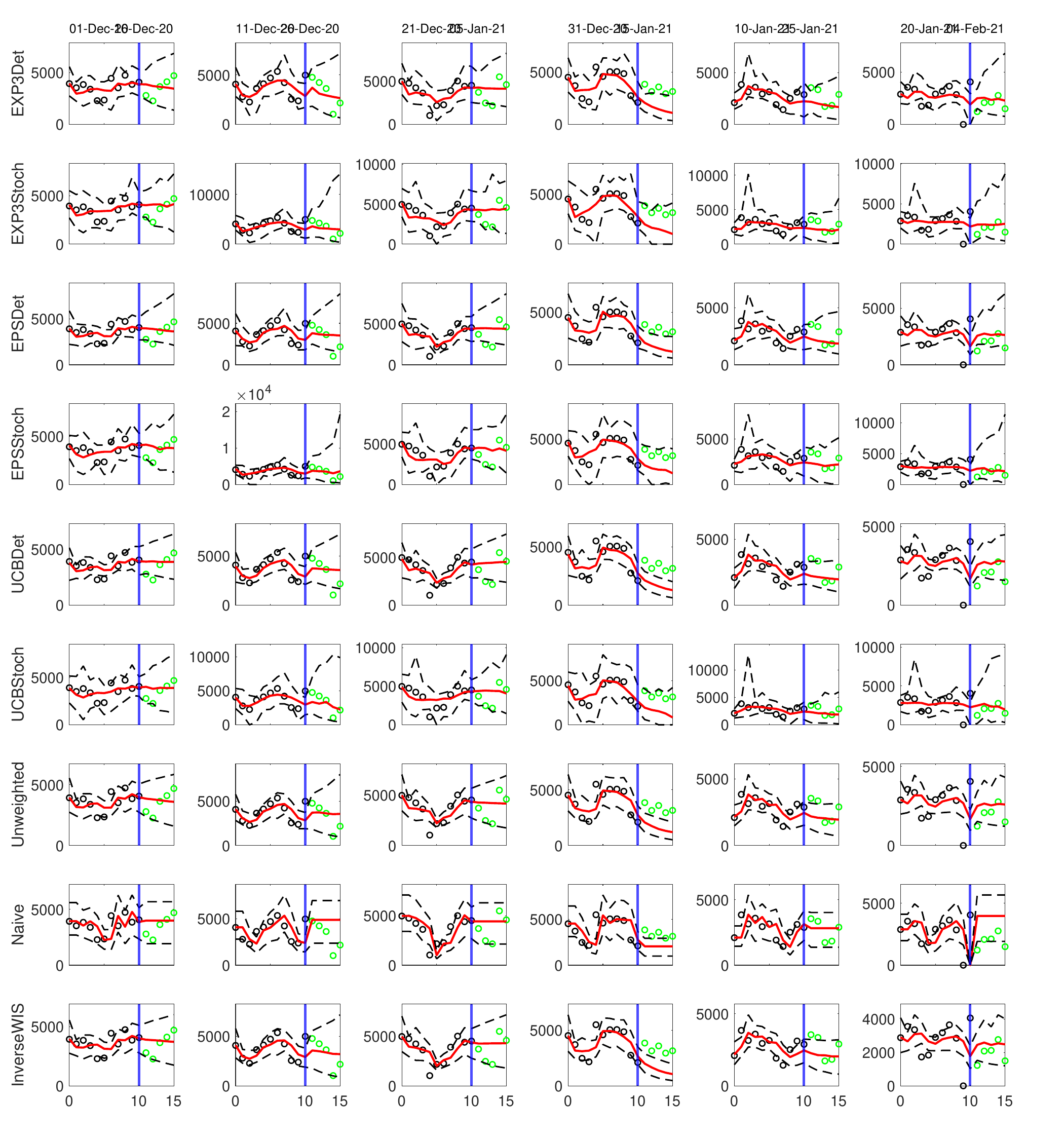}
\caption{\comparisonpanelcaption{2}{\fixedcalibration}{5}}
\label{fig:ensemble_fixed_wave2_fcst5}
\end{figure}
\begin{table}[H]
\centering
\scriptsize
\resizebox{\textwidth}{!}{%
\begin{tabular}{lrrrrrrrr}
\toprule
& \multicolumn{4}{c}{Calibration} & \multicolumn{4}{c}{Forecasting} \\
\cmidrule(lr){2-5} \cmidrule(lr){6-9}
Model & RMSE & WIS & 95\% PI Coverage (\%) & Mean 95\% PI Width & RMSE & WIS & 95\% PI Coverage (\%) & Mean 95\% PI Width \\
\midrule
SIR & 870.11 & 511.46 & 57.6\% & 1613.59 & 1317.29 & 817.93 & 66.7\% & 3095.14 \\
SEIR & 869.39 & 514.31 & 57.6\% & 1649.90 & 1280.84 & 781.99 & 66.7\% & 3102.65 \\
GLM & 1092.70 & 674.90 & 53.0\% & 1655.47 & 1858.53 & 1310.96 & 33.3\% & 2147.60 \\
Gompertz & 1049.76 & 667.29 & 45.5\% & 1567.70 & 1387.57 & 921.84 & 50.0\% & 2407.11 \\
Richards & 1046.60 & 650.15 & 53.0\% & 1665.86 & 1712.30 & 1214.36 & 40.0\% & 2274.17 \\
ARIMA & 1090.78 & 580.70 & 72.7\% & 2529.78 & 1234.73 & 733.59 & 76.7\% & 5166.68 \\
RWDrift & 1339.88 & 723.51 & 78.8\% & 3410.32 & 1714.31 & 1009.29 & 83.3\% & 5286.79 \\
SES & 1012.03 & 554.15 & 66.7\% & 2278.87 & 992.53 & 613.16 & 63.3\% & 2323.07 \\
Holt & 1165.93 & 576.34 & 86.4\% & 3726.14 & 1778.89 & 997.22 & 96.7\% & 7910.46 \\
ExpGrowth & 1340.17 & 721.52 & 78.8\% & 3490.60 & 1874.36 & 1115.81 & 83.3\% & 6024.44 \\
EXP3Det & 895.71 & 478.68 & 84.8\% & 2631.38 & 1175.03 & 705.99 & 70.0\% & 4052.00 \\
EXP3Stoch & 885.82 & 473.81 & 83.3\% & 3252.78 & 1196.30 & 694.90 & 96.7\% & 5662.14 \\
EPSDet & 912.82 & 503.01 & 75.8\% & 2274.80 & 1241.11 & 729.86 & 66.7\% & 3835.73 \\
EPSStoch & 888.10 & 471.03 & 89.4\% & 3253.85 & 1167.13 & 677.75 & 100.0\% & 5949.71 \\
UCBDet & 901.09 & 521.61 & 57.6\% & 1766.10 & 1228.08 & 761.88 & 56.7\% & 3004.81 \\
UCBStoch & 875.61 & 469.54 & 86.4\% & 3534.98 & 1173.63 & 698.56 & 96.7\% & 5671.02 \\
Unweighted & 909.01 & 512.41 & 63.6\% & 1895.73 & 1219.44 & 737.06 & 70.0\% & 3335.73 \\
Naive & 1174.40 & 636.56 & 75.8\% & 2568.81 & 1479.16 & 862.93 & 70.0\% & 3478.29 \\
InverseWIS & 880.27 & 496.66 & 65.2\% & 1845.86 & 1198.80 & 738.94 & 73.3\% & 3235.96 \\
\bottomrule
\end{tabular}%
}
\caption{\metricstablecaption{2}{\fixedcalibration}{5}}
\label{tab:ensemble_fixed_wave2_fcst5}
\end{table}
\paragraph{Ten-Day Forecasting Horizon}

\begin{figure}[H]
\centering
\includegraphics[width=\linewidth]{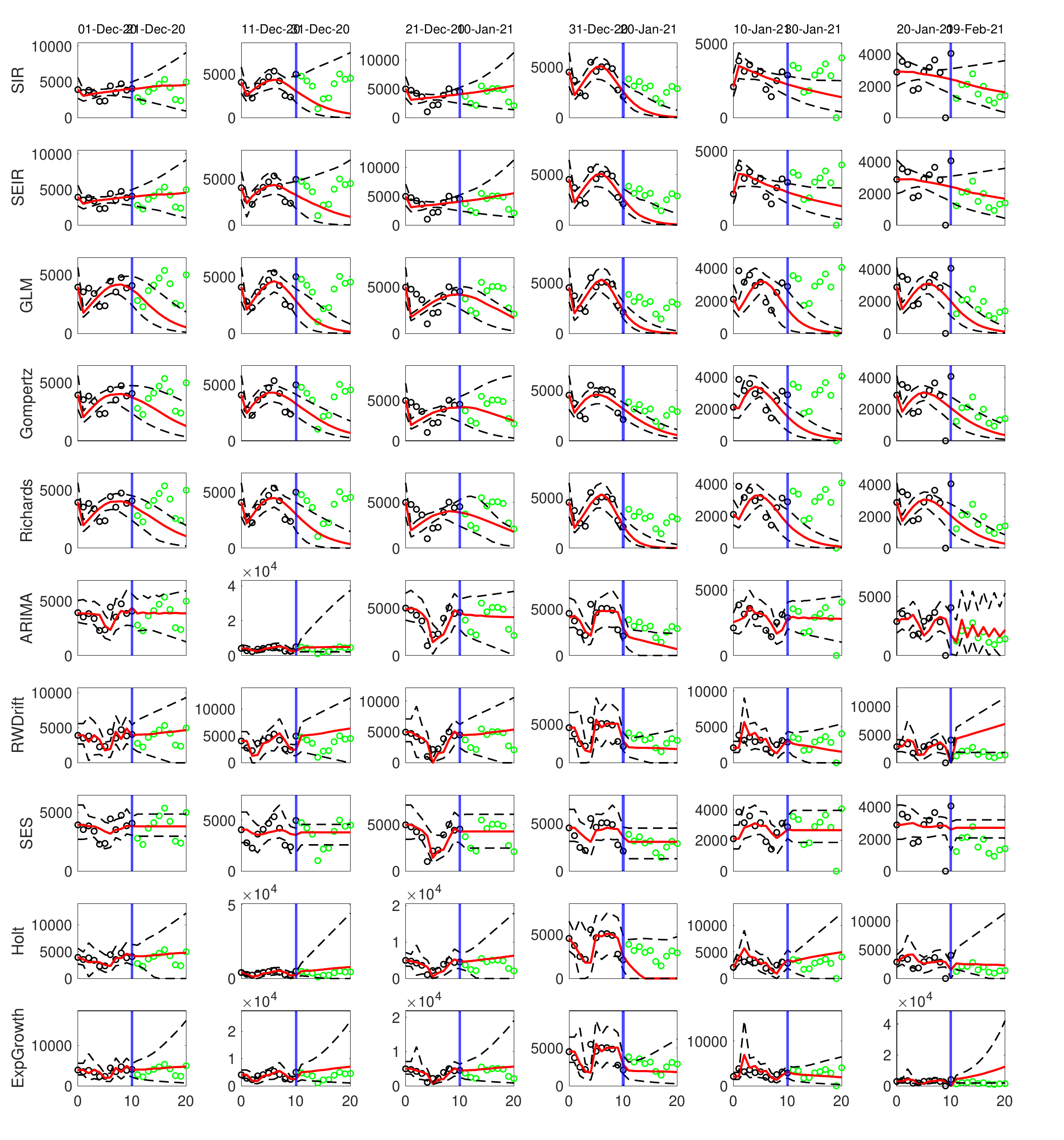}
\caption{\basepanelcaption{2}{\fixedcalibration}{10}}
\label{fig:base_fixed_wave2_fcst10}
\end{figure}
\begin{figure}[H]
\centering
\includegraphics[width=\linewidth]{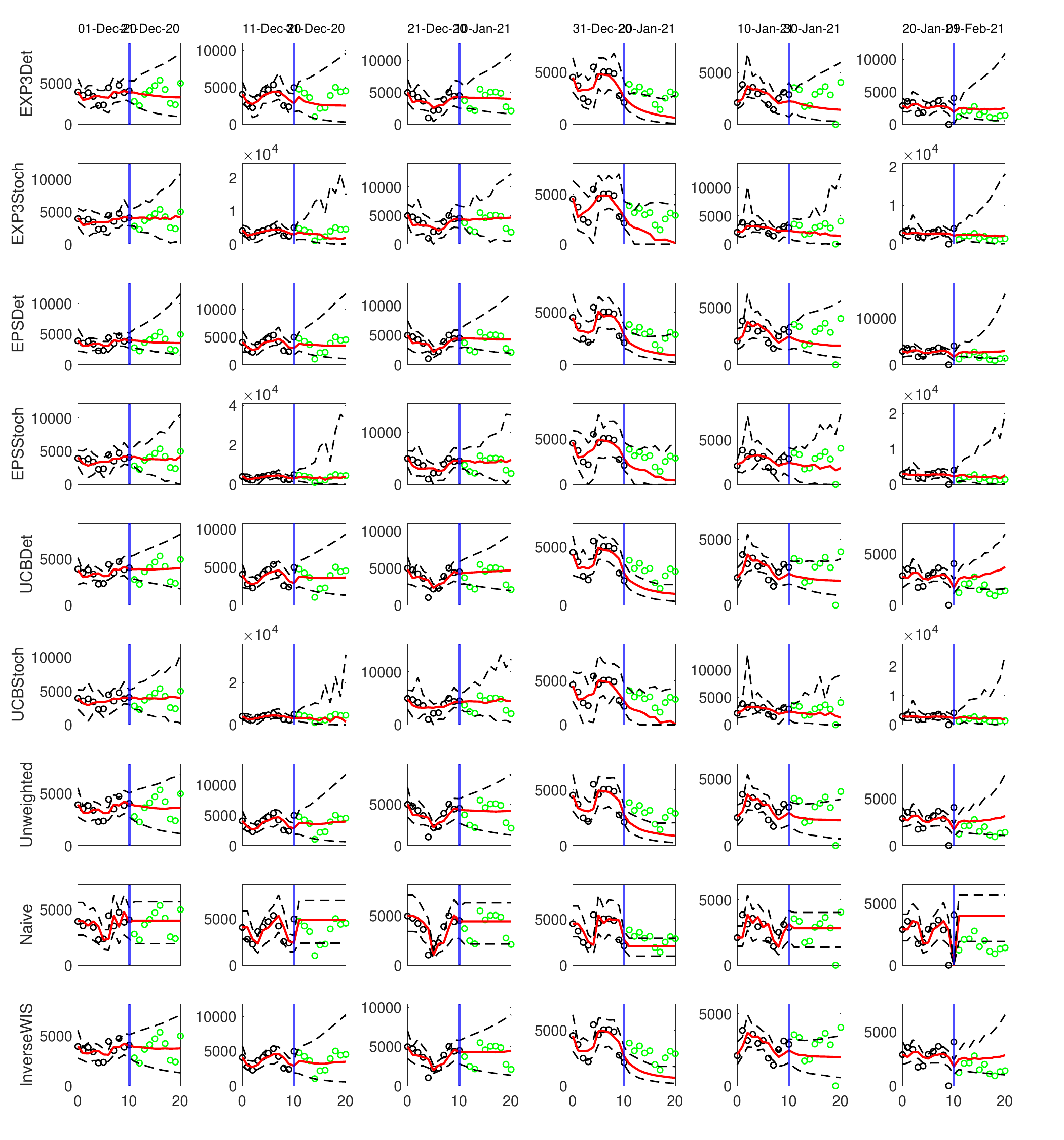}
\caption{\comparisonpanelcaption{2}{\fixedcalibration}{10}}
\label{fig:ensemble_fixed_wave2_fcst10}
\end{figure}
\begin{table}[H]
\centering
\scriptsize
\resizebox{\textwidth}{!}{%
\begin{tabular}{lrrrrrrrr}
\toprule
& \multicolumn{4}{c}{Calibration} & \multicolumn{4}{c}{Forecasting} \\
\cmidrule(lr){2-5} \cmidrule(lr){6-9}
Model & RMSE & WIS & 95\% PI Coverage (\%) & Mean 95\% PI Width & RMSE & WIS & 95\% PI Coverage (\%) & Mean 95\% PI Width \\
\midrule
SIR & 870.11 & 511.46 & 57.6\% & 1613.59 & 1658.16 & 1012.16 & 66.7\% & 3853.46 \\
SEIR & 869.39 & 514.31 & 57.6\% & 1649.90 & 1607.16 & 956.06 & 70.0\% & 3851.64 \\
GLM & 1092.70 & 674.90 & 53.0\% & 1655.47 & 2274.42 & 1721.59 & 18.3\% & 1701.21 \\
Gompertz & 1049.76 & 667.29 & 45.5\% & 1567.70 & 1801.71 & 1230.76 & 45.0\% & 2449.43 \\
Richards & 1046.60 & 650.15 & 53.0\% & 1665.86 & 2123.52 & 1584.37 & 26.7\% & 1874.73 \\
ARIMA & 1090.78 & 580.70 & 72.7\% & 2529.78 & 1266.20 & 714.88 & 81.7\% & 6783.22 \\
RWDrift & 1339.88 & 723.51 & 78.8\% & 3410.32 & 1991.21 & 1149.42 & 85.0\% & 6515.90 \\
SES & 1012.03 & 554.15 & 66.7\% & 2278.87 & 1129.08 & 694.88 & 58.3\% & 2323.07 \\
Holt & 1165.93 & 576.34 & 86.4\% & 3726.14 & 2005.78 & 1175.57 & 98.3\% & 11276.71 \\
ExpGrowth & 1340.17 & 721.52 & 78.8\% & 3490.60 & 2497.88 & 1454.11 & 83.3\% & 9881.39 \\
EXP3Det & 895.71 & 478.68 & 84.8\% & 2631.38 & 1362.66 & 772.49 & 78.3\% & 5263.97 \\
EXP3Stoch & 885.82 & 473.81 & 83.3\% & 3252.78 & 1468.37 & 784.63 & 96.7\% & 7917.87 \\
EPSDet & 912.82 & 503.01 & 75.8\% & 2274.80 & 1342.93 & 780.21 & 76.7\% & 5481.54 \\
EPSStoch & 888.10 & 471.03 & 89.4\% & 3253.85 & 1262.99 & 750.64 & 98.3\% & 8754.56 \\
UCBDet & 901.09 & 521.61 & 57.6\% & 1766.10 & 1385.69 & 821.64 & 60.0\% & 3748.33 \\
UCBStoch & 875.61 & 469.54 & 86.4\% & 3534.98 & 1398.35 & 774.21 & 96.7\% & 8229.20 \\
Unweighted & 909.01 & 512.41 & 63.6\% & 1895.73 & 1321.61 & 793.27 & 71.7\% & 4206.35 \\
Naive & 1174.40 & 636.56 & 75.8\% & 2568.81 & 1490.14 & 851.46 & 70.0\% & 3478.29 \\
InverseWIS & 880.27 & 496.66 & 65.2\% & 1845.86 & 1313.17 & 802.28 & 71.7\% & 4007.26 \\
\bottomrule
\end{tabular}%
}
\caption{\metricstablecaption{2}{\fixedcalibration}{10}}
\label{tab:ensemble_fixed_wave2_fcst10}
\end{table}
\begin{figure}[H]
\centering
\includegraphics[width=\linewidth]{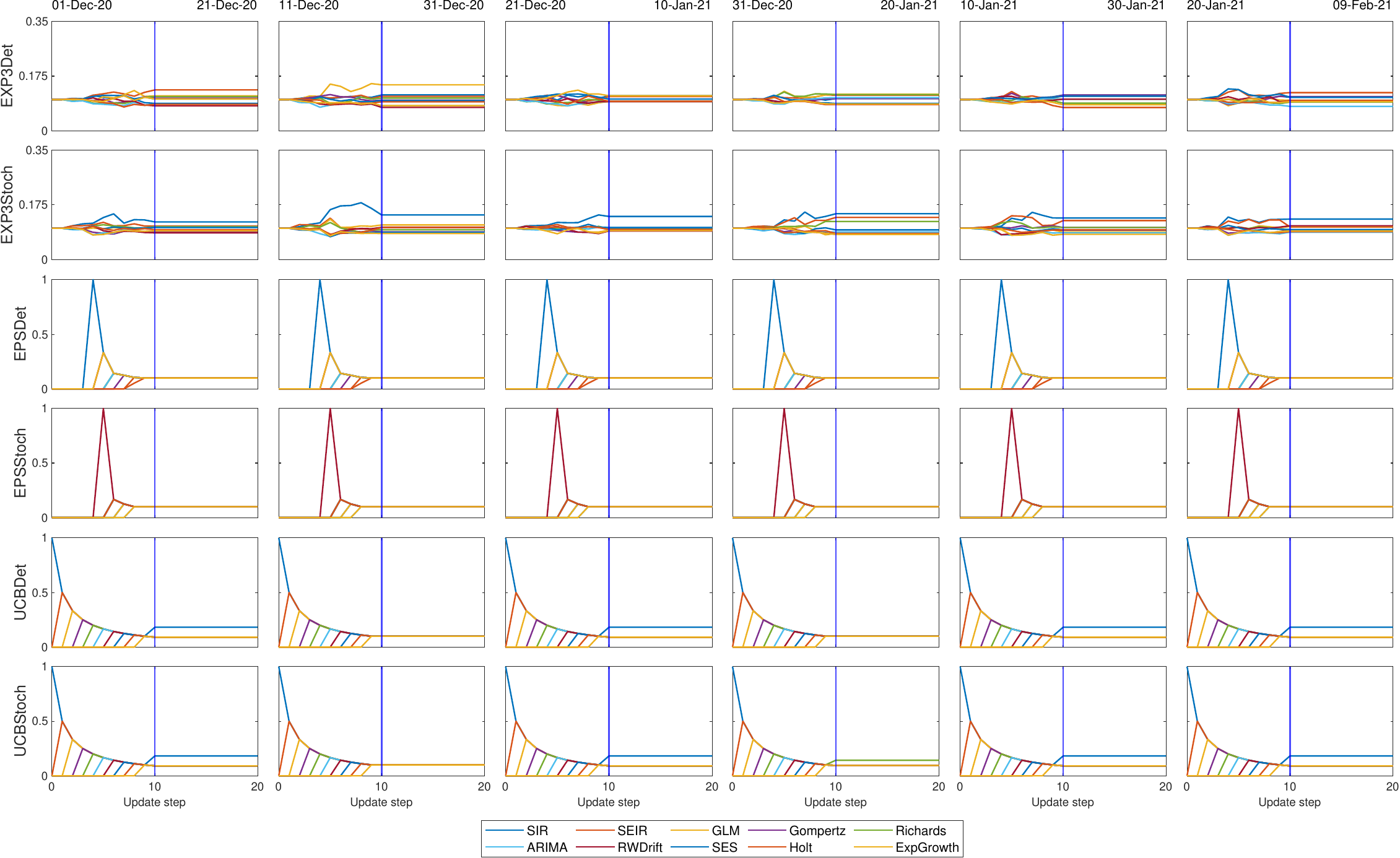}
\caption{\weightpanelcaption{2}{\fixedcalibration}}
\label{fig:weights_fixed_wave2}
\end{figure}
\subsubsection{Growing Calibration Period}
\paragraph{Ten-Day Forecasting Horizon}

\begin{figure}[H]
\centering
\includegraphics[width=\linewidth]{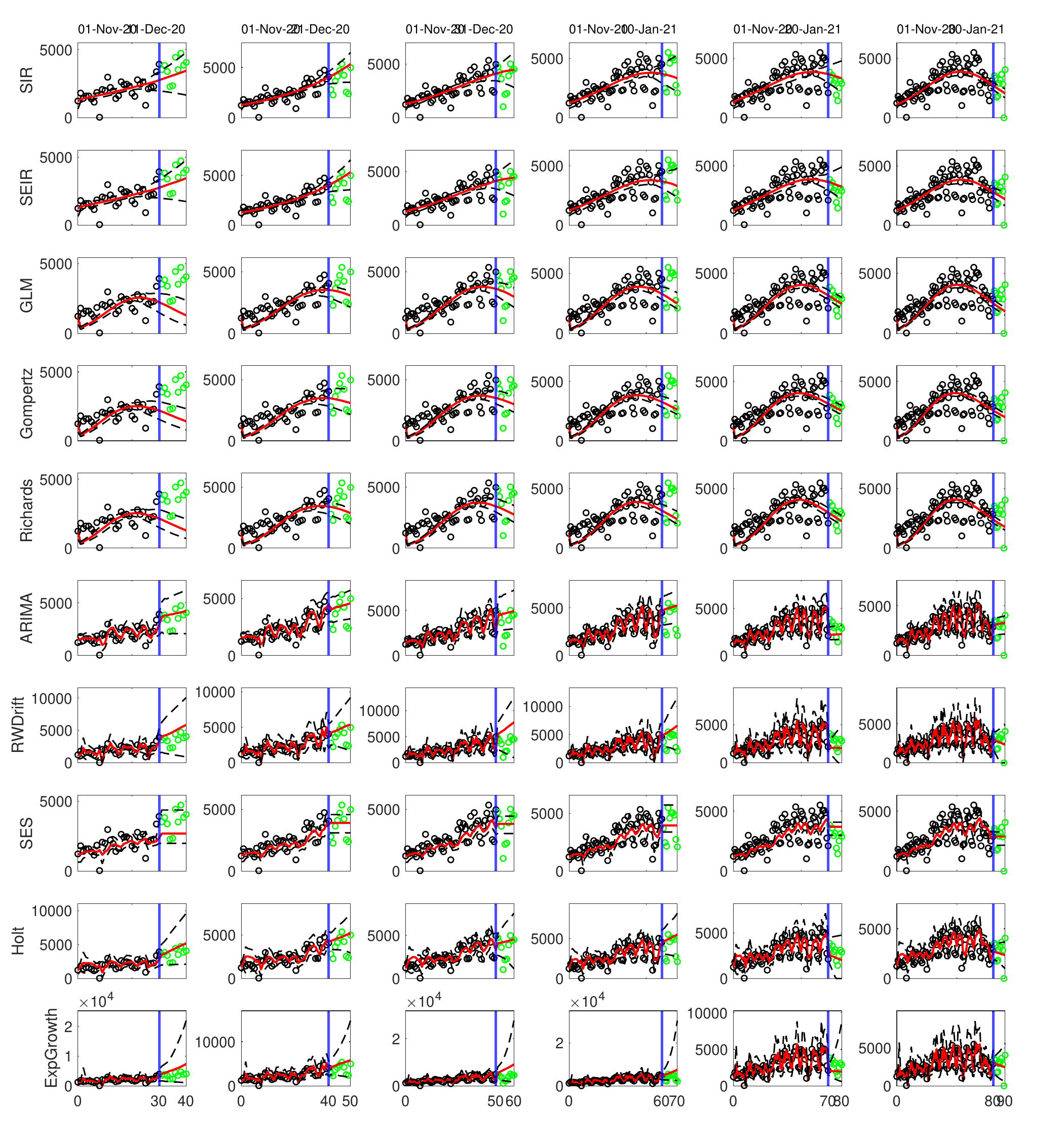}
\caption{\basepanelcaption{2}{\growingcalibration}{10}}
\label{fig:base_growing_wave2_fcst10}
\end{figure}
\begin{figure}[H]
\centering
\includegraphics[width=\linewidth]{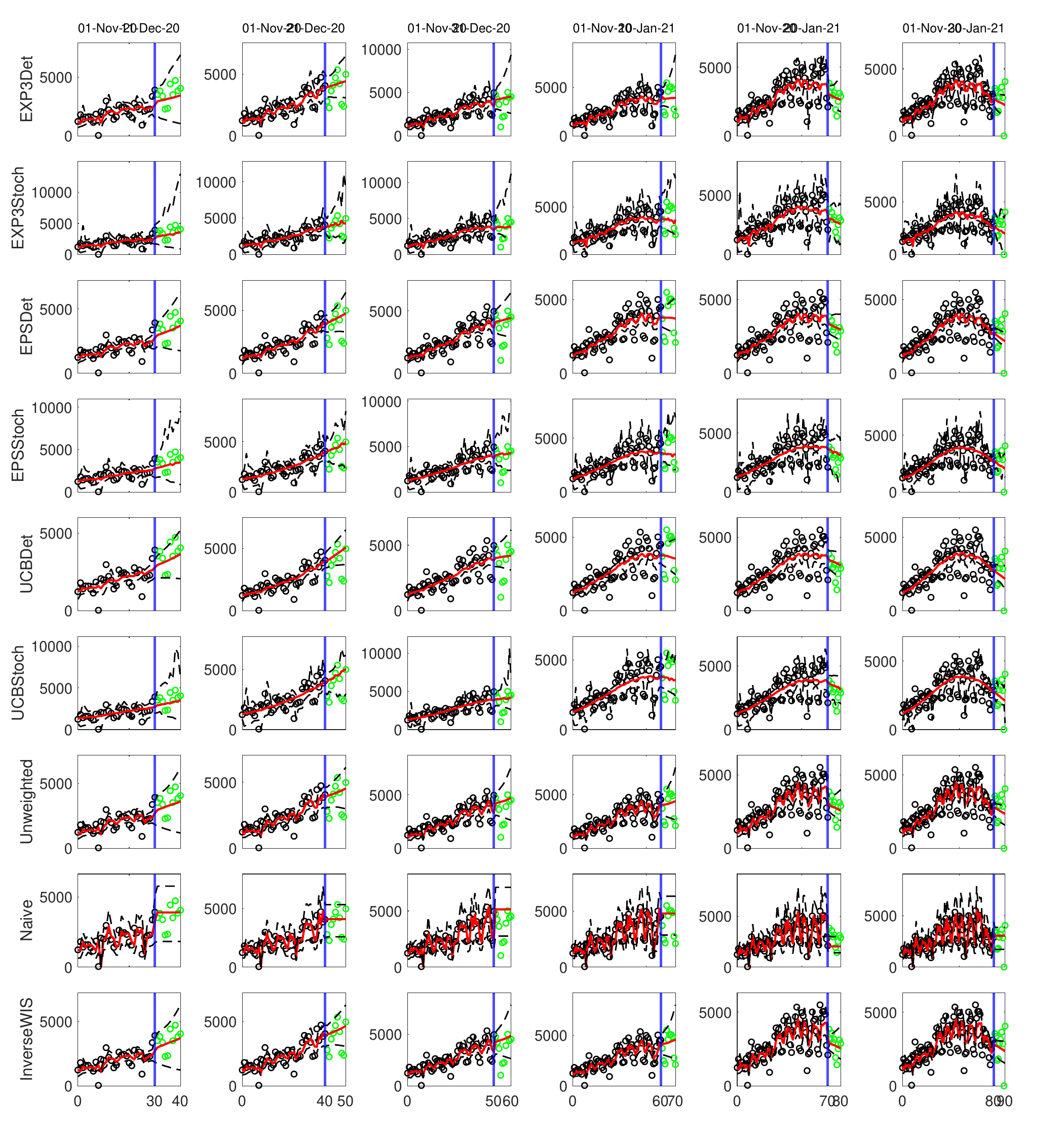}
\caption{\comparisonpanelcaption{2}{\growingcalibration}{10}}
\label{fig:ensemble_growing_wave2_fcst10}
\end{figure}
\begin{table}[H]
\centering
\scriptsize
\resizebox{\textwidth}{!}{%
\begin{tabular}{lrrrrrrrr}
\toprule
& \multicolumn{4}{c}{Calibration} & \multicolumn{4}{c}{Forecasting} \\
\cmidrule(lr){2-5} \cmidrule(lr){6-9}
Model & RMSE & WIS & 95\% PI Coverage (\%) & Mean 95\% PI Width & RMSE & WIS & 95\% PI Coverage (\%) & Mean 95\% PI Width \\
\midrule
SIR & 785.48 & 521.18 & 29.7\% & 603.45 & 1193.55 & 728.28 & 51.7\% & 1984.71 \\
SEIR & 785.95 & 521.51 & 30.5\% & 598.63 & 1178.98 & 716.22 & 51.7\% & 1970.24 \\
GLM & 924.60 & 684.77 & 23.1\% & 503.56 & 1355.40 & 992.44 & 25.0\% & 1168.44 \\
Gompertz & 930.22 & 682.21 & 24.2\% & 530.11 & 1333.19 & 978.02 & 26.7\% & 1142.85 \\
Richards & 957.06 & 708.00 & 23.8\% & 519.98 & 1407.86 & 1068.33 & 20.0\% & 1033.68 \\
ARIMA & 887.29 & 493.53 & 64.7\% & 1554.89 & 1270.64 & 689.51 & 66.7\% & 2705.14 \\
RWDrift & 974.01 & 528.85 & 73.8\% & 2347.18 & 1822.06 & 929.77 & 88.3\% & 5785.47 \\
SES & 872.81 & 557.10 & 42.9\% & 906.12 & 1181.44 & 757.69 & 48.3\% & 1634.08 \\
Holt & 929.92 & 499.28 & 71.2\% & 1898.75 & 1286.75 & 699.09 & 76.7\% & 3865.76 \\
ExpGrowth & 985.12 & 532.27 & 77.3\% & 2468.70 & 2149.55 & 1125.31 & 88.3\% & 8971.14 \\
EXP3Det & 793.02 & 455.11 & 60.2\% & 1305.23 & 1083.93 & 641.42 & 65.0\% & 2845.06 \\
EXP3Stoch & 803.36 & 436.26 & 76.0\% & 2059.41 & 1068.03 & 597.64 & 88.3\% & 4644.07 \\
EPSDet & 773.87 & 493.88 & 36.5\% & 687.84 & 1091.20 & 666.81 & 60.0\% & 2026.70 \\
EPSStoch & 784.52 & 446.29 & 69.0\% & 1787.60 & 1148.12 & 646.68 & 70.0\% & 3726.75 \\
UCBDet & 780.63 & 504.22 & 36.7\% & 741.56 & 1115.82 & 685.44 & 56.7\% & 1837.25 \\
UCBStoch & 786.64 & 474.00 & 58.2\% & 1401.46 & 1135.91 & 655.76 & 68.3\% & 2774.21 \\
Unweighted & 785.71 & 476.29 & 48.1\% & 900.96 & 1095.35 & 634.62 & 61.7\% & 2482.40 \\
Naive & 928.63 & 503.04 & 73.0\% & 2005.88 & 1306.38 & 703.26 & 70.0\% & 3042.48 \\
InverseWIS & 784.11 & 466.97 & 52.0\% & 958.18 & 1114.05 & 637.53 & 63.3\% & 2606.54 \\
\bottomrule
\end{tabular}%
}
\caption{\metricstablecaption{2}{\growingcalibration}{10}}
\label{tab:ensemble_growing_wave2_fcst10}
\end{table}
\paragraph{Thirty-Day Forecasting Horizon}

\begin{figure}[H]
\centering
\includegraphics[width=\linewidth]{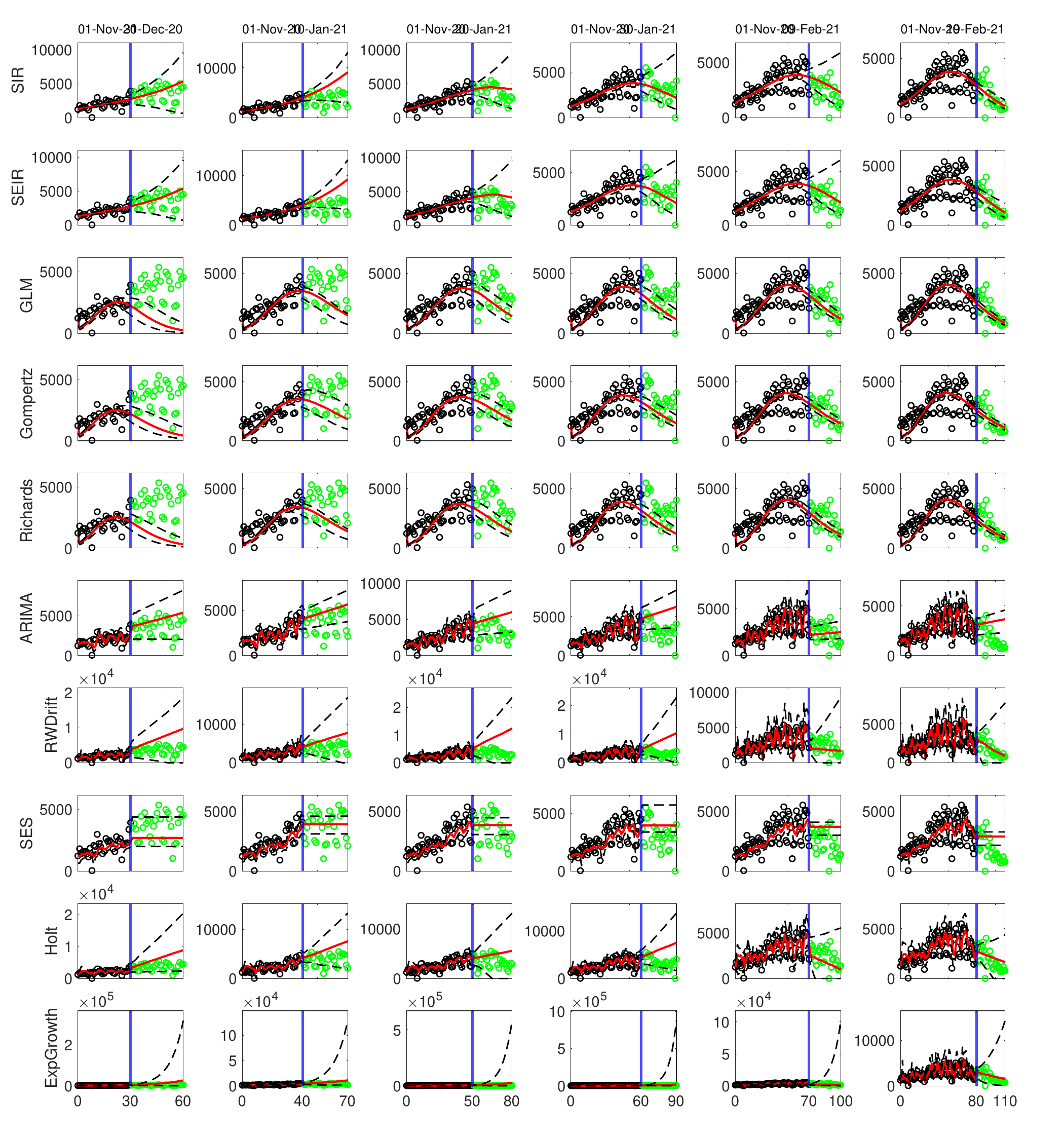}
\caption{\basepanelcaption{2}{\growingcalibration}{30}}
\label{fig:base_growing_wave2_fcst30}
\end{figure}
\begin{figure}[H]
\centering
\includegraphics[width=\linewidth]{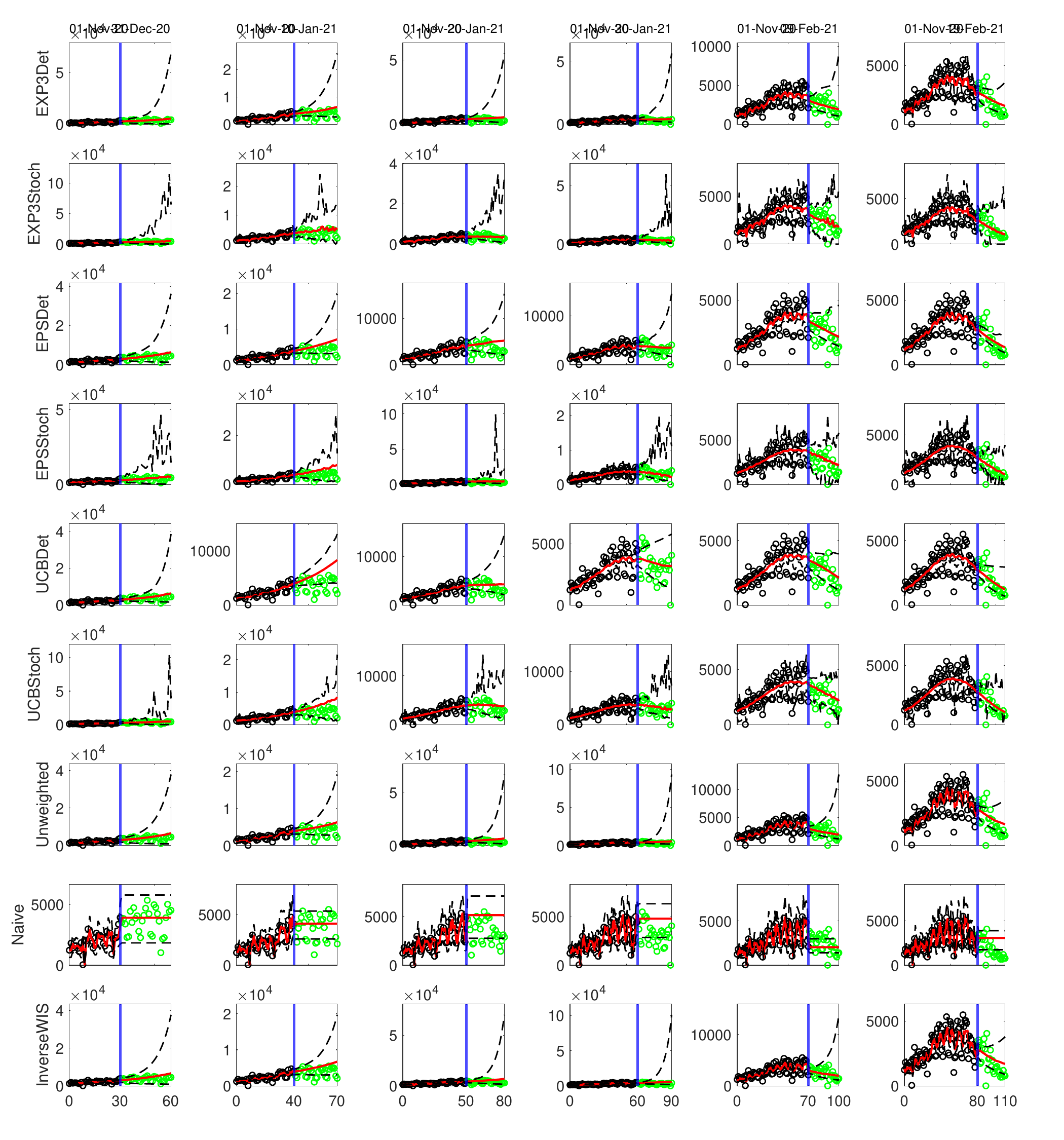}
\caption{\comparisonpanelcaption{2}{\growingcalibration}{30}}
\label{fig:ensemble_growing_wave2_fcst30}
\end{figure}
\begin{table}[H]
\centering
\scriptsize
\resizebox{\textwidth}{!}{%
\begin{tabular}{lrrrrrrrr}
\toprule
& \multicolumn{4}{c}{Calibration} & \multicolumn{4}{c}{Forecasting} \\
\cmidrule(lr){2-5} \cmidrule(lr){6-9}
Model & RMSE & WIS & 95\% PI Coverage (\%) & Mean 95\% PI Width & RMSE & WIS & 95\% PI Coverage (\%) & Mean 95\% PI Width \\
\midrule
SIR & 785.48 & 521.18 & 29.7\% & 603.45 & 1481.47 & 837.03 & 70.6\% & 3699.67 \\
SEIR & 785.95 & 521.51 & 30.5\% & 598.63 & 1479.77 & 828.45 & 71.1\% & 3681.96 \\
GLM & 924.60 & 684.77 & 23.1\% & 503.56 & 1563.90 & 1136.94 & 25.6\% & 1126.88 \\
Gompertz & 930.22 & 682.21 & 24.2\% & 530.11 & 1480.37 & 1053.05 & 27.8\% & 1169.00 \\
Richards & 957.06 & 708.00 & 23.8\% & 519.98 & 1611.08 & 1210.03 & 21.7\% & 960.88 \\
ARIMA & 887.29 & 493.53 & 64.7\% & 1554.89 & 1873.47 & 1120.21 & 53.3\% & 3263.02 \\
RWDrift & 974.01 & 528.85 & 73.8\% & 2347.18 & 3265.01 & 1606.52 & 94.4\% & 10211.41 \\
SES & 872.81 & 557.10 & 42.9\% & 906.12 & 1409.84 & 955.88 & 35.6\% & 1634.08 \\
Holt & 929.92 & 499.28 & 71.2\% & 1898.75 & 2193.63 & 1169.66 & 85.6\% & 6771.28 \\
ExpGrowth & 985.12 & 532.27 & 77.3\% & 2468.70 & 6268.20 & 3463.44 & 93.3\% & 80051.38 \\
EXP3Det & 793.02 & 455.11 & 60.2\% & 1305.23 & 1387.94 & 814.22 & 74.4\% & 10602.67 \\
EXP3Stoch & 803.36 & 436.26 & 76.0\% & 2059.41 & 1135.14 & 685.43 & 94.4\% & 12179.93 \\
EPSDet & 773.87 & 493.88 & 36.5\% & 687.84 & 1445.33 & 816.19 & 64.4\% & 5136.22 \\
EPSStoch & 784.52 & 446.29 & 69.0\% & 1787.60 & 1362.64 & 719.12 & 86.1\% & 8957.77 \\
UCBDet & 780.63 & 504.22 & 36.7\% & 741.56 & 1441.04 & 806.16 & 68.3\% & 4258.70 \\
UCBStoch & 786.64 & 474.00 & 58.2\% & 1401.46 & 1373.50 & 723.00 & 78.9\% & 6217.72 \\
Unweighted & 785.71 & 476.29 & 48.1\% & 900.96 & 1606.43 & 899.45 & 71.1\% & 10310.07 \\
Naive & 928.63 & 503.04 & 73.0\% & 2005.88 & 1532.27 & 890.72 & 62.8\% & 3042.48 \\
InverseWIS & 784.11 & 466.97 & 52.0\% & 958.18 & 1725.93 & 959.00 & 72.2\% & 10877.37 \\
\bottomrule
\end{tabular}%
}
\caption{\metricstablecaption{2}{\growingcalibration}{30}}
\label{tab:ensemble_growing_wave2_fcst30}
\end{table}
\begin{figure}[H]
\centering
\includegraphics[width=\linewidth]{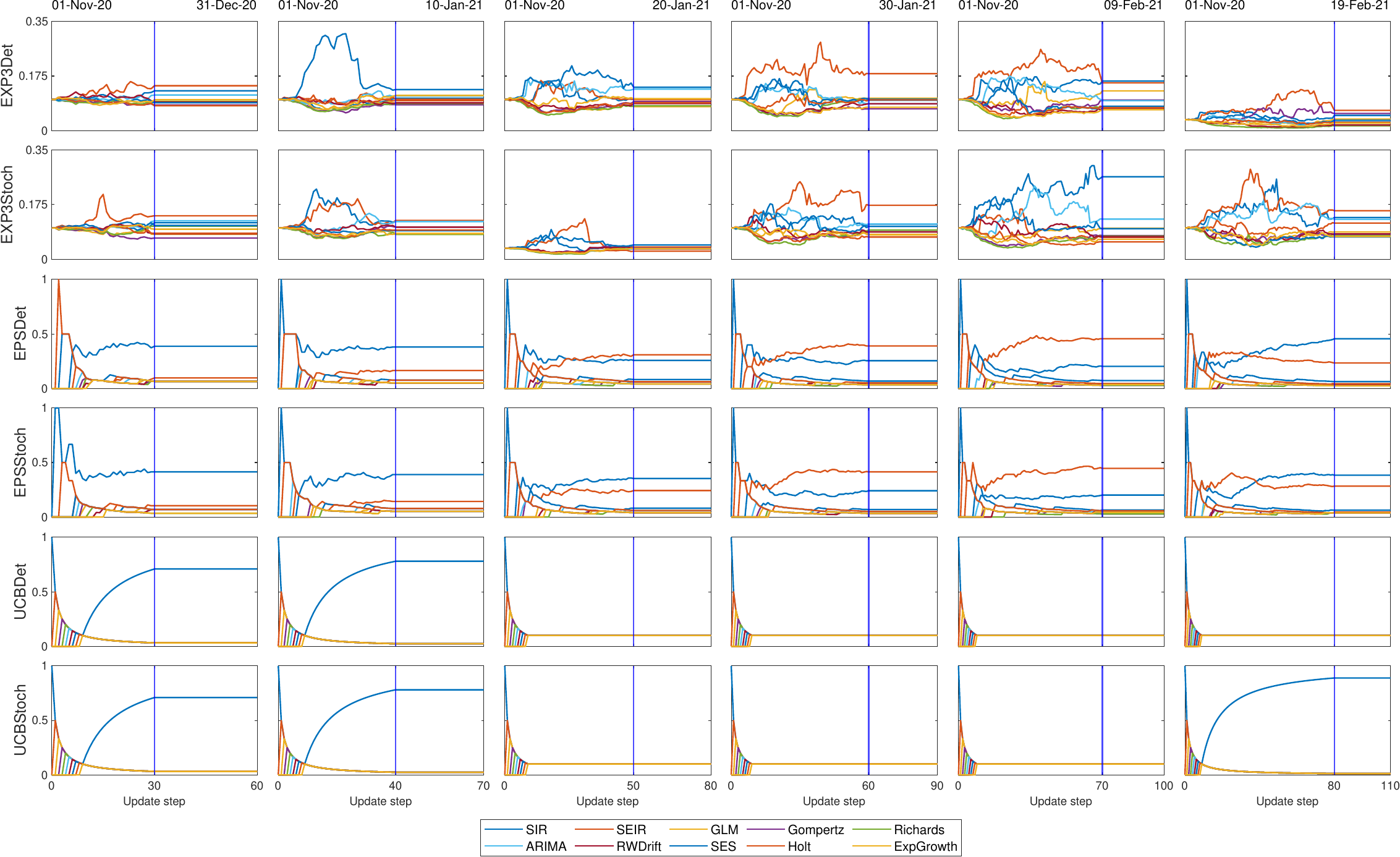}
\caption{\weightpanelcaption{2}{\growingcalibration}}
\label{fig:weights_growing_wave2}
\end{figure}
\subsection{Alabama third wave}
\subsubsection{Fixed Calibration Period}
\paragraph{Five-Day Forecasting Horizon}

\begin{figure}[H]
\centering
\includegraphics[width=\linewidth]{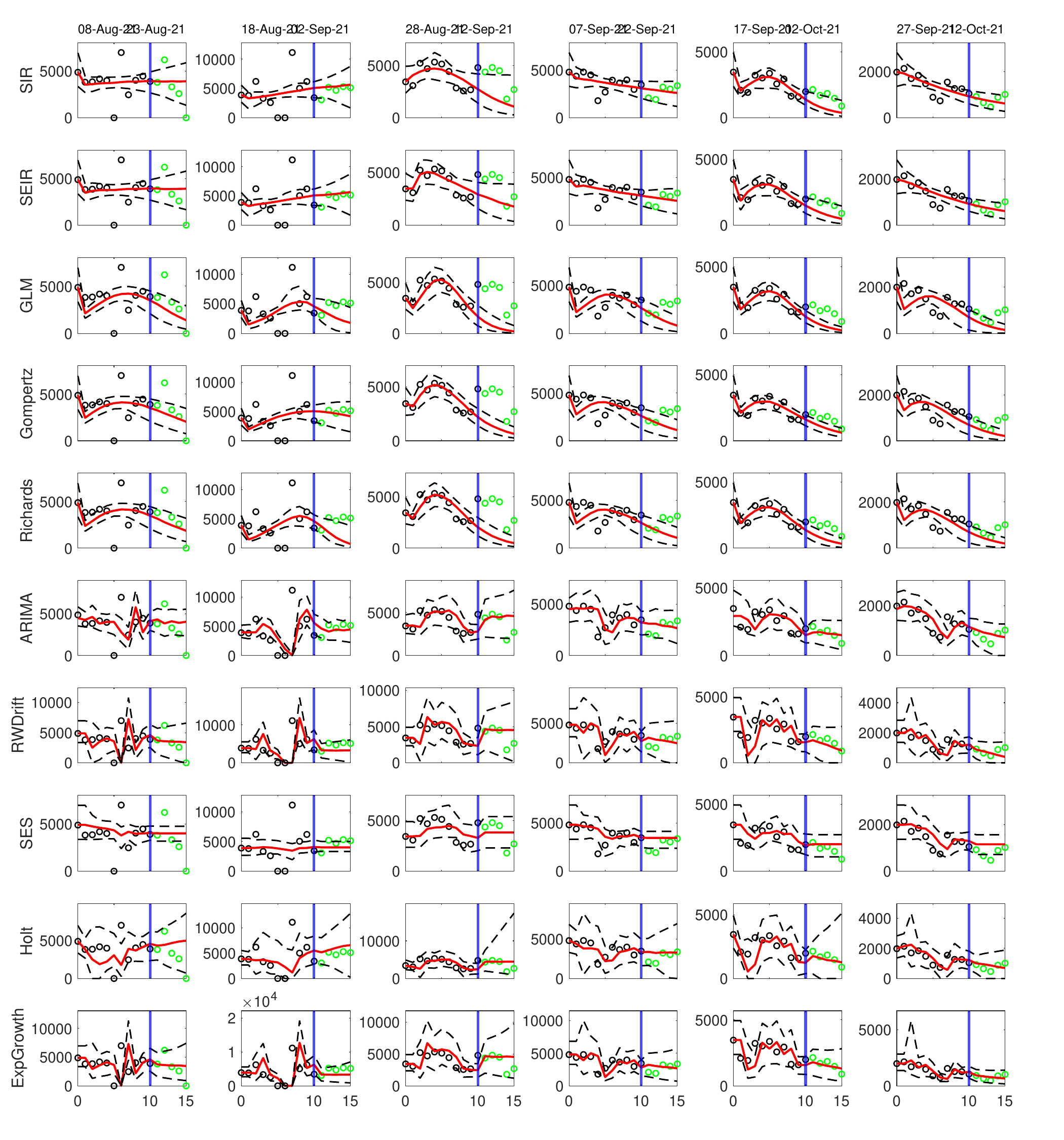}
\caption{\basepanelcaption{3}{\fixedcalibration}{5}}
\label{fig:base_fixed_wave3_fcst5}
\end{figure}
\begin{figure}[H]
\centering
\includegraphics[width=\linewidth]{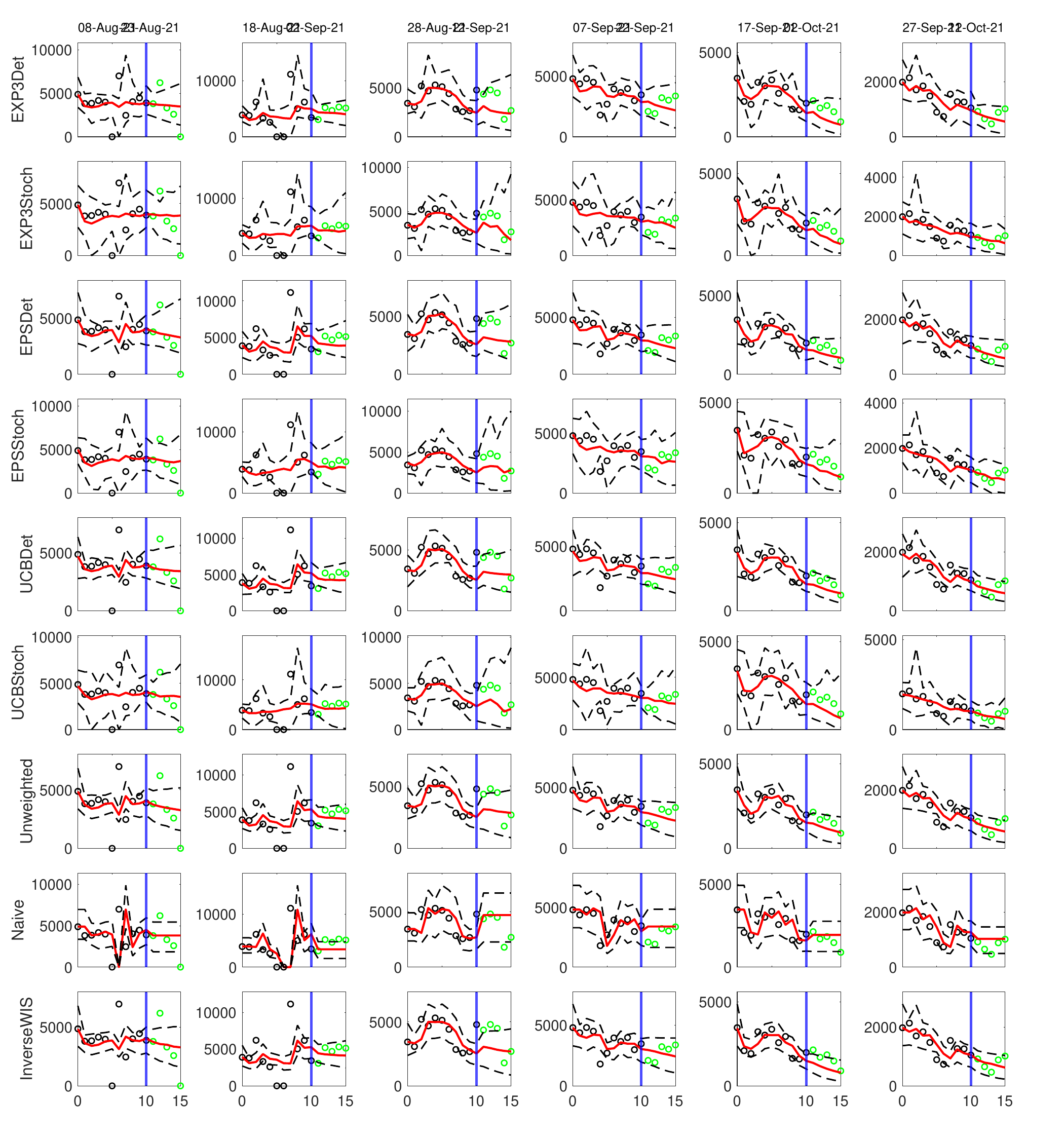}
\caption{\comparisonpanelcaption{3}{\fixedcalibration}{5}}
\label{fig:ensemble_fixed_wave3_fcst5}
\end{figure}
\begin{table}[H]
\centering
\scriptsize
\resizebox{\textwidth}{!}{%
\begin{tabular}{lrrrrrrrr}
\toprule
& \multicolumn{4}{c}{Calibration} & \multicolumn{4}{c}{Forecasting} \\
\cmidrule(lr){2-5} \cmidrule(lr){6-9}
Model & RMSE & WIS & 95\% PI Coverage (\%) & Mean 95\% PI Width & RMSE & WIS & 95\% PI Coverage (\%) & Mean 95\% PI Width \\
\midrule
SIR & 1136.60 & 656.95 & 68.2\% & 1570.22 & 1215.60 & 689.84 & 66.7\% & 2712.01 \\
SEIR & 1118.41 & 649.36 & 68.2\% & 1597.52 & 1110.83 & 644.92 & 70.0\% & 2628.53 \\
GLM & 1339.17 & 822.05 & 57.6\% & 1656.12 & 1809.83 & 1276.55 & 33.3\% & 1919.17 \\
Gompertz & 1253.88 & 765.36 & 51.5\% & 1491.66 & 1353.05 & 928.58 & 43.3\% & 1976.71 \\
Richards & 1275.27 & 773.93 & 56.1\% & 1587.56 & 1804.51 & 1198.66 & 26.7\% & 1851.61 \\
ARIMA & 1461.85 & 767.94 & 72.7\% & 2259.08 & 1027.59 & 558.25 & 90.0\% & 2906.76 \\
RWDrift & 1871.85 & 983.64 & 78.8\% & 3340.65 & 1099.18 & 572.42 & 96.7\% & 4309.08 \\
SES & 1294.33 & 725.30 & 72.7\% & 2066.64 & 1059.49 & 652.23 & 56.7\% & 1770.70 \\
Holt & 1484.80 & 702.86 & 86.4\% & 3688.83 & 1154.35 & 634.86 & 96.7\% & 6063.89 \\
ExpGrowth & 1869.52 & 986.17 & 75.8\% & 3376.22 & 1073.48 & 566.09 & 93.3\% & 4073.75 \\
EXP3Det & 1225.50 & 639.45 & 74.2\% & 2810.04 & 1029.90 & 551.38 & 86.7\% & 2940.23 \\
EXP3Stoch & 1207.66 & 633.23 & 84.8\% & 3377.76 & 940.21 & 530.54 & 93.3\% & 4330.64 \\
EPSDet & 1269.26 & 698.50 & 72.7\% & 2191.58 & 973.38 & 550.87 & 86.7\% & 2804.98 \\
EPSStoch & 1204.33 & 618.36 & 87.9\% & 3336.92 & 943.32 & 527.53 & 93.3\% & 4446.22 \\
UCBDet & 1250.61 & 709.23 & 66.7\% & 1701.91 & 951.35 & 528.51 & 83.3\% & 2302.96 \\
UCBStoch & 1169.10 & 614.12 & 86.4\% & 3565.05 & 983.67 & 535.09 & 93.3\% & 4484.62 \\
Unweighted & 1268.89 & 709.54 & 68.2\% & 1813.02 & 967.37 & 547.77 & 76.7\% & 2404.82 \\
Naive & 1710.09 & 911.93 & 68.2\% & 2521.95 & 1166.07 & 644.92 & 73.3\% & 2863.54 \\
InverseWIS & 1231.44 & 689.11 & 68.2\% & 1755.97 & 984.25 & 565.93 & 73.3\% & 2305.23 \\
\bottomrule
\end{tabular}%
}
\caption{\metricstablecaption{3}{\fixedcalibration}{5}}
\label{tab:ensemble_fixed_wave3_fcst5}
\end{table}
\paragraph{Ten-Day Forecasting Horizon}

\begin{figure}[H]
\centering
\includegraphics[width=\linewidth]{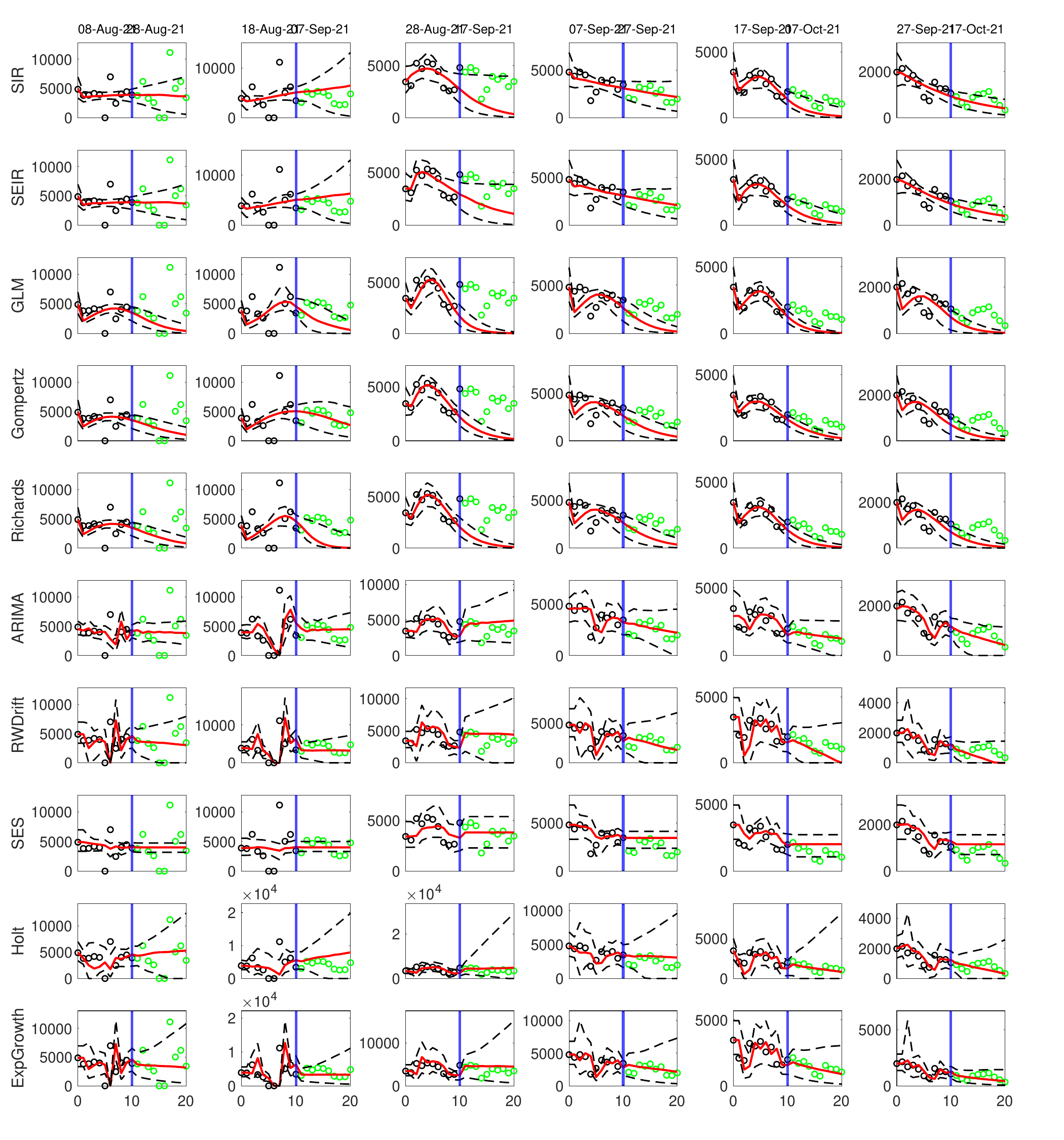}
\caption{\basepanelcaption{3}{\fixedcalibration}{10}}
\label{fig:base_fixed_wave3_fcst10}
\end{figure}
\begin{figure}[H]
\centering
\includegraphics[width=\linewidth]{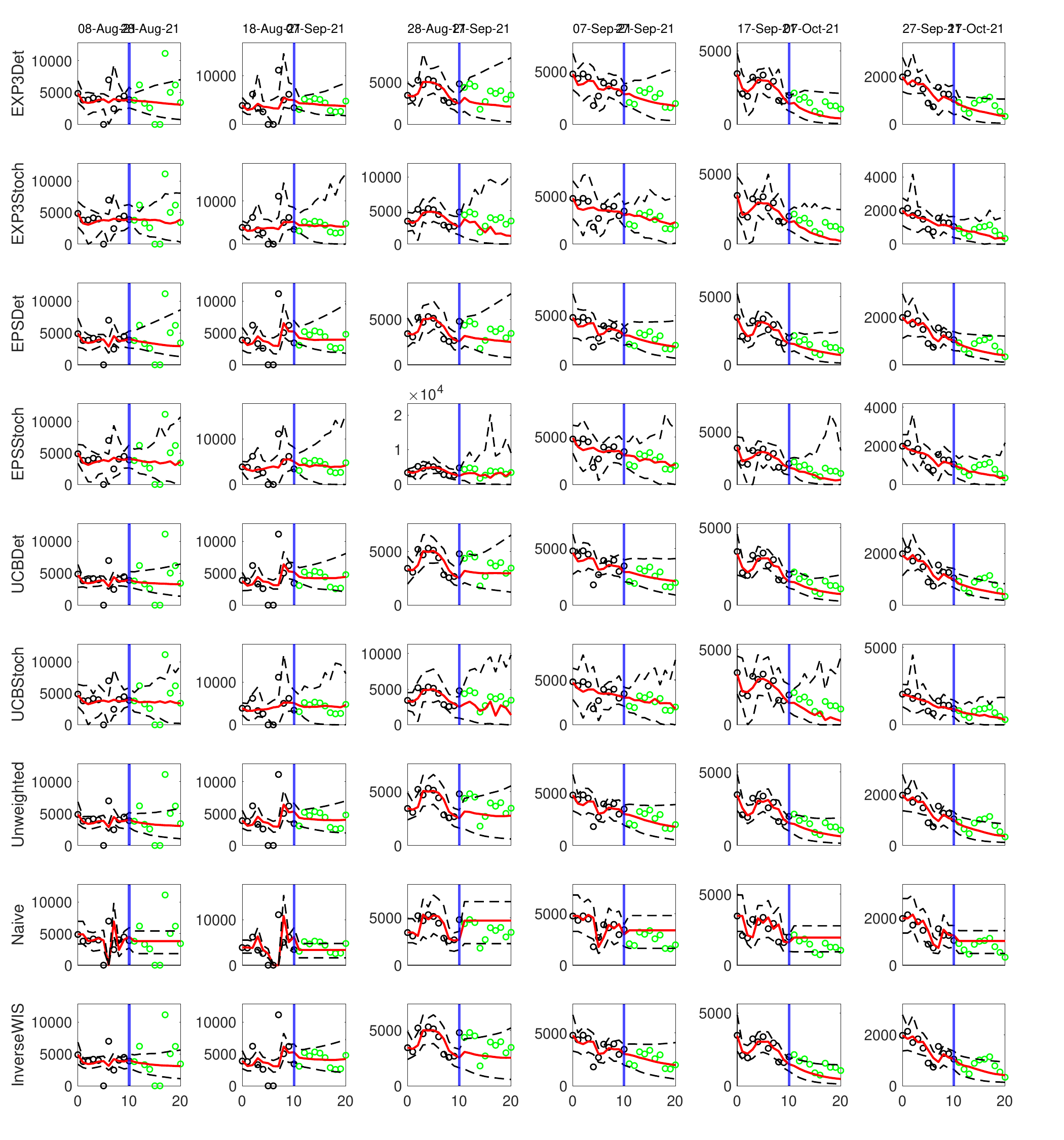}
\caption{\comparisonpanelcaption{3}{\fixedcalibration}{10}}
\label{fig:ensemble_fixed_wave3_fcst10}
\end{figure}
\begin{table}[H]
\centering
\scriptsize
\resizebox{\textwidth}{!}{%
\begin{tabular}{lrrrrrrrr}
\toprule
& \multicolumn{4}{c}{Calibration} & \multicolumn{4}{c}{Forecasting} \\
\cmidrule(lr){2-5} \cmidrule(lr){6-9}
Model & RMSE & WIS & 95\% PI Coverage (\%) & Mean 95\% PI Width & RMSE & WIS & 95\% PI Coverage (\%) & Mean 95\% PI Width \\
\midrule
SIR & 1136.60 & 656.95 & 68.2\% & 1570.22 & 1644.94 & 908.32 & 70.0\% & 3415.94 \\
SEIR & 1118.41 & 649.36 & 68.2\% & 1597.52 & 1516.86 & 859.93 & 68.3\% & 3330.50 \\
GLM & 1339.17 & 822.05 & 57.6\% & 1656.12 & 2295.47 & 1698.57 & 20.0\% & 1502.60 \\
Gompertz & 1253.88 & 765.36 & 51.5\% & 1491.66 & 1820.58 & 1308.61 & 31.7\% & 1863.08 \\
Richards & 1275.27 & 773.93 & 56.1\% & 1587.56 & 2282.15 & 1601.04 & 16.7\% & 1541.50 \\
ARIMA & 1461.85 & 767.94 & 72.7\% & 2259.08 & 1175.23 & 651.27 & 90.0\% & 3393.67 \\
RWDrift & 1871.85 & 983.64 & 78.8\% & 3340.65 & 1291.32 & 681.59 & 96.7\% & 4974.32 \\
SES & 1294.33 & 725.30 & 72.7\% & 2066.64 & 1246.00 & 782.80 & 55.0\% & 1770.70 \\
Holt & 1484.80 & 702.86 & 86.4\% & 3688.83 & 1528.95 & 893.86 & 95.0\% & 8853.32 \\
ExpGrowth & 1869.52 & 986.17 & 75.8\% & 3376.22 & 1183.46 & 636.88 & 93.3\% & 5378.12 \\
EXP3Det & 1225.50 & 639.45 & 74.2\% & 2810.04 & 1243.01 & 674.15 & 88.3\% & 3635.91 \\
EXP3Stoch & 1207.66 & 633.23 & 84.8\% & 3377.76 & 1273.06 & 686.84 & 93.3\% & 5384.41 \\
EPSDet & 1269.26 & 698.50 & 72.7\% & 2191.58 & 1166.76 & 646.79 & 90.0\% & 3565.76 \\
EPSStoch & 1204.33 & 618.36 & 87.9\% & 3336.92 & 1143.38 & 658.70 & 93.3\% & 6062.03 \\
UCBDet & 1250.61 & 709.23 & 66.7\% & 1701.91 & 1131.48 & 636.53 & 85.0\% & 2772.80 \\
UCBStoch & 1169.10 & 614.12 & 86.4\% & 3565.05 & 1238.82 & 687.32 & 93.3\% & 5614.52 \\
Unweighted & 1268.89 & 709.54 & 68.2\% & 1813.02 & 1160.17 & 652.03 & 80.0\% & 2773.23 \\
Naive & 1710.09 & 911.93 & 68.2\% & 2521.95 & 1334.56 & 743.75 & 75.0\% & 2863.54 \\
InverseWIS & 1231.44 & 689.11 & 68.2\% & 1755.97 & 1190.38 & 678.54 & 76.7\% & 2668.22 \\
\bottomrule
\end{tabular}%
}
\caption{\metricstablecaption{3}{\fixedcalibration}{10}}
\label{tab:ensemble_fixed_wave3_fcst10}
\end{table}
\begin{figure}[H]
\centering
\includegraphics[width=\linewidth]{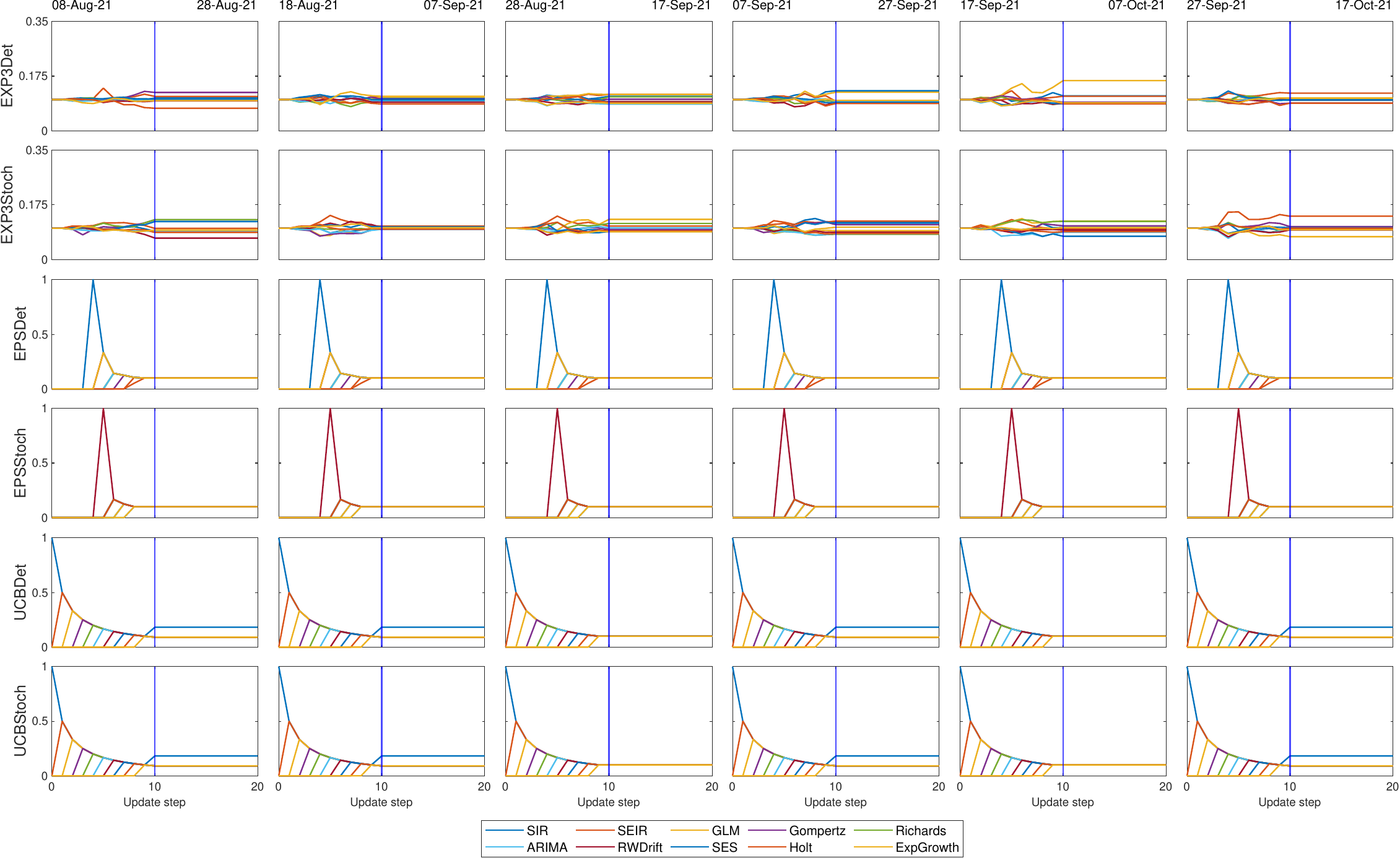}
\caption{\weightpanelcaption{3}{\fixedcalibration}}
\label{fig:weights_fixed_wave3}
\end{figure}
\subsubsection{Growing Calibration Period}
\paragraph{Ten-Day Forecasting Horizon}

\begin{figure}[H]
\centering
\includegraphics[width=\linewidth]{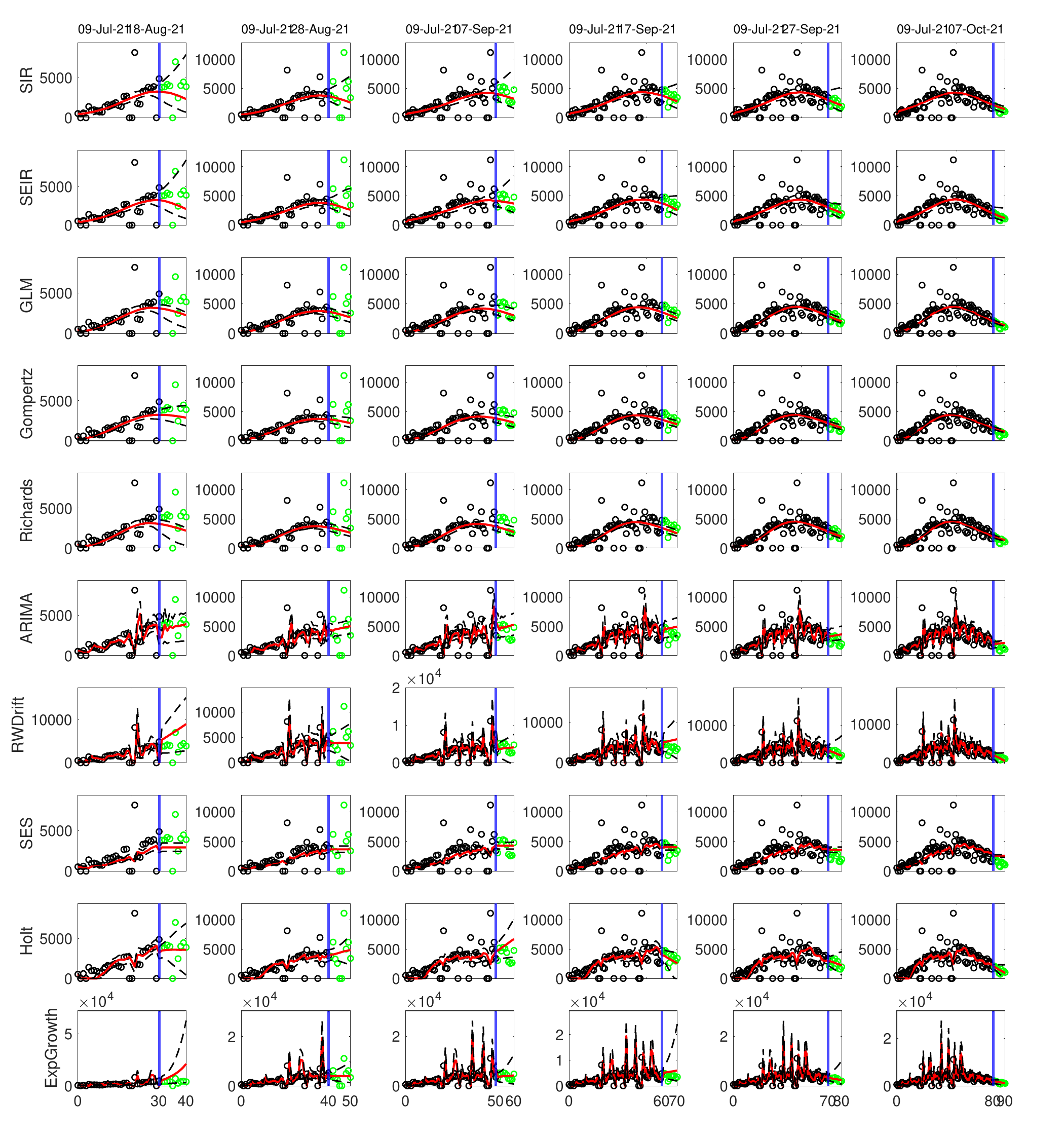}
\caption{\basepanelcaption{3}{\growingcalibration}{10}}
\label{fig:base_growing_wave3_fcst10}
\end{figure}
\begin{figure}[H]
\centering
\includegraphics[width=\linewidth]{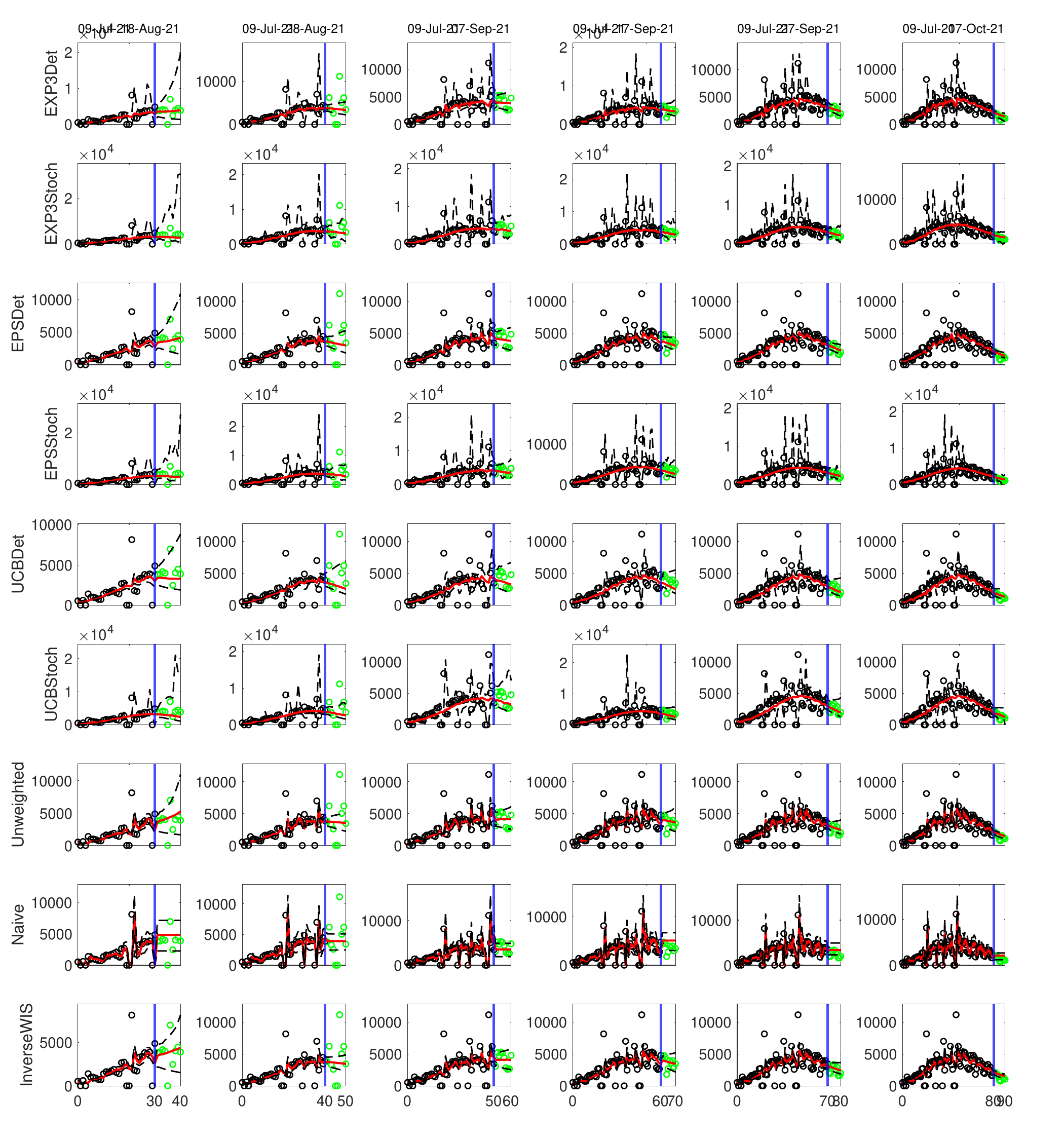}
\caption{\comparisonpanelcaption{3}{\growingcalibration}{10}}
\label{fig:ensemble_growing_wave3_fcst10}
\end{figure}
\begin{table}[H]
\centering
\scriptsize
\resizebox{\textwidth}{!}{%
\begin{tabular}{lrrrrrrrr}
\toprule
& \multicolumn{4}{c}{Calibration} & \multicolumn{4}{c}{Forecasting} \\
\cmidrule(lr){2-5} \cmidrule(lr){6-9}
Model & RMSE & WIS & 95\% PI Coverage (\%) & Mean 95\% PI Width & RMSE & WIS & 95\% PI Coverage (\%) & Mean 95\% PI Width \\
\midrule
SIR & 1654.54 & 900.71 & 38.7\% & 939.88 & 1393.92 & 774.46 & 83.3\% & 3546.20 \\
SEIR & 1651.86 & 876.73 & 46.2\% & 1067.73 & 1386.46 & 779.51 & 78.3\% & 3258.30 \\
GLM & 1664.79 & 968.33 & 17.6\% & 539.28 & 1420.91 & 965.77 & 31.7\% & 1246.08 \\
Gompertz & 1664.74 & 963.64 & 18.7\% & 555.10 & 1392.41 & 917.82 & 35.0\% & 1214.38 \\
Richards & 1680.53 & 990.83 & 19.9\% & 577.41 & 1471.77 & 1022.42 & 28.3\% & 1200.54 \\
ARIMA & 2206.20 & 1097.80 & 47.2\% & 1552.74 & 1628.12 & 974.88 & 50.0\% & 2524.80 \\
RWDrift & 2679.53 & 1252.39 & 61.5\% & 2289.66 & 1880.65 & 1048.53 & 88.3\% & 5882.08 \\
SES & 1845.04 & 1086.75 & 17.5\% & 650.06 & 1609.33 & 1153.07 & 21.7\% & 915.10 \\
Holt & 1896.65 & 1024.38 & 45.5\% & 1256.30 & 1464.25 & 814.14 & 78.3\% & 3240.74 \\
ExpGrowth & 3948.48 & 1813.52 & 51.1\% & 2905.20 & 2812.05 & 1519.70 & 86.7\% & 9613.72 \\
EXP3Det & 1767.97 & 890.21 & 48.7\% & 2012.54 & 1314.16 & 750.86 & 76.7\% & 3295.40 \\
EXP3Stoch & 1687.75 & 835.69 & 69.8\% & 3010.08 & 1350.07 & 745.20 & 90.0\% & 5532.10 \\
EPSDet & 1747.18 & 937.14 & 32.7\% & 853.14 & 1313.67 & 767.19 & 66.7\% & 2356.22 \\
EPSStoch & 1666.62 & 837.06 & 69.2\% & 2866.97 & 1357.80 & 752.95 & 85.0\% & 4402.00 \\
UCBDet & 1701.76 & 894.86 & 42.9\% & 1305.13 & 1363.09 & 769.31 & 81.7\% & 3129.62 \\
UCBStoch & 1673.79 & 854.49 & 59.7\% & 2189.75 & 1396.53 & 788.06 & 81.7\% & 3987.74 \\
Unweighted & 1876.93 & 1003.79 & 37.0\% & 904.70 & 1345.87 & 775.03 & 71.7\% & 2539.82 \\
Naive & 2572.26 & 1208.53 & 57.6\% & 1988.08 & 1656.56 & 960.94 & 66.7\% & 3051.14 \\
InverseWIS & 1812.23 & 977.90 & 30.6\% & 794.96 & 1319.28 & 771.34 & 65.0\% & 2188.20 \\
\bottomrule
\end{tabular}%
}
\caption{\metricstablecaption{3}{\growingcalibration}{10}}
\label{tab:ensemble_growing_wave3_fcst10}
\end{table}
\paragraph{Thirty-Day Forecasting Horizon}

\begin{figure}[H]
\centering
\includegraphics[width=\linewidth]{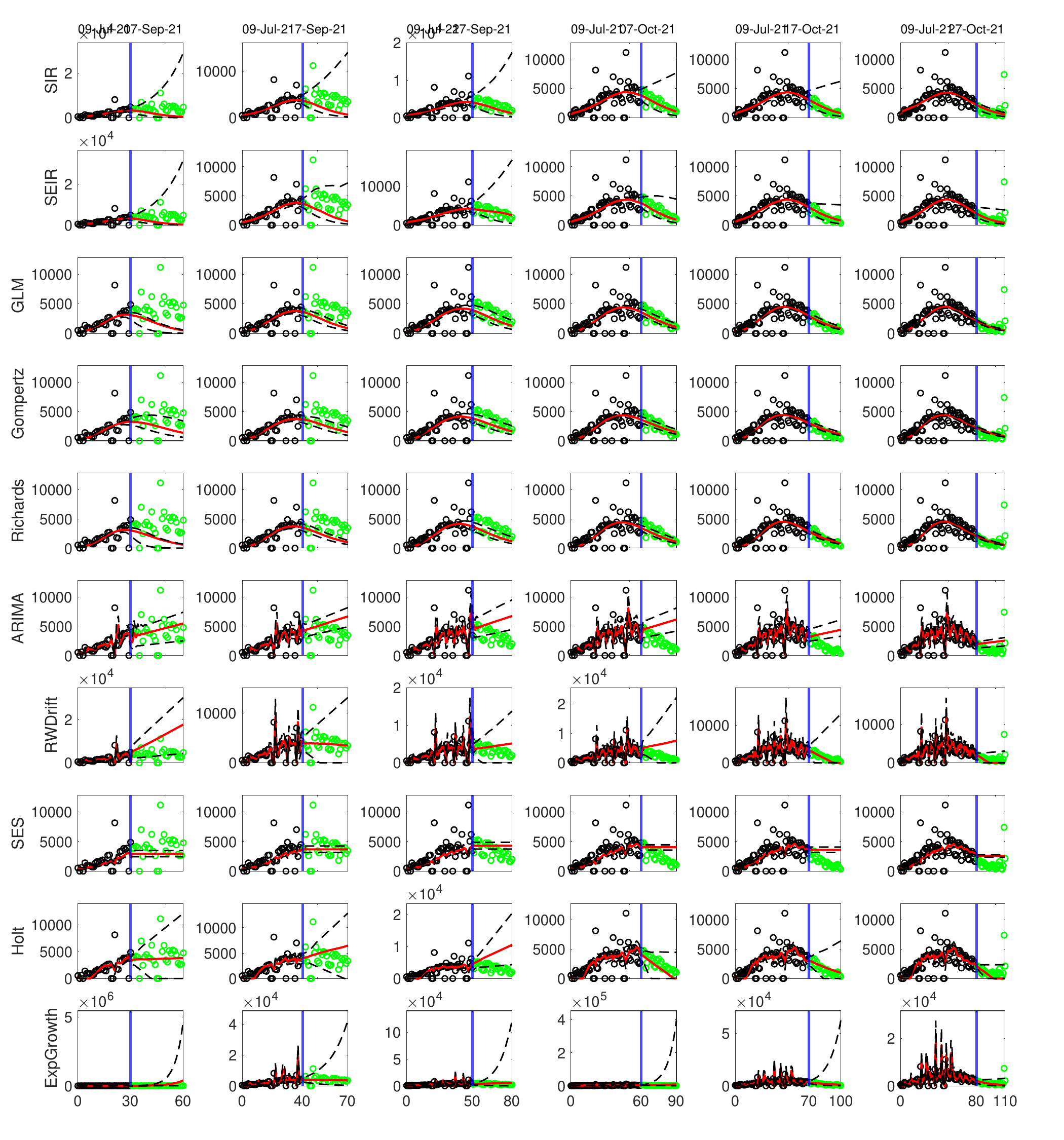}
\caption{\basepanelcaption{3}{\growingcalibration}{30}}
\label{fig:base_growing_wave3_fcst30}
\end{figure}
\begin{figure}[H]
\centering
\includegraphics[width=\linewidth]{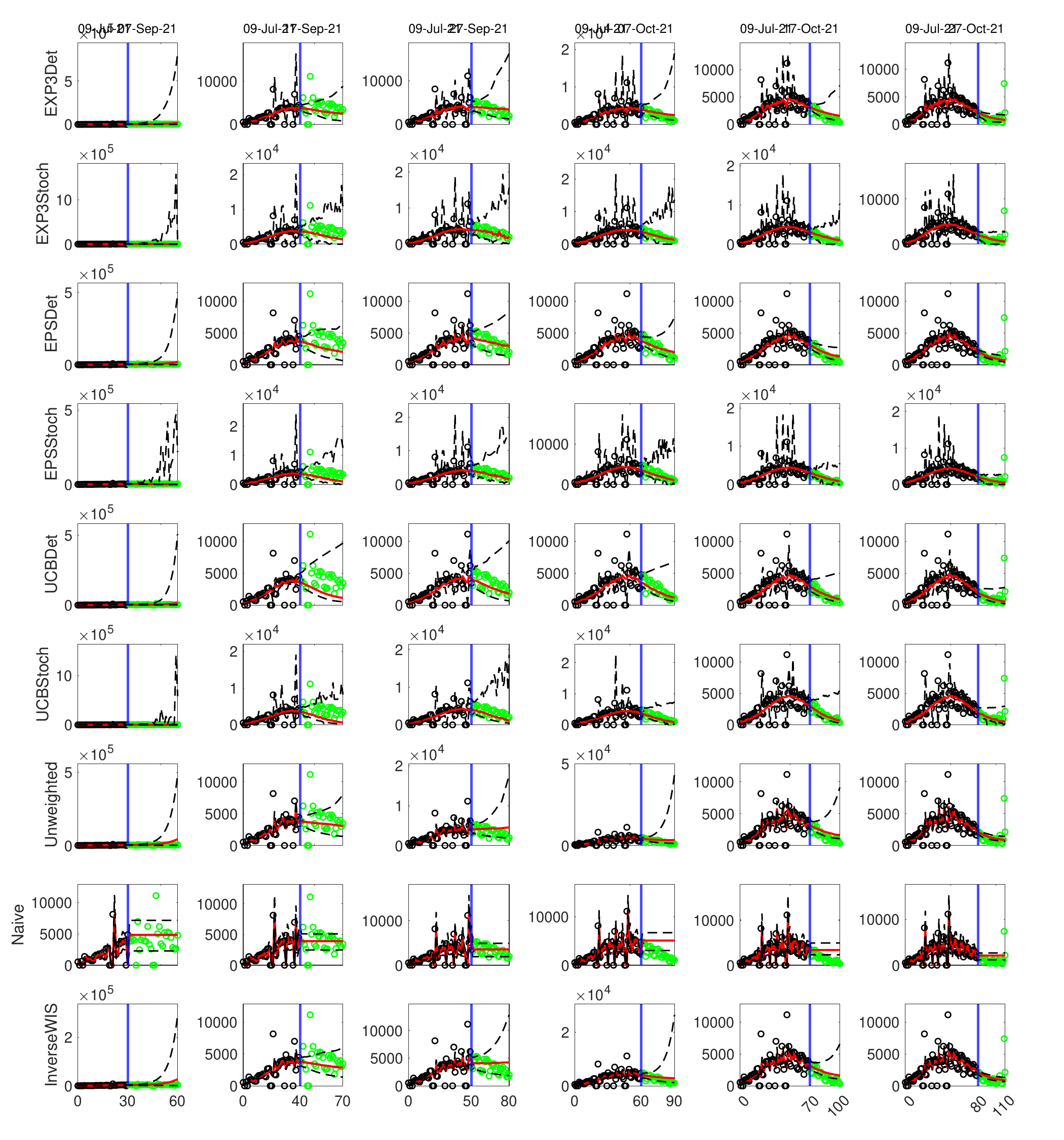}
\caption{\comparisonpanelcaption{3}{\growingcalibration}{30}}
\label{fig:ensemble_growing_wave3_fcst30}
\end{figure}
\begin{table}[H]
\centering
\scriptsize
\resizebox{\textwidth}{!}{%
\begin{tabular}{lrrrrrrrr}
\toprule
& \multicolumn{4}{c}{Calibration} & \multicolumn{4}{c}{Forecasting} \\
\cmidrule(lr){2-5} \cmidrule(lr){6-9}
Model & RMSE & WIS & 95\% PI Coverage (\%) & Mean 95\% PI Width & RMSE & WIS & 95\% PI Coverage (\%) & Mean 95\% PI Width \\
\midrule
SIR & 1654.54 & 900.71 & 38.7\% & 939.88 & 1628.98 & 861.14 & 90.0\% & 6774.54 \\
SEIR & 1651.86 & 876.73 & 46.2\% & 1067.73 & 1624.09 & 851.82 & 90.6\% & 6075.95 \\
GLM & 1664.79 & 968.33 & 17.6\% & 539.28 & 1628.74 & 1097.72 & 33.3\% & 1109.99 \\
Gompertz & 1664.74 & 963.64 & 18.7\% & 555.10 & 1492.81 & 920.99 & 35.0\% & 1312.97 \\
Richards & 1680.53 & 990.83 & 19.9\% & 577.41 & 1657.25 & 1142.86 & 31.1\% & 1018.50 \\
ARIMA & 2206.20 & 1097.80 & 47.2\% & 1552.74 & 2505.57 & 1659.79 & 34.4\% & 2944.98 \\
RWDrift & 2679.53 & 1252.39 & 61.5\% & 2289.66 & 3106.13 & 1666.15 & 91.1\% & 9829.02 \\
SES & 1845.04 & 1086.75 & 17.5\% & 650.06 & 1962.82 & 1452.31 & 16.1\% & 915.10 \\
Holt & 1896.65 & 1024.38 & 45.5\% & 1256.30 & 2140.29 & 1172.03 & 83.3\% & 5825.38 \\
ExpGrowth & 3948.48 & 1813.52 & 51.1\% & 2905.20 & 25219.24 & 11435.75 & 86.1\% & 163662.98 \\
EXP3Det & 1767.97 & 890.21 & 48.7\% & 2012.54 & 1957.09 & 1176.02 & 81.1\% & 30440.78 \\
EXP3Stoch & 1687.75 & 835.69 & 69.8\% & 3010.08 & 1458.04 & 953.94 & 95.0\% & 37588.55 \\
EPSDet & 1747.18 & 937.14 & 32.7\% & 853.14 & 2012.18 & 1187.12 & 75.6\% & 16342.27 \\
EPSStoch & 1666.62 & 837.06 & 69.2\% & 2866.97 & 1502.70 & 857.18 & 93.3\% & 22503.54 \\
UCBDet & 1701.76 & 894.86 & 42.9\% & 1305.13 & 1815.26 & 1013.17 & 87.8\% & 16244.70 \\
UCBStoch & 1673.79 & 854.49 & 59.7\% & 2189.75 & 1646.44 & 949.98 & 89.4\% & 23777.13 \\
Unweighted & 1876.93 & 1003.79 & 37.0\% & 904.70 & 3589.10 & 1673.71 & 66.7\% & 18790.41 \\
Naive & 2572.26 & 1208.53 & 57.6\% & 1988.08 & 1954.52 & 1191.65 & 52.2\% & 3051.14 \\
InverseWIS & 1812.23 & 977.90 & 30.6\% & 794.96 & 2543.95 & 1240.16 & 63.9\% & 12126.10 \\
\bottomrule
\end{tabular}%
}
\caption{\metricstablecaption{3}{\growingcalibration}{30}}
\label{tab:ensemble_growing_wave3_fcst30}
\end{table}
\begin{figure}[H]
\centering
\includegraphics[width=\linewidth]{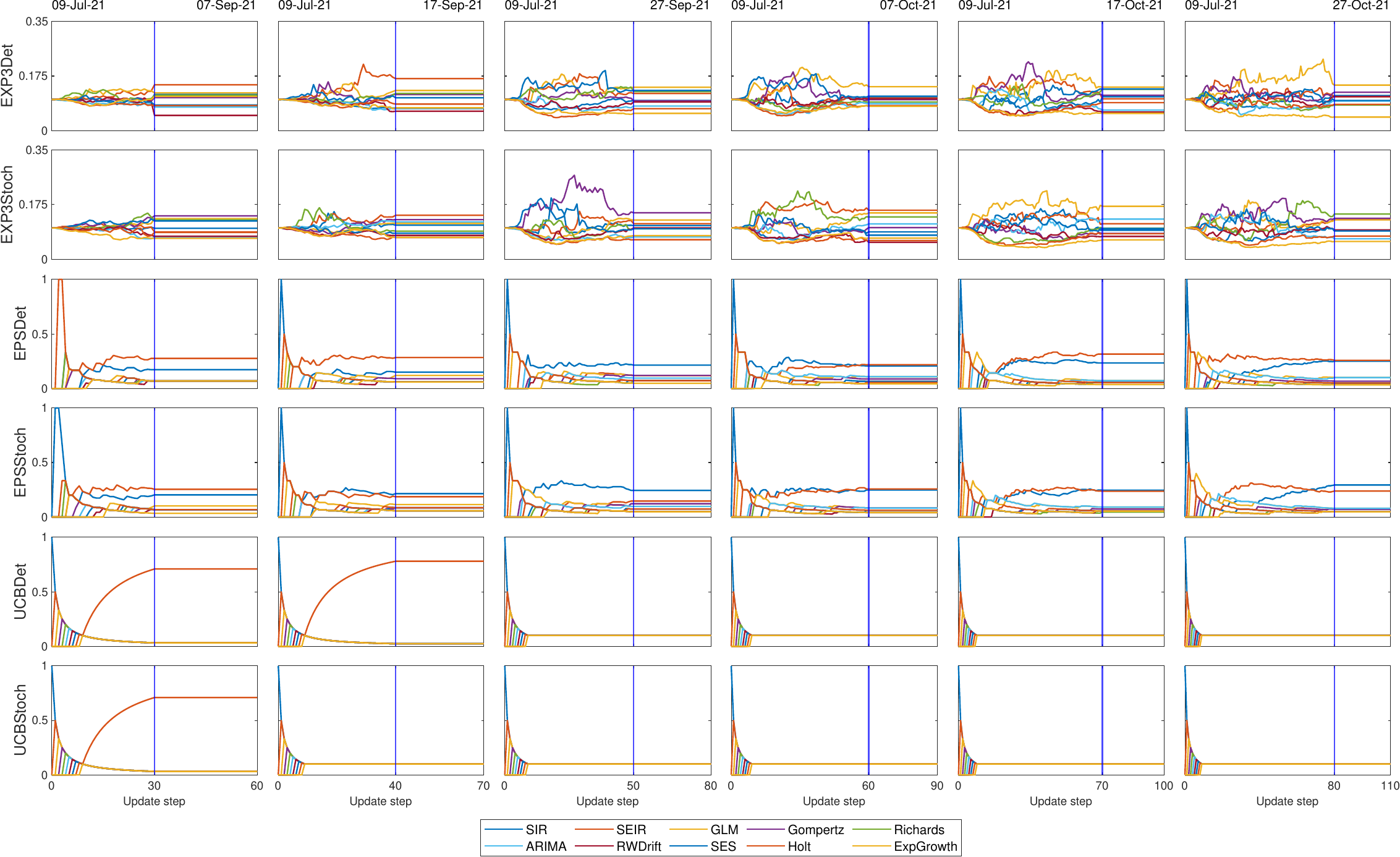}
\caption{\weightpanelcaption{3}{\growingcalibration}}
\label{fig:weights_growing_wave3}
\end{figure}

\end{document}